\documentclass[12pt,a4paper,twoside,fleqn,openright]{book}
\usepackage{barbu_usepackage_book}
\usepackage{barbu_usepackage_book_commandes}

\usepackage[english]{babel}
\usepackage{microtype}
\usepackage{libertine}
\usepackage{parskip}
\usepackage{float}

\fancypagestyle{fancy}{
\fancyhf{} 
\fancyfoot[C]{\bfseries -\thepage - } 
\fancyfoot[R]{Gibert \couleur{\textit{et al}.}} 
\fancyhead[LO]{\bfseries\rightmark} 
\renewcommand{\headrulewidth}{0.4pt} 
}

\fancypagestyle{plain}{
\fancyhf{} 
\fancyfoot[C]{\bfseries -\thepage - }
\fancyfoot[R]{Gibert \couleur{\textit{et al}.}} 
\renewcommand{\headrulewidth}{0pt} 
}

\DeclareGraphicsExtensions{.jpg,.mps,.pdf,.png,.eps} 

\definecolor{DARKRED}{rgb}{0.545,0.0,0.0}
\definecolor{DARKGREEN}{rgb}{0.0078,0.4314,0.5059}
\definecolor{LIGHTGREEN}{rgb}{0,0.6706,0.7412}
\definecolor{LIGHTBLUE}{rgb}{0,0.6,0.8667}
\definecolor{LIGHTORANGE}{rgb}{1,0.6,0.20}
\definecolor{LIGHTGRAY}{rgb}{0.6314,0.7804,0.8784}

\hypersetup{
    colorlinks=true,                          
    linkcolor=LIGHTBLUE, 
    citecolor=DARKGREEN, 
    urlcolor=blue,
    linktoc=all
}

\author{Gibert Dominique, Lopes Fernando, Courtillot Vincent et Boulé Jean-Baptiste}
\title{Information Theory \\ Signal Analysis and Processing \\ Inverse Problem}
\date{}

\numberwithin{equation}{chapter}
\numberwithin{figure}{chapter}

\renewcommand\thesection{\arabic{section}}

\numberwithin{equation}{section}

\makeindex
\graphicspath{{figures/pdf/}}

\begin{document}
\dominitoc

\begingroup
\thispagestyle{empty}
\begin{tikzpicture}[remember picture,overlay]
    
 \fill[top color=blue!70!black!20, bottom color=cyan!20!blue!60] 
        (current page.south west) rectangle (current page.north east);

    \node[opacity=0.3,anchor=south west, at=(current page.south west), 
    inner sep=0pt, outer sep=0pt] {\includegraphics[width=\paperwidth,height=\paperheight,keepaspectratio]{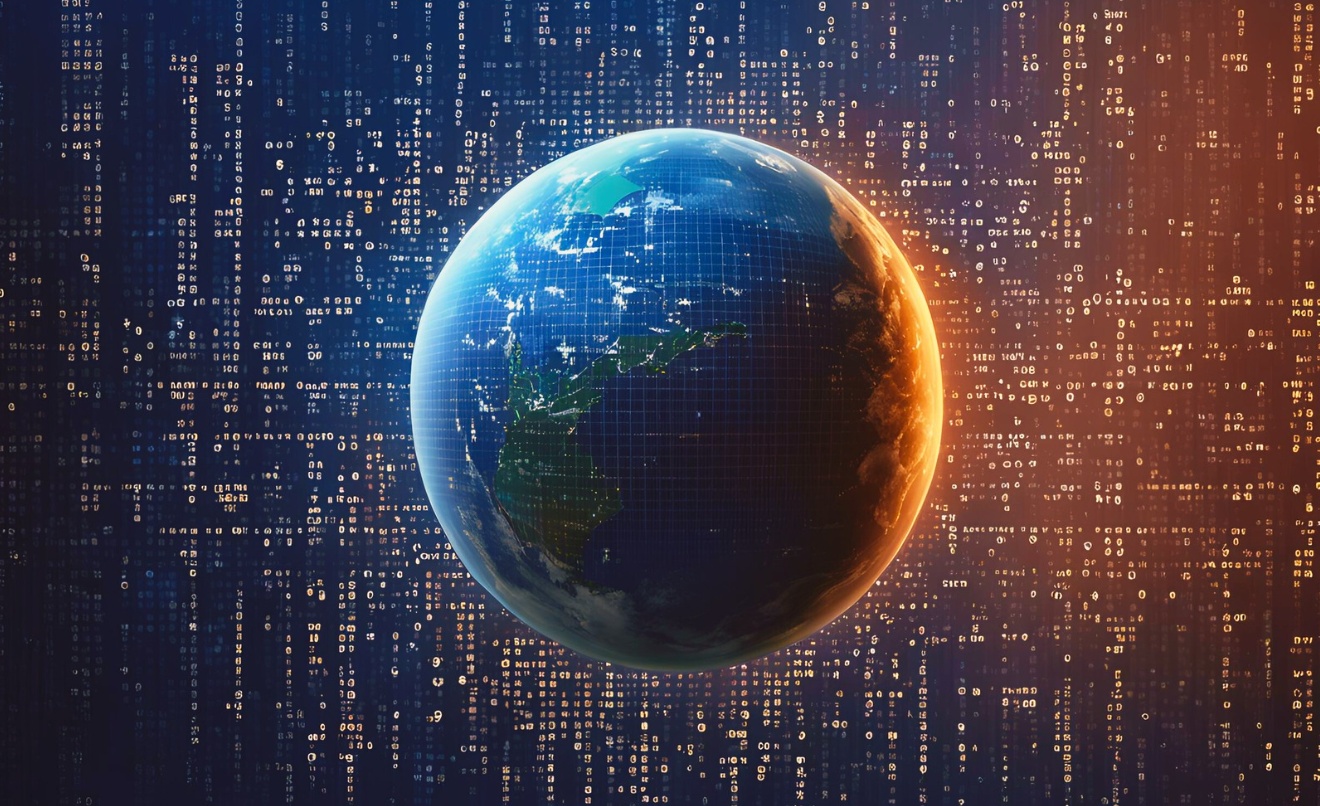}};

    \draw (current page.center) node [inner sep=1cm]{
        \Huge\centering\bfseries\color{black}\parbox[c][][t]{\paperwidth}{
            \centering \Huge Information Theory\\[15pt]
            {\Large Signal Analysis and Processing }\\[15pt]
            {\Large Inverse Problem}\\[20pt]
            {\small Gibert Dominique, Lopes Fernando, Courtillot Vincent et Boulé Jean-Baptiste}
        }
    }; 
\end{tikzpicture}
\vfill
\endgroup
\pagestyle{plain}

\tableofcontents
\dominitoc[n]    
\nomtcrule			

\chapter*{Introduction}
\pagestyle{plain}
Signal processing is a field that is difficult to describe in a few words. However, throughout these pages, we will see that the essential goal of signal processing techniques is to \couleur{separate a message from noise}.

This definition assumes that we know what the desired message is or what noise needs to be eliminated, which necessitates relying on \apriori considerations borrowed from the physics or chemistry of the problems being addressed. Without a fine understanding of the issues under study, the most sophisticated signal processing techniques in the world risk "spinning their wheels." It is important to understand the relationships between signal processing and chemistry and physics, as this is where the models used in signal processing find their justification. In our quest to isolate the message, we will see that it is often wise to transport the information carried by the signal into another "world," dual to the initial "world," where the information becomes more readable. The most well-known example is probably the transition from the time domain to the frequency domain \via the Fourier transform \index{Fourier}. The choice of the "host" world depends heavily on our \apriori knowledge of the problem at hand, and signal processing methods are like "glasses" through which we view a "landscape" of information where entities need to be recognized. Therefore, we will work on crafting glasses suited to our vision that provide the clearest possible images of the landscape. But that is not enough: you can process beautiful images of California obtained by the \couleur{SPOT} satellite\footnote{Satellite for Earth Observation. A family of French Earth observation satellites, initially launched between 1985 and 2002, and later between 2012 and 2014. URL: \href{http://www.intelligence-airbusds.com/en/99-spotmaps-high-resolution-colour-satellite-images}{http://www.intelligence-airbusds.com/en/99-spotmaps-high-resolution-colour-satellite-images}} and recognize very nice "roads" but not the slightest fault if you do not have this concept in mind. It is not the role of signal processing to conceptualize the entities to be recognized in the landscape of information, but that of other disciplines such as geology, chemistry, and physics. A first piece of advice: \couleur{learn a bit of signal processing and a lot of other things!}

The choice of the dual space in which to transport the information contained in the signal critically depends on the models adopted to represent the signals. This modeling, explicit or implicit, allows signal processing techniques to be integrated into the theory of inverse problems. This approach is very beneficial for understanding the importance of choosing signal models and for clarifying the notion of resolution. Some classic signal processing problems, such as deconvolution, directly fall under the theory of inverse problems and are better understood in this context.

In practice, the array of available signal processing techniques allows for a progressive approach and gradually clarifies the understanding of a particular signal. In all cases, the physics of the phenomena causing the signal provides valuable insights into the nature of the message to be extracted. For example, in satellite altimetry, a geophysicist aiming to study the geoid will seek to correct the undulations of the sea surface for their temporal variability, which is precisely the signal of interest to the oceanographer studying ocean currents. This antagonism of objectives can be illustrated in all branches of global physics: \couleur{one's signal may be another's noise}. We touch here on a very general human principle. The signals studied in geophysics are extremely varied and require a vast array of processing methods. As a result, many techniques are employed by geophysicists, and sometimes, when the standard array is no longer sufficient, some of them develop new methodologies that prove to be very broad in scope. This is the case with methods based on the criterion of entropy maximization \index{entropy maximization}, for example, or the case of wavelets \index{wavelet}. There are also rediscoveries such as the "Sompi" method \shortcite{kumazawa1990theory}, which closely resembles the method invented by Baron de \textit{Prony}—itself close to Fourier analysis \shortcite{hauer1990initial}—in 1795!

This course should be considered an introduction aimed at raising awareness among geophysicists dealing with signals. I have followed a classic approach based on the Fourier transform, from which I develop a number of "selected pieces" chosen either for their universal character (sampling, the uncertainty principle, \etc) or for their great practical utility (linear filtering, spectral analysis, \etc). The choice of the Fourier transform is both simple and in line with what is generally done in the literature on signal processing. Nevertheless, it remains debatable as sine and cosine functions, which have an unbounded support, do not always have a physical meaning. However, these functions have the immense advantage of being the \couleur{eigenfunctions of most of the major partial differential equations in mathematical physics} expressed in Cartesian coordinates. This is what makes them successful, along with plane harmonic waves in seismology. But the Earth is round, we drill cylindrical wells, and Cartesian coordinates are not always the best suited. We then have to abandon them along with sine and cosine functions, which give way to spherical harmonics, Bessel functions \index{Bessel}, \etc. Many geophysical signals must therefore be processed using models other than the Fourier transform (wavelets, spherical harmonics, \etc), but many points covered in this course (sampling, aliasing, duality, \etc) remain valid and adaptable to these function bases. Some readers will undoubtedly find this course scandalously incomplete. This is the result of a simple principle to which I have adhered unfailingly: I only discuss techniques that I have personally used. It seemed indispensable to me, in a course with a practical aim, to adopt such a principle because merely reading the specialized literature generally does not provide a precise idea of the operational character of the theories developed there. This concern to help the reader form a personal opinion is concretized by the fact that they can recreate all the figures in this book using the \href{https://fr.mathworks.com/products/matlab.html}{Matlab\textregistered} functions accompanying the book. It is, of course, possible to change the initial parameter values to test the limits of the presented techniques. These functions can also be used to carry out a number of additional practical exercises.
\chapter{\titrechap{The Fourier transform}}
\minitoc
\pagestyle{fancy}
\section{Definition of the Fourier Transform}
Almost all works on signal processing are based on the Fourier transform, which associates with a function $f\left(t\right)$ its Fourier transform $F(u)$. The expressions we have adopted for the direct and inverse Fourier transforms are those used by Bracewell in his book \shortcite{bracewell1986fourier}. They have the advantage of being symmetric and easy to remember.
\begin{equation}
	F\left(u\right)=\int_{-\infty}^{+\infty}f\left(t\right)\exp\left(-2i\pi ut\right)dt
	\label{fourier_directe}
\end{equation}

where $u\in\Bbb{R}$ is referred to as the frequency. For a wide class of functions, $f\left(t\right)$, the above integral equation is invertible and the original function can be reconstructed using the inverse Fourier transform,
\begin{equation}
	f\left(t\right)=\int_{-\infty}^{+\infty}F\left(u\right)\exp\left(+2i\pi ut\right)du
   \label{fourier_inverse}
\end{equation}

Many signal processing operations involve computing the Fourier transform of the signal, inspecting it, applying a series of simple operations to it, and finally reconstructing the processed signal by computing an inverse Fourier transform. Faced with this approach, a novice\footnote{That is, someone who dares to ask the right questions\thinspace !} often wonders: "\danger{Why this Fourier transform? Why not my Zébulon-Klack transform}", defined by the following relation,
\begin{equation}
	ZK\left(\zeta,\chi\right)=\int_{-\infty}^{+\infty}\arctan\left[f^{2}\left(t\right)\right]\left(\cosh\zeta+\sinh\chi\right)dt,
	\label{zebulon}
\end{equation}

\danger{of which I am very proud?}". I have never had to use the \danger{Zébulon-Klack} transform, but there may be a domain in mathematical physics where it is quite useful. Why not, since it is precisely in mathematical physics that the Fourier transform finds its justification. Ultimately, things are not as definitive as they might first appear, and it is important to explore the domain where the Fourier transform proves to be useful.

\section{Mathematical Physics}
The language of physics is constructed using mathematics. The laws of physics are expressed in the form of equations, which physicists spend considerable time solving within various contexts of complexity. The same law can be presented in very different mathematical forms. For instance, Newton's law of universal gravitation \index{Newton} can be written as
\begin{equation}
\left\Vert \overrightarrow{f}\right\Vert =G\frac{m_{1}m_{2}}{r^{2}}
\label{newton}
\end{equation}

where $G$ is the universal gravitational constant, and $\overrightarrow{f}$ is the mutual attraction force between the two masses $m_{1}$ and $m_{2}$ separated by the distance $r$. However, one can also describe the law of universal gravitation using Poisson's equation \index{Poisson},
\begin{equation}
\nabla^{2}\Phi=-4\pi G\rho
\label{poisson}
\end{equation}

where $\rho$ represents the mass density of the material in the considered region, and $\Phi$ is a potential whose gradient provides the gravitational attraction. A similar approach can be applied to the laws of electromagnetism, \etc \couleur{Poisson}'s equation (\ref{poisson}) is a partial differential equation that allows for a local formulation of gravitation within the framework of field theory. This local expression of physical laws is generally more satisfying to the mind as it removes the "magical" notion of action at a distance. I do not intend to delve further into this fascinating subject; interested readers may profitably consult the works of \shortciteN{feynman1980nature}\footnote{"The Nature of Physics"} or \shortciteN{Thom91}\footnote{"Predicting is not Explaining"}. This topic is often present in non-local formulations which are extensively used for practical reasons (e.g., geometric optics and ray theory in seismology). Partial differential equations are ubiquitous in physics, and it is remarkable that a few of these equations cover a vast range of mathematical physics, as illustrated by the chapter "\couleur{The Same Equations Have the Same Solutions}" in the physics course by \shortciteN{feynman2013cours}. We shall mention only, the Laplace equation,
\begin{equation}
\nabla^{2}\psi=0
\label{laplace}
\end{equation}

	the wave equation,
\begin{equation}
\nabla^{2}\psi-\frac{1}{c^{2}}\frac{\partial^{2}}{\partial t^{2}}\psi=0
\label{eq_ondes}
\end{equation}
\newpage

	and the diffusion equation,
\begin{equation}
\nabla^{2}\psi-\frac{1}{\kappa}\frac{\partial}{\partial t}\psi=0.
\label{eq_diffusion}
\end{equation}

	The solution of these equations, that is, finding the field $\psi$ while considering boundary conditions, initial conditions, etc., can only be achieved numerically in complex cases. Simple cases can be handled analytically through methods such as separation of variables and Green's functions \index{Green} (see, for instance, the books by \shortciteN{Morse53}). In this chapter and the one on convolution, we will see that these two techniques bestow a particular status upon the Fourier transform, though not upon the Zébulon-Klack transform!

\section{Orthogonal Functions}
The method of separation of variables, pioneered by Bernoulli \index{Bernoulli} in the mid-18th century \shortcite{bernoulli1753}, involves selecting a coordinate system (Cartesian, spherical, cylindrical, etc.) in which the unknown field is expressed as the product of functions, each depending on only one coordinate,
\begin{equation}
\psi(x,y,z,t)=f(x)\cdot g(y)\cdot h(z)\cdot s(t)
\label{sep_variables}
\end{equation}

When this solution form is substituted into the partial differential equation to be solved, it results in a system of differential equations coupled by arbitrary constants, referred to as "separation constants." The analytical form of the partial differential equation and the resulting differential equations depends on the choice of coordinate system. For example, in Cartesian coordinates $\left(x,y,z\right)$, the wave equation is
\begin{equation}
\frac{\partial^{2}}{\partial x^{2}}\psi+\frac{\partial^{2}}{\partial y^{2}}\psi+\frac{\partial^{2}}{\partial z^{2}}\psi-\frac{1}{c^{2}}\frac{\partial^{2}}{\partial t^{2}}\psi=0
\label{eq_ondes_cart}
\end{equation}

whereas in spherical coordinates $(r,\theta,\phi)$, it is written as
\begin{equation}
\frac{1}{r^{2}}\frac{\partial}{\partial r}(r^{2}\frac{\partial\psi}{\partial r})+\frac{1}{r^{2}\sin\theta}\frac{\partial}{\partial\theta}(\sin\theta\frac{\partial\psi}{\partial\theta})+\frac{1}{r^{2}\sin^{2}\theta}\frac{\partial^{2}\psi}{\partial\phi^{2}}-\frac{1}{c^{2}}\frac{\partial^{2}}{\partial t^{2}}\psi=0
\label{eq_ondes_spher}
\end{equation}

In all cases, the coupled differential equations can be expressed in the form of a Sturm-Liouville \index{Sturm-Liouville} equation,
\begin{equation}
\frac{d}{dt}[l(t)\frac{\partial P}{\partial t}]+[m(t)+u\cdot n(t)]P(t)=0
\label{eq_strum_liouville}
\end{equation}

where $u$ is the separation constant and the functions $l(t)$, $m(t)$, and $n(t)>0$ are determined by the chosen coordinate system. The solutions, $P$, are as numerous as the allowed values for the separation constant, denoted as $P(t\mid u)$. These solutions are known as the eigenfunctions of the differential equation. They have the important property of being mutually orthogonal, meaning they satisfy
\begin{equation}
\int n(t)P(t\mid u_{1})P^{*}(t\mid u_{2})dt=0:\textrm{if}; u_{1}\neq u_{2}
\end{equation}

where $^{*}$ denotes the complex conjugate and a weighted inner product with the function $n(t)$ is used. The bounds of the integral above depend on the range of the solutions being sought. The solution, $s(t)$, is a linear combination of all the particular solutions, which are the eigenfunctions,
\begin{equation}
s(t)=\int_{\Omega}S(u)P(t,u)du
\end{equation}

where $\Omega$ is the set of permissible values for $u$. The coefficients, $S(u)$, in this linear combination are adjusted according to the boundary conditions and initial conditions that the field $\psi$ must satisfy. These coefficients indicate "how much" of each eigenfunction $P(t\mid u)$ is involved in the "composition" of the function $s(t)$. To better understand their role, one can compare this situation to the more classical context of vector analysis, where a vector is decomposed into a basis. In this case, $s(t)$ plays the role of the vector to be decomposed, the eigenfunctions $P(t\mid u)$ are analogous to the basis vectors\footnote{which are generally infinite in number}, and $S(u)$ can be considered as the function providing the components of $s(t)$ in the basis $\left\{ P\left(t\mid u\right);u\in\Omega\right\} $. For the problem of constructing the solution $\psi$ to be well-posed, the partial differential equation to be solved must be accompanied by boundary and/or initial conditions that uniquely determine the components $S(u)$ by forming the inner product between the field expression at the boundaries and the basis functions\footnote{the example of potential field extension provided later illustrates this computation.}. This point is discussed very clearly in the books by \shortciteN{Morse53}. Each function, $f$, $g$, $h$, and $s$ in the expression of the field $\psi$ can thus be written as a linear combination of the eigenfunctions of the corresponding Sturm-Liouville equation.

\section{The Fourier Transform}

\subsection{Theoretical Foundations and Definitions}

The prominent role of the Fourier transform in signal processing is justified by the fact that many partial differential equations in physics lead to Sturm-Liouville differential equations where $n(t)=1$, and whose eigenfunctions are the $\cos$ and $\sin$ functions. The solutions then take the form
\begin{equation}
s(t)=\int_{0}^{+\infty}S_{\cos}(u)\cos(2\pi ut)du+\int_{0}^{+\infty}S_{\sin}(u)\sin(2\pi ut)du
\end{equation}

where the coefficients $S_{\cos}$ and $S_{\sin}$ are known as the Fourier coefficients. \couleur{Joseph Fourier (1768–1830)}, born in Auxerre, submitted his first paper on polynomial root approximations to the Académie des Sciences in 1789. After spending several years in Egypt, he was appointed Prefect in Grenoble in 1802. In 1807, he presented a paper on heat propagation to the Académie des Sciences. His major work \shortcite{fourier1822}, "Théorie analytique de la chaleur" (Analytical Theory of Heat), was published in 1822, and a few months later, he was appointed perpetual secretary of the Académie des Sciences. It is worth noting that Fourier became interested in statistics as early as 1798 and was recognized by the Académie des Sciences from 1816 as a specialist in insurance, statistics, and probability. Although some sums of trigonometric series had been calculated by Euler \index{Euler} (1707–1783), the history of trigonometric series can be traced back to the solution of the vibrating strings problem \shortcite{bernoulli1753}. The question of representing an arbitrary function, possibly discontinuous, by a trigonometric series quickly arose—a representation that the leading mathematicians of the time (1750) deemed impossible. It was not until fifty years later that Fourier addressed this issue while working on his analytical theory of heat. His initial work (1807) concerned only trigonometric series, and it was in 1812 that he introduced the Fourier integral. \couleur{He is credited with the notation $\int_{a}^{b}$}.

The variable $u \geq 0$ is called the frequency. This result is highly significant, indicating that \couleur{in many physical problems, the solutions can be expressed as a linear combination of cosine and sine functions}. It is possible to modify the expression of the solution above to match the form of the Fourier transform we encountered at the beginning of this chapter, eq. (\ref{fourier_directe}). The calculation is straightforward and uses the following Euler identities,
\begin{equation}
\cos(2\pi ut)=\dfrac{\exp(2i\pi ut)+\exp(-2i\pi ut)}{2}
\end{equation}

and
\begin{equation}
\sin(2\pi ut)=\dfrac{\exp(2i\pi ut)-\exp(-2i\pi ut)}{2i}
\end{equation}

Some algebraic manipulations then yield
\begin{equation}
s(t)=\int_{-\infty}^{+\infty}S(u)\exp(2i\pi ut)du,
\end{equation}

where, this time, the frequency $u$ can take negative values. The function $S\left(u\right)$ is called the Fourier transform of $s(t)$ and is given by
\begin{equation}
	\left\{ 
	\begin{array}{llll}
       S(u) & = & [S_{\cos}(u)-iS_{\sin}(u)]/2 & u\geq0\\
	   S(u) & = & [S_{\cos}(-u)+iS_{\sin}(-u)]/2 & u\leq0
	\end{array}\right\}
\end{equation}

\begin{figure}[!h]
\begin{center}
	\tcbox[colback=white]{\includegraphics[width=16cm]{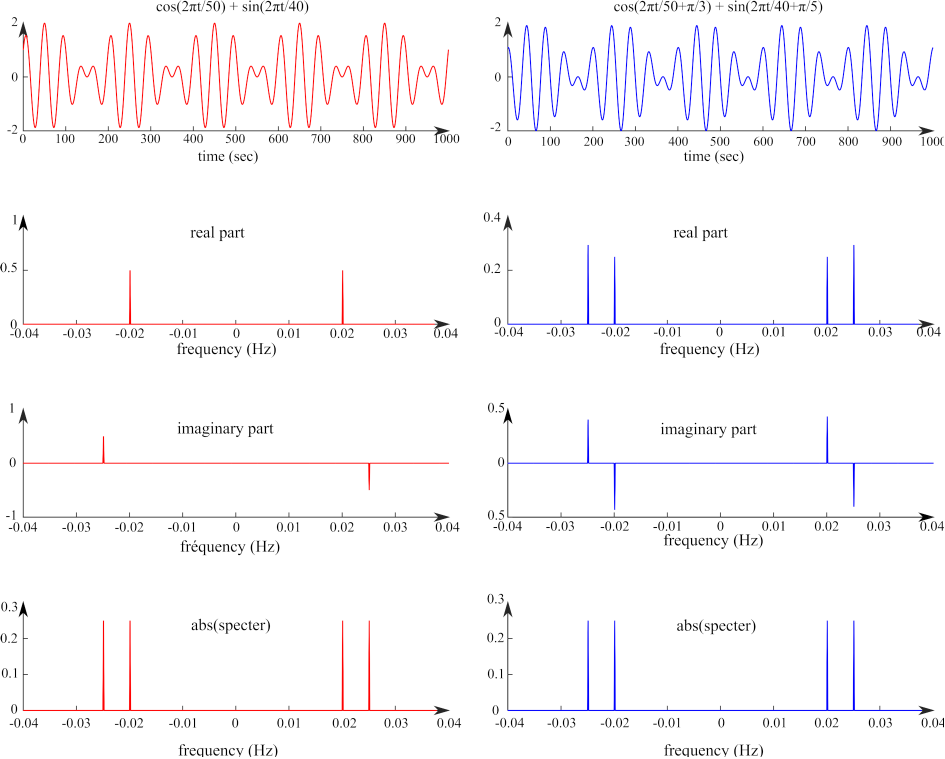}}
\end{center}
\caption{Examples of Fourier transforms of simple signals. On the left, the real part of the transform corresponds to the frequency carried by the cosine component of the signal, and the imaginary part corresponds to the sine component. On the right, phase shifts cause the two frequencies present in the signal to decompose into the real and imaginary parts of the Fourier transform.}
\label{fourier01}
\end{figure}

$S(u)$ can be computed using the orthogonality property of the eigenfunctions,
\begin{equation}
	\int_{-\infty}^{+\infty}\exp(2i\pi ut)\exp(-2i\pi vt)dt=\delta(u-v)
\end{equation}

where $\delta(t\neq0)=0$. We find
\begin{equation}
	S(u)=\int_{-\infty}^{+\infty}s(t)\exp(-2i\pi ut)dt
\end{equation}

which is the expression for the direct Fourier transform. The function \couleur{fourier\_01.m} computes the Fourier transforms of simple signals and illustrates the role of the real and imaginary parts (Figure \ref{fourier01}).

\newpage

\subsection{Notations}
We will employ two notations. The first, implicitly adopted up to this point, represents functions using lowercase letters in their original physical space (\eg $f(t)$) and their images in the Fourier dual space using uppercase letters (\eg $F(u)$). The second notation will be used only when the first is not applicable, representing the direct and inverse transformation operations by $\mathcal{F}$ and $\mathcal{F}^{-1}$, respectively. Therefore, we have:
\begin{equation}
\mathcal{F}f(t) = F(u)
\end{equation}

and
\begin{equation}
\mathcal{F}^{-1}F(u) = f(t)
\end{equation}

\subsection{Example: Extension of Potential Fields}
This example will introduce an initial application of the Fourier transform in geophysics: the extension, either upwards or downwards, of potential fields such as the Earth's magnetic field or the gravitational field. In the atmosphere, which we will approximate as a vacuum, these two fields satisfy Laplace's equation:
\begin{equation}
\nabla^{2}v = 0
\label{eq_laplace}
\end{equation}

where $v$ is the field to be extended from a surface on which it is assumed to be perfectly known. The field in question may be a potential or a component of a geophysical field such as gravity or the magnetic field. Let us restrict ourselves to a two-dimensional Cartesian geometry where Laplace's equation is written as:
\begin{equation}
\frac{\partial^{2}}{\partial x^{2}}v(x,z) + \frac{\partial^{2}}{\partial z^{2}}v(x,z) = 0
\end{equation}

By separating variables, we seek a solution of the form:
\begin{equation}
v(x,z) = l(x)m(z),
\end{equation}

which, when substituted into equation (\ref{eq_laplace}), yields:
\begin{equation}
\frac{1}{l(x)}\frac{d^{2}}{dx^{2}}l(x) = -\frac{1}{m(z)}\frac{d^{2}}{dz^{2}}m(z)
\end{equation}

This equation must be satisfied for all pairs $(x,z)$, which is only possible if each term is equal to a real constant, \danger{the famous separation constant $\alpha$},

\begin{equation}
\frac{d^{2}}{dx^{2}}l(x) - \alpha l(x) = 0
\end{equation}

and
\begin{equation}
\frac{d^{2}}{dz^{2}}m(z) + \alpha m(z) = 0
\end{equation}

The separation of variables has transformed the initial partial differential equation into a system of two coupled differential equations. If $\alpha > 0$, we find:
\begin{equation}
l(x, \alpha > 0) = L_{+}(\alpha)\exp(+\sqrt{\alpha}x) + L_{-}(\alpha)\exp(-\sqrt{\alpha}x),
\end{equation}

and if $\alpha \leq 0$,
\begin{equation}
l(x, \alpha \leq 0) = L_{\cos}(\alpha)\cos(\sqrt{|\alpha|}x) + L_{\sin}(\alpha)\sin(\sqrt{|\alpha|}x).
\end{equation}

Identical solutions are obtained for $m(z)$, though the sign of the constant $\alpha$ should be reversed:
\begin{equation}
m(z, \alpha \leq 0) = M_{+}(\alpha)\exp(+\sqrt{|\alpha|}z) + M_{-}(\alpha)\exp(-\sqrt{|\alpha|}z)
\end{equation}

and,
\begin{equation}
m(z, \alpha > 0) = M_{\cos}(\alpha)\cos(\sqrt{\alpha}z) + M_{\sin}(\alpha)\sin(\sqrt{\alpha}z).
\end{equation}

In the most general case, the solution $v(x,z)$ is a linear combination of the solutions above for all possible values of the separation constant $\alpha$. However, not all obtained solutions are necessarily physically acceptable. For instance, consider the specific case of calculating a field in the half-space $z \geq 0$ with sources located entirely in the half-space $z < 0$. In such a configuration, physical considerations indicate that $v \to 0$ as $z \to +\infty$, which eliminates the solutions $M_{+}(\alpha)\exp(+\sqrt{|\alpha|}z)$, $M_{\cos}(\alpha)\cos(\sqrt{\alpha}z)$, and $M_{\sin}(\alpha)\sin(\sqrt{\alpha}z)$. Ultimately, the acceptable solutions are:

\begin{equation}
l(x, \alpha \leq 0) = L_{\cos}(\alpha)\cos(\sqrt{|\alpha|}x) + L_{\sin}(\alpha)\sin(\sqrt{|\alpha|}x),
\end{equation}

and
\begin{equation}
m(z, \alpha \leq 0) = M_{-}(\alpha)\exp(-\sqrt{|\alpha|}z).
\end{equation}

The most general solution that can be constructed is therefore:
\begin{equation}
v(x,z) = \int_{-\infty}^{0} \left[ V_{\cos}(\alpha)\cos(\sqrt{|\alpha|}x) + V_{\sin}(\alpha)\sin(\sqrt{|\alpha|}x) \right] \exp(-\sqrt{|\alpha|}z) , d\alpha
\end{equation}

This expression resembles the Fourier transform discussed at the beginning of this chapter. The resemblance becomes clearer by performing the variable change $\sqrt{|\alpha|} \rightarrow 2\pi u$ and using Euler's identities to switch to complex notation:
\begin{equation}
v(x,z) = \int_{-\infty}^{+\infty} V(u) \exp(2i\pi ux) \exp(-2\pi |u|z) , du
\end{equation}

At $z = 0$, the expression is exactly the same as the inverse Fourier transform:
\begin{equation}
v(x,0) = \int_{-\infty}^{+\infty} V(u) \exp(2i\pi ux) , du
\end{equation}

Thus, by direct Fourier transform, we have:
\begin{equation}
V(u) = \int_{-\infty}^{+\infty} v(x,0) \exp(-2i\pi ux) , dx
\end{equation}

We are now able to write the complete chain of calculations for extending, upwards, a known potential field at $z = 0$:
\begin{equation}
v(x,0) \stackrel{\mathcal{F}}{\longmapsto} V(u) \longmapsto V(u) \exp(-2\pi |u|z) \stackrel{\mathcal{F}^{-1}}{\longmapsto} v(x,z)
\label{algoProlon01}
\end{equation}

\begin{figure}[H]
\begin{center}
\tcbox[colback=white]{\includegraphics[width=16cm]{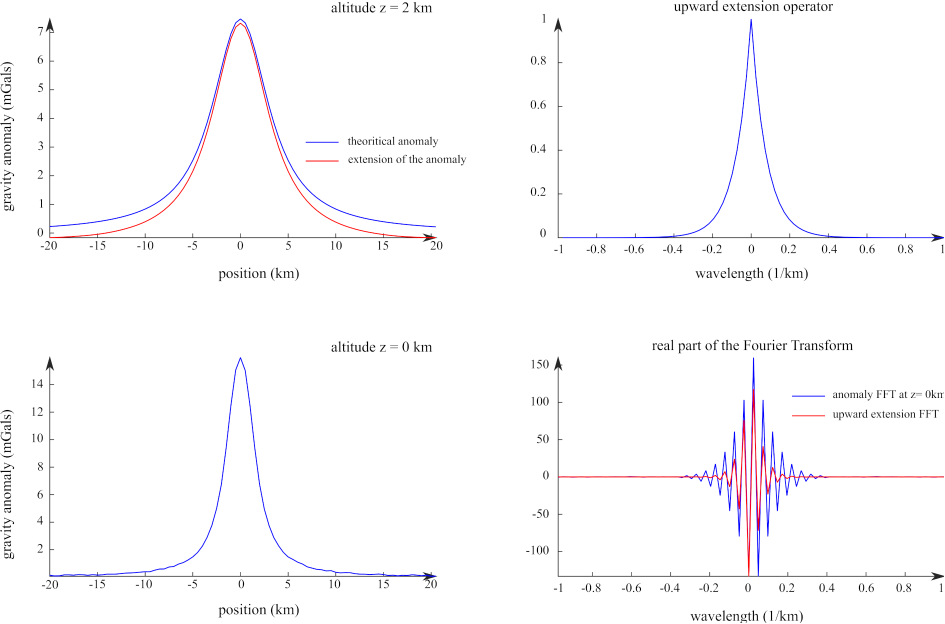}}
\end{center}
\caption{Illustration of the upward extension of a gravity anomaly. The discrepancy between the theoretical anomaly and the extended anomaly is due to numerical inaccuracies resulting from sampling.}
\label{prolonDemo01}
\end{figure}

\begin{figure}[H]
\begin{center}
\tcbox[colback=white]{\includegraphics[width=16cm]{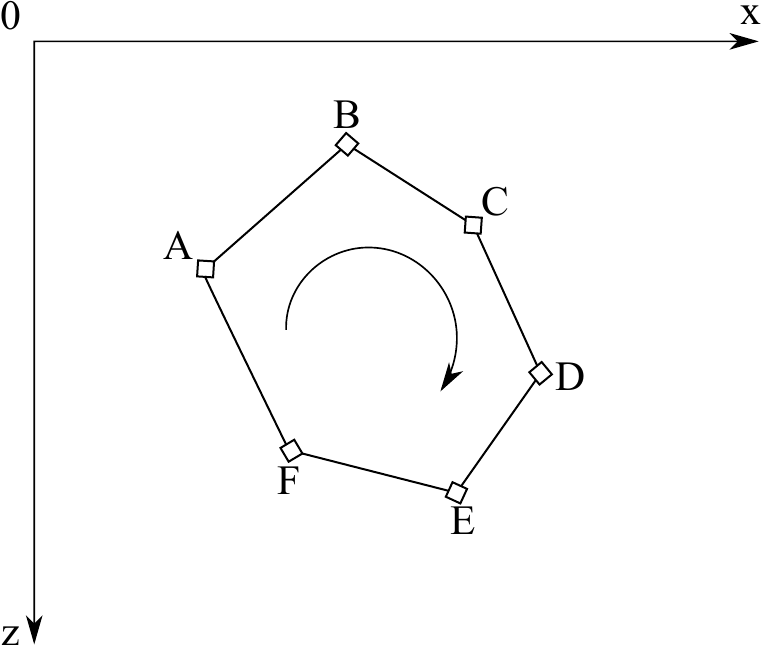}}
\end{center}
\caption{Polygon ABCDEF, of infinite dimension along Oy, used to calculate the theoretical magnetic or gravity anomaly produced by the perturbing body.}
\label{talwani}
\end{figure}

To illustrate this, the program \couleur{prolonDemo01.m} calculates the upward extension of a gravity anomaly obtained using the \couleur{talwani.m} function and produces Figure (\ref{prolonDemo01}). The method of \shortciteN{talwani1959rapid}\index{Talwani} allows, as in magnetism, for the calculation of the theoretical gravity anomaly of any body, such as a polygon, as shown in Figure (\ref{talwani}). This anomaly is the sum of the horizontal (X${i}$) and vertical (Z${i}$) contributions — using the notation from Talwani's paper — from each of the $n$ sides of the polygon ABCDEF:

\begin{equation}
V = 2G\rho \sum_{i=1}^{n} Z_i
\end{equation}

and
\begin{equation}
H = 2G\rho \sum_{i=1}^{n} H_i
\end{equation}

where $G$ is the universal gravitational constant and $\rho$ is the volumetric density of the object.

We will now describe the processing chain of the code \couleur{prolonDemo01.m}. This chain consists of three stages. The first stage involves calculating the Fourier transform (\warning{line 27}), $V(u)$, of the field measured at $z = 0$. The second stage is the application of the extension operator, which entails computing the product of $V(u)$ and the function $\exp(-2\pi |u|z)$ (\warning{line 34}). The final stage involves calculating the inverse Fourier transform of this product to obtain the field at the desired altitude $z$ (\warning{line 35}). Similar calculations can be performed for other potential field transformations, such as computing horizontal or vertical derivatives, downward extension, pole reduction of magnetic anomalies, etc. In all cases, the first and last stages of the processing chain involve Fourier transforms as long as the coordinate system is Cartesian. If the coordinates are spherical or cylindrical, the functions $\cos$ and $\sin$ are replaced by Legendre or Bessel functions, and the chain no longer includes Fourier transforms. The example we have examined is representative of what physicists do when processing signals: they perform calculations based on pure mathematics and physical considerations, which then provide solid theoretical foundations justifying the subsequent signal processing operations. In such an approach, computing the Fourier transform of the measured field does not "come out of a hat," but is justified by physical theory. In my opinion, this approach is the only one that can be beneficial. When signal processing operations lack genuine theoretical justification, they "often lead to poor results"! Within this intellectual framework, the role of signal processing is to \danger{master and implement, on incomplete and noisy data, a series of numerical calculations that best reproduce those of the underlying theory}.

\subsection{Break: The Hartley Transform\index{Hartley}}
At the beginning of this section, we saw that a real function, $s(t)$, can be expressed as follows,
\begin{equation}
s(t)=\int_{0}^{+\infty}S_{\cos}(u)\cos(2\pi ut)du+\int_{0}^{+\infty}S_{\sin}(u)\sin(2\pi ut)du
\label{fourier_hartley}
\end{equation}

From this, one can arrive at the classical expression for the Fourier transform through Euler’s identities and some algebraic manipulations. In the expression (\ref{fourier_hartley}), the functions $S_{\cos}$ and $S_{\sin}$ are real, and the Fourier transform, which is a complex function, has symmetry properties,
\begin{equation}
S(-u)=S^{*}(u)
\end{equation}

that render negative frequencies redundant. It is legitimate to question the utility of complicating matters by using a complex Fourier transform when "half of it" is unnecessary. If you are averse to this complexity, you might be attracted to the transform introduced by Ralph Hartley (\shortciteNP{hartley1942more}), which \couleur{Ronald Bracewell}\footnote{Ronald Newbold Bracewell (July 22, 1921 – August 12, 2007) was an Australian astronomer and physicist involved in the SETI program.} ardently supports. This transform is only applicable to real functions and can be easily derived from the above expression using the following elementary properties,
\begin{eqnarray}
\cos\left(-2\pi ut\right) & = & \cos\left(+2\pi ut\right)\\
\sin\left(-2\pi ut\right) & = & -\sin\left(+2\pi ut\right)
\end{eqnarray}

Thus, we have,
\begin{equation}
\int_{0}^{+\infty}S_{\cos}(u)\cos(2\pi ut)du=\int_{-\infty}^{+\infty}H_{\cos}(u)[\cos(2\pi ut)+\sin(2\pi ut)]du,
\end{equation}

where we defined,
\begin{equation}
H_{\cos}(u)=\frac{1}{2}S_{\cos}(|u|)
\end{equation}

Similarly,
\begin{equation}
\int_{0}^{+\infty}S_{\sin}(u)\sin(2\pi ut)du=\int_{-\infty}^{+\infty}H_{\sin}(u)[\cos(2\pi ut)+\sin(2\pi ut)]du,
\end{equation}

where
\begin{equation}
H_{\sin}(u\geq0)=\frac{1}{2}S_{\sin}(u);\text{and}; H_{\sin}(u\leq0)=-\frac{1}{2}S_{\sin}(-u)
\end{equation}

Finally, the signal $s(t)$ can be written as,
\begin{equation}
s(t)=\int_{-\infty}^{+\infty}H(u)\text{cas}(2\pi ut)du
\end{equation}

which we shall call the inverse Hartley transform, where the \text{cas} function is given by,
\begin{equation}
\text{cas}(2\pi ut)\equiv\cos(2\pi ut)+\sin(2\pi ut)
\end{equation}

and where the Hartley transform, $H(u)=H_{\cos}(u)+H_{\sin}(u)$, can be obtained via the direct transform of $s(t)$,
\begin{equation}
H(u)=\int_{-\infty}^{+\infty}s(t)\text{cas}(2\pi ut)dt
\end{equation}

In addition to the fact that it uses only real functions, the Hartley transform possesses symmetry properties that allow the construction of very fast numerical transformation algorithms. These algorithms are at least as fast, and certainly simpler, than specialized fast Fourier transform algorithms for real signals. Furthermore, the basis functions, $\text{cas}$, are real functions that have almost the same interesting properties in mathematical physics as the $\cos$ and $\sin$ functions.

\section{Fourier Series}
\subsection{Theoretical Foundations and Definitions}
The solution of partial differential equations sometimes involves conditions on the boundaries of a finite domain. For example, this occurs when studying the Earth's normal modes and assuming that the normal stresses on its surface are zero. In such cases, the values that certain separation constants can take are no longer real numbers but integers. This is known as mode selection. The solution then becomes,
\begin{equation}
s(t)=S_{0}+\sum_{n=1}^{\infty}S_{\cos,n}\cos\left(\frac{2\pi nt}{T}\right)+S_{\sin,n}\sin\left(\frac{2\pi nt}{T}\right)
\label{serie_fourier}
\end{equation}
where $T$ is the duration (or length) of the domain between the boundaries where conditions are imposed. The above equation is called a Fourier series\index{Fourier series}, and the coefficients $S_{\cos,n}$ and $S_{\sin,n}$ can be computed using the orthogonality properties of the eigenfunctions,
\begin{equation}
	\int_{0}^{T}\cos(\frac{2\pi mt}{T})\sin(\frac{2\pi nt}{T})dt=0\;\;\forall(m,n)
\end{equation}

\begin{equation}
	\int_{0}^{T}\cos(\frac{2\pi mt}{T})\cos(\frac{2\pi nt}{T})dt=\left\{ 
	\begin{array}{ll}
0 & m\neq n\\
T/2 & m=n
	\end{array}\right)
\end{equation}

and,
\begin{equation}
	\int_{0}^{T}\sin(\frac{2\pi mt}{T})\sin(\frac{2\pi nt}{T})dt=\left\{ 
	\begin{array}{ll}
0 & m\neq n\\
T/2 & m=n
	\end{array}\right)
\end{equation}

The Fourier coefficients are then given by,
\begin{equation}
S_{0}=\frac{1}{T}\int_{0}^{T}s(t)dt
\end{equation}

\begin{equation}
S_{\cos,n}=\frac{2}{T}\int_{0}^{T}s(t)\cos\left(\frac{2\pi nt}{T}\right)dt
\end{equation}

and,
\begin{equation}
S_{\sin,n}=\frac{2}{T}\int_{0}^{T}s(t)\sin\left(\frac{2\pi nt}{T}\right)dt
\end{equation}

Just as with the Fourier transform discussed in the previous section, more compact forms can be obtained using Euler's identities and complex notation,
\begin{equation}
s(t)=\sum_{n=-\infty}^{+\infty}S_{n}\exp\left(\frac{2i\pi nt}{T}\right)
\end{equation}

where,
\begin{equation}
S_{n}=\frac{1}{T}\int_{0}^{T}s(t)\exp\left(-\frac{2i\pi nt}{T}\right)dt
\end{equation}

Note that this time the sum extends over $n \in \mathbb{Z}$.

\subsection{Example: Vibrations of a Taut String}
We will focus on calculating the small amplitude vibrations, $v\left(x,t\right)$, of a taut string fixed at its ends. The partial differential equation relevant to this problem is the wave equation for one spatial dimension:
\begin{equation}
\frac{\partial^{2}}{\partial x^{2}}v(x,t)-\frac{1}{c^{2}}\frac{\partial^{2}}{\partial t^{2}}v(x,t)=0
\end{equation}

where $c$ is the wave propagation speed. Assuming,
\begin{equation}
v(x,t)=l(x)m(t)
\end{equation}

the separation of variables provides,
\begin{equation}
\frac{1}{l(x)}\frac{d^{2}}{dx^{2}}l(x)=\frac{1}{c^{2}m(t)}\frac{d^{2}}{dt^{2}}m(t)
\end{equation}

which must be satisfied for all pairs $(x,t)$. Introducing the separation constant $\alpha$, we obtain the system,
\begin{equation}
\frac{d^{2}}{dx^{2}}l(x)-\alpha l(x)=0
\end{equation}

and,
\begin{equation}
\frac{d^{2}}{dt^{2}}m(t)-\alpha c^{2}m(t)=0
\end{equation}

If $\alpha>0$, we find,
\begin{equation}
l(x,\alpha>0)=L_{+}(\alpha)\exp(+\sqrt{\alpha}x)+L_{-}(\alpha)\exp(-\sqrt{\alpha}x),
\end{equation}

and,
\begin{equation}
m(t,\alpha>0)=M_{+}(\alpha)\exp(+\sqrt{\alpha}ct)+M_{-}(\alpha)\exp(-\sqrt{\alpha}ct)
\end{equation}

When $\alpha\leq0$,
\begin{equation}
l(x,\alpha\leq0)=L_{\cos}(\alpha)\cos(\sqrt{|\alpha|}x)+L_{\sin}(\alpha)\sin(\sqrt{|\alpha|}x),
\end{equation}

and,
\begin{equation}
m(t,\alpha\leq0)=M_{\cos}(\alpha)\cos(\sqrt{|\alpha|}ct)+M_{\sin}(\alpha)\sin(\sqrt{|\alpha|}ct).
\end{equation}

Physical considerations specific to the problem must now be used to select acceptable solutions. We will only consider undamped vibrations, which allows us to eliminate the evanescent solutions $m(t,\alpha>0)$ and, consequently, $l(x,\alpha>0)$. The solutions corresponding to $$\alpha\leq0$$ are acceptable but must be subject to the boundary conditions of the string, which we will assume are located at $x=0$ and $x=L$. At these points, the vibrations must vanish, and the acceptable solutions must satisfy,
\begin{equation}
l(0,\alpha\leq0)=l(L,\alpha\leq0)=0
\end{equation}

which is only satisfied by,
\begin{equation}
L_{\sin}(\alpha)\sin(\sqrt{|\alpha|}x)
\end{equation}

when,
\begin{equation}
\sqrt{\left|\alpha\right|}=\frac{k\pi}{L};\text{with}; k \in \mathbb{N}^{*}
\end{equation}

The boundary condition of the string prevents a continuous variation of $\alpha$, and only discrete values are permitted. This is called mode selection. Ultimately, the most general acceptable solution is of the form,
\begin{equation}
v(x,t)=\sum_{k=1}^{+\infty}\sin\left(\frac{k\pi x}{L}\right)\left[V_{\cos,k}\cos\left(\frac{k\pi ct}{L}\right)+V_{\sin,k}\sin\left(\frac{k\pi ct}{L}\right)\right]
\end{equation}

where the coefficients $V_{\cos,k}$ and  $V_{\sin,k}$ need to be determined. This can be done by assuming the shape and velocity of the string at time $t=0$. For example, if,
\begin{equation}
v(x,0);\text{known and},;\left.\frac{\partial}{\partial t}v(x,t)\right|_{t=0}=0
\end{equation}

we have $V_{\sin,k}=0$, due to the initial velocity condition being zero, and $V_{\cos,k}$ such that,
\begin{equation}
v(x,0)=\sum_{k=1}^{+\infty}V_{\cos,k}\sin\left(\frac{k\pi x}{L}\right)
\end{equation}

This expression is a Fourier series, and the coefficients,
\begin{equation}
V_{\cos,k}=\frac{2}{L}\int_{0}^{L}v(x,0)\sin\left(\frac{k\pi x}{L}\right)dx
\end{equation}

The acceptable solution given the initial conditions is therefore,
\begin{equation}
v(x,t)=\sum_{k=1}^{+\infty}V_{\cos,k}\sin\left(\frac{k\pi x}{L}\right)\cos\left(\frac{k\pi ct}{L}\right)
\end{equation}

Note that this solution can be written as,
\begin{equation}
\left.v(x)\right|{t=t{0}}=\sum_{k=1}^{+\infty}V_{k}\left(t_{0}\right)\sin\left(\frac{k\pi x}{L}\right)
\end{equation}

where we have introduced the time-varying Fourier coefficients,
\begin{equation}
V_{k}\left(t_{0}\right)\equiv V_{\cos,k}\cos\left(\frac{k\pi ct_{0}}{L}\right)
\end{equation}

which indicates that at any time $t=t_{0}$, the shape of the string is a Fourier series. Similarly,
\begin{equation}
\left.v(t)\right|{x=x{0}}=\sum_{k=1}^{+\infty}V_{k}\left(x_{0}\right)\cos\left(\frac{k\pi ct}{L}\right)
\end{equation}

where we have defined,
\begin{equation}
V_{k}\left(x_{0}\right)\equiv V_{\cos,k}\sin\left(\frac{k\pi x_{0}}{L}\right)
\end{equation}

indicates that the vibrations at any point $x=x_{0}$ on the string are also a Fourier series with frequencies dependent on the length of the string\footnote{Hence the famous question posed by Mark Kac: "Can we hear the shape of a drum?" \shortcite{kac1966can}}.

\section{Properties of the Fourier Transform}
The Fourier transform has many properties, which are listed in the book by \shortciteN{bracewell1986fourier}. Here, we will only mention those that will be frequently used in the following sections of this book.

\subsection{Linearity}
This property is a direct consequence of the linearity of function integration:
\begin{equation}
\mathcal{F}\alpha f(t)+\beta g(t)=\alpha\mathcal{F}f(t)+\beta\mathcal{F}g(t),
\end{equation}

where $\alpha$ and $\beta$ are constants.

\subsection{Symmetries}
The symmetry properties of the Fourier transform are very useful for deducing and verifying certain results. Consider, for example, a real and even function,
\begin{equation}
f_{p}\left(-t\right)=f_{p}\left(t\right)\in\mathbb{R}
\end{equation}

\begin{figure}[H]
\begin{center}
\tcbox[colback=white]{\includegraphics[width=16cm]{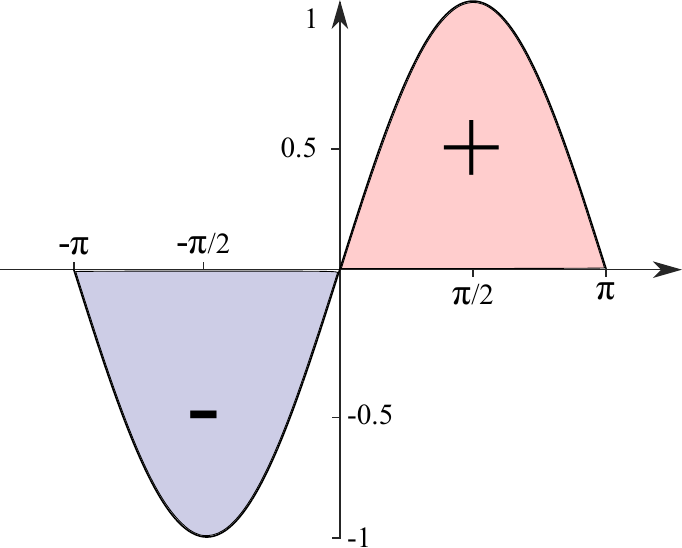}}
\end{center}
\caption{Sine function between -$\pi$ and $\pi$. In blue and red, the areas (negative and positive) of the sine function illustrating the integration of the sine over a period. The sum of these two areas is zero.}
\label{int_sin}
\end{figure}

The Fourier transform of such a function is given by,
\begin{equation}
 \begin{split}
	F_{p}(u) & =  \int_{-\infty}^{+\infty}f_{p}(t)\exp(-2i\pi ut)dt\\
 & =  \int_{-\infty}^{+\infty}f_{p}(t)\cos(-2\pi ut)dt+i \underbrace{\int_{-\infty}^{+\infty}f_{p}(t)\sin(-2\pi ut)dt}_{=0}\\
 & =  \int_{-\infty}^{+\infty}f_{p}(t)\cos(-2\pi ut)dt\\
 & =  F_{p}(-u)
\end{split}
\end{equation}

where it is verified that $F_{p}(u)$ is even and real. Indeed, since $f_p(t)$ is an even function and sin($-2\pi ut$) is an odd function, their product is an odd function, whose integral over the period is zero. To illustrate this, let’s take $f_p(t)$ constant and equal to 1; it is indeed an even function. We then find ourselves in the trivial case shown in Figure (\ref{int_sin}), where we sum two "signed" areas that cancel out. Similar calculations show that a real odd function, $f_{i}(-t)=-f_{i}(t)$, has a purely imaginary and odd Fourier transform. Thus, we can say that \couleur{the Fourier transform preserves parity}.

Any real function $f(t)$ can always be written as the sum of an odd function,
\begin{equation}
f_{i}(t)=[f(t)-f(-t)]/2
\end{equation}

and an even function,
\begin{equation}
f_{p}(t)=[f(t)+f(-t)]/2
\end{equation}

The linearity of the Fourier transform then establishes that the transform,
\begin{equation}
F(u)=F_{p}(u)+F_{i}(u)
\end{equation}

is complex and satisfies,
\begin{equation}
F(-u)=F^{*}(u)
\end{equation}

where $^{*}$ denotes the complex conjugate. \couleur{The information corresponding to negative frequencies is redundant} as it can be deduced from the information about positive frequencies. This property is utilized in numerical analysis, where specialized Fourier transform programs for real functions are found. All symmetries are summarized in the following formulas,
\begin{figure}[!h]
\centering
\tcbox[colback=white]{\includegraphics[width=16cm]{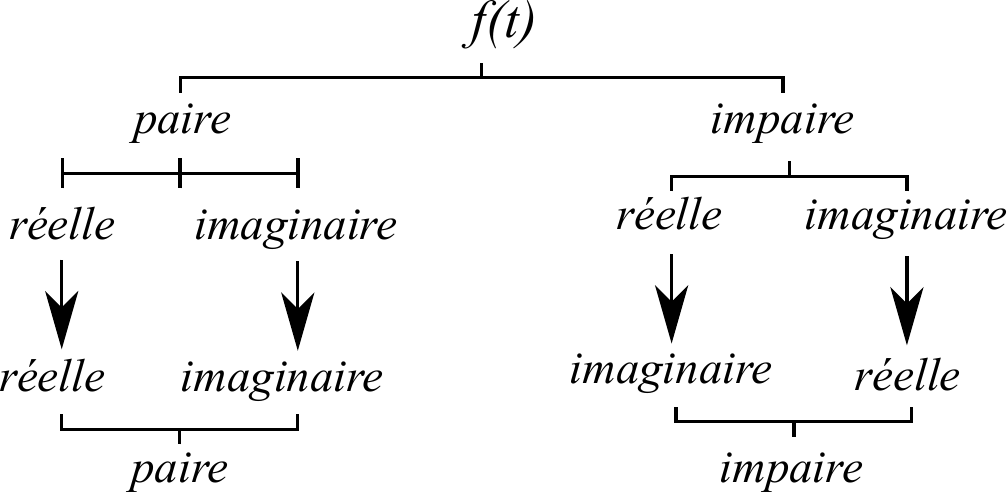}}
\caption{Symmetries of the Fourier transform for different types of functions}
\label{TFrelation}
\end{figure}

\subsection{Similarity}\index{similarity theorem}
\label{similitude}
This property, which is easily demonstrated by performing a change of variable in the integral defining the Fourier transform, expresses the duality \index{Fourier transform duality} that exists between a function and its Fourier transform:
\begin{equation}
\mathcal{F}\left[f\left(\alpha t\right)\right]\left(u\right)=\frac{1}{\left|\alpha\right|}F\left(u/\alpha\right).
\end{equation}

\couleur{This property indicates that the narrower the temporal support of a function, the wider the frequency support of its Fourier transform}. Denis Gabor\index{Gabor} first illustrated this in his famous 1946 paper, \textit{Theory of Communication} (\shortciteNP{gabor1946theory}), by introducing the notion of Heisenberg boxes\index{Heisenberg boxes}. We will not delve into the details of these boxes, also called time-frequency atoms when dealing with time-frequency transforms such as the Fourier transform or wavelet transforms; we will simply describe these boxes. For more details, readers are encouraged to refer to Chapter 4 of Stéphane Mallat's book, \textit{A Wavelet Tour of Signal Processing} (\shortciteNP{mallat1999wavelet}).

A brief preliminary discussion is necessary before describing these atoms. The linear operator $L$, whatever it may be, associates to any function $g\in\mathbb{L}^2(\mathbb{R})$ the following value:
\begin{equation}
Lg(\gamma) = \int_{-\infty}^{+\infty} g(t) \phi^{*}{\gamma}(t) dt = <g,\phi{\gamma}>
\end{equation}

The Parseval's theorem provides the following extension to the above expression:
\begin{equation}
Lg(\gamma) = \int_{-\infty}^{+\infty} g(t) \phi^{}{\gamma}(t) dt = \dfrac{1}{2\pi}\int{-\infty}^{+\infty} \hat{g}(u) \hat{\phi}^{}_{\gamma}(u) du
\label{heisenberg_box}
\end{equation}

With these two relations established, we can now briefly describe these boxes. A Fourier atom $\phi_{\gamma}$ is constructed using a window $f$ that can be translated in time by $t'$ and also modulated in frequency by $u$, giving:
\begin{equation}
\phi_{\gamma} (t)= \exp(iut) f(t-t')
\end{equation}

Relation (\ref{heisenberg_box}) shows that the information contained in $ <g,\phi_{\gamma}>$ depends only on the spread of $\phi_{\gamma}$ in time and frequency:
\begin{equation}
|\phi_{\gamma}|^2 = \int_{-\infty}^{+\infty} |\phi_{\gamma}(t)|^2 dt = 1
\end{equation}

$|\phi_{\gamma}(t)|^2$	 can be interpreted as a probability density centered at:
\begin{equation}
t_{\gamma} = \int_{-\infty}^{+\infty} t |\phi_{\gamma}(t)|^2 dt
\end{equation}

and whose spread  $\sigma^2_t(\gamma)$ is measured by the variance:
\begin{equation}
\sigma^2_t(\gamma) = \int_{-\infty}^{+\infty} (t-t_{\gamma})^2 |\phi_{\gamma}(t)|^2 dt
\end{equation}

Plancherel's formula ensures the following relation:
\begin{equation}
\int_{-\infty}^{+\infty} |\hat{\phi}{\gamma}(u)|^2 du = 2\pi |\phi{\gamma}|^2,
\label{plancherel_energie}
\end{equation}

Thus, we can naturally write the median frequency and the spread of the box in frequency as follows:
\begin{equation}
u_{\gamma} = \dfrac{1}{2\pi}\int_{-\infty}^{+\infty} u |\hat{\phi}_{\gamma}(u)|^2 du
\end{equation}

and:
\begin{equation}
\sigma^2_u(\gamma) = \dfrac{1}{2\pi}\int_{-\infty}^{+\infty} (u-u_{\gamma})^2 |\hat{\phi}_{\gamma}(t)|^2 du
\end{equation}

\begin{figure}[H]
\centering
\tcbox[colback=white]{\includegraphics[width=16cm]{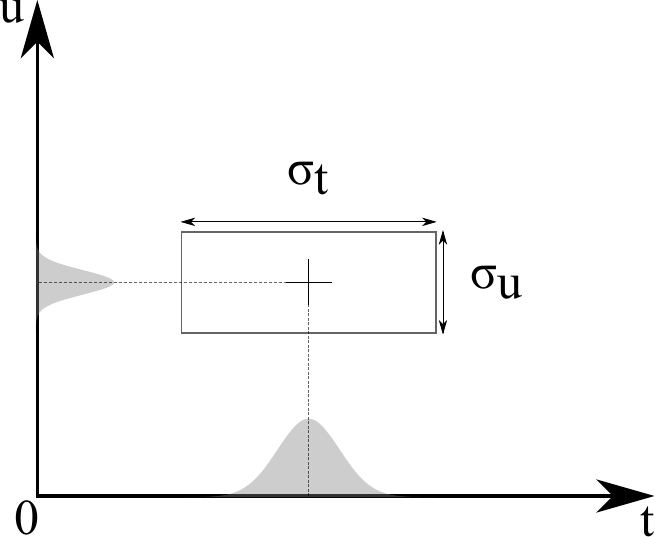}}
\caption{Heisenberg box schematizing the time-frequency duality of a Fourier atom}
\label{heisenberg_fig}
\end{figure}

We then obtain a rectangle (Figure (\ref{heisenberg_fig})) whose area is given by the product of the variances in frequency and time. Heisenberg's uncertainty theorem shows that the area of this rectangle is greater than or equal to 1/2, so:
\begin{equation}
\sigma_t\sigma_u \geqslant 1/2.
\label{incertitude_heis}
\end{equation}

Thus, it is clear that \couleur{the narrower the temporal support of a function, the wider the frequency support of its transform}.

\subsection{Translation}
This property is also derived by a simple change of variable. It expresses that translating a signal results in a phase shift in the Fourier dual space:
\begin{equation}
\mathcal{F}f(t-t_{0})=\exp(-2i\pi ut_{0})\mathcal{F}f(t)
\end{equation}

Reversing the application, it indicates that a frequency shift is equivalent to a time modulation:
\begin{equation}
\exp(2i\pi u_{0}t)f(t)=\mathcal{F}^{-1}F(u-u_{0})
\end{equation}

\subsection{Differentiation}
This property allows for the easy determination of the Fourier transforms of derivatives of a function. For the first derivative:
\begin{equation}
\begin{split}
	\mathcal{F}[\frac{d}{dt}f(t)](u) & =  \mathcal{F}[\lim_{\xi\downarrow0}{f(t+\xi)-f(t)}{\xi}](u)\\
 & =  \lim_{\xi\downarrow0}\frac{\mathcal{F}[f(t+\xi)-f(t)](u)}{\xi}\\
 & =  \lim_{\xi\downarrow0}\frac{[\exp(2i\pi u\xi)-1]F(u)}{\xi}\\
 & =  2i\pi uF(u)
	\end{split}
\end{equation}

The generalization to the n$^{ith}$ derivative is immediate:
\begin{equation}
	\mathcal{F}[\frac{d^{n}}{dt^{n}}f(t)](u)=(2i\pi u)^{n}F(u)
\end{equation}

Note also that the right-hand side of this expression remains valid when $n$ is not an integer but is a positive real number. This allows for the definition of the notion of non-integer differentiation of a function, which is useful for studying fractals and abrupt variations that occur in certain signals. Non-integer derivatives are also useful for studying wave propagation in highly heterogeneous media where properties vary randomly.

\section{Multidimensional Fourier Transforms}
\subsection{Example: Extension of Potential Fields}
The example of extending potential fields seen previously in the two-dimensional case can be extended to three dimensions. The calculations naturally lead to a two-dimensional Fourier transform. In the three-dimensional case, the potential must satisfy:
\begin{equation}
\frac{\partial^{2}}{\partial x^{2}}v(x,y,z)+\frac{\partial^{2}}{\partial y^{2}}v(x,y,z)+\frac{\partial^{2}}{\partial z^{2}}v(x,y,z)=0
\end{equation}

Assuming that the sources are located in the lower half-space, a similar reasoning to that used for the two-dimensional case leads to an acceptable solution:
\begin{equation}
v(x,y,z) = \int\int_{-\infty}^{+\infty}V (u_{x},u_{y})\exp[2i\pi(u_{x}x+u_{y}y)] \times\exp(-2\pi z\sqrt{u_{x}^{2}+u_{y}^{2}})du_{x}du_{y}
\end{equation}

Knowledge of the field in the plane $z=0$ provides:
\begin{equation}
V(u_{x},u_{y})=\int\int_{-\infty}^{+\infty}v(x,y,0)\exp[-2i\pi(u_{x}x+u_{y}y)]dxdy
\end{equation}

which is a two-dimensional Fourier transform. The inverse transform is given by:
\begin{equation}
v(x,y,0)=\int\int_{-\infty}^{+\infty}V(u_{x},u_{y})\exp[+2i\pi(u_{x}x+u_{y}y)]du_{x}du_{y}.
\end{equation}

The processing chain for the three-dimensional extension is the same as for the two-dimensional extension; it suffices to replace the one-dimensional Fourier transforms with their two-dimensional versions.

\subsection{General Definitions}
The generalization to $n$ dimensions leads to:
\begin{equation}
V(\overrightarrow{u})=\idotsint_{\Bbb{R}^{n}}v(\overrightarrow{x})\exp[-2i\pi\overrightarrow{u}\bullet\overrightarrow{x}]\overrightarrow{dx}
\end{equation}

for the direct Fourier transform, and
\begin{equation}
v(\overrightarrow{x})=\idotsint_{\Bbb{R}^{n}}V(\overrightarrow{u})\exp[+2i\pi\overrightarrow{u}\bullet\overrightarrow{x}]\overrightarrow{du}
\end{equation}

for the inverse Fourier transform, where $\bullet$ denotes the dot product.

\subsection{Sign Conventions in Space-Time}
The multidimensional Fourier transform we have defined is applicable to both spatial coordinates and time. However, it is wise to adopt a sign convention that differentiates the time dimension from the spatial dimensions.

\begin{equation}
V(\overrightarrow{u},u_{t})=\idotsint_{\Bbb{R}^{4}}v(\overrightarrow{x},t)\exp[-2i\pi(\overrightarrow{u}\bullet\overrightarrow{x}-u_{t}t)]\overrightarrow{dx}dt
\end{equation}

for the direct Fourier transform, and
\begin{equation}
v(\overrightarrow{x},t)=\idotsint_{\Bbb{R}^{4}}V(\overrightarrow{u},u_{t})\exp[+2i\pi(\overrightarrow{u}\bullet\overrightarrow{x}-u_{t}t)]\overrightarrow{du}du_{t}
\end{equation}

for the inverse Fourier transform. This definition of the Fourier transform is frequently used in seismology.
\chapter{\titrechap{Convolution and Correlation}}
\minitoc
\section{Convolution}
The convolution of two functions, $f(t)$ and $g(t)$, is defined by the integral:
\begin{equation}
[f*g](t)\equiv\int_{-\infty}^{+\infty}f(\tau)g(t-\tau)d\tau=\int_{-\infty}^{+\infty}f(t-\tau)g(\tau)d\tau
\label{equ_conv}
\end{equation}

where we use the classic notation $*$ for the convolution operator. Convolution is frequently encountered in signal processing because it appears in:
\begin{itemize}[label=$\bullet$]
\item linear systems theory,
\item Green's function theory when solving partial differential equations,
\item probability theory, where it is used to compute the distribution of sums of independent random variables.
\end{itemize}

The origins of convolution are as fundamental as those of the Fourier transform, and we will see that these two mathematical operations have remarkable properties with respect to each other. Before establishing these main properties, and as we did for the Fourier transform, we will first explore the "domain" of convolution.

\subsection{Where Do We Encounter Convolutions?}
\subsubsection{Temporal Convolution}
The concept of temporal convolution is closely related to the notions of \couleur{linearity and time (or space) invariance (or stationarity)}. One of the main tasks of physicists is to study systems through which signals pass. A system is characterized by a functional $\mathcal{G}$ that associates an input signal $e(t)$ with an output signal $s(t)$,
\begin{equation}
e(t)\longmapsto_{\mathcal{G}}s(t)\equiv\mathcal{G}[e(t)](t)
\end{equation}

The system in question can be the very object of the study, and its characteristics can be examined by injecting specific signals and observing the results. This approach is used when emitting electromagnetic or elastic waves into the Earth to study its structure (see Figure \ref{terre_ondes}). In other cases, it is the input signal that interests the physicist, and the system serves as a \couleur{pair of glasses} through which the phenomenon $e(t)$ is viewed. This occurs whenever measurements are made using an instrument, whether it is an astronomer looking at the sky through a telescope or a geophysicist recording ground vibrations with a seismometer. Many problems in experimental physics are of this nature, and their solutions are more or less easy to find depending on the complexity of the systems involved. The simplest systems one can imagine are linear time-invariant systems. These very simple systems arise in problems where the underlying physics is linear or as first-order approximations of nonlinear systems. A linear system satisfies the following relationships:
\begin{figure}[H]
\begin{center}
\tcbox[colback=white]{\includegraphics[width=16cm]{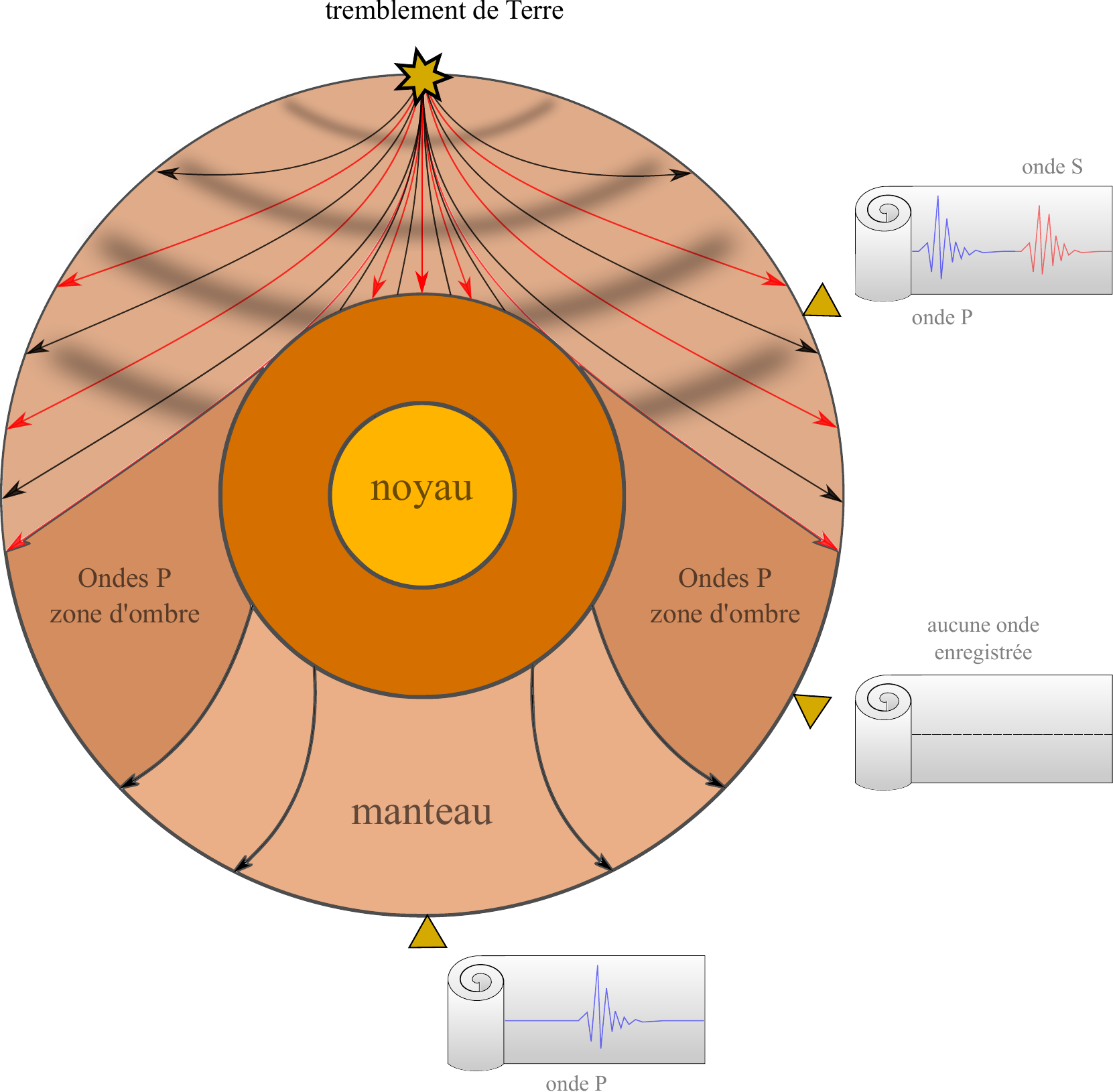}}
\end{center}
\caption{Diagram of Earth's structure obtained from seismic wave tomography. The yellow triangles represent the seismometers used by geophysicists.}
\label{terre_ondes}
\end{figure}

\begin{equation}
\mathcal{G}[\alpha_{1}e_{1}(t)+\alpha_{2}e_{2}(t)](t)=\alpha_{1}\mathcal{G}[e_{1}(t)](t)+\alpha_{2}\mathcal{G}[e_{2}(t)](t),
\end{equation}

and
\begin{equation}
s(t)=\mathcal{G}[e(t)](t)\Longrightarrow s(t-\xi)=\mathcal{G}[e(t-\xi)](t)
\end{equation}

These two properties allow us to establish that if the input signal consists of two signals of the same shape, with different amplitudes, occurring at different times, then,
\begin{equation}
f_{1}.e(t-t_{1})+f_{2}.e(t-t_{2})\longmapsto_{\mathcal{G}}f_{1}.s(t-t_{1})+f_{2}.s(t-t_{2})
\end{equation}

Of course, this can be generalized further,
\begin{equation}
\sum f_{i}.e(t-t_{i})\longmapsto_{\mathcal{G}}\sum f_{i}.s(t-t_{i})
\end{equation}

and even, in the limiting case where the input signals are infinitesimally close, forming a \textit{continuum},
\begin{equation}
\int_{-\infty}^{+\infty}f(\tau)e(t-\tau)d\tau\longmapsto_{\mathcal{G}}\int_{-\infty}^{+\infty}f(\tau)s(t-\tau)d\tau
\end{equation}

The integrals above are convolution integrals. Suppose now that the signals $e(t-\tau)$ in the left integral are impulses, $\delta(t-\tau)$, as brief as we want\footnote{The limit process, that is, an infinitely brief impulse, is discussed in the section on the Dirac impulse.}. In this case, somewhat like representing a function by a juxtaposition of \couleur{sticks} of different heights, the integral\footnote{Which we will revisit as the "sampling formula" in the section on the Dirac impulse.} becomes,
\begin{equation}
	\int_{-\infty}^{+\infty}f(\tau)\delta(t-\tau)d\tau=f(t)
\end{equation}

Let,
\begin{equation}
	g(t)\equiv\mathcal{G}[\delta(t)](t)
\end{equation}

be the system's impulse response. Then,
\begin{equation}
\mathcal{G}[f(t)](t)=\int_{-\infty}^{+\infty}f(\tau)g(t-\tau)d\tau
\end{equation}

This expression shows that \couleur{the response of a linear and time-invariant system is equal to the convolution product of the input signal with the system's impulse response.} The system is entirely characterized by its impulse response. The time-dependent system cannot respond before being excited, and its impulse response is causal, that is, such that,
\begin{equation}
	g(t<0)=0
\end{equation}

\begin{figure}[H]
\centering
\tcbox[colback=white]{\includegraphics[width=16cm]{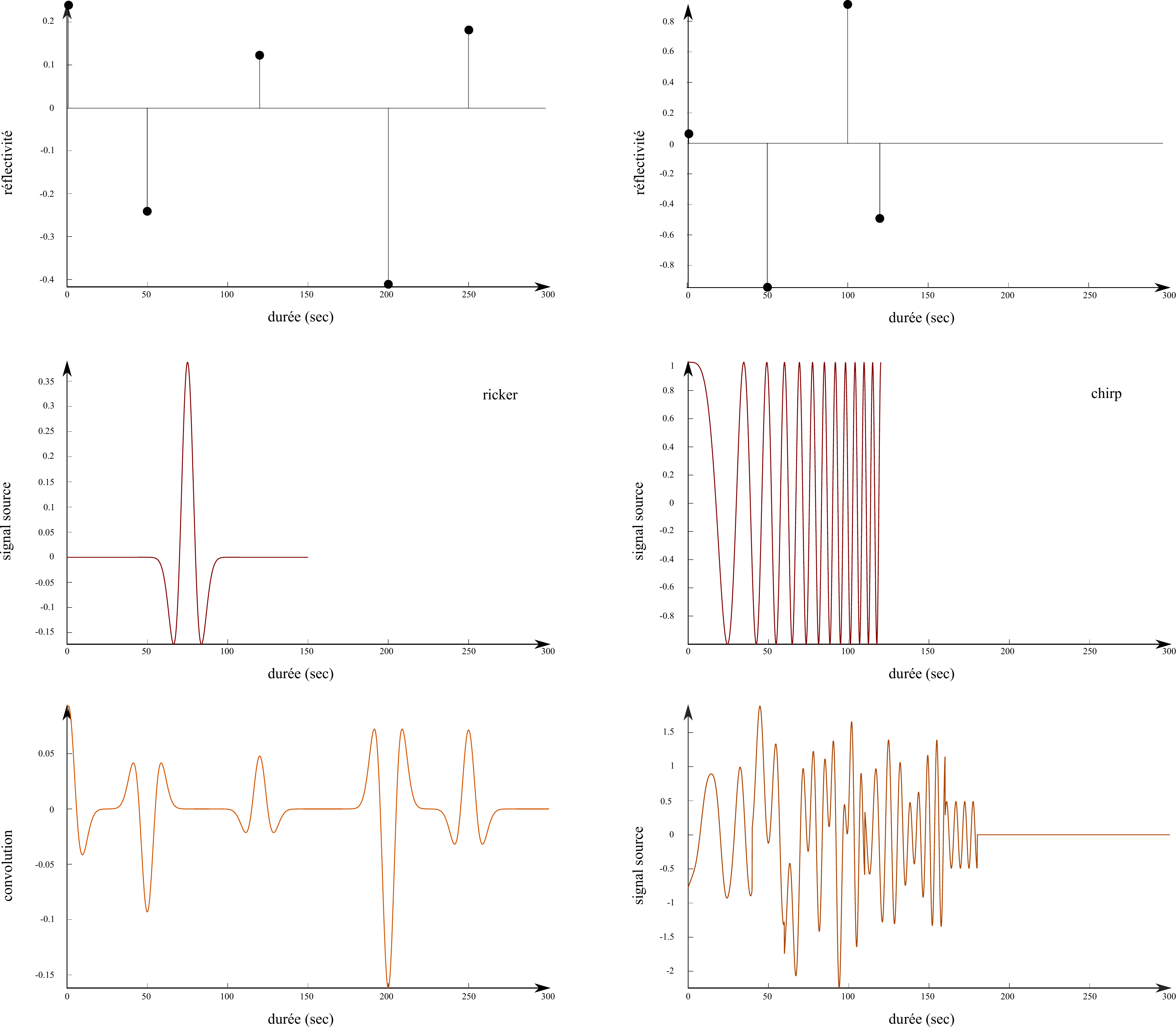}}
\caption{Temporal convolutions. This example is frequently encountered in seismics; the two signals at the top are the impulse responses of the subsurface. This is what would be obtained if we were capable of emitting a Dirac impulse using seismic sources. In practice, seismic sources are Ricker impulses (left center) or sweeps or chirps (right center). The resulting seismic traces (bottom) are the convolution of the impulse response (top) with the sources (middle). Note that the \couleur{sweep}, which has a long duration, completely masks the events present in the impulse response. How to retrieve them? See the section on correlation and Figure \ref{book2b}.}
\label{convolution_01}
\end{figure}

Many physical systems can be reasonably well represented by linear time-invariant systems. This is the case for many electronic circuits, optical setups, and mechanical assemblies. In seismology, the Earth is often considered an elastic medium and, therefore, linear and invariant. This approximation forms the basis for interpreting seismic recordings. To illustrate our points, we invite the reader to use the program \couleur{ex\_convolution.m} in which the convolution of \couleur{Ricker} \index{ricker} and \couleur{chirp} \index{chirp} is performed on random reflectivities. Figure (\ref{convolution_01}) provides an example.

The \couleur{Ricker} wavelet, sometimes called the \couleur{Mexican hat}, is defined by the following relation,
\begin{equation}
		r(t) = (1-2\pi^2 f^2 t^2) e^{-\pi^2 f^2 t^2} 
\end{equation}

It is also found in the form,
\begin{equation}
		r(t) = \dfrac{2}{\sqrt{3\sigma}\pi^{1/4}}(1-\dfrac{t^2}{\sigma^2})e^{-\dfrac{t^2}{2*\sigma^2}}
\end{equation}

The function \couleur{ricker.m} provides an implementation of the Ricker wavelet. The \couleur{chirp}, which in English means "tweet", is a pseudo-periodic signal of duration $T$, modulated in frequency ($\Delta f$) around a carrier frequency ($f_0$) and also modulated in amplitude. The function \couleur{chirp\_lin.m} provides an implementation of a particular case where the frequency ramp is linear and the envelope modulation remains constant. This signal is defined as follows,
\begin{equation}
		c(t) = \Re\{Ae^{2\pi i (f_0+\dfrac{\Delta f}{2T}.t-\dfrac{\Delta f}{2}).t}\} \quad \text{avec} \quad A=1 \quad \text{et} \quad \forall t \in \{0,T\}
\end{equation} 

\subsection{Spatial Convolution}
Spatial convolutions, which involve functions depending on spatial coordinates, are very common as they appear in the theory of \couleur{Green's functions} (a theory extensively covered in a dedicated book by \shortciteNP{roach1982green}) applied to solving partial differential equations. The Green's function represents the field created by a point source: a point mass in gravitation, a point charge in electrostatics, a dipole in magnetostatics, etc. For example, the gravitational potential created by a unit point mass located at the origin is given by:
\begin{equation}
	g(x,y,z)=\frac{G}{\sqrt{x^{2}+y^{2}+z^{2}}}
\end{equation}

As you know, the potential $v(x,y,z)$ of multiple masses is equal to the sum of the potentials caused by each mass, \ie,
\begin{equation}
v(x,y,z)=\iiint_{\Bbb{R}^{3}}\rho(\xi,\theta,\zeta)g(x-\xi,y-\theta,z-\zeta)d\xi d\theta d\zeta
\end{equation}

This is a three-dimensional spatial convolution where $\rho(\xi,\theta,\zeta)$ is the spatial mass density distribution. Analogous expressions are also encountered in potential theory, electromagnetism, etc. Spatial convolutions are also seen in seismic tomography for weakly diffracting media where the \couleur{Born approximation}\index{Born approximation} can be applied. This approximation originates from quantum mechanics for very sparse scattering potentials (\shortciteNP{cohen1998quantum}). In first-order Born approximation, only the incident wave and the waves scattered by a single interaction, a single scatterer, are considered and studied (\shortciteNP{hudson1981use}). More generally, it pertains to perturbation theory in mathematics.

\subsection{Convolution and Probability}
Convolution appears in probability theory as follows. Let $\alpha$ and $\beta$ be two independent random variables with respective probability densities $\alpha$ and $\beta$. The probability density $g(\gamma)$ of the sum $\gamma=\alpha+\beta$ is given by the convolution:
\begin{equation}
	g(\gamma)=\int_{-\infty}^{+\infty}a(\gamma-\xi)b(\xi)d\xi.
\end{equation}

	We will see later that this property, combined with the Central Limit Theorem, explains why the normal distribution holds a special place in statistics.

\subsection{Properties of Convolution}
\subsubsection{Commutativity, Associativity, Distributivity}
Convolution is commutative:
\begin{equation} 
	f*g=g*f
\end{equation}

associative:
\begin{equation}
	(f*g)*h=f*(g*h)
\end{equation}

and also distributive with respect to addition:
\begin{equation}
	f*\left(g+h\right)=f*g+f*h
\end{equation}

These properties are immediate consequences of those of integration and are easily established. Note, however, that it is due to the fact that one of the functions is "flipped" -- meaning that the integration variable appears with a negative sign (see equation \ref{equ_conv}) -- in the convolution integral that convolution is commutative. \couleur{Without this flipping, commutativity does not hold.}

\subsection{Fourier Transform of a Convolution}
The Fourier transform of a convolution product is obtained by explicitly writing out the following integrals:
\begin{equation}
	\begin{split}
\mathcal{F}[f*g](u) & =  \int_{-\infty}^{+\infty}\left\{ \int_{-\infty}^{+\infty}f(\xi)g(t-\xi)d\xi\right\} \exp(-2i\pi ut)dt\\
  & =  \int_{-\infty}^{+\infty}f(\xi)\left\{ \int_{-\infty}^{+\infty}g(t-\xi)\exp(-2i\pi ut)dt\right\} d\xi\\
 & =  \int_{-\infty}^{+\infty}f(\xi)\mathcal{F}[g(t-\xi)](u)d\xi\\
 & =  G(u)\int_{-\infty}^{+\infty}f(\xi)\exp(-2i\pi u\xi)d\xi\\
 & =  G(u)F(u)
 \end{split}
\end{equation}

\couleur{The Fourier transform of a convolution product is equal to the product of the Fourier transforms (Plancherel's theorem)}\index{Plancherel's theorem},
\begin{equation}
	\mathcal{F}\left[(f*g)\left(t\right)\right]\left(u\right)=F\left(u\right)G\left(u\right).
\end{equation}

The dual of the previous theorem indicates that:
\begin{equation}
	\mathcal{F}\left[f(t)g(t)\right](u)=\left[F*G\right](u).
\end{equation}

Applying this theorem to the specific case where $g(t)=f^{*}(t)$, we obtain:
\begin{equation}
	\begin{split}
\mathcal{F}\left[f(t)f^{*}(t)\right](u) & =  \mathcal{F}\left[\left|f(t)\right|^{2}\right](u)\\
 & =  F(u)*F^{*}(-u)
\end{split}
\end{equation}

which can be written as:
\begin{equation}
	\int_{-\infty}^{+\infty}\left|f(t)\right|^{2}\exp(-2i\pi ut)dt=\int_{-\infty}^{+\infty}F(v)F^{*}(u-v)dv
\end{equation}

By setting $u=0$:
\begin{equation}
	\int_{-\infty}^{+\infty}\left|f(t)\right|^{2}dt=\int_{-\infty}^{+\infty}\left|F(u)\right|^{2}du
\end{equation}

This important relation is known as the \couleur{Rayleigh-Parseval theorem}\index{Rayleigh-Parseval theorem}; it indicates that \couleur{the energy of the signal is conserved by the Fourier transform}. The simple form of the Fourier transform of a convolution product has significant consequences. From an analytical perspective, the simplification is substantial since one transitions from an integral formulation to a straightforward product of functions. This property, combined with the fact that convolution is a frequently encountered mathematical operation, greatly enhances the role of the Fourier transform in signal processing. Many calculations are simpler when performed \via the Fourier transform. For example, as seen in probability theory, the probability density $p_{s}\left(x\right)$ of a sum of $N$ independent random variables is given by the convolution chain:
\begin{equation} 
	p_{s}(x)=p_{1}(x)*p_{2}(x)*\cdots*p_{N}(x)
\end{equation}

which, after Fourier transform, becomes:
\begin{equation}
	P_{s}(u)=P_{1}(u)\times P_{2}(u)\times\cdots\times P_{N}(u),
\end{equation}

where the Fourier transforms $P_{i}(u)$ are called the characteristic functions of the probability densities $p_{i}(x)$.

\subsection{Differentiation of a Convolution}
We have:
\begin{equation}
	\begin{split}
\mathcal{F}^{-1}\left\{ \mathcal{F}\left[\frac{d}{dt}\left[f*g\right](t)\right](u)\right\} (t) & =  \mathcal{F}^{-1}\left[2i\pi uF\left(u\right)G\left(u\right)\right]\left(t\right)\\
 & =  \mathcal{F}^{-1}\left[2i\pi uF(u)\right](t)*\mathcal{F}^{-1}\left[G\left(u\right)\right]\left(t\right)\\
 & =  \left(\frac{d}{dt}f(t)\right)*g(t)
\end{split}
\end{equation}

and also:
\begin{equation}
	\begin{split}
\mathcal{F}^{-1}\left\{ \mathcal{F}\left[\frac{d}{dt}\left[f*g\right]\left(t\right)\right]\left(u\right)\right\} \left(t\right) & =  \mathcal{F}^{-1}\left[2i\pi uF\left(u\right)G\left(u\right)\right]\left(t\right)\\
 & =  \mathcal{F}^{-1}\left[F\left(u\right)\right]\left(t\right)*\mathcal{F}^{-1}\left[2i\pi uG\left(u\right)\right]\left(t\right)\\
 & =  f\left(t\right)*\left(\frac{d}{dt}g\left(t\right)\right)
\end{split}
\end{equation}

which simplifies to:
\begin{equation}
\frac{d}{dt}\left[f*g\right](t)=\left(\frac{d}{dt}f(t)\right)*g(t)=f(t)*\left(\frac{d}{dt}g(t)\right),
\end{equation}

\couleur{which should not be confused with the differentiation of a simple product of functions.}
\section{Correlation}
The cross-correlation of two functions $f(t)$ and $g(t)$ is defined by,
\begin{equation}
	\begin{split}
r_{f,g}(l) & =  f(t)\Diamond g(t)\\
 & \equiv  \int_{-\infty}^{+\infty}f^{*}(t)g(t+l)dt\\
 & =  \int_{-\infty}^{+\infty}f^{*}(t-l)g(t)dt\\
 & =  f^{*}(-t)*g(t)
\end{split}
\end{equation}

and can be interpreted as a convolution where one of the functions is not "reversed." The variable $l$ represents the time shift between the function and its replica. \couleur{Cross-correlation is not commutative},
\begin{equation}
	r_{f,g}(l)=f^{*}(-t)*g(t)\neq f(t)*g^{*}(-t)=r_{g,f}(l)
\end{equation}

The Fourier transform of the cross-correlation is easily calculated using the theorems discussed earlier,
\begin{equation}
	\begin{split}
R_{f,g}(u) & \equiv  \mathcal{F}\left[r_{f,g}(l)\right](u)\\
 & =  \int_{-\infty}^{+\infty}\int_{-\infty}^{+\infty}f^{*}(t)g(t+l)\exp(-2i\pi ul)dtdl\\
 & =  \int_{-\infty}^{+\infty}f^{*}(t)dt\int_{-\infty}^{+\infty}g(t+l)\exp(-2i\pi ul)dl\\
 & =  G(u)\int_{-\infty}^{+\infty}f^{*}(t)\exp(2i\pi ut)dt\\
 & =  G(u)\left[\int_{-\infty}^{+\infty}f(t)\exp(-2i\pi ut)dt\right]^{*}\\
 & =  F^{*}(u)G(u),
\end{split}
\end{equation}

where the property $z_{1}^{*}z_{2}=(z_{1}z_{2}^{*})^{*}$ has been used to transition from the fourth to the fifth line. Note that,
\begin{equation}
	F^{*}(u)G(u)=\left[F(u)G^{*}(u)\right]^{*}
\end{equation}

which implies,
\begin{equation}
	r_{f,g}(l)=r_{g,f}(-l)
\end{equation}

The autocorrelation is such that its Fourier transform is,
\begin{equation}
	R_{f,f}(u)=\left|F(u)\right|^{2}
\end{equation}

\couleur{The energy spectrum of a function is equal to the Fourier transform of the autocorrelation of the function}. This relationship between autocorrelation and the energy spectrum is known as the \couleur{Wiener-Khinchin theorem}\index{Wiener-Khinchin theorem} when $f(t)$ is a stochastic process\footnote{For more details, see the chapter on stochastic processes.}. In analytical calculations, such processes are generally defined by their autocorrelation function, and the Wiener-Khinchin theorem allows the deduction of the energy spectrum, though it does not provide information about the phase.

\begin{figure}[H]
\centering
\tcbox[colback=white]{\includegraphics[width=16cm]{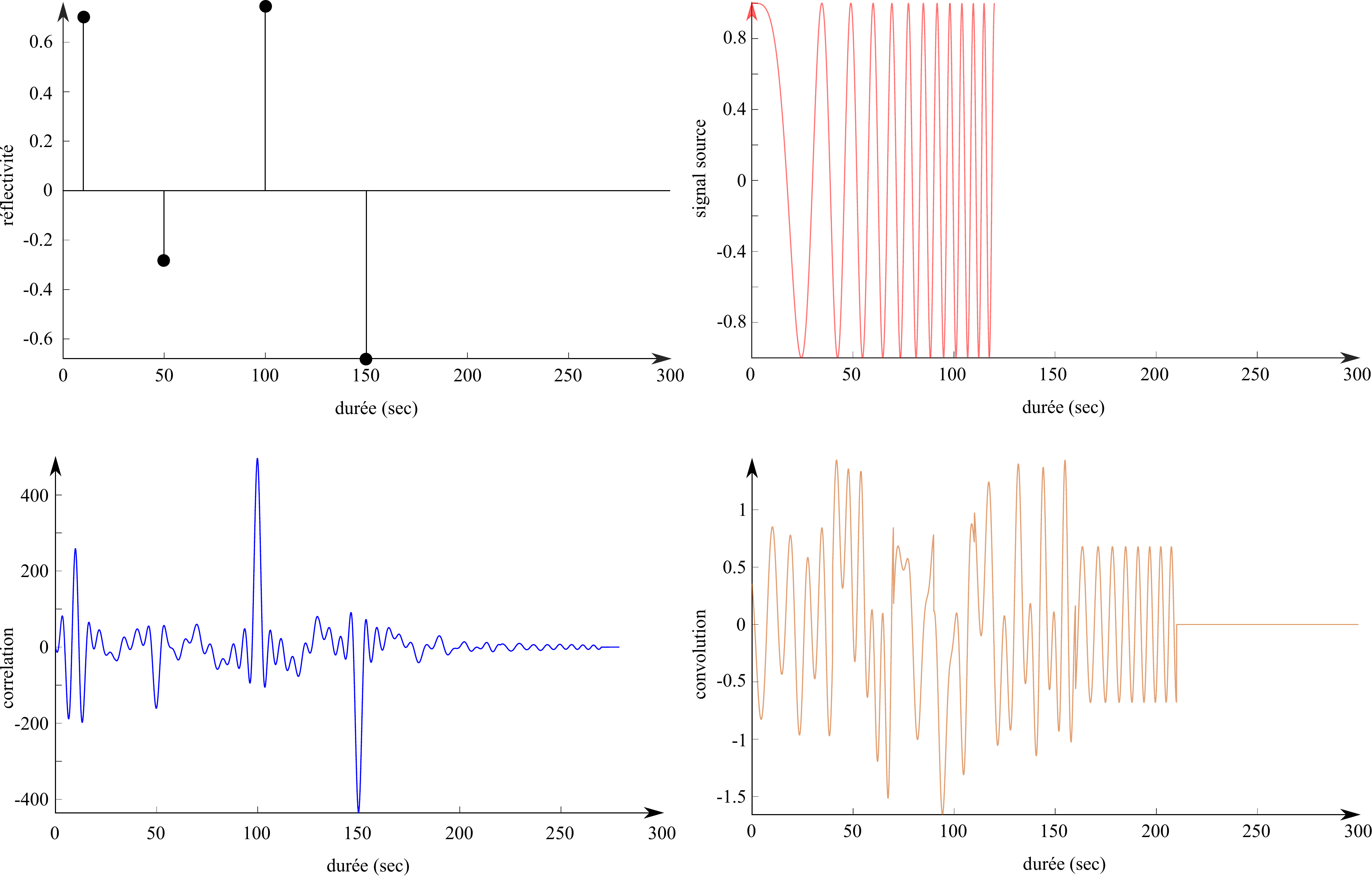}}
\caption{Example of cross-correlation. The convolution of an ideal seismic impulse response (top left) with a sweep (top right) emitted by a vibratory truck produces a trace (bottom right) in which arrivals are indistinguishable. The cross-correlation (bottom left) between this trace and the sweep helps to better discern the arrivals. This operation is routinely performed in seismic surveys when vibratory trucks are used as sources. Its effectiveness is due to the very specific shape (frequency sweep) of the sweep.}
\label{correlation_01}
\end{figure}

Cross-correlation (Figure \ref{correlation_01}) represents the power or energy if the two functions  $f$ and $g$ are physically associated, such as: intensity and voltage (power), magnetic and electric fields (Poynting vector), or force and velocity.

\chapter{\titrechap{The Hilbert Transform}}
\minitoc
\section{Definition}
The linear system whose transfer function -- that is, the Fourier transform of the impulse response -- is given by
\begin{equation}
	G(u)=-i\ \sgn(u)
\end{equation}

has the sole effect of advancing the phases by $\pi/2$ and is called a quadrature filter. The impulse response (\ref{rep_imp_hilbert}) allows us to obtain the system's response to an input $e(t)$.
\begin{equation}
	g(t)=\frac{1}{\pi t}
	\label{rep_imp_hilbert}
\end{equation}

This response is generally expressed in the following form,
\begin{equation}
	s(t)=\frac{1}{\pi}\int_{-\infty}^{+\infty}\frac{e(\tau)}{t-\tau}d\tau.
	\label{transformee_hilbert}
\end{equation}

By definition, $s(t)$ is called the \couleur{Hilbert}\index{Hilbert transform} transform of $e(t)$, in honor of David Hilbert (1862-1943), born in Konigsberg where he lived. He studied and began his career there until 1895, when he moved to Göttingen. His research covered a vast range of topics, including number theory, the theory of proof, algebraic geometry, variational calculus, and integral equations. His work on the development of arbitrary functions into series of orthogonal functions is particularly relevant for this course. We will use the following notation,
\begin{equation}
	F_{Hi}(t)\equiv\mathcal{H}[f](t)\equiv\frac{1}{\pi t}*f(t).
\end{equation}

\begin{figure}[H]
\begin{center}
\tcbox[colback=white]{\includegraphics[width=16cm]{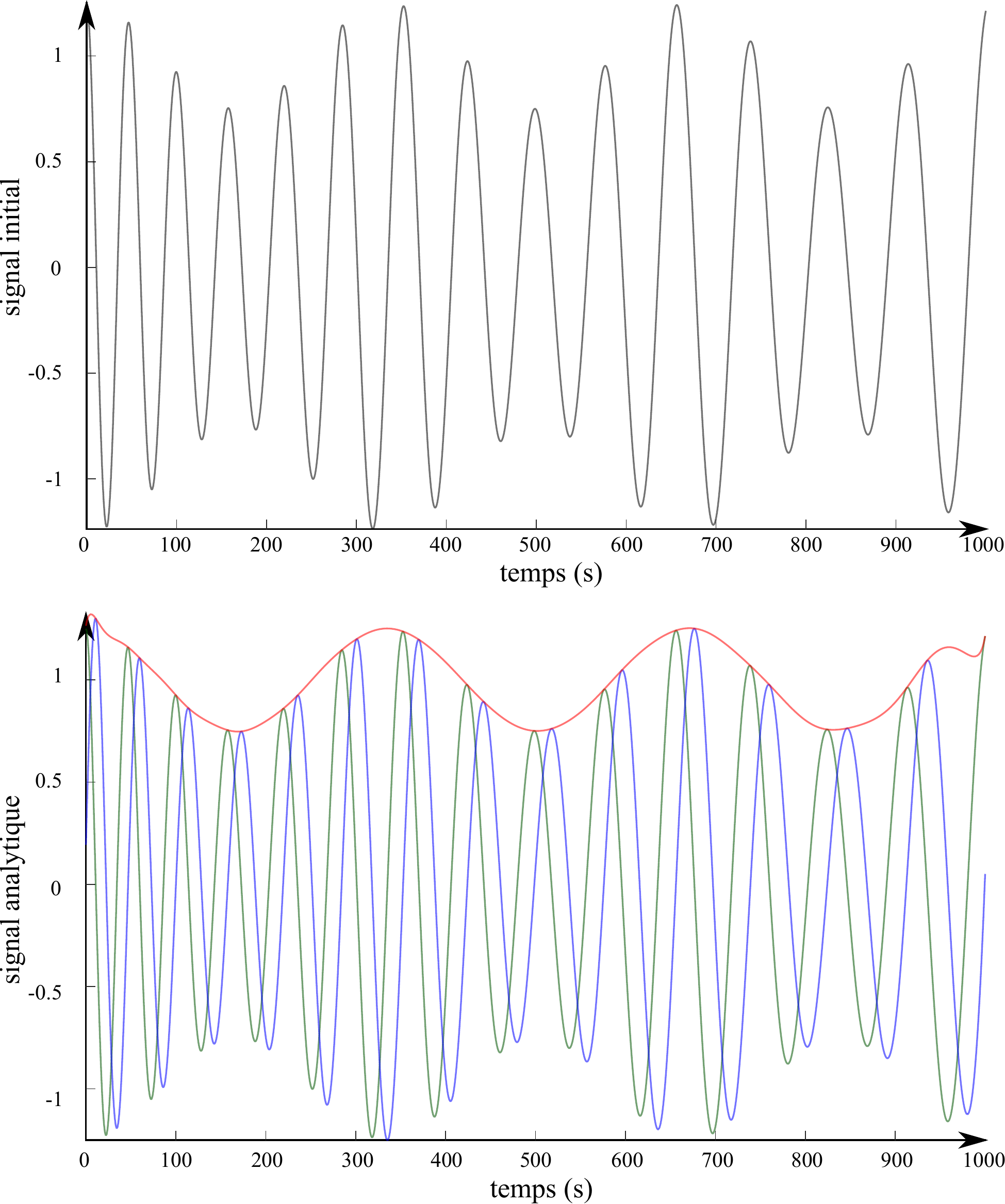}}
\end{center}
\caption{\textbf{Envelope Calculation by Hilbert Transform}. The initial signal (top, in gray) has an analytic function which is a complex function whose real part (bottom, in green) is equal to the initial signal itself, and whose imaginary part (bottom, in blue) is the Hilbert transform of the initial signal. The envelope (bottom, in red) is obtained by calculating the magnitude of the analytic signal. }
\label{hilbert_01}
\end{figure}

The \couleur{Hilbert} transform is used when studying causal signals. Non-stationary signals are often analyzed via their analytic signal, which is computed using the \couleur{Hilbert} transform. This transform therefore allows us to compute the analytic signal,
\begin{equation}
	f_{a}(t)\equiv f(t)+i\mathcal{H}[f](t)
\end{equation}

associated with $f(t)$. The magnitude of the analytic signal provides the envelope of $f(t)$ (figure \ref{hilbert_01}). An analytic signal is the complex equivalent of a real signal where all positive and zero frequencies are doubled, and negative frequencies are canceled. The program \couleur{ex\_hilbert\_transform.m} performs this computation and produces the images in figure (\ref{hilbert_01}). It uses the subfunction \couleur{hilbert\_transform.m}, which allows the user to choose either the \couleur{hilbert} function -- native to Matlab\textregistered -- or to more explicitly develop the Hilbert transform algorithm.

\section{Formulae: Hilbert Transforms}
\begin{equation}
	\cos(2\pi u_{0}t)\longmapsto+\sin(2\pi u_{0}t)
\end{equation}
\begin{equation}
	\sin(2\pi u_{0}t)\longmapsto-\cos(2\pi u_{0}t)
\end{equation}
\begin{equation}
	\frac{\sin(t)}{t}\longmapsto\frac{1-\cos(t)}{t}
\end{equation}
\begin{equation}
	\frac{1}{(1+t^{2})}\longmapsto\frac{t}{(1+t^{2})}
\end{equation}
\begin{equation}
	\exp(-\alpha|t|)\cos(2\pi u_{0}t)\longmapsto\exp(-\alpha|t|)\sin(2\pi u_{0}t)
\end{equation}
\chapter{\titrechap{Useful Functions in Fourier Analysis}}
\minitoc
\section{Catalogue of Useful Functions}
The previous chapters introduced us to the \couleur{Fourier} transformation through mathematical physics, that is, from an idealistic perspective where we did not question the feasibility of performing the calculations we developed using real signals. To delve deeper, it is now necessary to establish a link between this idealistic viewpoint and practical application. This link consists of the more or less rigorous answers to the inevitable questions that arise when dealing with real signals. While one can indeed pose numerous questions, the following are ubiquitous:

\begin{itemize}[label=$\bullet$]
    \item \warning{Infinity, present in $F\left(u\right)=\int_{-\infty}^{+\infty}f\left(t\right)\exp\left(-2i\pi ut\right)dt$, does not exist in the computer since the signal I have is of finite duration. How will this integral be evaluated? What errors will I incur?}
    \item \warning{Truncating the above integral is not sufficient because, even for a limited duration, I do not know the signal at all times but only at certain instances. What do I lose by not knowing the signal densely? How are my calculations affected?}
    \item \warning{What I measure is not the signal of interest but the signal plus noise, which consists of measurement errors and other unwanted signals. What is the impact of this noise on my calculations?}
\end{itemize}

The quality of the answers to these questions directly controls the analytical power of the different methods that will be used. Before discussing these answers in detail in the following chapters, it is necessary to have a set of "tools" that will allow us to "mathematize" the questions we pose. These tools will be functions or distributions that act as "scissors," "switches," "cameras," etc. With these tools, we will be able to mathematically articulate the transition from the ideal of mathematical physics to the reality of numerical processing. Some of the functions we will consider in this chapter are not functions in the strict sense and can only be rigorously manipulated in the sense of distributions, which the Anglo-Saxons call \couleur{"generalized functions"}. We will emphasize their physical significance and how these mathematical entities appear as physical limits\footnote{See, for example, the discussion concerning the \couleur{Dirac} impulse.}.

\section{Window (the "scissors")}
\begin{figure}[!h]
	\begin{center}
		\tcbox[colback=white]{\includegraphics[width=16cm]{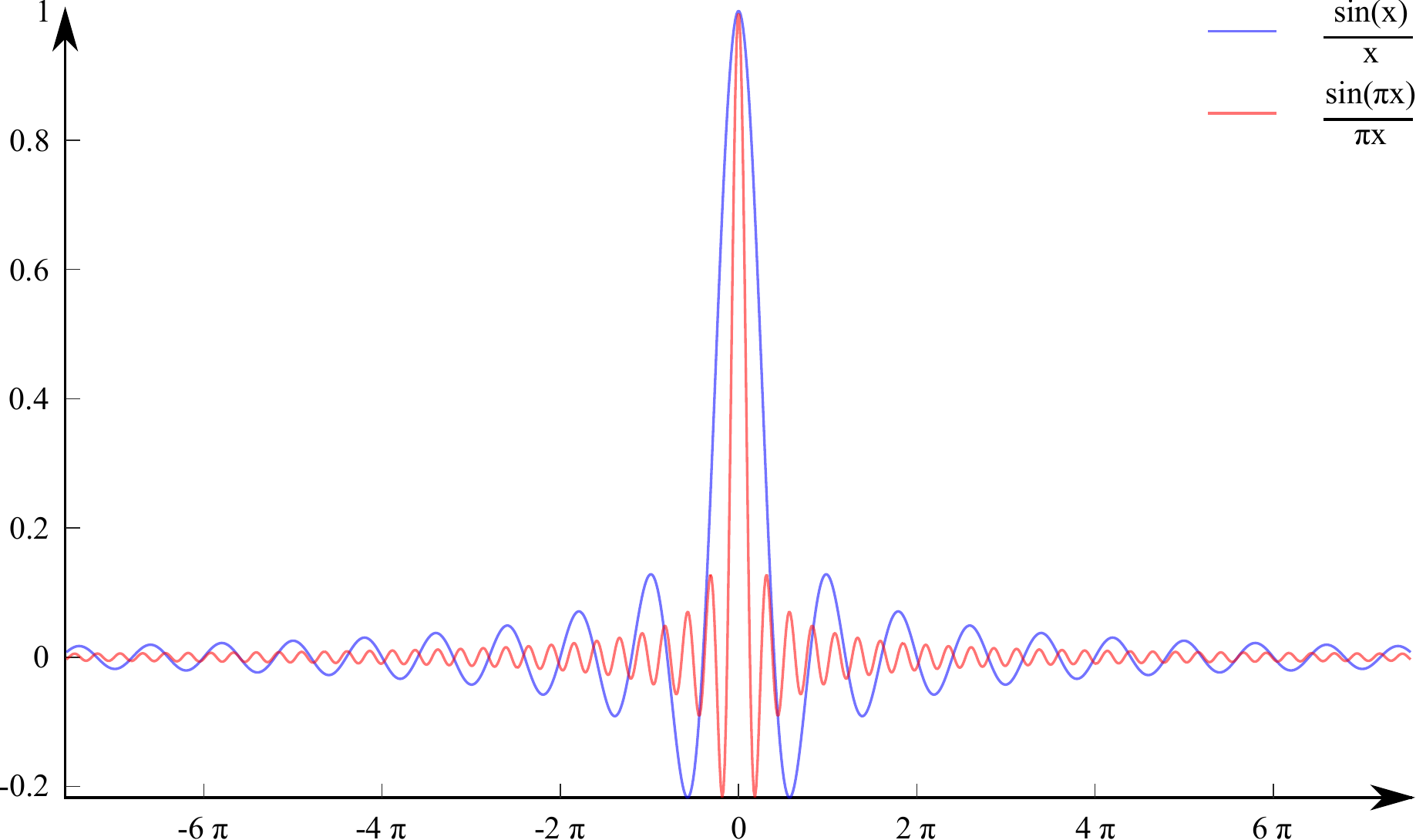}}
	\end{center}
\caption{Sine cardinal functions. In blue, the most commonly known unnormalized form, and in red, the normalized form.}
	\label{sin_cardinal}
\end{figure}

The window $\Pi(t)$, also known as the "rectangular" or "boxcar" function, is defined by,
\begin{equation}
	\Pi(t)= 
	\begin{cases}
		0 & |t|>1/2\\
		1 & |t|\leq1/2
	\end{cases}
\end{equation}

is one of the fundamental functions that we will continually use to symbolize the truncation of signals. We can thus view it as the pair of "scissors" in the toolbox that we are filling. Its \couleur{Fourier} transform,
\begin{equation}
	\mathcal{F}[\Pi(t)](u)=\frac{\sin(\pi u)}{\pi u}\equiv\textrm{sinc}(u)
\end{equation}

is easy to obtain by direct integration. The $\textrm{sinc}$ function, called "sine cardinal," is illustrated in figure (\ref{sin_cardinal}), and will be discussed in more detail in chapter (\ref{sinus_cardinal}). Note the presence of $\pi$ in its expression, in accordance with the definition by \shortciteN{bracewell1986fourier}, which many authors overlook. The \couleur{Fourier} transform of a window with width $T$, amplitude $\alpha$, and centered at $t_{0}$ is obtained by using the theorems seen in the previous chapter,
\begin{equation}
	\begin{split}
	\mathcal{F}[\alpha\Pi(\frac{t-t_{0}}{T})](u) & =  \alpha\mathcal{F}\left[\Pi(\frac{t-t_{0}}{T})\right](u)\\
 & =  \alpha\exp(-2i\pi ut_{0})\mathcal{F}[\Pi(\frac{t}{T})](u)\\
 & =  \alpha T\exp(-2i\pi ut_{0})\textrm{sinc}(uT)
	\end{split}
\end{equation}

Without encroaching too much on the following chapters, it is good to justify the use of the window function now. One might think that these scissors are unnecessary to express the fact that a signal is known only for a limited duration, and it is simpler to write that the \couleur{Fourier} transform of such a signal is,
\begin{equation}
\mathcal{F}[truncated\; signal](u)=\int_{beginning}^{end}(signal)\times\exp(-2i\pi ut)dt
\end{equation}

where we simply take the endpoints of the signal's observation interval as the limits of the integral. This calculation is correct and provides the same result as one would obtain using the window, but it has a major drawback: it implies a redefinition of the \couleur{Fourier} transform operator. Such redefinition is rigorously discouraged, which is why it is better to write,
\begin{equation}
	\mathcal{F}[truncated\; signal](u)=\mathcal{F}[(signal)\times\Pi(of\; the\; appropriate\; duration)](u)
\end{equation}

where we can use the symbolic notation $\mathcal{F}$ since we retain the initial definition of the \couleur{Fourier} transform.

\section{Cardinal Sine}\label{sinus_cardinal}
We have already encountered this function, which is the \couleur{Fourier} transform of the window function. It plays a role in the interpolation and filtering of signals. The cardinal sine, defined by,
\begin{equation}
	\textrm{sinc}(u)\equiv\frac{\sin(\pi u)}{\pi u}
\end{equation}

is such that,
\begin{equation}
	\begin{cases}
		\textrm{sinc}(0)  =  1\\
		\ \\
		\textrm{sinc}(n) =  0 \qquad (n\in\Bbb{Z})\\
		\ \\
		\int_{-\infty}^{+\infty}\textrm{sinc}(t)dt  =  1
	\end{cases}
\end{equation}

Using the duality properties of the \couleur{Fourier} transform, it is directly shown that,
\begin{equation}
	\mathcal{F}[\textrm{sinc}(t)](u)=\Pi(u)
\end{equation}

The importance of the cardinal sine comes from the fact that its \couleur{Fourier} transform is zero outside the interval $\left[-1/2;1/2\right]$. We will see, in the chapter on filtering, that convolution by a cardinal sine is a low-pass filtering. We will also see, in the chapter on sampling, that the cardinal sine allows, under certain conditions, the interpolation of signals for which only discrete values are known. Finally, note that for signals $f(t)$ such that $F(u)=0$ outside the interval $[-1/2;1/2]$ we have,
\begin{equation}
	F(u)\Pi(u)=F(u)
\end{equation}

which, after inverse \couleur{Fourier} transform, gives,
\begin{equation}
	[f*\textrm{sinc}](t)=f(t)
\end{equation}

For such signals with bounded spectra, the cardinal sine is the identity element of convolution.

\section{Triangle}
\begin{figure}[!h]
	\begin{center}
		\tcbox[colback=white]{\includegraphics[width=16cm]{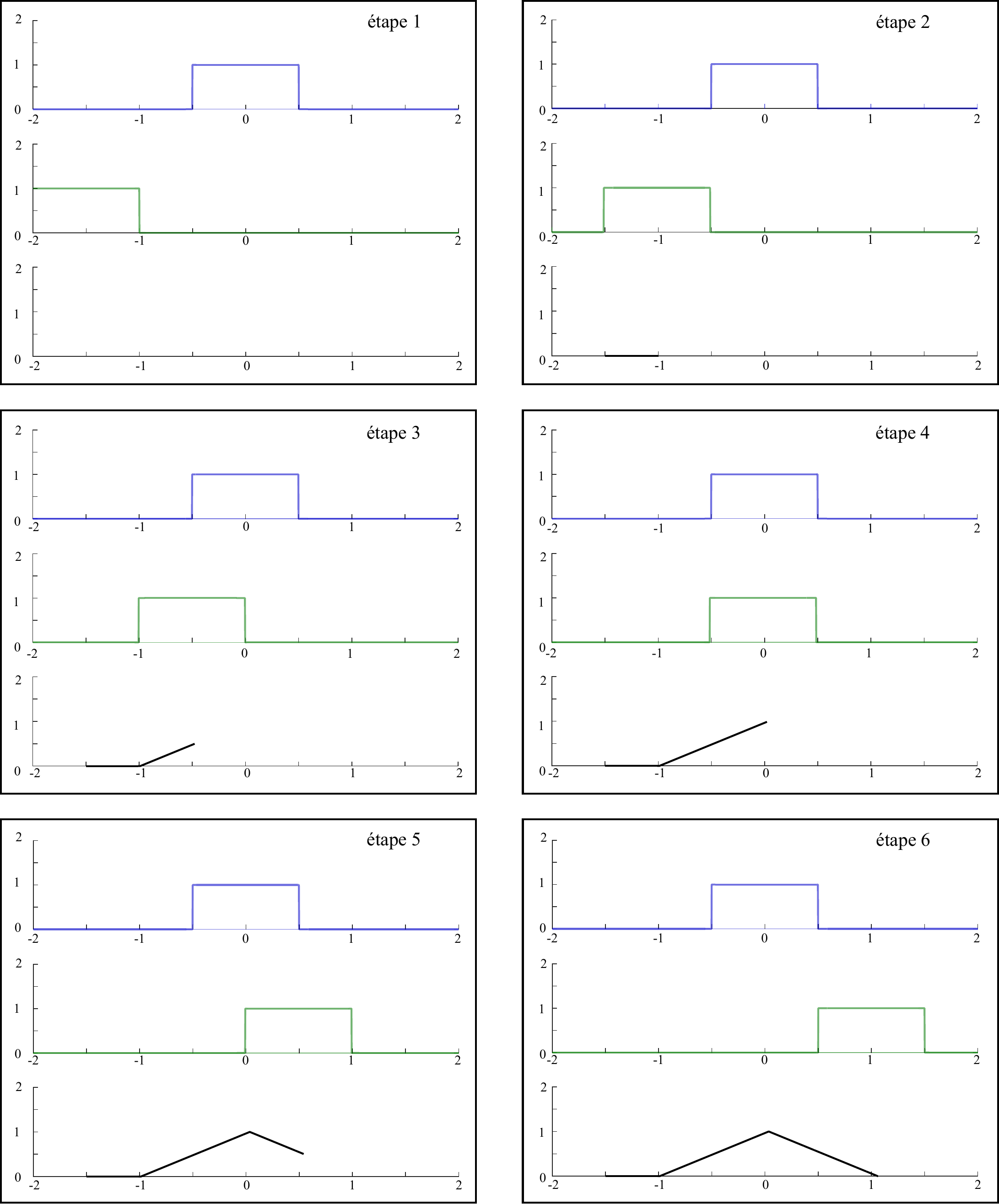}}
	\end{center}
\caption{From top to bottom, and from left to right, different stages of the self-convolution product of a rectangular function (blue curve) by itself (green curve). As can be seen, the result is the triangle function (black curve).}
	\label{convo_porte}
\end{figure}

The triangle function defined by the following relation,
\begin{equation}
	\Lambda(t)=\left\{ 
	\begin{array}{lll}
		0 & si & |t|>1\\
		1-|t| & si & |t|\leq1
	\end{array}\right.
\end{equation}

frequently appears in calculations as it is the self-convolution of the window function (Figure \ref{convo_porte}). The program \couleur{ex\_autoconv\_fenetre.m} demonstrates the result of the self-convolution product of a rectangular function in the form of an animation. This observation immediately shows that,
\begin{equation}
	\begin{split}
		\mathcal{F}\left[\Lambda\left(t\right)\right]\left(u\right) & =  \mathcal{F}\left\{ \left[\Pi*\Pi\right]\left(t\right)\right\} \left(u\right)\\
 & = \textrm{sinc}^{2}\left(u\right)
	\end{split}
\end{equation}

\section{Exponential Functions}
\subsection{Exponential Decaying to Infinity}
First, let's consider the case of the function $\exp\left(-|t|\right)$, which often appears in the analysis of signals generated by damping or diffusion processes. Its \couleur{Fourier} transform is computed by direct integration. Let us first note that,
\begin{equation}
	\begin{split}
	\mathcal{F}\left[\exp\left(-|t|\right)\right]\left(u\right) & =  	\int_{-\infty}^{+\infty}\exp\left(-\left|t\right|\right)\exp\left(-2i\pi ut\right)\textrm{d}t\\
 & =  \int_{-\infty}^{+\infty}\exp\left(-|t|\right)\cos\left(-2i\pi ut\right)\textrm{d}t\\
 & =  2\int_{0}^{+\infty}\exp\left(-|t|\right)\cos\left(-2i\pi ut\right)\textrm{d}t\\
 & =  2\Re\left[\int_{0}^{+\infty}\exp\left(-|t|\right)\exp\left(-2i\pi ut\right)\textrm{d}t\right]
	\end{split}
\end{equation}

Therefore, we have,
\begin{equation}
\begin{split}
\mathcal{F}\left[\exp\left(-|t|\right)\right]\left(u\right) & =  2\Re\left[\int_{0}^{+\infty}\exp\left[\left(-2i\pi u-1\right)t\right]\textrm{d}t\right]\\
 & =  2\Re\left(\frac{1}{2i\pi u+1}\right)\\
 & =  \frac{2}{\left(2\pi u\right)^{2}+1}.
 \end{split}
\end{equation}

We will revisit this Fourier transform when we study \couleur{Butterworth} filters.

\subsection{Gaussian}
The case of the Gaussian function $\exp\left(-\pi t^{2}\right)$ is interesting for different reasons. This function is important in statistics where it represents the \couleur{Gauss} distribution underlying the least squares methods. We have already seen the role of convolution in probability theory. Moreover, the Gaussian is often used in distribution theory and plays a part in defining \couleur{Heisenberg}'s uncertainty principle (see chapter \ref{similitude}), whose implications we will explore in signal processing. The calculation of the \couleur{Fourier} transform is clever,
\begin{equation}
	\begin{split}
\mathcal{F}\left[\exp\left(-\pi t^{2}\right)\right]\left(u\right) & =  \int_{-\infty}^{+\infty}\exp\left(-\pi t^{2}\right)\exp\left(-2i\pi ut\right)\textrm{d}t\\
 & =  \exp\left(-\pi u^{2}\right)\int_{-\infty}^{+\infty}\exp\left[-\pi\left(t+iu\right)^{2}\right]\textrm{d}t\\
 & =  \exp\left(-\pi u^{2}\right)\int_{-\infty}^{+\infty}\exp\left(-\pi\xi^{2}\right)\textrm{d}\xi\\
 & =  \exp\left(-\pi u^{2}\right),
	\end{split}
\end{equation}

where we have used the property $\int_{-\infty}^{+\infty}\exp\left(-\pi\xi^{2}\right)\textrm{d}\xi=1$. \warning{Note that the Gaussian function is its own Fourier transform}.

\section{\couleur{Dirac} Delta Function (the "photo")}
The Dirac delta function, $\delta(t)$, is named in honor of \couleur{Paul Dirac} (1902-1984) who was born in Bristol and studied electrical engineering at the University. In 1923, he moved to Cambridge as a student and published, two years later, his paper on the fundamental laws of quantum mechanics (\shortciteNP{dirac1925fundamental}). This work was based on recent results by \couleur{Heisenberg} and extended them significantly. Other papers followed, and in 1933, at the age of 31, \couleur{Dirac} received the Nobel Prize in Physics. In 1937, he married \couleur{Margit Wigner}, sister of the eminent physicist \couleur{Eugen Wigner}. Between 1930 and 1940, \couleur{Dirac} focused on developing quantum electrodynamics; his last papers concerned general relativity. It was in 1926 that \couleur{Dirac} introduced his famous "function" $\delta(t)$, which is zero everywhere except at the origin and has an integral equal to 1, to represent a unit impulse at $t=0$ with no effect for $t\neq0$. $\delta(t)$ is not a function in the usual sense, as a function that is zero outside the origin has an integral of zero. The Dirac delta function was empirically manipulated for a long time until it found a rigorous mathematical justification within the framework of distribution theory developed by \couleur{Laurent Schwartz} in 1950 (\shortciteNP{schwartz1950theorie}). 

This function is not a classical function and can only be formally defined in the sense of distributions. It is difficult to enumerate all the roles played by this distribution, which is encountered in numerous calculations. The attribute "photo" attached to the Dirac delta function is there to remind us that it allows us, thanks to the sampling formula, to mathematically express the fact \warning{"that we sample the value of a signal at a given instant"}. But the Dirac delta function is more than that, as we will see. Historically, the notion of an impulse was introduced by physicists before mathematicians invented distributions. It should be noted that the impulse is in line with other physical idealizations such as point mass, point charge, infinitely thin layers, etc., which are easily manageable in calculations but physically unrealizable. We will approach the Dirac delta function in this way: as an ideal that we can never exactly achieve but can approximate closely enough to be useful. In this context, "sufficiently close" is reached when we can no longer measure the duration of the impulse, or when the response time of the excited system is so much longer than the duration of the excitation that it doesn't matter. Thus, according to this definition, the same \warning{stimulus} may or may not be considered an impulse; it will be up to you to judge based on the overall characteristics of the excited system. When dealing with an impulse, it is not important to specify its duration, and we will write that the excitation occurred at the origin of time,
\begin{equation}
	\delta\left(t\right)=0 \ \forall \ t\neq0
\end{equation}

However, the integral of the impulse represents what the excited system will dissipate and must be defined,
\begin{equation}
	\int_{-\infty}^{+\infty}\delta\left(t\right)dt=1
\end{equation}

\begin{figure}[H]
	\begin{center}
		\tcbox[colback=white]{\includegraphics[width=16cm]{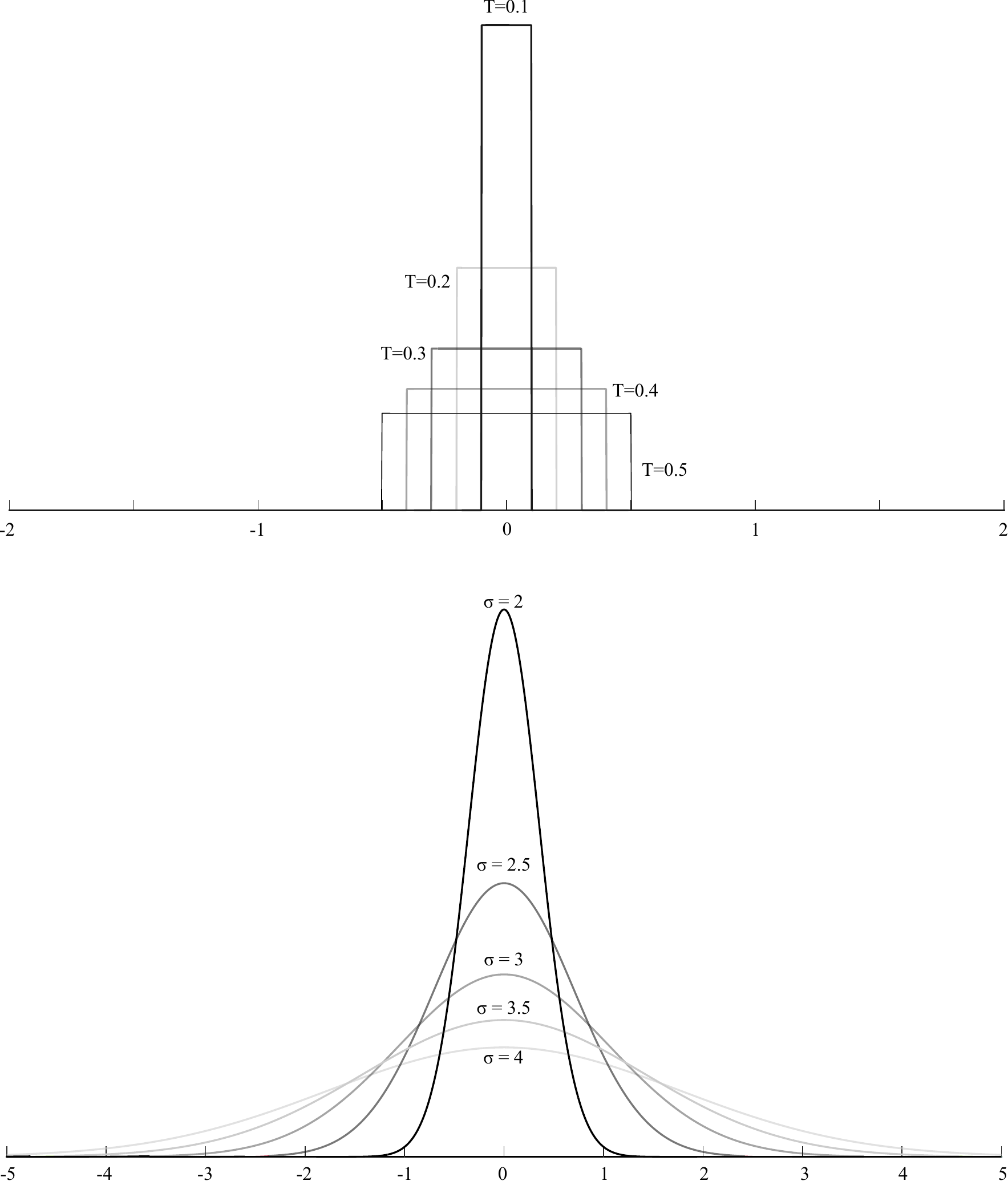}}
	\end{center}
\caption{Top: Evolution of the window function as $T$ approaches 0. Bottom: Evolution of the Gaussian function as the standard deviation $\sigma$ approaches 0. As can be seen in both cases, the temporal support decreases, the functions thin out and tend to infinity, thus approaching Dirac distributions.}
	\label{porte_gaussienne}
\end{figure}

The fundamental properties of the Dirac delta function can be established by representing $\delta(t)$ as the limit of classical functions localized around the origin (see figure \ref{porte_gaussienne} obtained with the program \couleur{porte\_gaussienne\_vers\_dirac.m}). One can, for example, use the window function,
\begin{equation}
	\delta\left(t\right)=\lim_{T\downarrow0}\frac{1}{T}\Pi\left(\frac{t}{T}\right)
\end{equation}

where the Gaussian function,
\begin{equation}
	\delta\left(t\right)=\lim_{\sigma\downarrow0}\frac{1}{\sqrt{\pi}\sigma}\exp\left(-\frac{t^{2}}{\sigma^{2}}\right)
\end{equation}

Let's demonstrate the sampling formula using this type of representation of the Dirac delta function,
\begin{equation}
	\begin{split}
	\int_{-\infty}^{+\infty}\delta\left(t\right)f\left(t\right)\textrm{d}t & = 	\lim_{T\downarrow0}\int_{-\infty}^{+\infty}\frac{1}{T}\Pi\left(\frac{t}{T}\right)f\left(t\right)\textrm{d}t\\
 & =  \lim_{T\downarrow0}\frac{1}{T}\int_{-T/2}^{+T/2}f\left(t\right)\textrm{d}t\\
 & =  \lim_{T\downarrow0}\frac{1}{T}\int_{-T/2}^{+T/2}\left[f\left(0\right)+tf^{\left(1\right)}\left(0\right)+\frac{t^{2}}{2}f^{\left(2\right)}\left(0\right)+\cdots\right]\textrm{d}t\\
 & =  \lim_{T\downarrow0}\left[f\left(0\right)+\frac{T^{2}}{24}f^{\left(2\right)}\left(0\right)+\cdots\right]\\
 & =  f\left(0\right).\nonumber 
	\end{split}
\end{equation}

By generalizing this result, it is easy to establish that the \couleur{Dirac} delta function is the identity element of convolution,
\begin{equation}
	\left[\delta*f\right]\left(t\right)=f\left(t\right)
\end{equation}

We deduce the translation formula,
\begin{equation}
	\delta\left(t-t_{0}\right)*f\left(t\right)=f\left(t-t_{0}\right)
\end{equation}

The \couleur{Plancherel} theorem immediately provides the Fourier transform of the Dirac delta function,
\begin{equation}
\mathcal{F}\left[\delta*f\right]\left(u\right)=F\left(u\right)\Longrightarrow\mathcal{F}\left[\delta\left(t\right)\right]\left(u\right)=1.
\end{equation}

Among the many properties of the Dirac delta function, we cite,
\begin{equation}
	\delta\left(-t\right)=\delta\left(t\right)
\end{equation}
\begin{equation}
	t\delta\left(t\right)=0
\end{equation}
\begin{equation}
	f\left(t\right)\delta\left(t-t_{0}\right)=f\left(t_{0}\right)\delta\left(t_{0}-t\right)
\end{equation}

and
\begin{equation}
	\delta\left(\alpha t\right)=\frac{1}{\left|\alpha\right|}\delta\left(t\right)
\end{equation}

The last property, which means that the Dirac delta function is a homogeneous distribution, is useful when dealing with Dirac combs that we will see later. It ensures the consistency of the Fourier transform of the Dirac delta function,
\begin{equation}
	\begin{split}
\delta\left(\alpha t\right) & =  \mathcal{F}^{-1}\left\{ \mathcal{F}\left\{ \delta\left(\alpha t\right)\right\} \right\} \left(t\right)\\
 & =  \mathcal{F}^{-1}\left[\frac{1}{\left|\alpha\right|}\right]\left(t\right)\\
 & =  \frac{1}{\left|\alpha\right|}\delta\left(t\right)
	\end{split}
\end{equation}

\section{Sign Function}
This function is defined by,
\begin{equation}
	\textrm{sgn}\left(t\right)=
		\begin{cases}
			+1 \qquad si \qquad t>0\\
			\ 0 \qquad \ si \qquad t=0\\
			-1 \qquad si \qquad t<0
		\end{cases}
\end{equation}

It has a \couleur{Fourier} transform that can only be calculated in the sense of distributions. To do this, let's introduce the function,
\begin{equation}
	\phi_{T}\left(t\right)=\exp\left(-\frac{\left|t\right|}{T}\right)\textrm{sgn}\left(t\right)
\end{equation}

of which the sign function is a limiting case,
\begin{equation}
	\lim_{T\rightarrow\infty}\phi_{T}\left(t\right)=\textrm{sgn}\left(t\right).
\end{equation}

We have,
\begin{equation}
	\begin{split}
\mathcal{F}\left[\phi_{T}\left(t\right)\right]\left(u\right) & =  \int_{-\infty}^{+\infty}\exp\left(-\frac{\left|t\right|}{T}\right)\textrm{sgn}\left(t\right)\exp\left(-2i\pi ut\right)\textrm{d}t\\
 & =  -\int_{-\infty}^{0}\exp\left(\frac{t}{T}\right)\exp\left(-2i\pi ut\right)\textrm{d}t+\int_{0}^{+\infty}\exp\left(-\frac{t}{T}\right)\exp\left(-2i\pi ut\right)\textrm{d}t\\
 & = \frac{-4i\pi u}{1/T^{2}-\left(2i\pi u\right)^{2}}
	\end{split}
\end{equation}

When $T \rightarrow +\infty$, we obtain a limiting \couleur{Fourier} transform which is that of the sign function,
\begin{equation}
	\mathcal{F}\left[\textrm{sgn}\left(t\right)\right]\left(u\right)=\frac{1}{i\pi u}
\end{equation}

\section{Heaviside Distribution (the switch)}
The Heaviside step function is named after \couleur{Oliver Heaviside} (1850-1925) who was born in London. In his youth, he was interested in experiments on electricity, and he published his first scientific paper at the age of twenty-two. His work concerned the transmission of electrical signals in transatlantic cables. A controversy arose between him and the specialists who did not believe in his technique of reducing attenuation by using inductances judiciously placed along the cable. It was during this time that \couleur{Heaviside} established the telegrapher's equation,$\frac{1}{C}\frac{\partial^{2}V}{\partial x^{2}}=L\frac{\partial^{2}V}{\partial t^{2}}+R\frac{\partial V}{\partial t}$ where $C$, $L$, and $R$ are the capacitance, inductance, and resistance of the line, respectively. Mainly concerned with the problems of transmitting electromagnetic signals over long distances, he predicted, simultaneously with \couleur{A.E. Kennelly} from Harvard University, the existence of the ionosphere. His studies on transient signals led him to develop a clever mathematical formalism that was a precursor to our current symbolic calculus, using \couleur{Fourier} and \couleur{Laplace} transforms. It was in this context that he invented the famous distribution now bearing his name,
\begin{equation}
	\textrm{H}\left(t\right)=
		\begin{cases}
			+1 \qquad \ si \qquad t>0\\
			1/2 \qquad si \qquad t=0\\
			\ 0 \qquad  \ \ si \qquad t<0
		\end{cases}
\end{equation}

This distribution is essential for representing discontinuities such as those caused by the opening or closing of a circuit and for selecting the causal part of a signal. Additionally, convolution with $\textrm{H}\left(t\right)$ allows for integrating a signal,
\begin{equation}
	\begin{split}
		\left[\textrm{H}*f\right]\left(t\right) & = 		\int_{-\infty}^{+\infty}f\left(\xi\right)\textrm{H}\left(t-\xi\right)\textrm{d}\xi\\
 & =  \int_{-\infty}^{t}f\left(\xi\right)\textrm{d}\xi,
	\end{split}
\end{equation}

from which we deduce that,
\begin{equation}
	\frac{\textrm{d}}{\textrm{d}t}\left(\textrm{H}*f\right)=\frac{\textrm{d}}{\textrm{d}t}\textrm{H}*f=f,
\end{equation}

which shows that,
\begin{equation}
	\frac{\textrm{d}}{\textrm{d}t}\textrm{H}\left(t\right)=\delta\left(t\right)
\end{equation}

A rigorous demonstration of this result requires the use of distributions,
\begin{equation}
	\begin{split}
	\frac{\textrm{d}}{\textrm{d}t}\textrm{H}\left(t\right) & =  \lim_{h\downarrow0}\frac{\textrm{H}\left(t+h\right)-\textrm{H}\left(t\right)}{h}\\
 & =  \lim_{h\downarrow0}\frac{1}{h}\Pi\left(\frac{t+h/2}{h}\right)\\
 & =  \delta\left(t\right)
	\end{split}
\end{equation}

Letting,
\begin{equation}
	\textrm{H}\left(t\right)=\frac{1}{2}\left[1+\textrm{sgn}\left(t\right)\right]
\end{equation}

the results from the previous sections immediately provide,
\begin{equation}
\mathcal{F}\left[\textrm{H}\left(t\right)\right]\left(u\right)=\frac{1}{2}\delta\left(u\right)+\frac{1}{2i\pi u}
\end{equation}

The second term on the right-hand side is equal to the inverse of the differentiation operator we encountered in the first chapter; it is the integration operator such that,
\begin{equation}
	\begin{split}
\mathcal{F}\left[\int_{-\infty}^{+\infty}f\left(\xi\right)\textrm{d}\xi\right]\left(u\right) & =  \mathcal{F}\left[\textrm{H}*f\right]\left(u\right)\\
 & =  \frac{1}{2i\pi u}F\left(u\right)
	\end{split}
\end{equation}

"Neglecting" the \couleur{Dirac} impulse in this expression is equivalent to ignoring a potential constant of integration."

\section{\couleur{Dirac} Comb (the camera)}
This distribution is extremely important for describing signal sampling. It is defined as a sequence of \couleur{Dirac} impulses occurring at a cadence of \(\tau = 1\),
\begin{equation}
	\textrm{shah}\left(t\right)\equiv\sum_{n=-\infty}^{+\infty}\delta\left(t-n\right)
\end{equation}

The main properties of this distribution are:
\begin{equation}
	\textrm{shah}\left(t+n\right)=\textrm{shah}\left(t\right)\;\left(n\in\Bbb{Z}\right)
\end{equation}

which indicates that $\textrm{shah}$ is periodic with period 1,
\begin{equation}
	\int_{n-1/2}^{n+1/2}\textrm{shah}\left(t\right)dt=1\;\left(n\in\Bbb{Z}\right)
\end{equation}

and
\begin{equation}
	\textrm{shah}\left(t\right)=0\;\left(t\notin\Bbb{Z}\right)
\end{equation}

which are directly established from the fundamental properties of the \couleur{Dirac} impulse. Furthermore,
\begin{equation}
	\textrm{shah}\left(\frac{t}{\tau}\right)=\left|\tau\right|\sum_{n=-\infty}^{+\infty}\delta\left(t-n\tau\right)
\end{equation}

This last relation is demonstrated using the homogeneity property of the \couleur{Dirac} impulse,
\begin{equation}
	\begin{split}
\textrm{shah}\left(\frac{t}{\tau}\right) & =  \sum_{n=-\infty}^{+\infty}\delta\left(\frac{t}{\tau}-n\right)\\
 & =  \sum_{n=-\infty}^{+\infty}\delta\left(\frac{t-n\tau}{\tau}\right)\\
 & =  \left|\tau\right|\sum_{n=-\infty}^{+\infty}\delta\left(t-n\tau\right)
	\end{split}
\end{equation}

The \couleur{Fourier} transform of the comb function can be calculated using a trick involving writing \(\textrm{shah}(t)\), which is 1-periodic, as a \couleur{Fourier} series,
\begin{equation}
	\textrm{shah}\left(t\right)=\sum_{n=-\infty}^{+\infty}\alpha_{n}\exp\left(2i\pi nt\right)
\end{equation}

where the coefficients are,
\begin{equation}
	\begin{split}
\alpha_{n} & =  \int_{-1/2}^{1/2}\textrm{shah}\left(t\right)\exp\left(-2i\pi nt\right)\textrm{d}t\\
 & =  \int_{-1/2}^{1/2}\delta\left(t\right)\exp\left(-2i\pi nt\right)\textrm{d}t\\
 & =  \left.\exp\left(-2i\pi nt\right)\right|_{t=0}\\
 & =  1 
	\end{split}
\end{equation}

Thus,
\begin{equation}
	\textrm{shah}\left(t\right)=\sum_{n=-\infty}^{+\infty}\exp\left(+2i\pi nt\right)
\end{equation}

The \couleur{Fourier} transform of the comb is then,
\begin{equation}
	\begin{split}
\mathcal{F}\left[\textrm{shah}\left(t\right)\right]\left(u\right) & =  \sum_{n=-\infty}^{+\infty}\mathcal{F}\left[\delta\left(t-n\right)\right]\left(u\right)\\
 & =  \sum_{n=-\infty}^{+\infty}\exp\left(-2i\pi nu\right)\\
 & =  \sum_{n=-\infty}^{+\infty}\exp\left(+2i\pi nu\right)\\
 & = \textrm{shah}\left(u\right)
	\end{split}
\end{equation}

\danger{The \couleur{Dirac} comb is its own \couleur{Fourier} transform.}

\section{Sine and Cosine Functions}
The calculation of the \couleur{Fourier} transforms of these functions involves distributions, and we have:
\begin{equation}
	\begin{split}
\mathcal{F}\left[\cos\left(2\pi u_{0}t\right)\right]\left(u\right) & =  \mathcal{F}\left[\frac{\exp\left(-2i\pi u_{0}t\right)+\exp\left(+2i\pi u_{0}t\right)}{2}\right]\left(u\right)\\
 & =  \frac{1}{2}\int_{-\infty}^{+\infty}\exp\left[-2i\pi\left(u+u_{0}\right)t\right]\textrm{d}t+\frac{1}{2}\int_{-\infty}^{+\infty}\exp\left[-2i\pi\left(u-u_{0}\right)t\right]\textrm{d}t\\
 & =  \frac{1}{2}\delta\left(u+u_{0}\right)+\frac{1}{2}\delta\left(u-u_{0}\right).
	\end{split}
\end{equation}

An analogous reasoning yields,
\begin{equation}
	\mathcal{F}\left[\sin\left(2\pi u_{0}t\right)\right]\left(u\right)=\frac{i}{2}\delta\left(u+u_{0}\right)-\frac{i}{2}\delta\left(u-u_{0}\right)
\end{equation}

\section{Form: Fourier Transforms}
\begin{equation}
	t\exp\left(-t\right)\textrm{H}\left(t\right)\longmapsto\frac{1}{\left(1+2i\pi u\right)^{2}}
\end{equation}
\begin{equation}
	\frac{1}{2}\Pi\left(t+1\right)+\frac{1}{2}\Pi\left(t-1\right)\longmapsto\cos\left(2\pi 	u\right)\textrm{sinc}\left(u\right)
\end{equation}
\begin{equation}
	\Pi\left(t\right)*\textrm{sgn}\left(t\right)\longmapsto-i\frac{\textrm{sinc}\left(u\right)}{\pi u}
\end{equation}
\begin{equation}
	\Delta\left(t\right)*\textrm{sgn}\left(t\right)\longmapsto-i\frac{\textrm{sinc}^{2}\left(u\right)}{\pi u}
\end{equation}
\begin{equation}
	\frac{1}{\sqrt{\left|t\right|}}\longmapsto\frac{1}{\sqrt{\left|u\right|}}
\end{equation}
\begin{equation}
	\left|\cos\left(\pi t\right)\right|\longmapsto\frac{1}{2}\textrm{shah}\left(u\right)\left[\textrm{sinc}\left(u+\frac{1}{2}\right)+\textrm{sinc}\left(u-\frac{1}{2}\right)\right]
\end{equation}
\begin{equation}
	\exp\left(-\pi t^{2}\right)\cos\left(2\pi t\right)\longmapsto\frac{1}{2}\exp\left[-\pi\left(u+1\right)^{2}\right]+\frac{1}{2}\exp\left[-\pi\left(u-1\right)^{2}\right]
\end{equation}

\begin{equation}
	\textrm{J}_{1}\left(2\pi t\right)/2t\longmapsto\sqrt{1-u^{2}}\Pi\left(u/2\right)
\end{equation}

\begin{equation}
	\textrm{J}_{0}\left(2\pi t\right)\longmapsto\pi^{-1}\Pi\left(u/2\right)\left(1-u^{2}\right)^{-1/2}
\end{equation}
\begin{equation}
	\tanh\left(\pi t\right)\longmapsto-i\textrm{cosech}\left(\pi u\right)
\end{equation}
\chapter{\titrechap{Sampling}}
\minitoc
\section{Sampling\index{Sampling}}
We will now address a very important part of the course, and what we will see in this chapter constitutes one of the "launching pads" necessary for practical applications in signal processing. Analog signal processing, or continuous processing, is becoming increasingly rare, although it should be noted here that analog modification of signals still exists at the sensor level; however, digital signal processing is becoming more frequent due to the increasing power of computers and the flexibility allowed by digital processing, which permits operations that are unachievable by analog means, \eg, non-causal filtering. Sampling is an essential step in digital signal processing; for a signal to be "digested" by the computer, it must be presented as a finite sequence (\id, of limited duration) of values (\id, discrete) coded on a certain number of bits. The operations of truncation, discretization, and quantization will modify the theoretical expressions we have seen so far (\eg, the bounds of the \couleur{Fourier} integral will not be infinite) and the role of this chapter is to examine the main effects of sampling and their impact on the theoretical expressions seen so far.

\subsection{Signal truncation}
In many cases, the signal we wish to study is not known in its entirety, but only for a limited duration. The question that then arises is to what extent the sample we possess is representative of the total, unknown signal. We can represent the truncation of a signal using the window,

\begin{equation}
	s_{T}\left(t\right)=s\left(t\right)\Pi\left(\frac{t-t_{0}}{T}\right)
\end{equation}

where $s_{T}\left(t\right)$ is the truncated part of the total signal $s\left(t\right)$. By transitioning into the dual space of \couleur{Fourier},
\begin{equation}
	S_{T}\left(u\right)=T\times S\left(u\right)*\left[\exp\left(-2i\pi ut_{0}\right)\textrm{sinc}\left(uT\right)\right]
\end{equation}

which shows that the \couleur{Fourier} transform of the truncated signal is a degraded version of that of the total signal. The degradation results from the convolution by $\textrm{sinc}$, which has the effect of "mixing" the values of $S\left(u\right)$. When the observation period is long, the central lobe of the sinc function is very narrow, and the degradation is minimal; however, according to the similarity principle, if the recording window is short, the central lobe is wide and the frequency resolution is poor. To better understand this, let us consider the simple case $s\left(t\right)=\cos\left(2\pi u_{0}t\right)$ for which $s_{T}\left(t\right)=\cos\left(2\pi u_{0}t\right)\Pi\left(t/T\right)$. We then have,
\begin{equation}
	S_{T}\left(u\right)=\left(T/2\right)\left\{ \textrm{sinc}\left[\left(u-u_{0}\right)T\right]+\textrm{sinc}\left[\left(u+u_{0}\right)T\right]\right\}
	\label{eq_troncature}
\end{equation}

that is to say, the two Dirac impulses of $S\left(u\right)$ are replaced by two sinc functions, which no longer allow for an infinitely precise determination of the frequency $u_{0}$. By analogy with the resolution of an optical instrument, we can define the frequency resolution as being equal to the half-width of the central lobe of the sinc functions,
\begin{equation}
	\delta u\approx\frac{1}{T}
\end{equation}

\begin{figure}[H]
	\begin{center}
		\tcbox[colback=white]{\includegraphics[width=16cm]{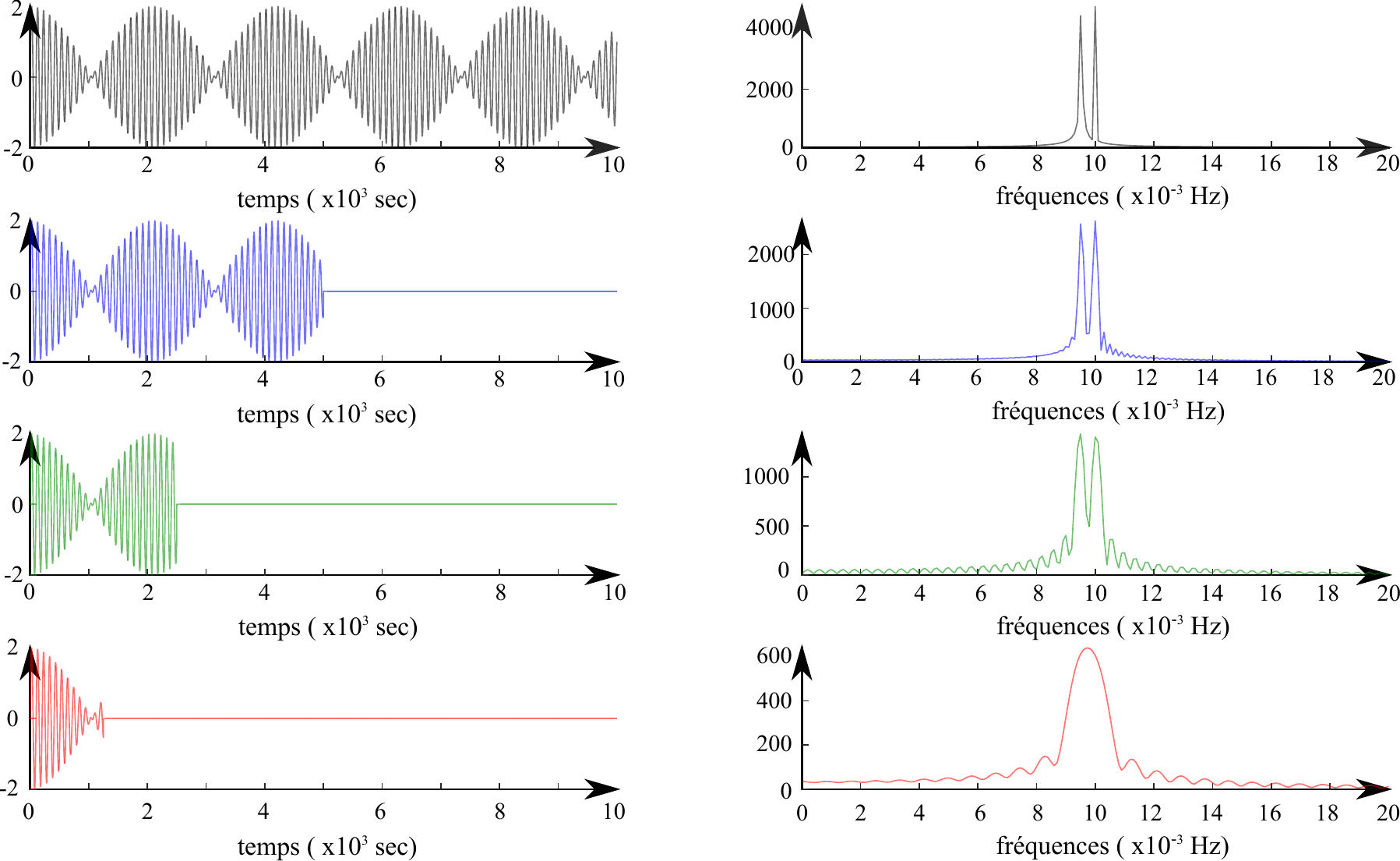}}
	\end{center}
	\caption{Illustration of the effects of truncation on frequency resolution $\delta u$. The two frequencies become indistinguishable when the truncation no longer reveals the presence of beats in the signal. At that point, the truncated signal can be interpreted as a single damped sinusoid instead of two sinusoids producing beats. We indeed find that the amplitude of the Fourier transform is also modified and corresponds to half the duration of the truncation window (\cf equation \ref{eq_troncature}).}
	\label{ex_troncature}
\end{figure}

\warning{The longer the observation period, the better the resolution.} The function \couleur{ex\_troncature.m} illustrates the influence of truncation on frequency resolution. The results are shown in the figure \ref{ex_troncature}.

\subsection{Discretization}
\subsubsection{Spectral duplication}
Discretization involves replacing the continuous signal $s\left(t\right)$ with the sequence of values taken by the signal at multiples of the discretization interval $\tau$. The thus discretized signal constitutes a set of discrete values that can be represented by,
\begin{equation}
	s_{\tau}\left(t\right)=s\left(t\right)\textrm{shah}\left(\frac{t}{\tau}\right)
\end{equation}

\warning{This representation, which uses the product of a function with a distribution, is not very rigorous and only makes sense when it appears under an integral.} This representation allows us to compute the Fourier transform of the discretized signal,
\begin{equation}
	\begin{split}
	S_{\tau}\left(u\right) & =  \tau S\left(u\right)*\textrm{shah}\left(\tau u\right)\\
 & = \sum_{n=-\infty}^{+\infty}S\left(u\right)*\delta\left(u-\frac{n}{\tau}\right)\\
 & =  \sum_{n=-\infty}^{+\infty}S\left(u-\frac{n}{\tau}\right)
	\end{split}
\end{equation}

which shows us that $S_{\tau}\left(u\right)$ consists of an infinite number of duplicates of $S\left(u\right)$, spaced at intervals of $u_{e}=\tau^{-1}$. \warning{The Fourier transform of a discretized signal is therefore a periodic function with period $\tau^{-1}$.}

\subsection{Correct Discretization: \couleur{Shannon} Interpolation\index{Shannon's Theorem}}
This theorem, established by \couleur{Claude Shannon}\index{Claude Shannon} (1916-2001) while he was an engineer at Bell Laboratories \shortcite{shannon1951mathematical}, forms the foundation of discrete signal processing and information theory. If the signal has a bounded spectrum, meaning that $S\left(u\right)=0$ when $\left|u\right|>u_{c}$, the duplicates will not overlap if the sampling frequency is such that,
\begin{equation}
	u_{e}>2u_{c}
\end{equation}

that is,
\begin{equation}
	u_{c}<u_{N}
\end{equation}
where the \couleur{Nyquist} frequency\index{Nyquist Frequency} $u_{N}=u_{e}/2$. This condition, known as the \couleur{Shannon} sampling theorem, intuitively expresses the fact that the period of a periodic phenomenon can only be determined if the phenomenon is observed more than twice per period. \warning{Note that strictly sampling twice per period is insufficient; sample $\sin\left(2\pi t\right)$ from $t=0$ and you will see!} When this condition is satisfied, it is possible to recover the Fourier transform of the total signal,

\begin{equation}
	S\left(u\right)=S_{\tau}\left(u\right)\Pi\left(\frac{u}{u_{e}}\right)
\end{equation}

whence,
\begin{equation}
	s\left(t\right)=\tau^{-1}s_{\tau}\left(t\right)*\textrm{sinc}\left(t/\tau\right)
\end{equation}

that is,
\begin{equation}
	\begin{split}
	s\left(t\right) & =  \frac{1}{\tau}\left[s\left(t\right)\textrm{shah}\left(\frac{t}{\tau}\right)\right]*\textrm{sinc}\left(\frac{t}{\tau}\right)\\
 & =  \frac{1}{\tau}\left[\tau\sum_{n=-\infty}^{+\infty}s\left(t\right)\delta\left(t-n\tau\right)\right]*\textrm{sinc}\left(\frac{t}{\tau}\right)\\
 & =  \sum_{n=-\infty}^{+\infty}\left[s\left(t\right)\delta\left(t-n\tau\right)\right]*\textrm{sinc}\left(\frac{t}{\tau}\right)\\
 & =  \sum_{n=-\infty}^{+\infty}\int_{-\infty}^{+\infty}\textrm{sinc}\left(\frac{t-\xi}{\tau}\right)s\left(\xi\right)\delta\left(\xi-n\tau\right)d\xi\\
 & =  \sum_{n=-\infty}^{+\infty}s\left(n\tau\right)\textrm{sinc}\left(\frac{t-n\tau}{\tau}\right),				
 	\end{split}
 	\label{reconstruct_shannon}
\end{equation}

where the transition to the last line uses the sampling formula. The final equality is known as the "Shannon interpolation formula"\index{Shannon interpolation} and allows for the recovery of the continuous signal from the discrete series $s_{\tau}\left(t\right)$. It is verified that for $t=k\tau$, $\textrm{sinc}\left(k - \right)=\delta_{k}^{n}$ and that the interpolation formula correctly yields $s\left(k\tau\right)$.

\subsection{Incorrect Discretization: Spectral Aliasing}
\begin{figure}[H]
	\begin{center}
		\tcbox[colback=white]{\includegraphics[width=16cm]{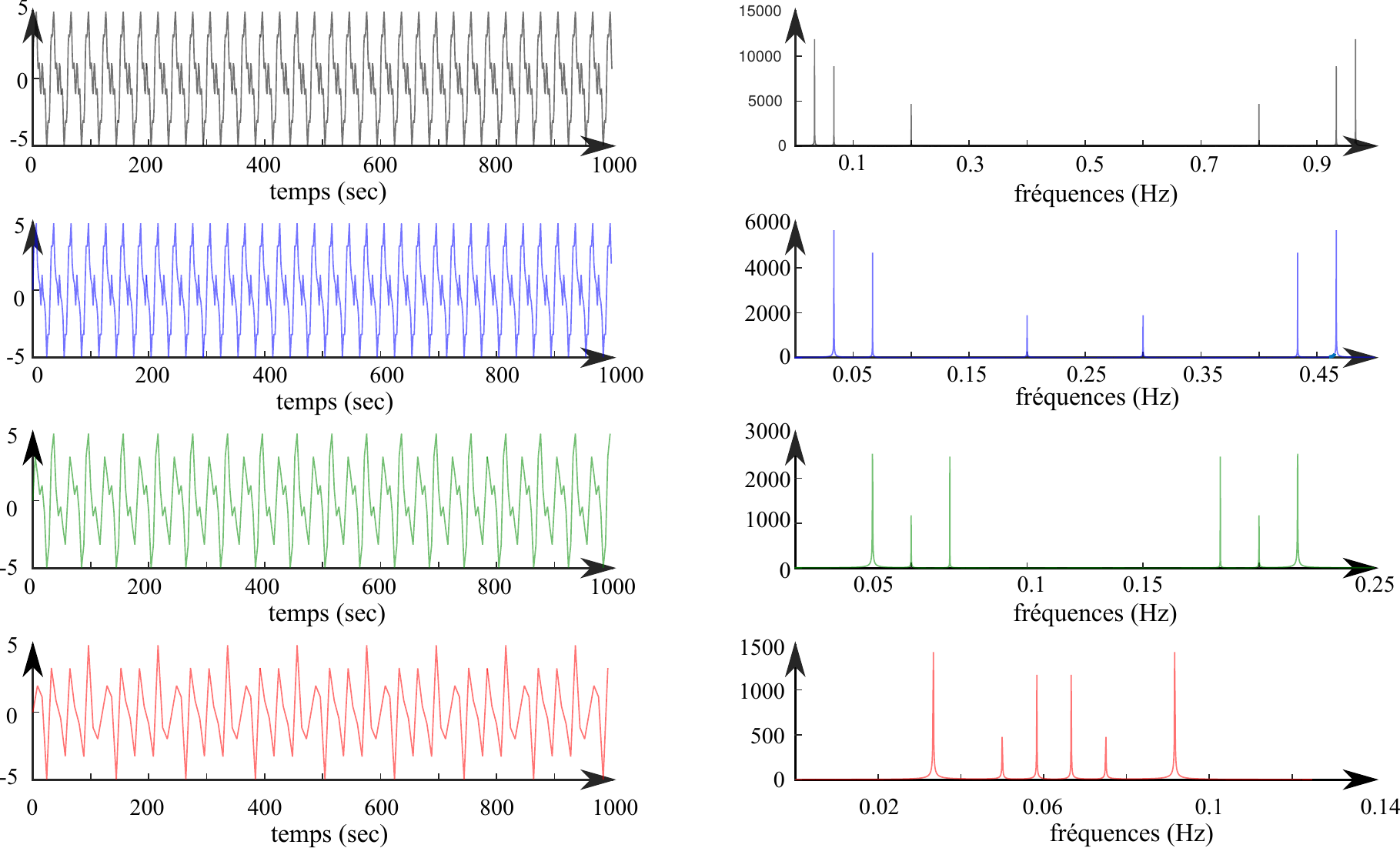}}
	\end{center}
	\caption{Illustration of the effects of discretization and spectral aliasing on the apparent frequencies of three sinusoids with frequencies of 0.2 Hz, 0.067 Hz, and 0.033 Hz. The sampling periods, from bottom to top (from black to red), are 1 second (1 Hz), \textcolor{blue}{2 seconds (0.5 Hz)}, \textcolor{DARKGREEN}{4 seconds (0.25 Hz)}, and \textcolor{red}{8 seconds (0.125 Hz)}. The initial signal (top left) is discretized with a one-second interval and results from the superposition of sinusoids, including two high-frequency ones, as shown by the magnitude of the Fourier transform of the signal (top right). Under-sampling with a two-second interval (blue curves) violates the Shannon condition and causes spectral aliasing. Under-sampling with a four-second interval (green curves) further distorts the spectral lines.
}
	\label{repliement}
\end{figure}

The phenomenon of spectral aliasing is an artifact that occurs when the discretization of a signal does not satisfy the \couleur{Shannon} sampling theorem. In this case,
\begin{equation}
	\tau^{-1}=u_{e}<2u_{c}
\end{equation}

and the \textit{duplicates} overlap. The function \couleur{ex\_repliement.m} illustrates the influence of discretization. The results are shown in Figure \ref{repliement}.

\begin{figure}[!h]
	\begin{center}
		\tcbox[colback=white]{\includegraphics[width=16cm]{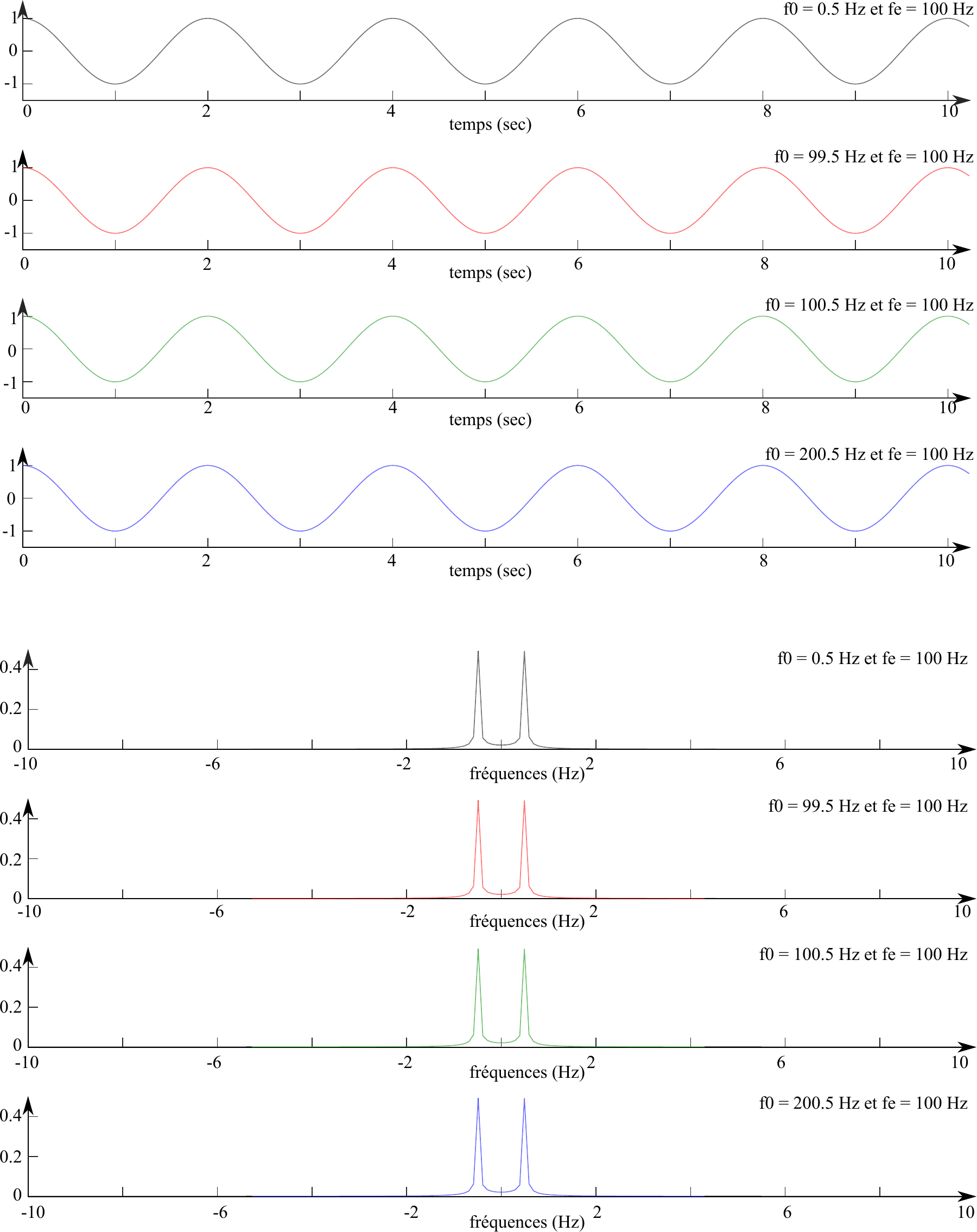}}
	\end{center}
	\caption{Example of spectral aliasing. At the top, the identical waveforms of four sinusoids with different frequencies, whose spectra are also identical because, in 3 out of 4 cases (red, green, and blue curves), we do not meet the \couleur{Shannon} sampling theorem.}
	\label{ex_shannon}
\end{figure}

It is thus impossible to recover $S\left(u\right)$ as we did previously (Figure \ref{repliement}). A sinusoidal signal with a frequency $u_{0}>u_{e}/2$ will be converted into a signal with an apparent frequency $u_{a}=u_{0}-mu_{e}$ where $m$ is the integer such that $\left|u_{a}\right|<u_{e}/2$. This phenomenon is analogous to a stroboscopic effect, where the apparent rotational speed of a mechanical part depends on the ratio between the actual rotational speed and the strobe light frequency. Spectral aliasing is a very serious problem because it transfers energy from high frequencies to low frequencies, resulting in an unacceptable spectrum (Figure \ref{repliement}). Before sampling a signal, one must either ensure that it does not contain significant energy outside the interval $\left[-u_{e}/2; +u_{e}/2\right]$, or filter the signal with a low-pass filter to remove high frequencies before the discretization process. You might think that aliasing can only occur during analog-to-digital conversions at the sensor level. This is incorrect, and experience shows that aliasing often occurs within the computer when, for practical reasons, "one only takes one point out of five because it will be sufficient and takes up less space"! A final example of spectral aliasing, which leads to incorrect interpretations, is shown in Figure \ref{ex_shannon}. This was obtained from the function \couleur{ex\_shannon.m}. As can be seen, we sampled at 100 Hz four sinusoids with frequencies of 0.5 Hz, 99.5 Hz, 100.5 Hz, and 200.5 Hz. Despite these different frequencies, the waveforms (top) are rigorously identical, and their respective \couleur{Fourier} spectra (bottom) suggest that these four signals are the same and beat at 0.5 Hz.

\begin{figure}[!h]
	\begin{center}
		\tcbox[colback=white]{\includegraphics[width=16cm]{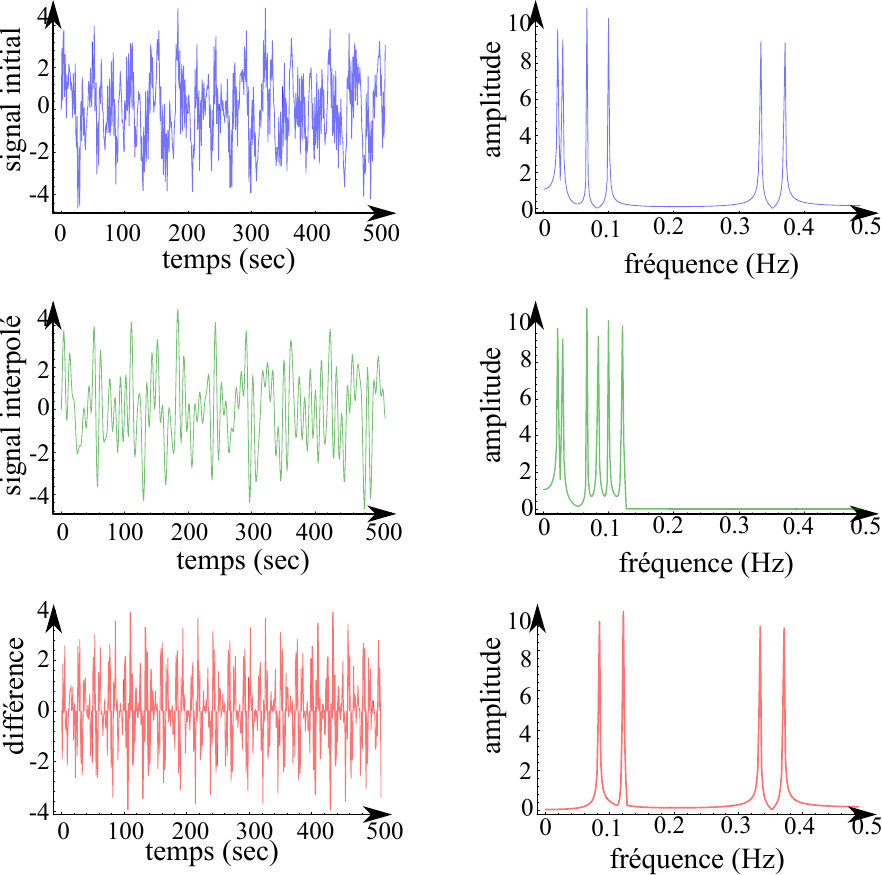}}
	\end{center}
	\caption{Loss of information due to spectral aliasing. The signal in the middle left is constructed by under-sampling the initial signal (top left) with a two-second interval followed by \couleur{Shannon} interpolation at a one-second interval. Although the constructed signal is sampled as finely as the initial signal, the aliasing that occurred during under-sampling has not been eliminated, as shown by the amplitude spectrum in the middle right. The difference (bottom right) between this spectrum and that of the initial signal (top right) shows that the interpolated signal no longer contains the original high frequencies between 0.3 and 0.4 Hz, which have been aliased around 0.1 Hz. This is also evident in the difference (bottom left) between the two signals, which contains the high-frequency oscillations missing in the interpolated signal.
}
	\label{interp_shannon}
\end{figure}

\paragraph{Vocabulary} -- In this book, we use the term \couleur{spectral duplication} to refer to the phenomenon that occurs when discretizing a signal using the comb function. The term \couleur{aliasing} is used to describe what happens when discretization does not satisfy the Shannon condition. In many texts, especially those written in English, \warning{you will encounter the term "aliasing," which has a dual meaning as it can refer either to spectral duplication or to aliasing}.

\subsection{Analog-to-Digital Conversion: Quantization}
	Quantization occurs during the analog-to-digital conversion, which provides a signal generally encoded in base 2. The smallest value that can be encoded is 1, and if the encoding is done with $n$ bits, the largest value is $2^{n}-1$. The encoding process will reduce the infinite number of possible values that the analog signal can take to a finite and relatively small number of digital values; we will see later that this process is accompanied by the generation of quantization noise. An encoding can be characterized by its dynamic range,
\begin{equation}
	\text{dynamique}\left(dB\right)\equiv20\log2^{n}
\end{equation}

For example, a 12-bit converter has a dynamic range of approximately 72 dB. This is the ratio between the smallest value and the largest value that can be converted.
\chapter{\titrechap{The Z-Transform}}
\minitoc
\section{The Utility of the $Z$-Transform}
Instead of a formal mathematical approach, we will present the $Z$-transform as a convenient notation for manipulating the \couleur{Fourier} transforms of signals discretized at a constant interval,
\begin{equation}
	s_{\tau}\left(t\right)=s\left(t\right)\textrm{shah}\left(\frac{t}{\tau}\right)
	\label{tz_01}
\end{equation}

for which,
\begin{equation}
	\begin{split}
	S_{\tau}\left(u\right) & = \tau\int_{-\infty}^{+\infty}\left\{ \sum_{n=-\infty}^{+\infty}s\left(t\right)\delta\left(t-n\tau\right)\right\} \exp\left(-2i\pi ut\right)dt\\
 & =  \tau\sum_{n=-\infty}^{+\infty}\int_{-\infty}^{+\infty}\left[s\left(t\right)\exp\left(-2i\pi ut\right)\right]\delta\left(t-n\tau\right)dt\\
 & =  \tau\sum_{n=-\infty}^{+\infty}s\left(n\tau\right)\exp\left(-2i\pi un\tau\right)
	\end{split}
	\label{tz_02}
\end{equation}

where the factor $\tau$ ensures correct scaling and equivalence between this expression and the continuous \couleur{Fourier} transform. The $Z$-transform is simply obtained by making the following variable change,
\begin{equation}
	Z\equiv\exp\left(-2i\pi u\tau\right)
\end{equation}

Some authors use the conjugate definition, $Z\equiv\exp\left(+2i\pi u\tau\right)$. In any case, this allows expression (\ref{tz_02}) to be rewritten in the form of a $Z$-transform,
\begin{equation}
	\begin{split}
S\left(Z\right) & =  \tau\sum_{n=-\infty}^{+\infty}s\left(n\tau\right)Z^{n}\\
 & =  \tau\left[\cdots+s\left(-\tau\right)Z^{-1}+s\left(0\right)+s\left(\tau\right)Z+s\left(2\tau\right)Z^{2}+\cdots\right]
\end{split}
\end{equation}

which is a polynomial in $Z$. In fact, this definition can be extended to any complex $Z$, but the choice we have made is appropriate because it allows for the equivalence between the $Z$-transform and the \couleur{Fourier} transform of discrete signals. The above expressions show that the $Z$-transform is merely a way of writing the $Z$, but the choice we have made is appropriate because it allows for the equivalence between the $Z$-transform and the \couleur{Fourier} transforms of discrete signals. When performing calculations involving the Fourier transforms of such signals, it is up to you to decide if using the $Z$-transform notation is useful or not. We encourage the reader to consult Jon \couleur{Claerbout}'s book, \textit{Fundamentals of Geophysical Data Processing} \shortcite{claerbout1985fundamentals}, to see numerous applications of the $Z$-transform. It is clear that all properties of the Fourier transform are preserved for the $Z$-transform, whose primary interest lies in the manipulation of discrete signals. For example, just as
\begin{equation}
	\exp\left(-2i\pi u\tau\right)S\left(u\right)
\end{equation}

est la transformée de \couleur{Fourier} du signal $s\left(t\right)$ retardé d'un pas de temps,
\begin{equation}
	ZS\left(Z\right)
\end{equation}

is the $Z$-transform of the discrete signal $s_{\tau}\left(t\right)$ delayed by the same time step: $Z$ can be considered as the unit delay operator. Similarly, the $Z$-transform of the convolution of two discrete signals is equal to the product of their $Z$-transforms.

\section{Formulary: $Z$-Transforms}
\begin{equation}
\left[\mathbf{a}_{0},a_{1},a_{2},\cdots\right]\longmapsto a_{0}+a_{1}Z+a_{2}Z^{2}+\cdots
\end{equation}

\begin{equation}
\left[1,1,1,\cdots\right]\longmapsto\frac{1}{1-Z}
\end{equation}

\begin{equation}
\left[0,1,1,\cdots\right]\longmapsto\frac{Z}{1-Z}
\end{equation}

\begin{equation}
\left[0,1,2,3,\cdots,n,\cdots\right]\longmapsto\frac{Z}{\left(1-Z\right)^{2}}
\end{equation}

\begin{equation}
\left[0,1,4,9,\cdots,n^{2},\cdots\right]\longmapsto\frac{Z\left(1+Z\right)}{\left(1-Z\right)^{3}}
\end{equation}

\begin{equation}
\left[0,1,8,27,\cdots,n^{3},\cdots\right]\longmapsto\frac{3Z^{2}\left(1+Z\right)}{\left(1-Z\right)^{4}}+\frac{Z\left(1+2Z\right)}{\left(1-Z\right)^{3}}
\end{equation}

\begin{equation}
\left[1,\exp\left(-\alpha\right),\exp\left(-2\alpha\right),\cdots\right]\longmapsto\frac{1}{1-\exp\left(-\alpha\right)Z}
\end{equation}

\begin{equation}
\left[0,1-\exp\left(-\alpha\right),1-\exp\left(-2\alpha\right),\cdots\right]\longmapsto\frac{Z\left(1-\exp\left(-\alpha\right)Z\right)}{\left(1-Z\right)\left(1-\exp\left(-\alpha\right)Z\right)}
\end{equation}

\begin{equation}
\left[0,\exp\left(-\alpha\right),4\exp\left(-2\alpha\right),\cdots\right]\longmapsto\frac{\left(1+\exp\left(-\alpha\right)Z\right)\exp\left(-\alpha\right)Z}{\left(1-\exp\left(-\alpha\right)Z\right)^{3}}
\end{equation}

\begin{equation}
\left[0,\exp\left(-\alpha\right),2\exp\left(-2\alpha\right),\cdots\right]\longmapsto\frac{\exp\left(-\alpha\right)Z}{\left(1-\exp\left(-\alpha\right)Z\right)^{2}}
\end{equation}

\chapter{\titrechap{The Discrete \couleur{Fourier} Transform}}
\minitoc
\section{The Discrete \couleur{Fourier} Transform}
\subsection{Discretization of the Fourier Transform}
Just as we discussed the issue of discretizing time-domain signals, we will now address the discretization of their \couleur{Fourier} transforms. To be correct, the discretization of time-domain signals must satisfy the \couleur{Shannon} criterion. Some remarks based on the duality of the \couleur{Fourier} transform will allow us to establish an equivalent rule without performing any calculations. We have seen that the discretization of a signal can only be rigorously achieved if its \couleur{Fourier} transform has limited support. By duality, we infer that the discretization of the Fourier transform can only be done for signals with limited temporal support. Poor temporal discretization leads to spectral aliasing: poor frequency sampling will cause aliasing of the signal. There will be no temporal aliasing if the frequency discretization satisfies the dual \couleur{Shannon} criterion.
\begin{equation}
	\nu<\frac{1}{T},
\end{equation}

where $T$ is the duration of the temporal support of the signal and $\nu$ is the frequency sampling interval. We have also seen that temporal discretization induces a periodicization of the \couleur{Fourier} transform: frequency discretization induces a temporal periodicization. Strictly speaking, it is only possible to discretize the signal and its \couleur{Fourier} transform without violating the \couleur{Shannon} criterion and its dual if the signal is periodic. In practice, a discrete Fourier transform will therefore always be the \couleur{Fourier} transform of a periodic signal. This limitation is severe and should never be forgotten. The previous reasoning allows us to derive the formula for the discrete Fourier transform. It can also be obtained using a more intuitive approach by revisiting the \couleur{Fourier} transform of a discretized signal,
\begin{equation}
	S_{\tau}\left(u\right)=\tau\sum_{n=-\infty}^{+\infty}s\left(n\tau\right)\exp\left(-2i\pi un\tau\right),
\end{equation}

that is, after truncating (centered at the origin) the signal to a duration $T$,
\begin{equation}
S_{\tau,T}\left(u\right)=T\textrm{sinc}\left(Tu\right)*\left[\tau\sum_{n=-\infty}^{+\infty}s\left(n\tau\right)\exp\left(-2i\pi un\tau\right)\right]
\end{equation}

The frequency resolution can be taken as half the width of the central lobe of the sinc function,
\begin{equation}
	\nu=\frac{1}{T}
\end{equation}

The range of useful frequencies being $\left[0;\tau^{-1}\right[$, the discrete frequencies are found to be
\begin{equation}
	u_{k}=k\nu\;\left(k=0,1,\cdots,N-1\right)
\end{equation}

where $N$ is the number of points in the truncated discrete signal. Under these conditions, the discrete \couleur{Fourier} transform is given by
\begin{equation}
	S_{\tau,T}\left(k\nu\right)=\tau\sum_{n=0}^{N-1}s\left(n\tau\right)\exp\left(\frac{-2i\pi kn}{N}\right)\;\left(k=0,1,\cdots,N-1\right)
\end{equation}

\subsection{The Fast \couleur{Fourier} Transform Algorithm}
\begin{figure}[H]
	\begin{center}
		\tcbox[colback=white]{\includegraphics[width=16cm]{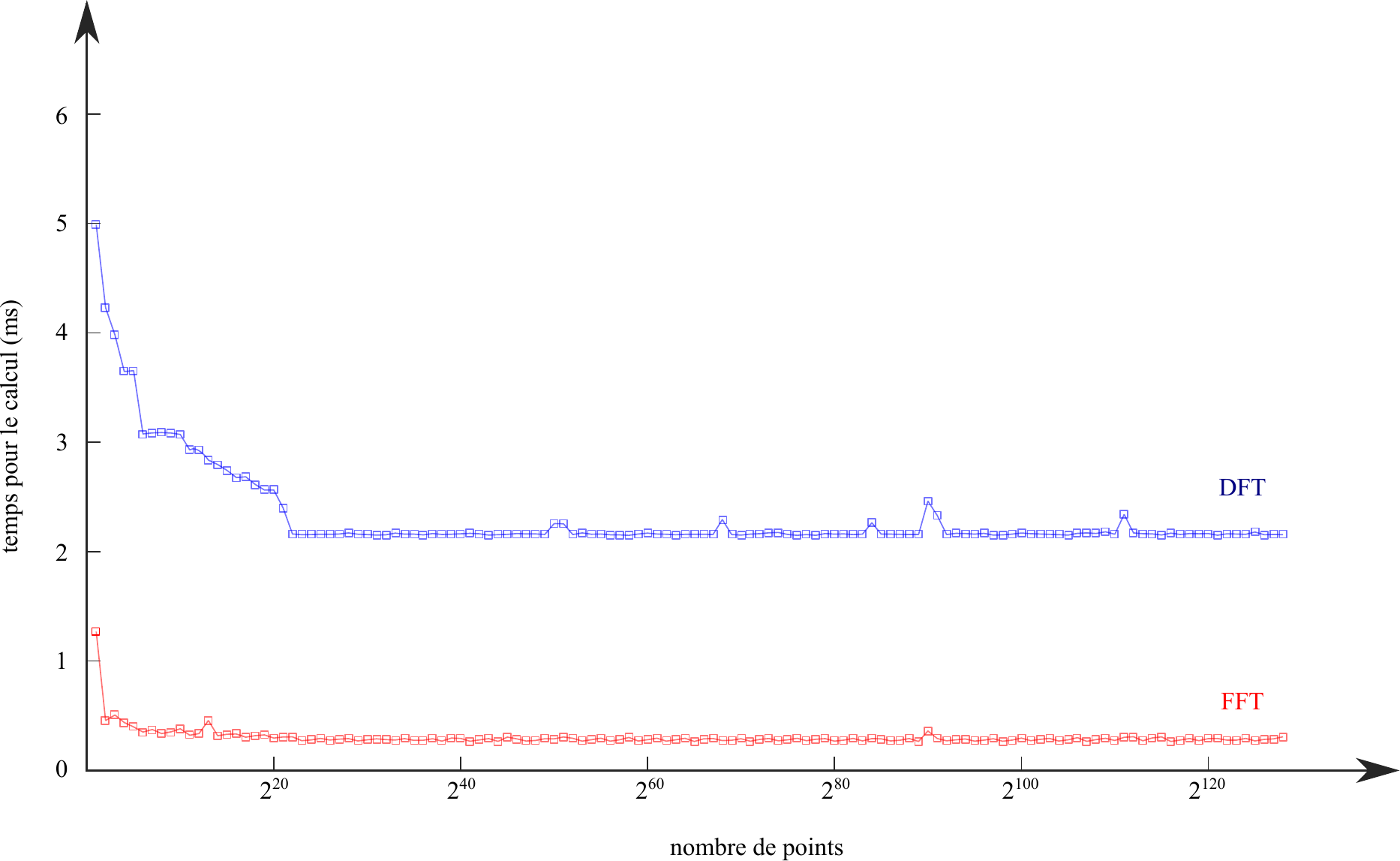}}
	\end{center}
	\caption{Comparison of computation times between the DFT and the FFT. The \couleur{Fourier} transform is performed on a signal consisting of $2^N$ points in red and $2^N-1$ points in blue. It is observed that the so-called fast algorithm is nearly 5 times faster.}
	\label{compa_fft_dft}
\end{figure}

The algorithm we will examine in this section was a revolution in numerical analysis and signal processing; it is considered one of the ten greatest algorithms of the 20\textsuperscript{th} century \shortcite{cipra2000best}. It allows for the rapid computation of discrete \couleur{Fourier} transforms of digitized signals, which, at the time of its discovery, made many previously impractical analysis methods feasible. To give you an idea of the algorithm's power, we will cite only the test results by \couleur{Jon Claerbout} \shortcite{Claerbout/92}, where his "slow" program—i.e., the one implementing the double loop of the discrete \couleur{Fourier} transform with $N^{2}$ iterations—takes 153 seconds to compute the discrete \couleur{Fourier} transform of a 1024-point signal, while the program using the fast algorithm takes only 0.7 seconds! The program \couleur{ex\_dft\_vs\_fft.m} compares computation times between a fast \couleur{Fourier} transform and a discrete \couleur{Fourier} transform on Matlab\textregistered. In the program, we use the \couleur{fft} function indiscriminately, but it is important to know that if the signal whose spectrum we want to compute does not have a dimension that is a power of 2, the classical algorithm implementing the double loop is used. Even though software and computers have made enormous advancements since \couleur{Claerbout}'s test, Figure (\ref{compa_fft_dft}) still shows up to a factor of 5 difference in computation time between the two algorithms. The fast Fourier transform has been widely used since the famous 1965 article by \couleur{James Cooley} and \couleur{John Tukey} \shortcite{cooley1965algorithm}, although the algorithm was originally conceived by \couleur{Carl Friedrich Gauss} in 1805, and has been adapted several times since, including notable work by \couleur{Cornelius Lanczos} in 1942 \shortcite{Lanczos1942}.

Let us denotes,
\begin{equation}
	W\equiv\exp\left(-2i\pi/N\right)
\end{equation}

the discrete \couleur{Fourier} transform then takes the form,
\begin{equation}
	S_{\tau,T}\left(k\nu\right)=\tau\sum_{n=0}^{N-1}s\left(n\tau\right)W^{kn}\;\left(k=0,1,\cdots,N-1\right)
\end{equation}

which, when adopting a matrix notation, becomes
\begin{equation}
	\left[
	\begin{array}{c}
		S\left(0\right)\\
		S\left(\nu\right)\\
		S\left(2\nu\right)\\
		S\left(3\nu\right)\\
		\vdots
	\end{array}
	\right]=\tau\left[
	\begin{array}{ccccc}
		1 & 1 & 1 & 1 & \cdots\\
		1 & W & W^{2} & W_{{}}^{3} & \cdots\\
		1 & W^{2} & W^{4} & W^{6} & \cdots\\
		1 & W^{3} & W^{6} & W^{9} & \cdots\\
		\vdots & \vdots & \vdots & \vdots & \ddots
	\end{array}\right].\left[
	\begin{array}{c}
		s\left(0\right)\\
		s\left(\tau\right)\\
		s\left(2\tau\right)\\
		s\left(3\tau\right)\\
		\vdots
	\end{array}\right]
	\label{fft_prod_01}
\end{equation}

The inverse matrix has the same structure and is obtained by replacing \( W \) with \( W^{-1} \), so
\begin{equation}
	\left[
	\begin{array}{c}
		s\left(0\right)\\
		s\left(\tau\right)\\
		s\left(2\tau\right)\\
		s\left(3\tau\right)\\
		\vdots
	\end{array}
	\right]=\tau\left[
	\begin{array}{ccccc}
		1 & 1 & 1 & 1 & \cdots\\
		1 & W^{-1} & W^{-2} & W_{{}}^{-3} & \cdots\\
		1 & W^{-2} & W^{-4} & W^{-6} & \cdots\\
		1 & W^{-3} & W^{-6} & W^{-9} & \cdots\\
		\vdots & \vdots & \vdots & \vdots & \ddots
	\end{array}\right].\left[
	\begin{array}{c}
		S\left(0\right)\\
		S\left(\nu\right)\\
		S\left(2\nu\right)\\
		S\left(3\nu\right)\\
		\vdots
	\end{array}\right],
	\label{fft_prod_02}
\end{equation}

which amounts to computing the inverse discrete \couleur{Fourier} transform,
\begin{equation}
s\left(n\tau\right)=\nu\sum_{k=0}^{N-1}S_{\tau,T}\left(k\nu\right)W^{-kn}\;\left(n=0,1,\cdots,N-1\right).
\end{equation}

Here we recognize that the matrix \( W \) is a \couleur{Vandermonde} matrix. The above matrix equations require \( N^2 \) multiplications and as many additions, which quickly becomes enormous, explaining the significance of the work by \couleur{Cooley} and \couleur{Tukey} in 1965. Their algorithm made it possible to compute discrete \couleur{Fourier} transforms very quickly by reducing the calculation of a discrete \couleur{Fourier} transform of length \( N \) to that of two transforms of length \( N/2 \). In fact, \couleur{J. Claerbout} notes that \couleur{Vern Herbert} of \textit{Chevron Standard Ltd.} had already programmed this as early as 1962. In practice, the discrete signal \( s(n\tau) \) is decomposed into two interleaved signals \( {}^{1}s \) and \( {}^{2}s \) such that
\begin{equation}
	\begin{cases}
		s^{^{\prime}}\left(n\right)  \equiv  s\left(2n\tau\right) \qquad \quad\left(n=0,1,\cdots,N/2-1\right)\\
		s^{^{\prime\prime}}\left(n\right)  \equiv  s\left[\left(2n+1\right)\tau\right]\;\;\left(n=0,1,		\cdots,N/2-1\right)
	\end{cases}
\end{equation}

We then obtain,
\begin{equation}
	\begin{split}
	S_{\tau,T}\left(k\nu\right) & =  \tau\sum_{n=0}^{N/2-1}s\left(2n\tau\right)W^{2kn}+\tau\sum_{n=0}^{N/2-1}s\left[\left(2n+1\right)\tau\right]W^{k\left(2n+1\right)}\\
 & =  \tau\sum_{n=0}^{N/2-1}s^{^{\prime}}\left(n\right)\left(W^{2}\right)^{kn}+\tau W^{k}\sum_{n=0}^{N/2-1}s^{^{\prime\prime}}\left(n\right)\left(W^{2}\right)^{nk}
	\end{split}
\end{equation}

for \( k = 0, 1, \ldots, N/2-1 \), the two sums are the discrete \couleur{Fourier} transforms of the interleaved signals,
\begin{equation} 
	S_{\tau,T}\left(k\nu\right)=\frac{S_{2\tau,T}^{^{\prime}}\left(k\nu\right)}{2}+\frac{S_{2\tau,T}^{^{\prime\prime}}\left(k\nu\right)}{2}W^{k}
\end{equation}

For \( k = 0, 1, \ldots, N/2-1 \), and setting \( k = l + N/2 \), the two sums can be written as
\begin{equation}
	\begin{split}
&S_{\tau,T}\left[\left(l+\frac{N}{2}\right)\nu\right]  \\ &=\tau\sum_{n=0}^{N/2-1}s^{^{\prime}}\left(n\right)\left(W^{2}\right)^{n\left(l+\frac{N}{2}\right)}+\tau W^{l+N/2}\sum_{n=0}^{N/2-1}s^{^{\prime\prime}}\left(n\right)\left(W^{2}\right)^{n\left(l+\frac{N}{2}\right)}\\
 &=  \tau\sum_{n=0}^{N/2-1}s^{^{\prime}}\left(n\right)W^{nN}\left(W^{2}\right)^{nl}+\tau W^{N/2}W^{l}\sum_{n=0}^{N/2-1}s^{^{\prime\prime}}\left(n\right)W^{nN}\left(W^{2}\right)^{nl}
	\end{split}
\end{equation}

but \( W^{lN} = 1 \) and \( W^{N/2} = -1 \), which allows us to obtain the simplified form
\begin{equation}
	S_{\tau,T}\left[\left(k+\frac{N}{2}\right)\nu\right]=\frac{S_{2\tau,T}^{^{\prime}}\left(k\nu\right)}{2}-\frac{S_{2\tau,T}^{^{\prime\prime}}\left(k\nu\right)}{2}W^{k}
\end{equation}

with \( k = 0, 1, \cdots, N/2-1 \). Thus, the computation of the discrete \couleur{Fourier} transform of a series with \( N \) values has been reduced to that of two transforms of interlaced series with \( N/2 \) values each. If \( N = 2^{p} \), this reduction can be performed \( p \) times, starting the process by computing the discrete \couleur{Fourier} transforms of \( N \) series containing only one value, then of series with 2 values, then 4, and so forth, up to the complete series. Overall, the number of operations is significantly reduced: the algorithm described enables the calculation of the transform of a series of \( N \) values with only \( 2Np \) operations, compared to \( 2N^{2} \) for the direct algorithm using the matrix form.

To illustrate what we have just discussed, the following subroutine, \texttt{TFR}, written in \textbf{Fortran} -- which stands for \textbf{For}mula \textbf{Tran}lator -- implements the Fast \couleur{Fourier} Transform (FFT) algorithm. The program computes the direct transform when the variable \texttt{dirinv=1} and the inverse transform when \texttt{dirinv=-1}. The complex values of the signal are provided in the array \texttt{signal}, and the number of values, \texttt{n}, must be an integer power of 2. Other programs can be found in some of the books cited in the bibliographic references (\shortciteN{Claerbout/92}, \shortciteN{claerbout1985fundamentals}, \shortciteN{Kanasewich/81}, and \shortciteN{Press/al/86}).
\newpage

\begin{lstlisting}
subroutine TFR(dirinv,signal,n)

   integer n,i,j,k,m,istep
   real dirinv,scale,arg
   complex signal(n),cmplx,cw,cdel,ct

   scale=1./sqrt(float(n))
   do i =1,n
       signal(i)=signal(i)*scale
   end do
   j=1
   k=1

   do i=1,n
       if(i.le.j) then
          ct=signal(j)
          signal(j)=signal(i)
          signal(i)=ct
       end if
       m=n/2
       do while (j.gt.m.and.m.gt.1)
           j=j-m
           m=m/2
       end do
       j=j+m
   end do

   do while(k.ge.n)
       istep=2*k
       cw=1.
       arg=dirinv*3.14159265/float(k)
       cdel=cmplx(cos(arg),sin(arg))
       do m=1,k
          do i=m,n,istep
             ct=cw*signal(i+k)
             signal(i+k)=signal(i)-ct
             signal(i)=signal(i)+ct
             cw=cw*cdel
          end do
          k=istep
       end do
   end do

   return
	
end
\end{lstlisting}

\warning{Go deeper \dots, but not too deep}
The relations (\ref{fft_prod_01}) and (\ref{fft_prod_02}) illustrate that the Discrete \couleur{Fourier} Transform (DFT) is essentially the product of a well-known \couleur{Vandermonde} matrix, which contains all the frequencies necessary for the decomposition (resp. reconstruction) of our signal $s$ (resp. $S$) via the coefficients of the sinusoids that facilitate these transformations. These coefficients are the unknowns. Therefore, we can view the direct (and inverse) Fourier transform as an inverse problem, which could be expressed using the notation from \couleur{William Menke}'s book, \textit{Geophysical Data Analysis: Discrete Inverse Theory} \shortcite{menke1984geophysical}, as follows for the direct transform,
\begin{equation}
		S = Ws
\end{equation}

and, 
\begin{equation}
		s = W^{-1}S
\end{equation}

for the inverse transform. We will not delve into the details of inverse problem theory (linear, nonlinear, gradient, conditioning, \etc), but will instead illustrate the relations (\ref{fft_prod_01}) and (\ref{fft_prod_02}) using the program \couleur{ex\_fourier\_coeff.m}, which calls the functions \couleur{fourier\_coeff.m} and \couleur{fourier\_reconstruct.m}.

\begin{figure}[H]
	\begin{center}
		\tcbox[colback=white]{\includegraphics[width=12cm]{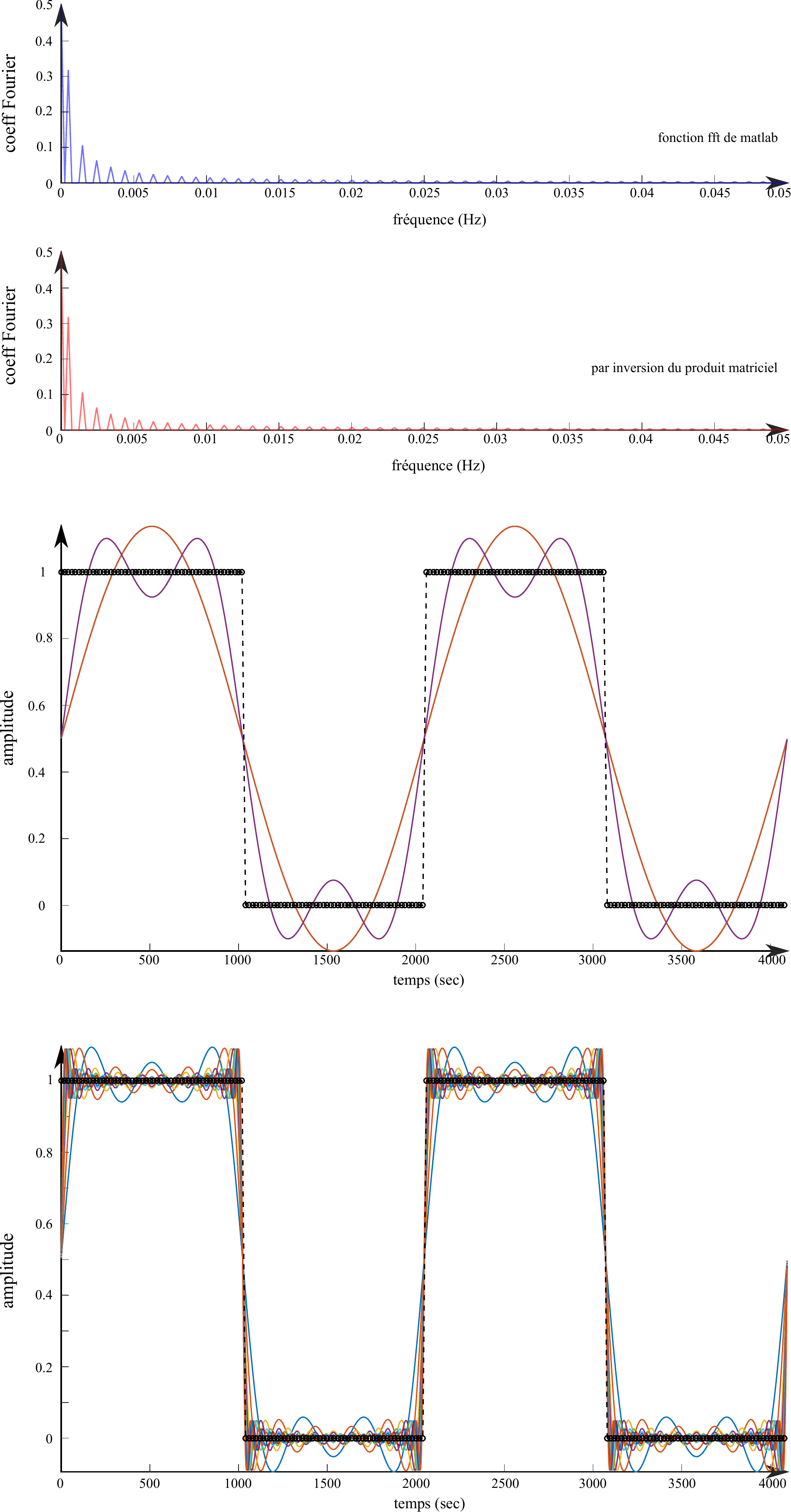}}
	\end{center}
	\caption{Direct and inverse \couleur{Fourier} transforms obtained by inverting the \couleur{Vandermonde} matrix.}
	\label{fourier_prod}
\end{figure}

Figure (\ref{fourier_prod}) illustrates the results. At the top, we have the \couleur{Fourier} spectra of a window function (shown in the two figures below in black line), ranging from 0 to the \couleur{Nyquist} frequency (0.05 Hz). The blue curve was obtained using Matlab's native \texttt{fft} function, while the red curve was obtained by inverting relation (\ref{fft_prod_01}). As can be seen, the two spectra are identical. The two figures below correspond to the reconstruction of the window function, which means inverting relation (\ref{fft_prod_02}), using a limited number of frequencies—i.e., not using all the frequencies previously calculated. For the calculation of $S$, it is evident that the number of frequencies to be computed, as previously discussed, must be at least equal to the number of points in the window $s$. For the inverse operation, and as we have also seen, to accurately reconstruct the original signal, we need to use at least as many frequencies as there are points in the signal. Here we illustrate two cases where, out of the 4096 points of the window $s$, thus requiring at least 4096 frequency values, we keep only 10 (center figure) and 100 (bottom figure). The sum of all reconstructed components should theoretically be exactly equal to the original signal (according to the \couleur{Plancherel} theorem, \cf relation \ref{plancherel_energie}, and \couleur{Shannon}'s reconstruction formula, \cf relation \ref{reconstruct_shannon}). However, the window function is one of the rare cases where this is not possible, which explains the apparent oscillations, also known as the \couleur{Gibbs} effect, because its spectrum is not of bounded support! Remember that this function has a cardinal sine as its Fourier transform, \cf figure (\ref{sin_cardinal}).

We could have intuitively predicted this, as ultimately, this function is discontinuous and transitions from 0 to 1 in an infinitesimally small amount of time. Invoking Heisenberg's uncertainty principle, \cf relation (\ref{incertitude_heis}), implies that the frequency required to describe this jump must be infinite, which is physically and numerically impossible.
\newpage
\chapter{\titrechap{Stochastic Processes}}
\minitoc
Until now, all the calculations we have performed have been within a deterministic framework, and we have always implicitly assumed that the signals we dealt with were perfectly known. In practice, this view is insufficient, and it is necessary to account for the fact that the signals being processed contain a certain amount of noise. Generally, the reasons given for the presence of noise include measurement uncertainties, electronic noise, \etc When signals are noisy, they can no longer be treated deterministically; a probabilistic approach must be adopted, in which the signal under study is considered as a sample drawn from the set of all possible signals. This set is called a stochastic process, with the particular signal being a realization of that process. This chapter does not provide a detailed exposition on stochastic processes; it is merely a general overview meant to introduce a few key terms. For a thorough presentation on the subject, we refer the reader to \couleur{Athanasios Papoulis}'s book, \textit{Probability, Random Variables, and Stochastic Processes} \shortcite{Papoulis/84}, listed in the bibliography.

\section{Definition of Stochastic Processes}
Such a process is characterized by its moments, among which the most useful are the mean,
\begin{equation}
	\mu_{x}\left(t\right)=\lim_{N\rightarrow\infty}\frac{1}{N}\sum_{n=1}^{N}x_{n}\left(t\right)
\end{equation}

and the autocorrelation,
\begin{equation}
	r_{x,x}\left(t,t+\tau\right)=\lim_{N\rightarrow\infty}\frac{1}{N}\sum_{n=1}^{N}x_{n}\left(t\right)x_{n}\left(t+\tau\right)
\end{equation}

where the signals \( x_{n}(t) \) represent realizations of the process \(\{ x \}(t)\). From a practical standpoint, completely arbitrary stochastic processes are not very useful because one rarely has a large number of realizations to calculate the aforementioned statistical attributes. It is generally to circumvent this difficulty that particularly simple processes are introduced, which we will briefly discuss below. \warning{Before discussing these, it is essential to keep in mind that these simplified processes have the immense advantage of being easily manipulable but also the significant drawback of often being too idealized to be realistic!} This certainly explains why many theoretically sophisticated signal processing methods are rarely applicable in practice, as the signals they are supposed to handle do not exist,
\begin{itemize}[label=$\bullet$]
	\item when the statistical moments of a process do not depend on time, the process is said to be stationary in the strict sense\index{strict stationarity}; 
	\item when only the mean and the autocorrelation are time-independent, the process is said to be second-order stationary or weakly stationary\index{second-order stationarity}.
\end{itemize}

It is worth noting right away that such processes are rare in practice, probably because they are information-poor: many geophysical signals owe their richness to their non-stationarity—the most illustrative example is certainly seismic signals. Consequently, it is primarily the noise itself that may be well described by stationary stochastic processes, rather than the signals as a whole. Among stationary processes are ergodic processes, where moments can be computed from a single sample by replacing ensemble sums with integrals over the values taken over time by a single realization,
\begin{equation}
	\mu_{x}=\lim_{T\rightarrow\infty}\frac{1}{T}\int_{0}^{T}x\left(t\right)dt
\end{equation}

for the mean and,
\begin{equation}
	r_{x,x}\left(\tau\right)=\lim_{T\rightarrow\infty}\frac{1}{T}\int_{0}^{T}x\left(t\right)x\left(t+\tau\right)dt
\end{equation}

for the autocorrelation. The energy spectrum of a stochastic process is computed using the \couleur{Wiener-Khinchine} theorem\index{Wiener-Khinchine theorem}, which we have already discussed,
\begin{equation}
	\left|X\left(u\right)\right|^{2}=\mathcal{F}\left[r_{x,x}\left(\tau\right)\right]\left(u\right).\end{equation}

\section{1/f Noise}
It is common for the energy spectra of geophysical signals to follow power law distributions,
\begin{equation} 
	E\left(u\right)=E_{0}u^{-\beta}
\end{equation}

where, in general, $\beta \in \left[0; 4\right]$. Such signals are referred to as "$1/f$ noises", and examples include the topography of young oceanic floors, geoid undulations, temporal variations of the Earth's magnetic field, \ldots  Such noises are invariant under scale changes, meaning that whether one contracts or dilates the time scale, the energy spectrum retains its power-law form with the same exponent. Thus, $1/f$ noises appear similar at all scales; they are statistically self-similar. There are numerous articles on this subject, such as the one by \couleur{Jérémy Kasdin} from 1995, \textit{Discrete Simulation of Colored Noise and Stochastic Processes and $1/f^{\alpha}$ Power Law Noise Generation} \shortcite{kasdin1995discrete}. Here, we will only mention three specific types of such noises.

\begin{itemize}[label=$\bullet$]
\item   \textbf{pink "noise"} has a constant energy per octave band, in contrast to white noise, whose spectrum is constant across all frequencies. For this type of noise, the coefficient $\beta$ has a value of 1;

\item \textbf{\couleur{brownian} "noise"}, also known as \couleur{brownian} motion\index{Brownian motion} in honor of the Scottish botanist \couleur{Robert Brown} (1773–1858), who first described in 1828 \shortcite{brown1828xxvii} the very irregular movements of large particles within pollen grains. For this type of noise, also referred to as red noise, the coefficient $\beta$ is equal to 2;

\item \textbf{black "noise"}, named by analogy to the thermal radiation of a black body, has a spectrum that decreases even more rapidly, and its coefficient $\beta$ is equal to 3.
\end{itemize}

\begin{figure}[H]
	\begin{center}
		\tcbox[colback=white]{\includegraphics[width=16cm]{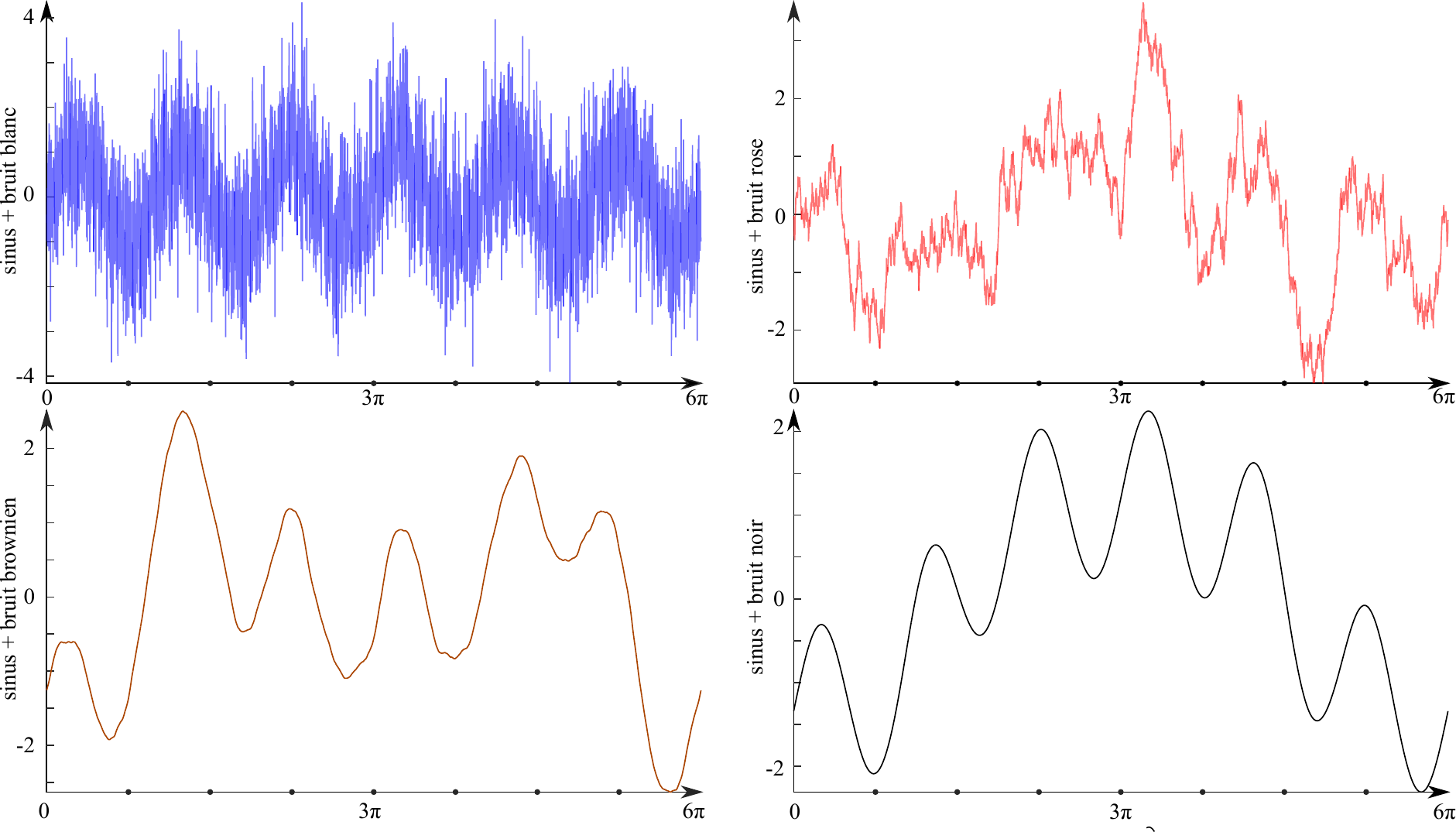}}
	\end{center}
	\caption{\textbf{1/$f$ Noise + Sine Wave}. These noises have identical phase spectra, which gives them correlated morphologies and allows us to observe that the black noise is a "smoother" version of the \textit{brownian} noise, which in turn is a "smoother" version of the pink noise, \etc}
	\label{bruit_sig}
\end{figure}

\begin{figure}[H]
	\begin{center}
		\tcbox[colback=white]{\includegraphics[width=16cm]{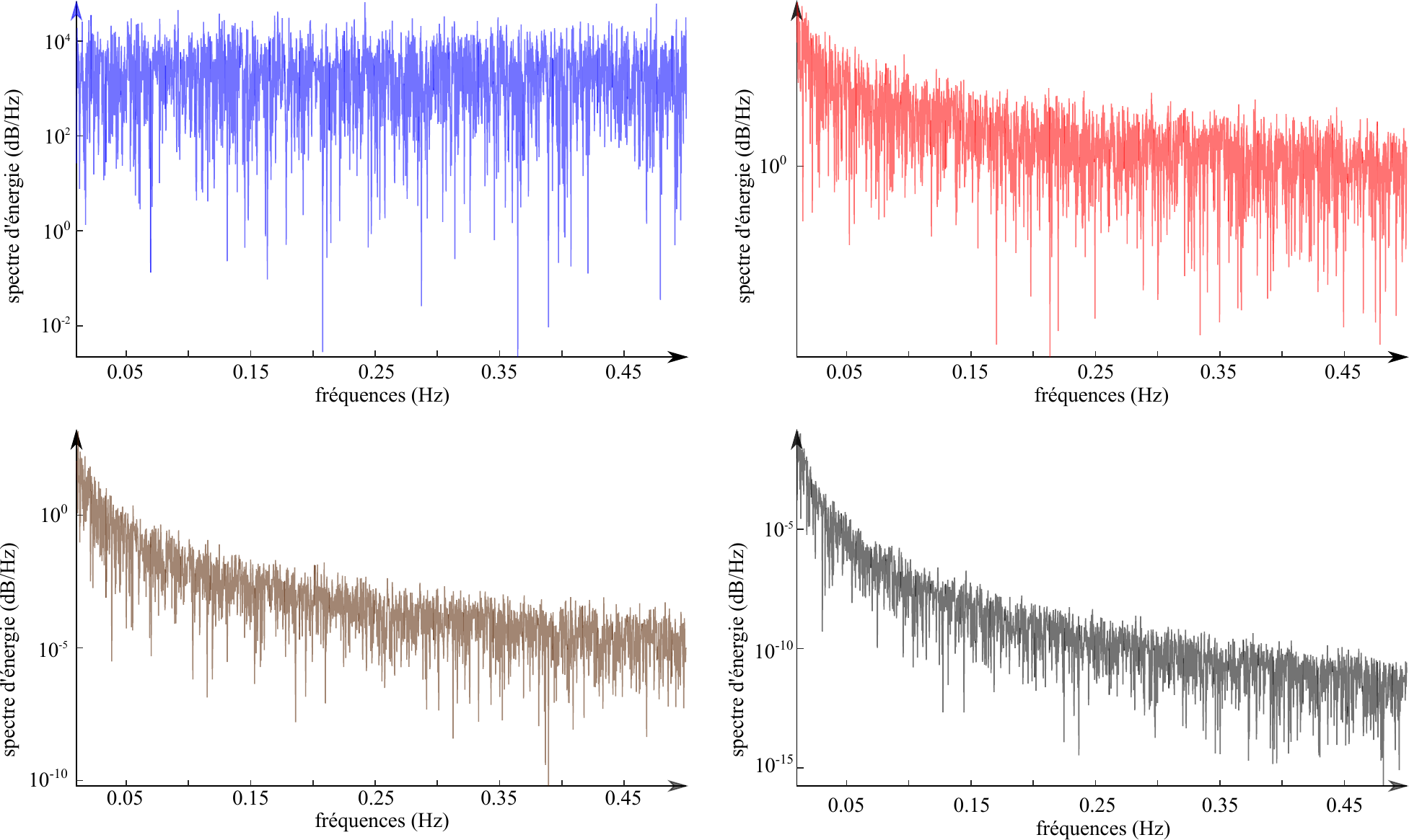}}
	\end{center}
	\caption{\textbf{1/$f$ Noise}. Representation of the energy spectra, in logarithmic scale, for white noise (blue curve), pink noise (red curve), \couleur{brownian} noise (brown curve), and black noise (black curve).
}
	\label{bruit_sig_02}
\end{figure}

To illustrate all that we have discussed, the program \couleur{ex\_bruit.m}, which utilizes the sub-function \couleur{fct\_bruit\_colore.m}, adds the various types of noise we have mentioned to a sinusoidal function. Figures (\ref{bruit_sig}) and (\ref{bruit_sig_02}) demonstrate the nature and effect of these noises on a given signal.

\section{White Noise}
\begin{figure}[H]
	\begin{center}
		\tcbox[colback=white]{\includegraphics[width=16cm]{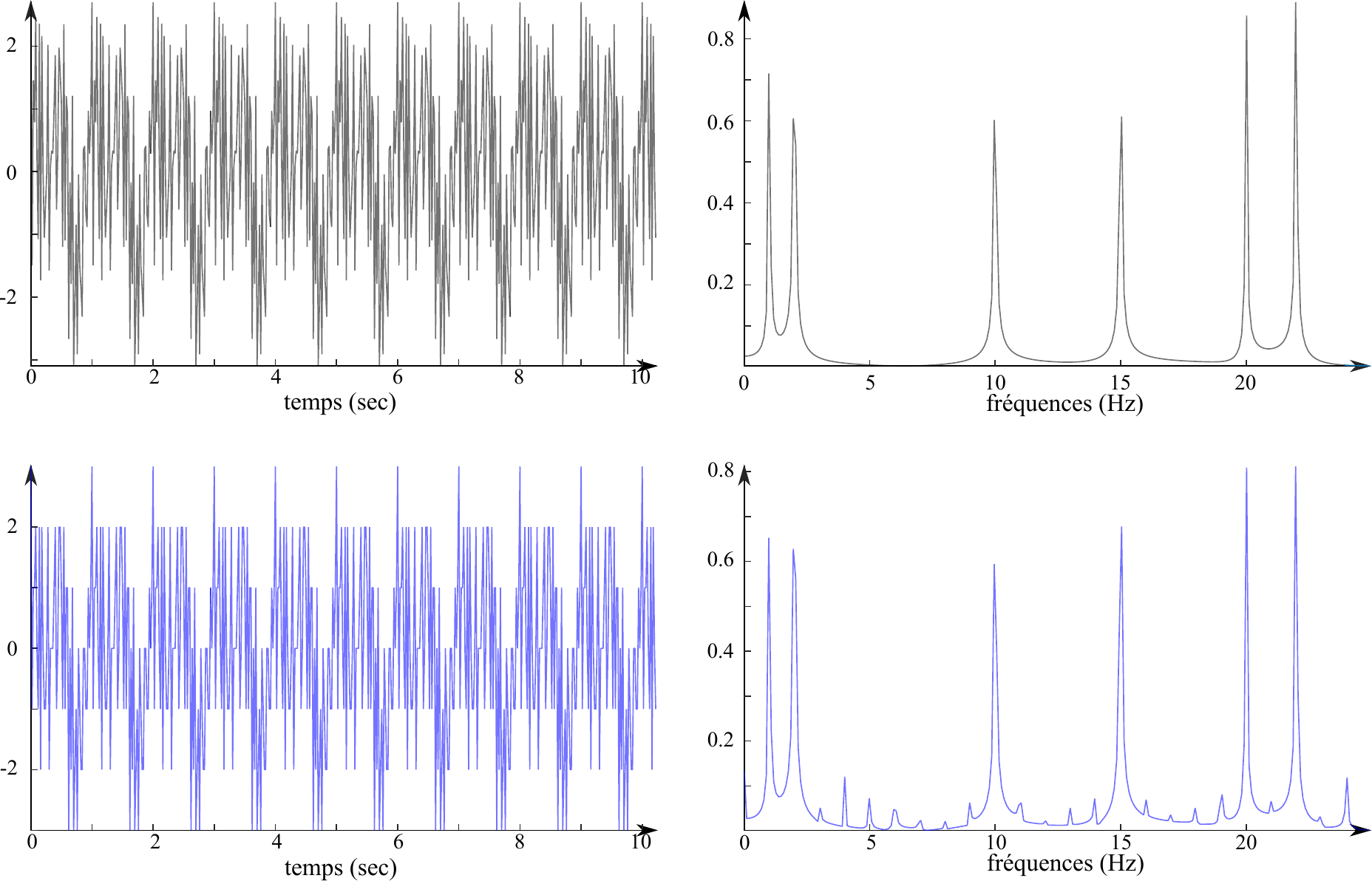}}
	\end{center}
	\caption{Quantization noise. The initial signal (top left). Its severe quantization (middle left) produces an amplitude spectrum (middle right) that is noisy compared to the initial spectrum (top right). Curves obtained with the program \couleur{ex\_bruit\_quantification.m}.}
	\label{bruit_quantif}
\end{figure}

\begin{figure}[H]
	\begin{center}
		\tcbox[colback=white]{\includegraphics[width=16cm]{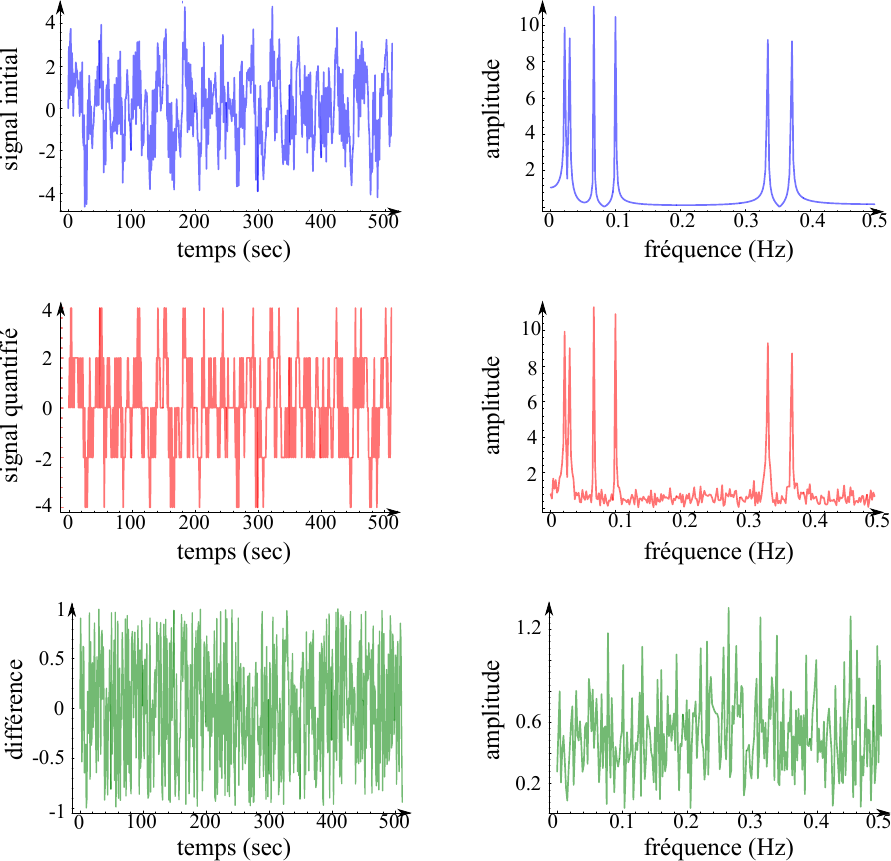}}
	\end{center}
	\caption{Quantization noise. The initial signal (top left) is still our "favorite" previously shown in several figures. Its severe quantization (middle left) yields an amplitude spectrum (middle right) that is noisy compared to the initial spectrum (top right). The difference between the two signals (bottom left) is the quantization noise, whose spectrum (bottom right) is close to that of white noise.}
	\label{bruit_quantif_02}
\end{figure}

White noise (blue curves in Figures \ref{bruit_sig} and \ref{bruit_sig_02}), $\left\{ b\right\} \left(t\right)$, is a stationary ergodic stochastic process in which successive values exhibit no correlation,

\begin{equation}
	r_{b,b}\left(\tau\right)=a^{2}\delta\left(\tau\right)
\end{equation}

where $a$ is a constant that sets the noise energy,
\begin{equation}
	\left|B\left(u\right)\right|^{2}=a^{2}
\end{equation}

which is uniformly distributed across the frequency axis (\cf blue curve in figure \ref{bruit_sig_02}), hence the term "white noise" by analogy with physical optics. With a correlation distance of zero, past values provide no information for predicting future values. A common example of nearly white noise is quantization noise (figures \ref{bruit_quantif} and \ref{bruit_quantif_02}), which is generated during coding operations where analog signals are transformed into digital signals through discretization. During such coding, analog values are assigned their digital counterparts, leading to a "rounding error," which constitutes the quantization noise $\left\{ q\right\} \left(t\right)$ whose probability density is approximately,
\begin{equation}
	\mathcal{P}\left\{ q\right\} =
	\begin{cases}
		1/\delta q \qquad si \qquad q\in\left[-\delta q/2;\delta q/2\right]\\
		\ \ 0 \qquad \quad si \qquad q\notin\left[-\delta q/2;\delta q/2\right]
	\end{cases} 
\end{equation}

where $\delta q$ is the quantization increment. The total energy (variance) of the noise,
\begin{equation}
	\begin{split}
	r_{q,q}\left(0\right) & =  \int_{-\infty}^{+\infty}\mathcal{P}\left\{ q\right\} q^{2}dq\\
 & =  \delta q^{2}/12
	\end{split}
\end{equation}

is uniformly distributed among the coefficients of the discrete \couleur{Fourier} transform of the sampled signal. Thus, if the series contains $256$ values and $\delta q = 0.10$, the average level of the energy spectrum is approximately $3 \times 10^{-6}$.

\section{Brownian Noise}
These noises are related to the "white" noise paradigm mentioned earlier. Even when limited to a few octaves, practically realizable "white" noises are very useful for describing stochastic processes such as those involved in \couleur{brownian} motion. If, in such motion, the increments $dx\left(t\right)$ are derived from "white" noise, then the position,
\begin{equation}
	x\left(t\right)=\int_{0}^{t}dx\left(\xi\right)d\xi
\end{equation}

will be a \couleur{brownian} noise (see Figures \ref{bruit_sig} and \ref{bruit_sig_02}, brown curves), such that,
\begin{equation}
	\left|X\left(u\right)\right|^{2}=X_{0}u^{-2}
\end{equation}

\warning{We will encounter \couleur{brownian} noises when the signal under study is the sum of random increments.}

\section{Pink Noise}
These noises (Figures \ref{bruit_sig} and \ref{bruit_sig_02}, red curves) are such that,
\begin{equation}
	E\left(u\right)=E_{0}u^{-1}
\end{equation}

\begin{figure}[H]
	\begin{center}
		\tcbox[colback=white]{\includegraphics[width=16cm]{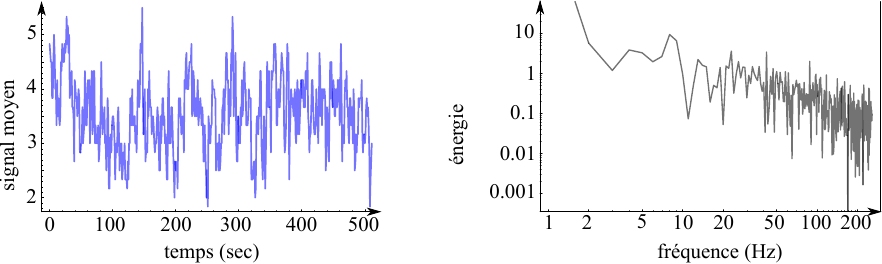}}
	\end{center}
	\caption{Pink noise generated by the superposition of simple processes with different time constants. In this example, the processes are six dice, with the first being rolled at each time increment, the second only every other time, the third every fourth time, \etc The average of the values displayed by the dice is calculated at each time increment, producing the noise on the left, whose amplitude spectrum (on the right) is reasonably pink. Thus, it is seen that the superposition of a small number of processes with different time constants easily produces pink noise. Such situations likely occur frequently in Nature.
}
	\label{bruit_rose}
\end{figure}

They are encountered in a wide range of situations, leading to the assertion that they play, with respect to $1/f$ noises, a role similar to that of the normal distribution with respect to statistical distributions. These noises have been noted for their aesthetic properties, and some authors have pointed out that many musical sounds exhibit "pink" spectra. Electronic noises generated by semiconductors are also "pink". Although not the only method, "pink" noises are easily created by superimposing relaxation processes with sufficiently different time constants (\cf Figure \ref{bruit_rose}).
\newpage 

\section{Black Noise}
These noises (figures \ref{bruit_sig} and \ref{bruit_sig_02}, brown curves) correspond to signals for which $\beta>2$, and they are often associated with "catastrophic" geophysical processes such as the floods of the Nile, whose level variations have a spectrum where $\beta=2.8$. "Black" noises have the particularity of possessing statistical persistence in accordance with the famous law of series; thus, the floods of the Nile occur in successive years as shown by \couleur{Harold Edwin Hurst} \shortcite{hurst1951long}. The Hurst exponent\index{Hurst exponent},
\begin{equation}
	H\equiv\frac{\log\left(R/\sqrt{S}\right)}{\log\left(T\right)}
\end{equation}

where $R$, $S$, and $T$ are respectively the maximum range, the variance, and the observation duration of the signal, allows for the measurement of the persistence of a statistical phenomenon. Moreover, $\beta=2H+1$.

\section{Stable Laws (\couleur{Gauss}, \couleur{Cauchy}, \etc)}
We have already reported that the probability density, $h$, of the sum of independent random variables is given by the convolution of the individual distributions $f$ and $g$. Therefore, in general, we have
\begin{equation}
	h\left(x+y\right)=f\left(x\right)*g\left(y\right)
\end{equation}

where the forms of the functions are \apriori arbitrary and different from one another. It is interesting to search for distributions that are invariant with respect to the above convolution; that is, distributions that yield the same distribution after convolution, up to a dilation and a translation. We seek functions,
\begin{equation}
	f_{x+y}\left(x+y\right)=f_{x}\left(x\right)*f_{y}\left(y\right)
\end{equation}

\begin{figure}[H]
	\begin{center}
		\tcbox[colback=white]{\includegraphics[width=16cm]{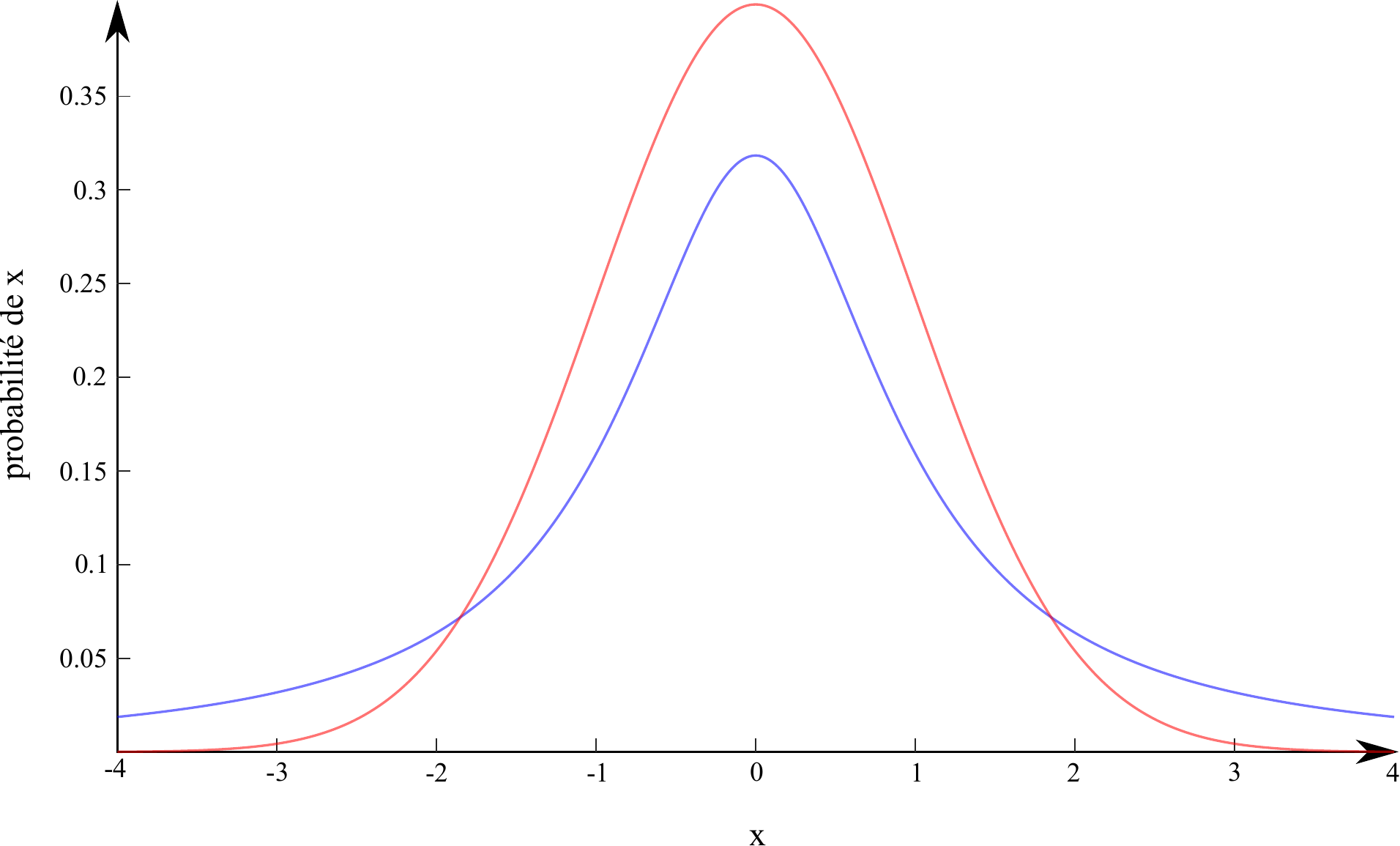}}
	\end{center}
	\caption{\textbf{Two stable laws}. The distributions of \couleur{Gauss} (red) and \couleur{Cauchy} (blue). Although these two distributions do not appear very different at first glance, the resulting statistical consequences are dramatically so, as you will see later in Figures \ref{book5a} and \ref{book5b}. These two curves were produced using the program \couleur{ex\_lois\_stables.m}, which calls the sub-functions \couleur{fct\_cauchy.m} and \couleur{fct\_normal.m}.}
	\label{lois_stables}
\end{figure}

\begin{figure}[H]
	\begin{center}
		\tcbox[colback=white]{\includegraphics[width=16cm]{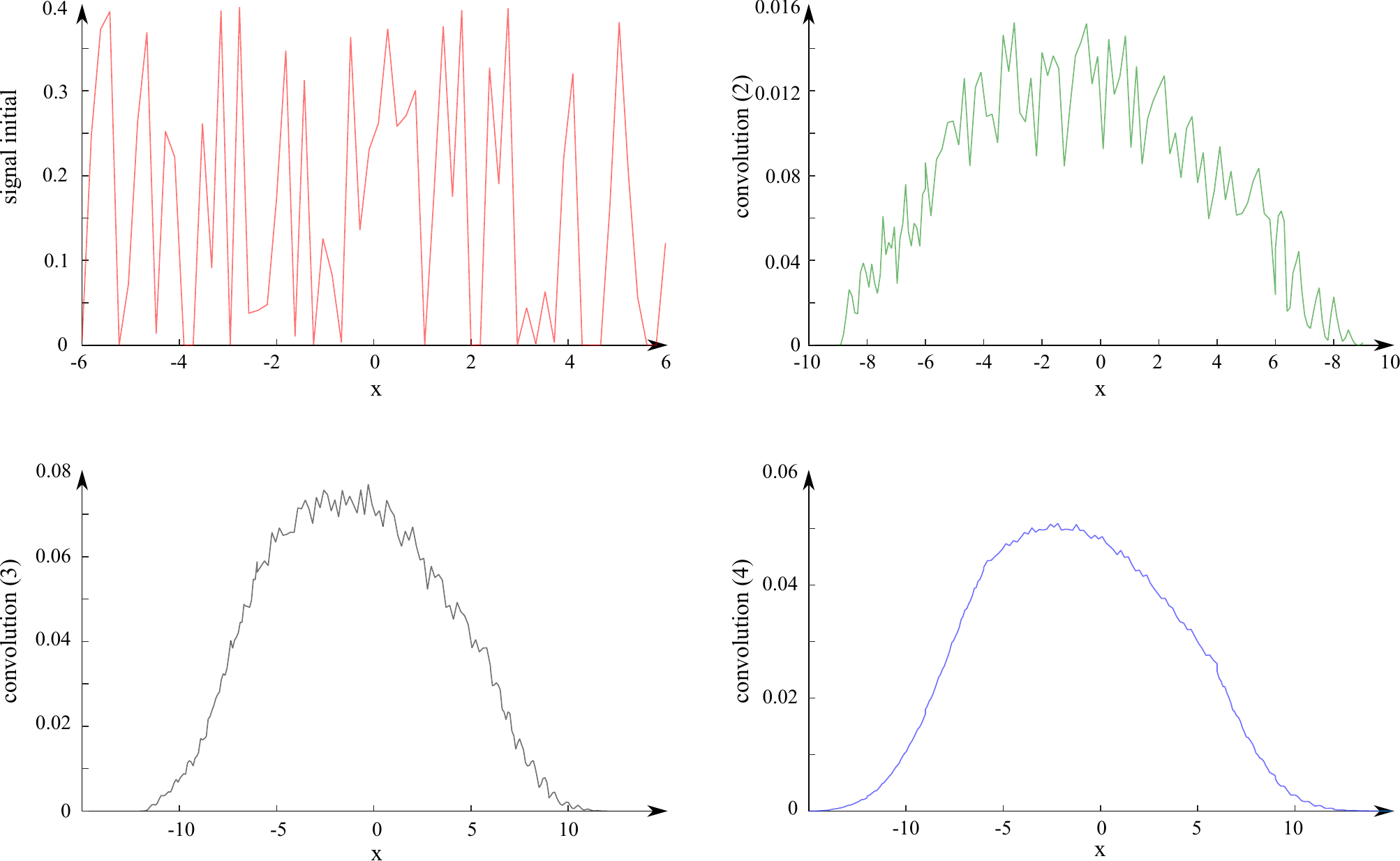}}
	\end{center}
	\caption{Illustration of the central limit theorem. The initial distribution (top left) has a finite variance and differs markedly from the \couleur{Gaussian} distribution with the same variance. Convolution of this initial distribution with itself (top right) yields a distribution that is already closer to the \couleur{Gaussian}. The triple autocorrelation (bottom left) and quadruple autocorrelation (bottom right) show that convergence is very rapid. Redo this figure by changing the initial distribution using the program \couleur{ex\_theoreme\_central\_limite.m}; you will see that the convergence is striking in almost all cases.}
	\label{theo_central}
\end{figure}

Such distributions have been termed 'stable laws' by the French mathematician \couleur{Paul Lévy} (1886--1971), and they serve as probabilistic attractors. The most well-known stable law is certainly \warning{the normal, or Gaussian, distribution} \index{normal law} (Figure \ref{lois_stables}),
\begin{equation}
	N\left(x\mid\mu_{x},\sigma_{x}^{2}\right)=\frac{1}{\sigma_{x}\sqrt{2\pi}}\exp\left[-\frac{\left(x-\mu_{x}\right)^{2}}{2\sigma_{x}^{2}}\right]
\end{equation}

where $\mu_{x}$ and $\sigma_{x}^{2}$ are the mean and the variance, respectively. We directly establish that,\begin{equation}
	\mathcal{P}\left\{ z=x+y\right\} =N\left(z\mid\mu_{z}=\mu_{x}+\mu_{y},\sigma_{z}^{2}=\sigma_{x}^{2}+\sigma_{y}^{2}\right)
\end{equation}

where it is always assumed that $x$ and $y$ are independent random variables. \warning{The convolution of normal laws is thus a normal law whose variance is equal to the sum of the variances and whose mean is equal to the sum of the means}. The normal law is well-known because it can be obtained as the limiting distribution of an infinite sum of independent variables whose distributions have finite variances; this is the consequence of the central limit theorem\index{central limit theorem} (Figure \ref{theo_central}).

It follows that the normal distribution is often used to describe the probabilistic behavior of physical measurements, with the reasoning being that these measurements incorporate a multitude of disturbances, whose sum is likely to conform to a normal statistic. While it is true that many autocorrelated distributions converge rapidly to the normal law, this should not be regarded as an absolute generalization, and there are cases where this is not the case.

The normal distribution is not the only stable law; in fact, there are infinitely many, including the \couleur{Cauchy} distribution\index{Cauchy law} (Figure \ref{lois_stables}, blue curve)
\begin{equation}
	C\left(x\mid m_{x},s_{x}\right)=\frac{s_{x}/\pi}{\left(x-m_{x}\right)^{2}+s_{x}^{2}}
\end{equation}

\begin{figure}[H]
	\begin{center}
		\tcbox[colback=white]{\includegraphics[width=16cm]{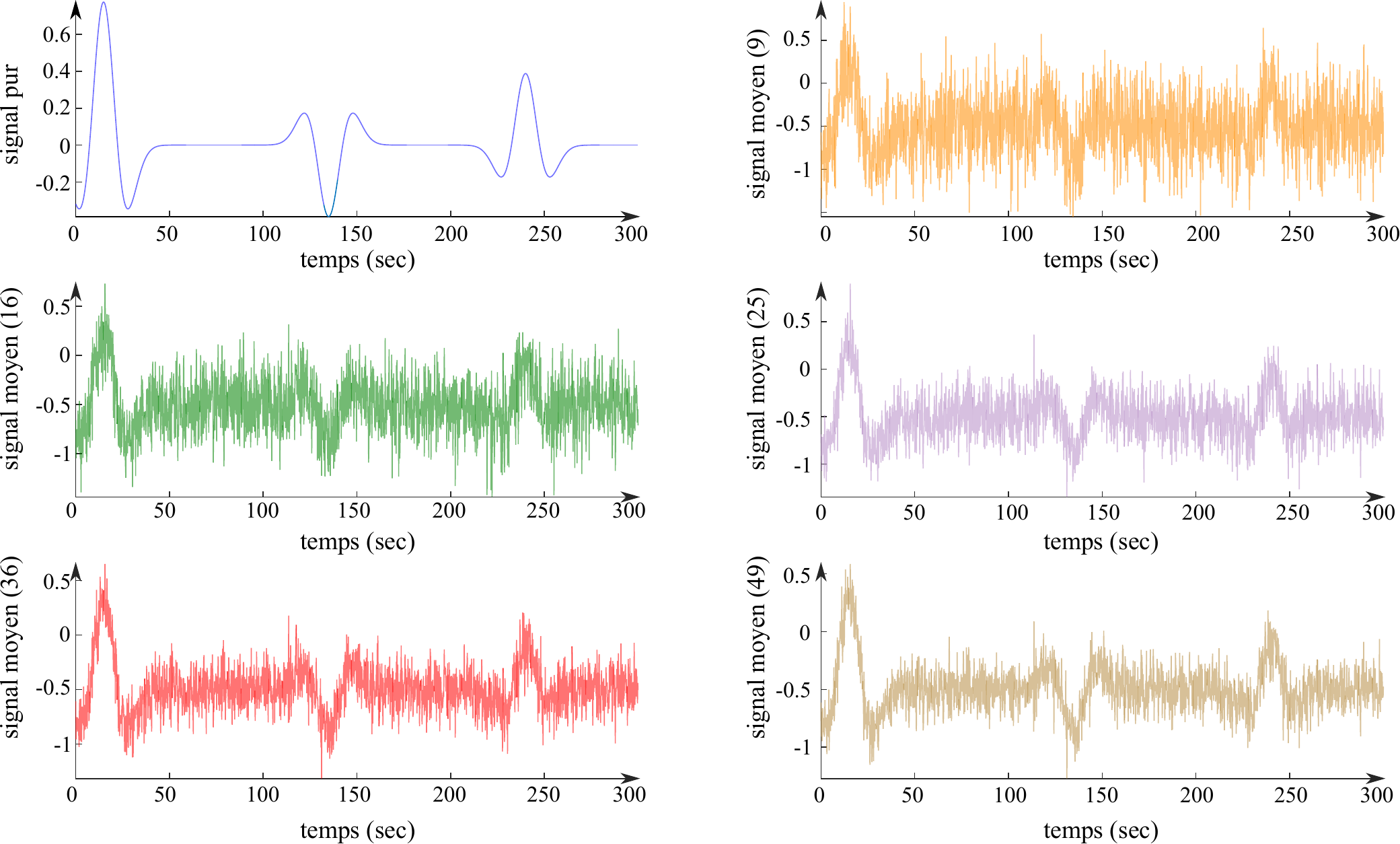}}
	\end{center}
	\caption{\textbf{Gaussian noise}. Addition of traces to increase the signal-to-noise ratio. In this example, the noise is white and \couleur{Gaussian}. The goal is to recover the initial signal (top left) by calculating the average of a certain number of noisy realizations. The average of 9 realizations (top right) allows for the identification of the first two events of the pure signal. Averages taken over more realizations (middle and bottom) improve the signal-to-noise ratio and allow for the recovery of the pure signal arrivals. An average taken over an infinite number of realizations converges stochastically to the pure signal.\label{bruit_gaussien}}
\end{figure}

\begin{figure}[H]
	\begin{center}
		\tcbox[colback=white]{\includegraphics[width=16cm]{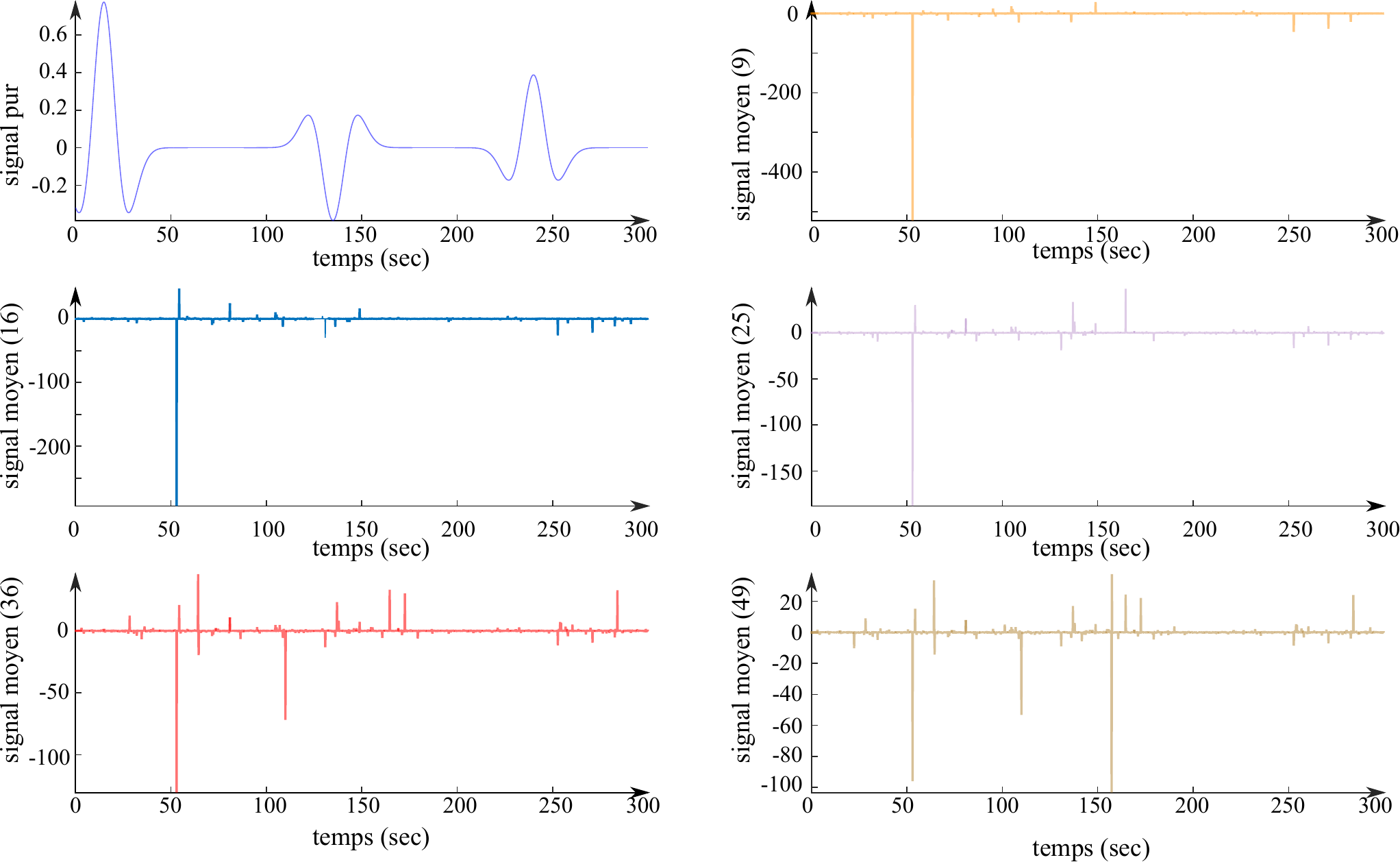}}
	\end{center}
	\caption{\textbf{Cauchy noise}. Addition of traces to increase the signal-to-noise ratio. The disaster becomes apparent when summing the traces; the signal-to-noise ratio increases dramatically!}
	\label{bruit_cauchien}
\end{figure}

It is \warning{interesting to note that the mean and variance are not defined for this distribution}. But that is not the worst part; in fact, it is easily shown that,
\begin{equation}
	\mathcal{P}\left\{ z=\frac{x+x}{2}\right\} =C\left(z\mid m_{z}=m_{x},s_{z}=s_{x}\right)
\end{equation}

that is, the mean of two variables following a \couleur{Cauchy} distribution also follows the same \couleur{Cauchy} distribution, and therefore exhibits the same dispersion around the median. This result is a significant issue because it indicates that, with respect to the \couleur{Cauchy} statistic, "unity does not strengthen" (Figures \ref{bruit_gaussien} and \ref{bruit_cauchien} obtained using the program \couleur{ex\_stack\_cauchy\_gauss.m}). This would not be problematic if the \couleur{Cauchy} distribution were not common; unfortunately, this is not the case. For example, the ratio of two independent variables with identical distributions follows the \couleur{Cauchy} statistic\footnote{Admittance calculators, coherence functions, and other transfer functions: beware!}

\chapter{\titrechap{Time-Frequency Duality}}
\minitoc
We will examine some correspondences that exist between a function and its \couleur{Fourier} transform. These correspondences provide a better understanding of the \couleur{Fourier} transformation and facilitate obtaining certain quick results about the function from its transform and \viceversa. For example, it is easily shown that,
\begin{equation}
	\int_{-\infty}^{+\infty}f\left(t\right)\mathrm{d}t=F\left(0\right)
\end{equation}

higher-order moments can be obtained by applying the "reverse differentiation theorem"\index{reverse differentiation theorem},
\begin{equation}
	\mathcal{F}\left[\left(-2i\pi t\right)^{n}f\left(t\right)\right]\left(u\right)=F^{\left(n\right)}\left(u\right)
\end{equation}

from which,
\begin{equation}
	\int_{-\infty}^{+\infty}t^{n}f\left(t\right)\mathrm{d}t=\frac{F^{\left(n\right)}\left(0\right)}{\left(-2i\pi\right)^{n}}
\end{equation}

Two particularly interesting cases are the first and second-order moments, which, when normalized by the zero-order moment, allow the calculation of the barycentric abscissa of a function,
\begin{equation}
	\left\langle t\right\rangle _{f}\equiv\frac{\int_{-\infty}^{+\infty}tf\left(t\right)\mathrm{d}t}{\int_{-\infty}^{+\infty}f\left(t\right)\mathrm{d}t}=-\frac{F^{\left(1\right)}\left(0\right)}{2i\pi F\left(0\right)}
\end{equation}

and the quadratic mean abscissa, which can also be viewed as a reduced moment of inertia,
\begin{equation}
	\left\langle t^{2}\right\rangle _{f}\equiv\frac{\int_{-\infty}^{+\infty}t^{2}f\left(t\right)\mathrm{d}t}{\int_{-\infty}^{+\infty}f\left(t\right)\mathrm{d}t}=-\frac{F^{\left(2\right)}\left(0\right)}{4\pi^{2}F\left(0\right)}
\end{equation}

The quadratic mean abscissa of a convolution product is easily calculated by noting that,
\begin{equation}
	\begin{split}
	\int_{-\infty}^{+\infty}t^{2}\left[f*g\right]\left(t\right)\mathrm{d}t & =  -\left.\frac{\mathcal{F}\left[f*g\right]^{\left(2\right)}\left(u\right)}{4\pi^{2}}\right|_{u=0}\\
 & =  -\left.\frac{F^{\left(2\right)}G+2F^{\left(1\right)}G^{\left(1\right)}+FG^{\left(2\right)}}{4\pi^{2}}\right|_{u=0}
	\end{split}
\end{equation}

and,
\begin{equation}
	\int_{-\infty}^{+\infty}\left[f*g\right]\left(t\right)\mathrm{d}t=F\left(0\right)G\left(0\right),
\end{equation}

from which the desired result follows,
\begin{equation}
	\begin{split}
\left\langle t^{2}\right\rangle _{f*g} & =  -\frac{1}{4\pi^{2}}\left(\frac{F^{\left(2\right)}\left(0\right)}{F\left(0\right)}+\frac{G^{\left(2\right)}\left(0\right)}{G\left(0\right)}+2\frac{F^{\left(1\right)}\left(0\right)}{F\left(0\right)}\frac{G^{\left(1\right)}\left(0\right)}{G\left(0\right)}\right)\\
 & =  \left\langle t^{2}\right\rangle _{f}+\left\langle t^{2}\right\rangle _{g}+2\left\langle t\right\rangle _{f}\times\left\langle t\right\rangle _{g}.	
	\end{split}
\end{equation}

If one of the two functions has its barycentric abscissa at the origin, we recover the rule of additivity of variances. We invite the reader to focus on the section related to this analogy in the work by \shortciteN{bracewell1986fourier},
\begin{equation}
	\left\langle t^{2}\right\rangle _{f*g}=\left\langle t^{2}\right\rangle _{f}+\left\langle t^{2}\right\rangle _{g}.
\end{equation}

\section{Measuring signal duration}
We have already considered the notion of the duration of a signal when introducing the \couleur{Dirac} impulse. This notion was unambiguous because we used the window function, whose bounded support defines the duration unequivocally. Signals with unbounded support have infinite duration; however, many of these signals have most of their energy concentrated in time, and it is then possible to associate a finite duration with them, which can be termed the effective duration\index{effective duration}. The challenge is to choose a method for calculating this duration; the simplest approach is to adopt the duration of a window that has the same ordinate at the origin and the same zero-order moment as the signal in question,
\begin{equation}
	\begin{split}
	D_{f} & \equiv  \frac{\int_{-\infty}^{+\infty}f\left(t\right)\mathrm{d}t}{f\left(0\right)}\\
 & =  \frac{F\left(0\right)}{\int_{-\infty}^{+\infty}F\left(u\right)\mathrm{d}u}\\
 & =  \frac{1}{D_{F}},
	\end{split}
\end{equation}

from which a first duality relationship follows,
\begin{equation}
	D_{f}\times D_{F}=1
\end{equation}

to be compared with the similarity theorem\index{similarity theorem} (see paragraph \ref{similitude}), which states that a dilation of the time axis corresponds to a contraction of the frequency axis. However, this definition of duration is not satisfactory as it is not invariant under translation; such invariance can be achieved by using the autocorrelation function, which is known to have a maximum at the origin. One then defines,
\begin{equation}
	\begin{split}
	D_{f\Diamond f} & \equiv  \frac{\int_{-\infty}^{+\infty}\left[f\Diamond f\right]\left(\tau\right)\mathrm{d}\tau}{\left[f\Diamond f\right]\left(0\right)}\\
 & =  \frac{\int_{-\infty}^{+\infty}f^{*}\left(\tau\right)\mathrm{d}\tau\int_{-\infty}^{+\infty}f\left(\tau\right)\mathrm{d}\tau}{\int_{-\infty}^{+\infty}f^{*}\left(\tau\right)f\left(\tau\right)\mathrm{d}\tau}\\
 & =  \frac{\left|F\left(0\right)\right|^{2}}{\int_{-\infty}^{+\infty}\left|F\left(u\right)\right|^{2}\mathrm{d}u}\\
 & =  \frac{1}{D_{\left|F\right|^{2}}}
	\end{split}
\end{equation}

where the infinite limits of the integral in the numerator of the first line allow, through a change of variable, a rewriting in the form of a product of two integrals. Thus, a new \warning{correlation-energy duality}\index{correlation-energy duality} relationship has been obtained,
\begin{equation}
	D_{f\Diamond f}\times D_{\left|F\right|^{2}}=1
\end{equation}

The duration of the autocorrelation has the drawback of involving the signal indirectly; therefore, several authors have introduced an alternative definition based on the calculation of the 'moment of inertia'\index{moment of inertia of a signal} and the 'center of gravity' of the signal\index{center of gravity of a signal},
\begin{equation}
	\left(D_{\left|f\right|^{2}}^{2}\right)^{2}\equiv\frac{\int_{-\infty}^{+\infty}t^{2}\left|f\left(t\right)\right|^{2}\mathrm{d}t}{\int_{-\infty}^{+\infty}\left|f\left(t\right)\right|^{2}\mathrm{d}t}-\left(\frac{\int_{-\infty}^{+\infty}t\left|f\left(t\right)\right|^{2}\mathrm{d}t}{\int_{-\infty}^{+\infty}\left|f\left(t\right)\right|^{2}\mathrm{d}t}\right)^{2}
\end{equation}

This definition is satisfactory in many cases and often yields results that align with intuition. We will see that adopting this definition leads to an uncertainty relation identical to that of \couleur{Heisenberg}\index{Heisenberg boxes} in quantum physics.

\section{The Uncertainty Principle in Signal Processing}
\subsection{Deterministic Approach}
This principle, universal in Physics, also applies in signal processing whenever the \couleur{Fourier} transform is involved, thus allowing an analogy with wave phenomena. Before addressing the uncertainty relation as it is known in Quantum Physics, let us consider the case of a truncated sinusoid using a window. In this case, the \couleur{Fourier} transform of the sinusoid of infinite duration, composed of two \couleur{Dirac} impulses, is replaced by two sinc functions. The intuition\footnote{In this regard, reading the passage concerning this issue in the "\textit{Feynman Lectures on Physics}" \shortcite{feynman2011feynman} is instructive.} suggests that the uncertainty in the frequency of the sinusoid is something like half the width of the central lobe of the sinc function,
\begin{equation}
	\delta u\approx\frac{1}{T}
\end{equation}

where $T$ is the duration of the truncated signal. This definition is identical to that used by optical scientists, who define the resolution of an optical instrument as the radius of the first dark ring in the diffraction pattern; it corresponds to the duration $D_{f}$ discussed in the previous section. Adopting the quadratic duration $D_{\left|f\right|^{2}}^{2}$ allows us to obtain the \couleur{Heisenberg} uncertainty relation—which has already been addressed in this book, \cf relation (\ref{incertitude_heis})—in honor of its originator.

\couleur{Werner Heisenberg} (1901-1976) was born in Würzburg (Germany) and studied theoretical physics in Munich (\couleur{Sommerfeld}) and Göttingen (\couleur{Born}). He published a dozen articles on atomic physics between 1922 and 1924, and during the same period (1923), he completed his doctoral thesis on turbulent flows. It was in 1924 that he met \couleur{Wolfgang Pauli} and \couleur{Niels Bohr}, and the following year he laid the foundations for a new mechanics of atomic systems. The continuation of this work led to the famous uncertainty principle (1927) \shortcite{heisenberg1927anschaulichen}, which earned him the Nobel Prize in 1932. After the war, \couleur{Heisenberg} reconstructed the \couleur{Max Planck} Institute of Physics; he then worked on a relativistic quantum field theory (with \couleur{Pauli}), a theory of nuclear structure motivated by the discovery of the neutron (1932), and a meson field theory (with \couleur{Yukawa} in 1935) which was confirmed by the discovery of the meson (1947) in cosmic rays. He briefly returned to the study of turbulent flows in 1948

Let us restrict ourselves to the case of a signal whose barycentric abscissa, as well as that of its \couleur{Fourier} transform, are at the origin. We then have,
\begin{equation}
	\begin{split}
\left(D_{\left|f\right|^{2}}^{2}\times D_{\left|F\right|^{2}}^{2}\right)^{2} & =  \frac{\int_{-\infty}^{+\infty}t^{2}\left|f\left(t\right)\right|^{2}\mathrm{d}t}{\int_{-\infty}^{+\infty}\left|f\left(t\right)\right|^{2}\mathrm{d}t}\times\frac{\int_{-\infty}^{+\infty}u^{2}\left|F\left(u\right)\right|^{2}\mathrm{d}u}{\int_{-\infty}^{+\infty}\left|F\left(u\right)\right|^{2}\mathrm{d}u}\\
 & =  \frac{\int_{-\infty}^{+\infty}\left|tf\left(t\right)\right|^{2}\mathrm{d}t}{\int_{-\infty}^{+\infty}\left|f\left(t\right)\right|^{2}\mathrm{d}t}\times\frac{\int_{-\infty}^{+\infty}\left|uF\left(u\right)\right|^{2}\mathrm{d}u}{\int_{-\infty}^{+\infty}\left|F\left(u\right)\right|^{2}\mathrm{d}u}\\
 & =  \frac{\int_{-\infty}^{+\infty}\left|tf\left(t\right)\right|^{2}\mathrm{d}t\int_{-\infty}^{+\infty}\left|f^{\left(1\right)}\left(t\right)\right|^{2}\mathrm{d}t}{4\pi^{2}\left(\int_{-\infty}^{+\infty}\left|f\left(t\right)\right|^{2}\mathrm{d}t\right)^{2}},
 \end{split}
\end{equation}

where we have used the \couleur{Parseval} theorem and then the differentiation relation. By employing the \couleur{Schwarz} inequality (\cf relation \ref{inegalite_schwarz}),
\begin{equation}
	\begin{split}
\left(D_{\left|f\right|^{2}}^{2}\times D_{\left|F\right|^{2}}^{2}\right)^{2} & \geq  \frac{\left(\int_{-\infty}^{+\infty}\left[tf^{*}\left(t\right)f^{\left(1\right)}\left(t\right)+tf\left(t\right)f^{\left(1\right)*}\left(t\right)\right]\mathrm{d}t\right)^{2}}{\left(4\pi\right)^{2}\left(\int_{-\infty}^{+\infty}\left|f\left(t\right)\right|^{2}\mathrm{d}t\right)^{2}}\\
 & =  \frac{\left(\int_{-\infty}^{+\infty}t\left[f\left(t\right)f^{*}\left(t\right)\right]^{\left(1\right)}\mathrm{d}t\right)^{2}}{\left(4\pi\right)^{2}\left(\int_{-\infty}^{+\infty}\left|f\left(t\right)\right|^{2}\mathrm{d}t\right)^{2}}\\
 & =  \frac{\left(\int_{-\infty}^{+\infty}f\left(t\right)f^{*}\left(t\right)\mathrm{d}t\right)^{2}}{\left(4\pi\right)^{2}\left(\int_{-\infty}^{+\infty}\left|f\left(t\right)\right|^{2}\mathrm{d}t\right)^{2}}\\
 & =  \frac{1}{\left(4\pi\right)^{2}}
\end{split}
\end{equation}

where the transition from the second to the third line involves integration by parts. A final evident simplification leads to the \couleur{Heisenberg} relation,
\begin{equation}
	D_{\left|f\right|^{2}}^{2}\times D_{\left|F\right|^{2}}^{2}\geq\frac{1}{4\pi}
\end{equation}

The equality is achieved (optimal time-frequency resolution) by the Gaussians that we will encounter in the chapter on wavelets. More generally, the uncertainty principle indicates that it is illusory to claim an infinitely good resolution simultaneously in time and frequency; \warning{the observation is unavoidable, the "\couleur{Dirac} monochromatic" does not exist!}

The \couleur{Schwarz} inequality is demonstrated as follows. Let $F\left(u\right)$ and $G\left(u\right)$, be two functions, and a real constant $\varepsilon$. We have,
\begin{equation}
	\int_{-\infty}^{+\infty}\left|F\left(u\right)+\varepsilon G\left(u\right)\right|^{2}\mathrm{d}u>0.\end{equation}

After expansion, this expression becomes,
\begin{equation}
	\int_{-\infty}^{+\infty}\left|F\left(u\right)\right|^{2}\mathrm{d}u+\varepsilon\int_{-\infty}^{+\infty}\left[F^{*}\left(u\right)G\left(u\right)+F\left(u\right)G^{*}\left(u\right)\right]\mathrm{d}u+\varepsilon^{2}\int_{-\infty}^{+\infty}\left|G\left(u\right)\right|^{2}\mathrm{d}u>0,
\end{equation}

which is a quadratic polynomial in $\varepsilon$ that, to remain always positive, must have a non-positive discriminant, that is to say, such that,
\begin{equation}
\left[\int_{-\infty}^{+\infty}\left[F^{*}\left(u\right)G\left(u\right)+F\left(u\right)G^{*}\left(u\right)\right]\mathrm{d}u\right]^{2}\leq4\int_{-\infty}^{+\infty}\left|F\left(u\right)\right|^{2}\mathrm{d}u\int_{-\infty}^{+\infty}\left|G\left(u\right)\right|^{2}\mathrm{d}u
\label{inegalite_schwarz}
\end{equation}

which is the sought inequality.

\section{Causal signal duality}
Causal signals, which are identically zero at negative times, are common in signal processing. They can be expressed in the form,
\begin{equation}
	f_{c}\left(t\right)=\textrm{H}\left(t\right)f\left(t\right)
\end{equation}

and thus we have,
\begin{eqnarray}
	F_{c}\left(u\right) & = & \frac{1}{2}\left[\delta\left(u\right)-\frac{i}{\pi u}\right]*F\left(u\right)\\
 & = & \frac{1}{2}\left[F\left(u\right)-i\left(\frac{1}{\pi u}\right)*F\left(u\right)\right]
\end{eqnarray}

Using the definition of the \couleur{Hilbert} transform, relation (\ref{transformee_hilbert}), one obtains the \couleur{Bayard-Bode} relation\index{Bayard-Bode relation},
\begin{equation}
	F_{c}\left(u\right)=\frac{1}{2}\left[F\left(u\right)-i\mathcal{H}\left[F\right]\left(u\right)\right],\end{equation}

which indicates that the \couleur{Fourier} transform of a causal signal has an imaginary part equal to the negative of the \couleur{Hilbert} transform of the real part. This property is used to rapidly compute the numerical \couleur{Hilbert} transform of signals using the fast \couleur{Fourier} transform algorithm.

\section{Minimum Delay Signals}
\subsection{Utility of Minimum Delay Signals}
The objective of this section is to provide some clarifications regarding a class of signals frequently encountered in geophysics, particularly in seismic deconvolution, known as "minimum delay signals" or "minimum phase signals." This class of signals is somewhat shrouded in mystery, and the numerous conversations we have had with "specialists" on the subject lead us to believe that a straightforward presentation of these signals is not without value. These signals can be introduced in an extremely formal manner\footnote{For example, in the work by \couleur{E.R. Robinson}, \textit{"Seismic Deconvolution"}.}, but we prefer to adopt an approach more connected to physical principles. The basic principle justifying the use of minimum delay signals is to observe that, when excited by a source of energy, physical systems arrange to reemit this energy as quickly as the laws describing their behavior allow. This mode of operation relies on the principles of least action, which form the foundation of physics. It turns out that this principle of optimal energy restitution can serve as an additional constraint, proving very useful for regularizing certain signal processing problems. In practice, it is necessary to have a "measure" of a signal's duration that allows quantifying the "rapidity" of energy restitution. We have seen that several choices are possible, and we will adopt the quadratic measure\footnote{Le $4\pi^{2}$  is included merely to simplify some of the expressions that will follow.}.
\begin{equation}
	R\left(f\right)\equiv4\pi^{2}\int_{0}^{+\infty}t^{2}\left|f\left(t\right)\right|^{2}\mathrm{d}t.
\end{equation}

Now, the problem we wish to solve is as follows: let $\left\{ f_{n},\left(t\right)\right\}$ be a collection of causal signals, all of which have the same amplitude spectrum $\left|F\left(u\right)\right|$. The objective is to find, within this collection, the unique signal $f_{\min}\left(t\right)$ such that,
\begin{equation}
	R\left(f_{\min}\right)\;\;\;\mathrm{MINIMUM}
\end{equation}

This signal is referred to as the minimum-delay signal associated with the collection. Given that the amplitude spectrum is fixed, the signals $f_{n}\left(t\right)$ differ by their phase spectra $\phi_{n}\left(u\right)$. It follows that the minimum-energy-delay constraint, which operates in the time domain, should be accompanied by a condition on the phase in the frequency domain. Hence the term "minimum-phase signal"\index{minimum-phase signal}.

\section{Case of Continuous Signals}
Let us begin with the case of continuous signals whose \couleur{Fourier} transform
\begin{equation}
	F_{n}\left(u\right)=\left|F\left(u\right)\right|\exp\left[i\phi_{n}\left(u\right)\right]
\end{equation}

The energy-delay measurement then takes the form,
\begin{equation}
	\begin{split}
	R\left(f_{n}\right) & =  \int_{-\infty}^{+\infty}\left|F_{n}^{\left(1\right)}\left(u\right)\right|^{2}\mathrm{d}u\\
 & =  \int_{-\infty}^{+\infty}\left|\left[\left|F\left(u\right)\right|^{\left(1\right)}+i\left|F\left(u\right)\right|\phi_{n}^{\left(1\right)}\left(u\right)\right]\exp\left[i\phi_{n}\left(u\right)\right]\right|^{2}\mathrm{d}u\\
 & =  \int_{-\infty}^{+\infty}\left[\left|F\left(u\right)\right|^{\left(1\right)}\right]^{2}\mathrm{d}u+\int_{-\infty}^{+\infty}\left|F\left(u\right)\right|^{2}\left[\phi_{n}^{\left(1\right)}\left(u\right)\right]^{2}\mathrm{d}u
	\end{split}
\end{equation}

Energy recovery is as fast as possible when the last term on the right-hand side is minimized,
\begin{equation} 		
\int_{-\infty}^{+\infty}\left[\phi_{\min}^{\left(1\right)}\left(u\right)\right]^{2}\mathrm{d}u\;\;\mathrm{MINIMUM}
\end{equation}

The signal with minimal energy-delay must also be of minimal phase variation, that is, as least dispersed as possible.

\section{Case of Discrete Signals}
Consider now a discrete signal comprising $L+1$ values,
\begin{equation}
	s=\left\{ \mathbf{s}_{0},s_{1},\cdots,s_{L}\right\} 
\end{equation}

The factorization of the $Z$-transform of this signal shows that it can be generated by convolving $L$ dipoles,\begin{equation}
	\left\{ \mathbf{s}_{0},s_{1},\cdots,s_{L}\right\} =\left\{ \alpha_{1},\beta_{1}\right\} *\left\{ \alpha_{2},\beta_{2}\right\} *\cdots*\left\{ \alpha_{L},\beta_{L}\right\}
\end{equation}

The amplitude spectrum of this signal is equal to the product of the amplitude spectra of the dipoles,
\begin{equation}
	\left|\mathcal{F}\left\{ \mathbf{s}_{0},s_{1},\cdots,s_{L}\right\} \right|=\prod_{l=1}^{L}\left|\alpha_{l}+\beta_{l}Z\right|
\end{equation}

Noting that,
\begin{equation}
	\left|\alpha_{l}+\beta_{l}Z\right|=\left|\beta_{l}+\alpha_{l}Z\right|
\end{equation}

we observe that the $2^{L}$ signals generated by convolving the $L$ dipoles, whether inverted or not, have $\left|\mathcal{F}\left\{ \mathbf{s}{0},s{1},\cdots,s_{L}\right\} \right|$ as their amplitude spectrum. Each dipole offers the alternative,
\begin{equation}
\left\{ \alpha_{l},\beta_{l}\right\} \;\; ou\;\;\left\{ \beta_{l},\alpha_{l}\right\} 
\end{equation}

depending on whether it is inverted or not. Among these two dipoles, the one with the largest absolute value for the first term is the minimum-delay dipole. For example, among,
\begin{equation}
	\left\{ 1,-2\right\} \;\; et\;\;\left\{ -2,1\right\} 
\end{equation}

it is the dipole $\left\{ -2,1\right\}$ that is of minimum delay. We then observe that among the $2^{L}$ signals that can be created from the initial dipoles, there is one that corresponds to the particular case where all the dipoles are of minimum delay. This signal, which we will denote by $\left\{ \mathbf{s}{0},s{1},\cdots,s_{L}\right\}{\min}$, is called the minimum-delay signal associated with $\left\{ \mathbf{s}{0},s_{1},\cdots,s_{L}\right\}$. It has the distinguishing feature of possessing an amplitude spectrum identical to that of the initial signal.

Now let us examine the phase spectrum of the minimum-delay signal. Knowing that,
\begin{equation}
	\arg\left(\mathcal{F}\left\{ \mathbf{s}_{0},s_{1},\cdots,s_{L}\right\} \right)=\sum_{l=1}^{L}\arg\left(\alpha_{l}+\beta_{l}Z\right)
\end{equation}

we are led to examine the phase spectra of the dipoles generating the signal. The phase spectrum of the minimum-delay dipole, $\left\{ \alpha_{l},\beta_{l}\right\} $, is given by,
\begin{equation}
	\begin{split}
\phi_{\min,l}\left(u\right) & \equiv  \arg\left(\alpha_{l}+\beta_{l}Z\right)\\
 & =  -\arctan\left[\frac{\beta_{l}\sin\left(2\pi u\tau\right)}{\alpha_{l}+\beta_{l}\cos\left(2\pi u\tau\right)}\right]
	\end{split}
\end{equation}

and,
\begin{equation}
	\begin{split}
	\phi_{\min,l}^{\left(1\right)}\left(u\right) & \equiv  \frac{\textrm{d}}{\textrm{d}u}\arg\left(\alpha_{l}+\beta_{l}Z\right)\\
 & =  -\frac{2\pi\tau\left[\beta_{l}^{2}+\alpha_{l}\beta_{l}\cos\left(2\pi u\tau\right)\right]}{\alpha_{l}^{2}+\beta_{l}^{2}+2\alpha_{l}\beta_{l}\cos\left(2\pi u\tau\right)}
	\end{split}
\end{equation}

An identical calculation applied to the inverted dipole\footnote{Also known as the maximum-delay dipole.}, $\left\{  \beta_{l},\alpha_{l}\right\} $, yields,
\begin{equation}
	\begin{split}
	\phi_{\max,l}^{\left(1\right)}\left(u\right) & \equiv  \frac{\textrm{d}}{\textrm{d}u}\arg\left(\beta_{l}+\alpha_{l}Z\right)\\
 & =  -\frac{2\pi\tau\left[\alpha_{l}^{2}+\alpha_{l}\beta_{l}\cos\left(2\pi u\tau\right)\right]}{\alpha_{l}^{2}+\beta_{l}^{2}+2\alpha_{l}\beta_{l}\cos\left(2\pi u\tau\right)}.\\
 & =  \phi_{\min,l}^{\left(1\right)}\left(u\right)-\frac{2\pi\tau\left(\alpha_{l}^{2}-\beta_{l}^{2}\right)}{\alpha_{l}^{2}+\beta_{l}^{2}+2\alpha_{l}\beta_{l}\cos\left(2\pi u\tau\right)}
	\end{split}
\end{equation}

Note that,
\begin{equation}
	\left|\alpha_{l}\right|>\left|\beta_{l}\right|\Longrightarrow\alpha_{l}^{2}+\alpha_{l}\beta_{l}\cos\left(2\pi u\tau\right)>0
\end{equation}

and,
\begin{equation}
	\alpha_{l}^{2}+\beta_{l}^{2}+2\alpha_{l}\beta_{l}\cos\left(2\pi u\tau\right)>0
\end{equation}

These inequalities reveal that,
\begin{equation}
	\phi_{\max,l}^{\left(1\right)}\left(u\right)<0
\end{equation}

that is, the phase $\phi_{\max,l}^{\left(1\right)}\left(u\right)$ of the maximum-delay dipole is a monotonically decreasing function. This is not the case for the phase of the minimum-delay dipole, which can be either increasing or decreasing. The triangular inequality
\begin{equation}
	\left|a+b\right|\leq\left|a\right|+\left|b\right|
\end{equation}

applied to the relation,
\begin{equation}
\phi_{\min,l}^{\left(1\right)}\left(u\right)=\phi_{\max,l}^{\left(1\right)}\left(u\right)+\frac{2\pi\tau\left(\alpha_{l}^{2}-\beta_{l}^{2}\right)}{\alpha_{l}^{2}+\beta_{l}^{2}+2\alpha_{l}\beta_{l}\cos\left(2\pi u\tau\right)}
\end{equation}

yields,
\begin{equation}
\left|\phi_{\min,l}^{\left(1\right)}\left(u\right)\right|\leq\left|\phi_{\max,l}^{\left(1\right)}\left(u\right)\right|+\frac{2\pi\tau\left(\alpha_{l}^{2}-\beta_{l}^{2}\right)}{\alpha_{l}^{2}+\beta_{l}^{2}+2\alpha_{l}\beta_{l}\cos\left(2\pi u\tau\right)}
\end{equation}

Since,
\begin{equation}
	\frac{2\pi\tau\left(\alpha_{l}^{2}-\beta_{l}^{2}\right)}{\alpha_{l}^{2}+\beta_{l}^{2}+2\alpha_{l}\beta_{l}\cos\left(2\pi u\tau\right)}>0
\end{equation}

\begin{equation}
	\left|\phi_{\min,l}^{\left(1\right)}\left(u\right)\right|<\left|\phi_{\max,l}^{\left(1\right)}\left(u\right)\right|
\end{equation}

Returning to the case of the complete signal, we have observed that the phase,
\begin{equation}
	\arg\left(\mathcal{F}\left\{ \mathbf{s}_{0},s_{1},\cdots,s_{L}\right\} \right)=\sum_{l=1}^{L}\phi_{l}\left(u\right)
\end{equation}

which immediately gives,
\begin{equation}
	\frac{\mathrm{d}}{\mathrm{d}u}\arg\left(\mathcal{F}\left\{ \mathbf{s}_{0},s_{1},\cdots,s_{L}\right\} \right)=\sum_{l=1}^{L}\phi_{l}^{\left(1\right)}\left(u\right)
\end{equation}

L'inégalité triangulaire permet d'obtenir que,
\begin{equation}
	\left|\frac{\mathrm{d}}{\mathrm{d}u}\arg\left(\mathcal{F}\left\{ \mathbf{s}_{0},s_{1},\cdots,s_{L}\right\} \right)\right|\leq\sum_{l=1}^{L}\left|\phi_{l}^{\left(1\right)}\left(u\right)\right|,
\end{equation}

and, using the results obtained for the minimum-delay dipole, it follows directly that,
\begin{equation}
	\left|\frac{\mathrm{d}}{\mathrm{d}u}\arg\left(\mathcal{F}\left\{ \mathbf{s}_{0},s_{1},\cdots,s_{L}\right\} _{\min}\right)\right|<\left|\frac{\mathrm{d}}{\mathrm{d}u}\arg\left(\mathcal{F}\left\{ \mathbf{s}_{0},s_{1},\cdots,s_{L}\right\} \right)\right|
\end{equation}

This inequality indicates that the minimum-delay signal associated with a collection of signals generated by $L$ dipoles is the signal with the slowest phase variation, that is, the signal with the least possible dispersion.

\section{The Cepstral Domain}\index{domaine cepstral}
The title of this subsection is not the result of typographical dyslexia, but indeed introduces one of the most intriguing aspects of the time-frequency duality. The "cepstral" domain \shortcite{oppenheim2004frequency} is the realm of homomorphic deconvolution\index{homomorphic deconvolution}, where "quefrency", "liftering", "sispha", and "alanysis" reign. It involves transforming a time-domain signal into another domain, analogous to time, using the properties of real and complex logarithms. The real cepstrum utilizes only the amplitude of the signal's spectrum, and by neglecting its phase, it becomes impossible to reconstruct all the initial information. With the complex logarithm \shortcite{oppenheim1965superposition}, it becomes possible to accurately reconstruct both the phase and amplitude of the original signal. The "cepstral transform" $\mathcal{C}(\tau)$ of a time-domain signal $y(t)$ is given by the following relation,
\begin{equation}
	\mathcal{C}(\tau) = \mathcal{F}(\mathtt{ln}(|\mathcal{F}(y(t))|))
	\label{eq_cepstre}
\end{equation}

This definition (\ref{eq_cepstre}), from an algorithmic perspective, can be expressed in the form,
\begin{equation}
	\begin{split}
   			&\hat{y}(u) = \mathcal{F}(y(t)) \\
   			&\mathcal{L}y(u) = \mathtt{ln}(r_{\hat{y}}) + \sqrt{-1}*\phi_{\hat{y}} \\
   			&\mathcal{C}(\tau) = \mathcal{R}(\mathcal{F}^{-1}(\mathcal{L}y))
   \end{split}
   \label{cepstre_algo}
\end{equation}

$\mathcal{F}$ and $\mathcal{F}^{-1}$ are the direct and inverse Fourier transforms, respectively, whose magnitude and phase are represented by $r$ and $\phi$.

The idea here is to use the fact that a recorded signal, such as a seismic wave, results from the convolution of a source (\eg a Ricker wavelet) with the impulse response of the medium (\eg a distribution of reflectors). Since cepstral analysis allows us to transition from the data space to a space where the cepstra of the two convolved elements are simply superimposed and added, if their supports are "sufficiently distant", it will be possible to separate and reconstruct either the propagated source or the medium's response by simply canceling out a part of the total cepstrum. One way to approach this is to consider the following situation: the wave propagates through a medium rich in reflectors. The more reflectors there are, the less the cepstral supports will be overlapping. The program \couleur{ex\_deconv\_homo.m} illustrates this concept."
\begin{figure}[H]
	\begin{center}
		\tcbox[colback=white]{\includegraphics[width=16cm]{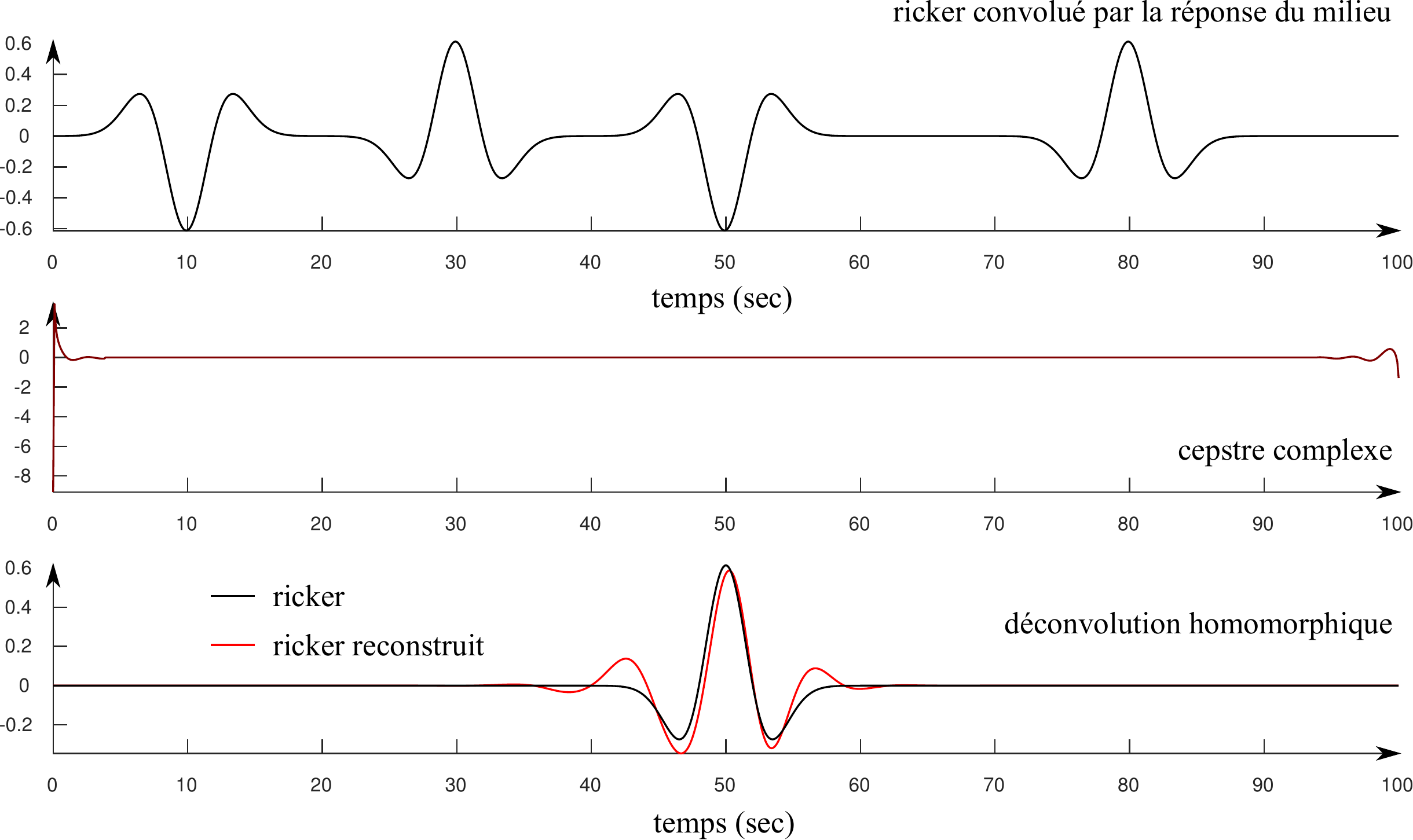}}
	\end{center}
	\caption{Homomorphic Deconvolution. The first figure illustrates the signal to be analyzed. It simply results from the convolution of a Ricker wavelet, shown in black in the bottom figure, with a sequence of positive (+1) or negative (-1) reflectors. The cepstrum, obtained using the relations (\ref{cepstre_algo}), is shown in the center in brown. By canceling out the part of this curve corresponding to the Green's function of the medium, it is possible to reconstruct the source through an inverse cepstral transform. The result of the deconvolution (red Ricker in the bottom figure) is superimposed on the original source (black Ricker). }
	\label{ex_cepstre}
\end{figure}
\chapter{\titrechap{Linear Filtering}}
\minitoc
This chapter deals exclusively with linear filtering applied to signals through a convolution operation \shortcite{kanasewich1981time}. We have already encountered this type of filtering when studying linear systems, where the output signal is a filtered version of the input signal. A linear filter is fully characterized by its transfer function, which is the \couleur{Fourier} transform of its impulse response. The magnitude of the transfer function, known as the gain, indicates which frequencies will be attenuated, preserved, or amplified. Traditionally, examining the gain allows filters to be classified as low-pass, high-pass, band-pass, or all-pass; however, geophysics also employs numerous filters that do not fit these categories, such as potential field extension operators, pole reduction filters, \etc The ideal low-pass filter is a rectangular function,
\begin{equation}
	\Pi\left(\frac{u}{2u_{b}}\right)
\end{equation}

with an impulse response that is a sinc function (\cf figure (\ref{sin_cardinal}))
\begin{equation}
	2u_{b}\textrm{sinc}\left(2u_{b}t\right)
\end{equation}

High-pass or band-pass filters can be constructed in a similar manner, and they all share the drawback of having an oscillatory impulse response with a decay that is slower the more abrupt the cutoff of their gain. These ideal filters are impractical and their discretization makes them perform poorly in practice. In particular, the "rectangular" filters exhibit oscillations near the edges known as the \couleur{Gibbs} phenomenon. Most of the time, the filters used have a real impulse response, and often it is required that \warning{they be additionally non-phase-shifting, which is not possible if the filter is causal}

\section{Filters and $Z$-Transforms}
We will see how the $Z$-transform\index{Z-transform} allows us to study and practically design digital filters applicable to sampled signals. Consider the discrete convolution,
\begin{equation}
	s_{n}\equiv s\left(n\tau\right)=\sum_{k=-\infty}^{+\infty}f_{k}e_{n-k}
\end{equation}

where $f_{k}$ is a discrete filter whose characteristics we wish to determine. After applying the $Z$-transform, the convolution becomes
\begin{equation}
	S\left(Z\right)=F\left(Z\right)E\left(Z\right)
\end{equation}

For example, filtering,
\begin{equation}
	s_{n}=\frac{1}{5}\sum_{k=-\infty}^{+\infty}e_{n-k}
\end{equation}

corresponding to a moving average over 5 values has the filter\footnote{In the following, we will denote the value of discrete filters at time zero in boldface.}
\begin{equation}
	f=\left\{ \frac{1}{5};\frac{1}{5};\frac{1}{5};\frac{1}{5};\frac{1}{5}\right\} 
\end{equation}

whose $Z$-transform is,
\begin{equation}
	F\left(Z\right)=\frac{1}{5}\left(Z^{-2}+Z^{-1}+1+Z+Z^{2}\right)
\end{equation}

It is then possible to calculate the gain of the filter, given by,
\begin{equation}
	\left|F\left(Z\right)\right|=\sqrt{F\left(Z\right)F^{*}\left(1/Z\right)}
\end{equation}

which, after expansion on the unit circle of \couleur{Fourier}, \ie, in the particular case where Z=e$^{2i\pi u\tau}$,
\begin{equation}
	\left|F\left(u\right)\right|=\sqrt{1+\frac{8}{5}\cos\left(2\pi u\tau\right)+\frac{6}{5}\cos\left(4\pi u\tau\right)+\frac{4}{5}\cos\left(6\pi u\tau\right)+\frac{2}{5}\cos\left(8\pi u\tau\right)}
	\label{ex_gain}
\end{equation}

The expression (\ref{ex_gain}) allows the calculation of the gain for any frequency within the \couleur{Shannon} interval $\left[-1/2\tau;+1/2\tau\right]$ and a comparison with the gain of the ideal continuous filter consisting of a rectangular pulse of 5 seconds duration when $\tau=1$.
\begin{figure}[H]
	\begin{center}
		\tcbox[colback=white]{\includegraphics[width=16cm]{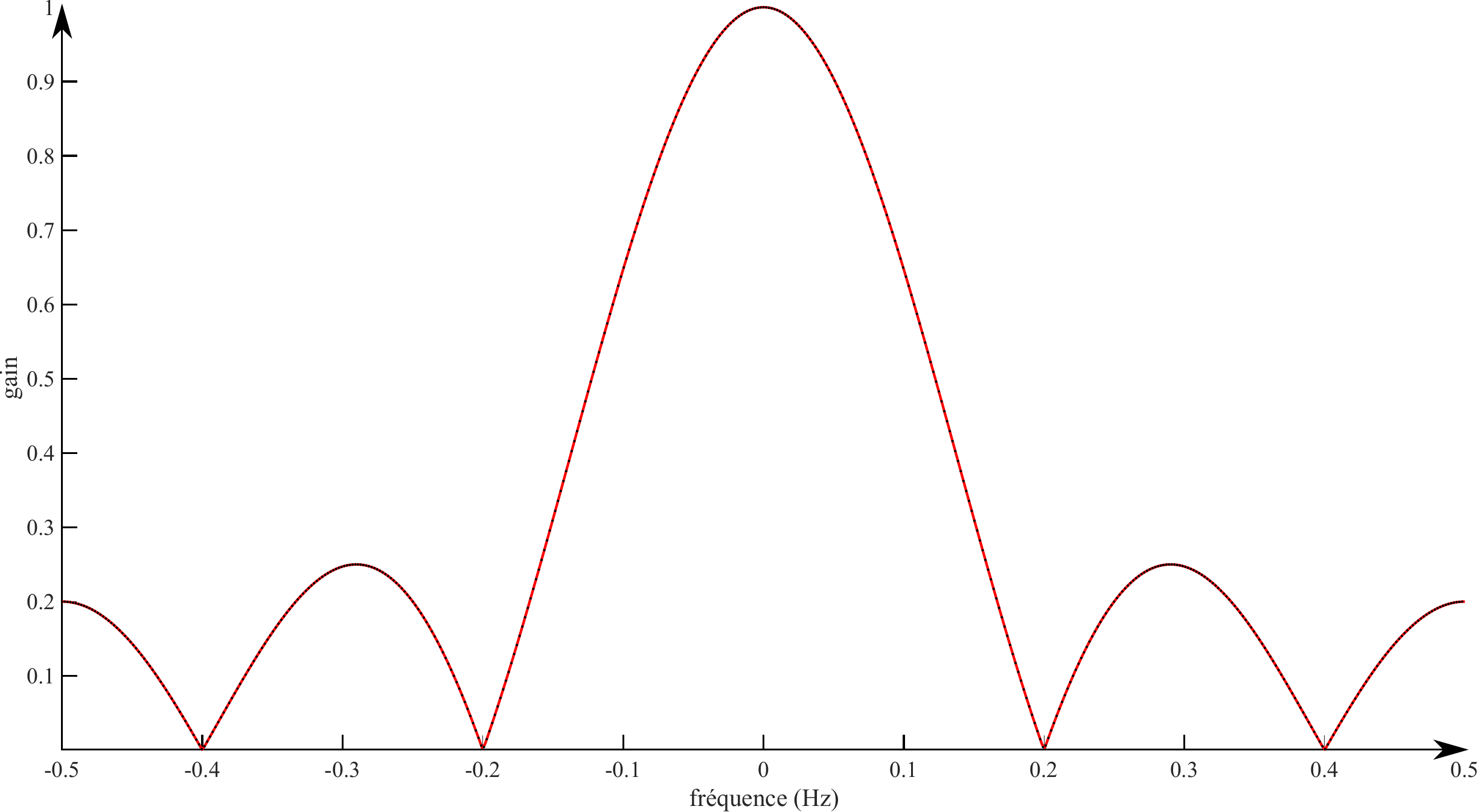}}
	\end{center}
	\caption{Gain of the discrete filter (1,1,1,1,1)/5. Comparison between the gain given by expression (\ref{ex_gain}), in red, and that of a perfect filter (black dashed lines) obtained by a 5-second duration rectangular pulse. (\couleur{ex\_gain\_transformeeZ.m}) \label{tz_gain}}
\end{figure}

One can take any filter,
\begin{equation}
	f=\left\{ +864;-144;+186;-55;-79;+4;+4\right\}
\end{equation}

calculate its $Z$-transform,
\begin{equation}
	F\left(Z\right)=864-144Z+186Z^{2}-55Z^{3}-79Z^{4}+4Z^{5}+4Z^{6}
\end{equation}

and factorize it,
\begin{equation}
	F\left(Z\right)=\left(4+Z\right)\left(4-Z\right)\left(-3+2iZ\right)\left(-3-2iZ\right)\left(2-Z\right)\left(3+Z\right)
\end{equation}

\warning{in order to decompose the initial filter as a cascade of dipoles}. This operation allows for easier study of the filter characteristics based on those of the dipoles; thus, the stability analysis of the overall filter can be performed. If one of the dipoles is numerically unstable, the entire filter will also be unstable. The filters we have just discussed consist of a sequence, more or less long, of numerical values that are convolved with the signal to be processed. This is why they are called finite impulse response filters\footnote{"FIR" in Anglo-Saxon terminology.}\index{finite impulse response filter} as opposed to infinite impulse response filters\footnote{"IIR" in Anglo-Saxon terminology.}\index{infinite impulse response filter} which we will now encounter.

\section{Operator and Filters in Numerical Analysis}
Finite difference operators are widely used filters in numerical analysis for solving partial differential equations. There is a whole range of filters that approximate the ideal operator to varying degrees or possess particular qualities (causal, anti-causal, \etc). Two widely used second derivative operators are,
\begin{equation}
	\left\{ 1;-2;1\right\} 
\end{equation}

and,
\begin{equation}
	\left\{ -\frac{1}{12};\frac{15}{12};-\frac{28}{12};\frac{15}{12};-\frac{1}{12}\right\}. 
\end{equation}

\begin{figure}[H]
	\begin{center}
		\tcbox[colback=white]{\includegraphics[width=16cm]{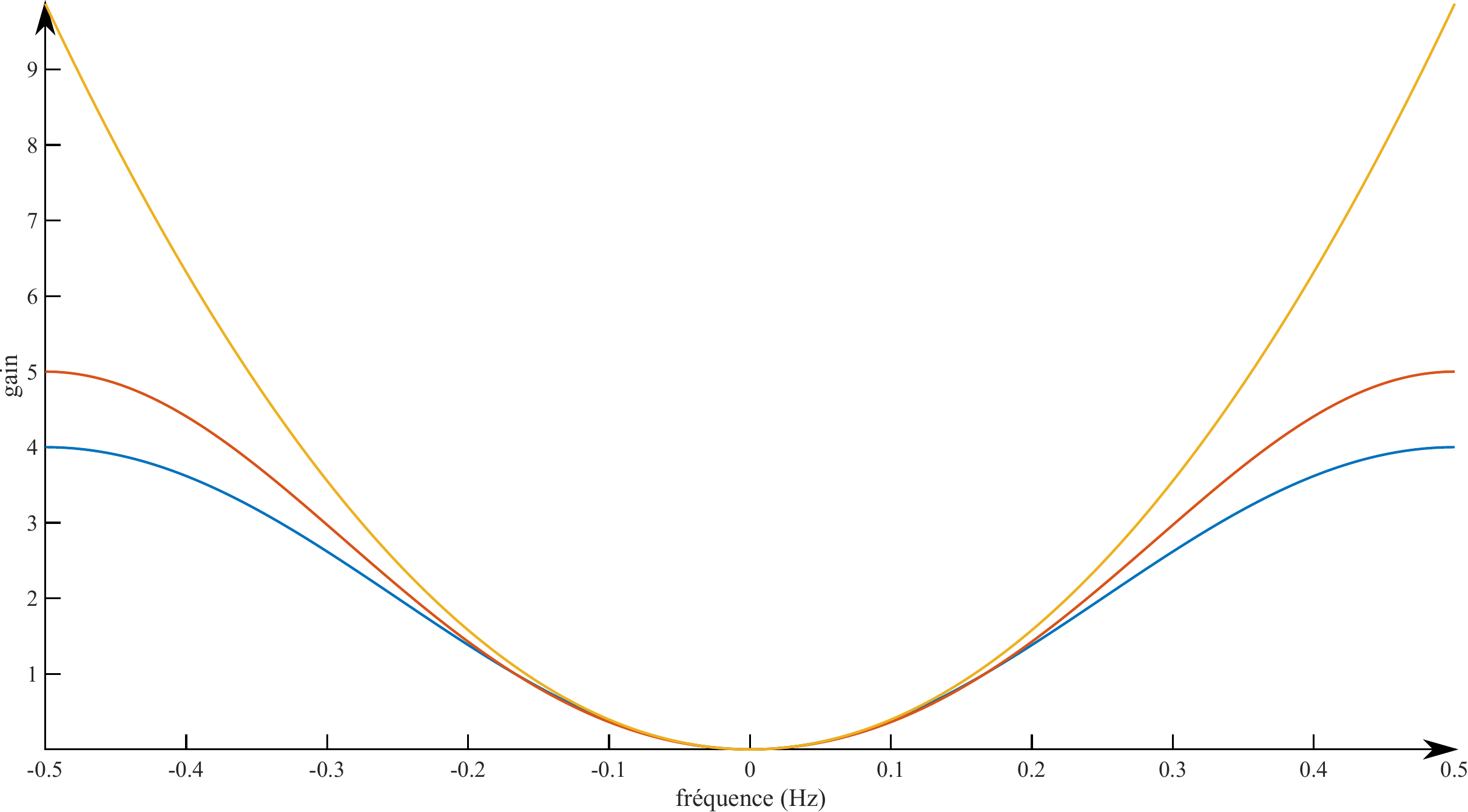}}
	\end{center}
	\caption{Gains of finite difference operators. This example shows the gains (solid lines) of the two second derivative operators, $\left\{ 1;-\textbf{2};1\right\}$ and $\left\{-\frac{1}{12};\frac{15}{12};-\frac{\textbf{28}}{\textbf{12}};\frac{15}{12};-\frac{1}{12}\right\}$. The yellow curve represents the gain of the ideal operator, $\left(2\pi u\right)^{2}$, which is best approximated by the 5-term filter (red curve). Both filters are non-phase-shifting as they are centered on the origin. The calculations assume a unit sampling step, and you will notice that proper use of the filters is only possible if limited to low frequencies, approximately Nyquist/5 for the less efficient operator, which requires sampling the signals with a finer step than that dictated by the \couleur{Shannon} rule. \label{tz_gains}}
\end{figure}

The $Z$-transform allows for the calculation of the gain and phase of these filters (\cf figure \ref{tz_gains}) and to deduce their characteristics. In particular, it is easy to define the frequency domain of validity for the discrete operator. The program \couleur{ex\_gain\_tz.m} produces the following results,

\section{Narrowband Filters}
\subsection{Recursiveness and Infinite Impulse Response}
The band-pass filter with the narrowest bandwidth is the one that only retains a particular frequency,
\begin{equation}
	F\left(u\right)=\delta\left(u+u_{0}\right)+\delta\left(u-u_{0}\right)
\end{equation}

\begin{equation}
	f\left(t\right)=\exp\left(-2i\pi u_{0}t\right)+\exp\left(+2i\pi u_{0}t\right)
\end{equation}

\begin{figure}[H]
	\begin{center}
		\tcbox[colback=white]{\includegraphics[width=16cm]{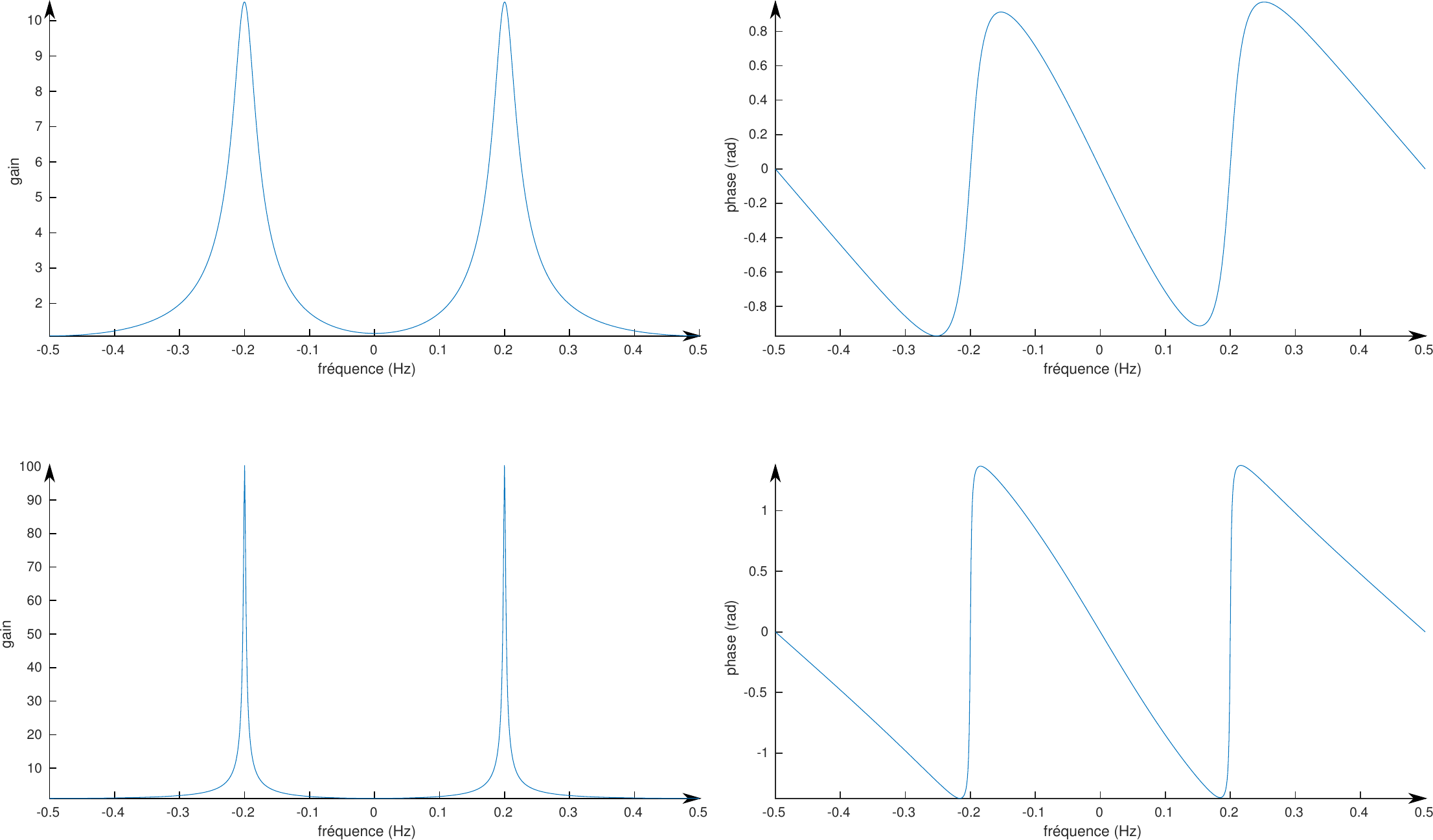}}
	\end{center}
	\caption{Gain and phase of the narrowband filter. The upper filter was constructed with $\epsilon=0.1$, and the lower one with $\epsilon=0.01$. (\couleur{ex\_bande\_etroite.m}) \label{bande_etroite}}
\end{figure}

A causal discretization of this filter provides,
\begin{equation}
	\begin{split}
	F\left(Z\right) & =  1+ZZ_{0}+\left(ZZ_{0}\right)^{2}+\left(ZZ_{0}\right)^{3}+\ldots\\
 &   +1+Z/Z_{0}+\left(Z/Z_{0}\right)^{2}+\left(Z/Z_{0}\right)^{3}+\ldots\\
 & =  1/\left(1-ZZ_{0}\right)+1/\left(1-Z/Z_{0}\right)
	\end{split}
\end{equation}

with $Z_{0}=\exp\left(-2i\pi u_{0}\tau\right)$. Note that writing the $Z$-transform of the filter as a ratio of polynomials allows for the manipulation of an infinite impulse response. The above expression shows that calculating the filter's gain will pose a numerical problem at $u=\pm u_{0}$. This is because the values for which the denominator of $F\left(Z\right)$ is zero, known as the poles\index{poles of the Z-transform}, lie on the unit circle. When $Z$ traverses this circle, the poles are encountered, leading to numerical issues. The desired filter is not realizable in its current form and must be modified to eliminate these numerical difficulties. The solution is to place the poles just off the unit circle so that the gain is no longer infinite. The trade-off is that the filter will no longer be as perfect as initially desired (\cf figure \ref{bande_etroite}). Thus, let us set
\begin{equation}
	Z_{0}^{^{\prime}}=\left(1-\epsilon\right)Z_{0}
\end{equation}

and,
\begin{equation}
	Z_{0}^{^{\prime\prime}}=\left(1-\epsilon\right)/Z_{0}
\end{equation}

with $\epsilon>0$, we then obtain the modified filter,
\begin{equation}
	\begin{split}
	F^{^{\prime}}\left(Z\right) & = & \frac{1}{1-ZZ_{0}^{^{\prime}}}+\frac{1}{1-ZZ_{0}^{^{\prime\prime}}}\\
 & = & \frac{\alpha_{0}+\alpha_{1}Z}{1+\beta_{1}Z+\beta_{2}Z^{2}}
	\end{split}
\end{equation}

with,
\begin{equation}
	\begin{cases}
		\alpha_{0} \qquad = \qquad 2\\
		\alpha_{1} \qquad = \qquad -2\left(1-\epsilon\right)\cos\left(2\pi u_{0}\tau\right)\\
		\beta_{1} \qquad = \qquad -2\left(1-\epsilon\right)\cos\left(2\pi u_{0}\tau\right)\\
		\beta_{2} \qquad = \qquad \left(1-\epsilon\right)^{2}
	\end{cases}
\end{equation}

The filtering operation can be expressed as a product of $Z$-transforms,
\begin{equation}
	S\left(Z\right)=E\left(Z\right)F^{^{\prime}}\left(Z\right)
\end{equation}

that is, using the filter's expression,
\begin{equation}
	S\left(Z\right)=E\left(Z\right)\left(\alpha_{0}+\alpha_{1}Z\right)-ZS\left(Z\right)\left(\beta_{1}+\beta_{2}Z\right)
\end{equation}

Réécrivons cette expression en développant chaque terme,
\begin{equation}
	\begin{array}{c}
		S\left(Z\right)\\
		\Updownarrow\\
		\left[
		\begin{array}{c}
			s_{0}\\
			+\\
			s_{1}Z\\
			+\\
			s_{2}Z^{2}\\
			+\\
			\vdots
		\end{array}\right]
	\end{array}=
	\begin{array}{c}
		\alpha_{0}E\left(Z\right)\\
		\Updownarrow\\
		\left[
		\begin{array}{c}
			\alpha_{0}e_{0}\\
			+\\
			\alpha_{0}e_{1}Z\\
			+\\
			\alpha_{0}e_{2}Z^{2}\\
			+\\
			\vdots
		\end{array}\right]
	\end{array}+
	\begin{array}{c}
		\alpha_{1}ZE\left(Z\right)\\
		\Updownarrow\\
		\left[
		\begin{array}{c}
			\\\\\alpha_{1}e_{0}Z\\
			+\\
			\alpha_{1}e_{1}Z^{2}\\
			+\\
			\vdots\end{array}\right]
		\end{array}-
		\begin{array}{c}
			\beta_{1}ZS\left(Z\right)\\
			\Updownarrow\\
			\left[
			\begin{array}{c}
				\\\\\beta_{1}s_{0}Z\\
				+\\
				\beta_{1}s_{1}Z^{2}\\
				+\\
				\vdots
			\end{array}\right]
		\end{array}-
		\begin{array}{c}
			\beta_{2}Z^{2}S\left(Z\right)\\
			\Updownarrow\\
			\left[
			\begin{array}{c}
				\\\\\\\\\beta_{2}s_{0}Z^{2}\\
				+\\
				\vdots
			\end{array}\right]
		\end{array}.
\end{equation}

Since the equality must hold for all $Z$, it is necessary that it holds individually for each power of $Z$, that is, for each term in the expression above. This leads to the recursive expressions,
\begin{equation}
	\begin{cases}
		s_{0} \qquad = \qquad \alpha_{0}e_{0}\\
		s_{1} \qquad = \qquad \alpha_{0}e_{1}+\alpha_{1}e_{0}-\beta_{1}s_{0}\\
		s_{n} \qquad = \qquad \alpha_{0}e_{n}+\alpha_{1}e_{n-1}-\beta_{1}s_{n-1}-\beta_{2}s_{n-2}
	\end{cases}
\end{equation}

\begin{figure}[H]
	\begin{center}
		\tcbox[colback=white]{\includegraphics[width=16cm]{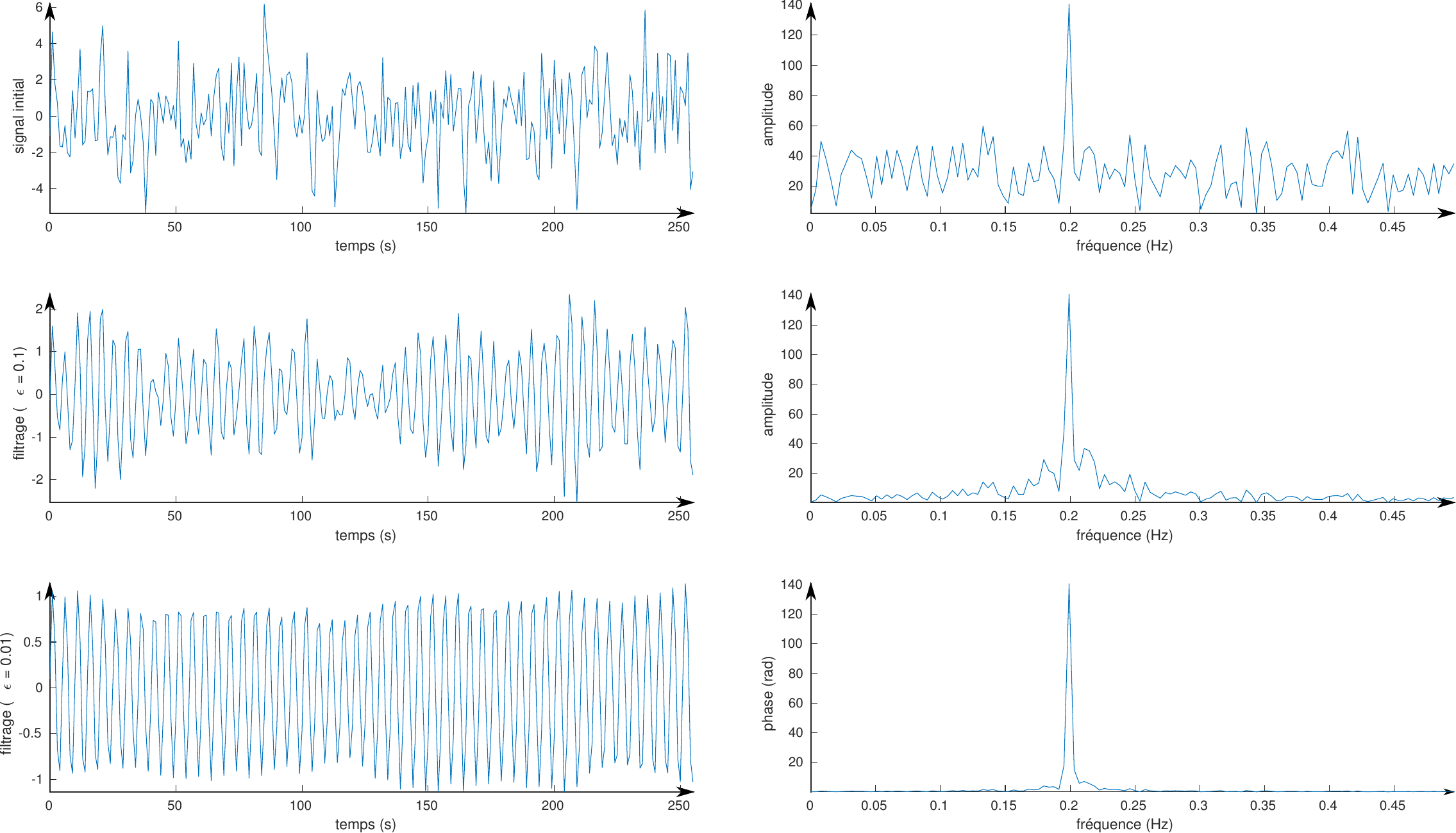}}
	\end{center}
	\caption{Applications of the narrowband filter. The initial signal (top left) consists of a pure frequency (0.20 Hz) and Gaussian white noise with unit variance. The amplitude spectrum (top right) clearly shows the spectral component and the noise level. The first filtering attempt (middle left) removes a significant portion of the noise (middle right). The filter used is the one at the top of Figure \ref{bande_etroite}. The second attempt (bottom left) was performed with the filter closer to the ideal shown at the bottom of Figure \ref{bande_etroite}. The spectral analysis of the filtered signal (bottom right) shows that the noise has indeed been further eliminated. However, the filtered signal (bottom left) shows a very disturbing boundary effect. The more or less significant nature of these boundary effects can be understood by examining the impulse response of the filters (\cf Figure \ref{be_repimp}).\label{bande_etoite_eps}}
\end{figure}

\begin{figure}[H]
	\begin{center}
		\tcbox[colback=white]{\includegraphics[width=16cm]{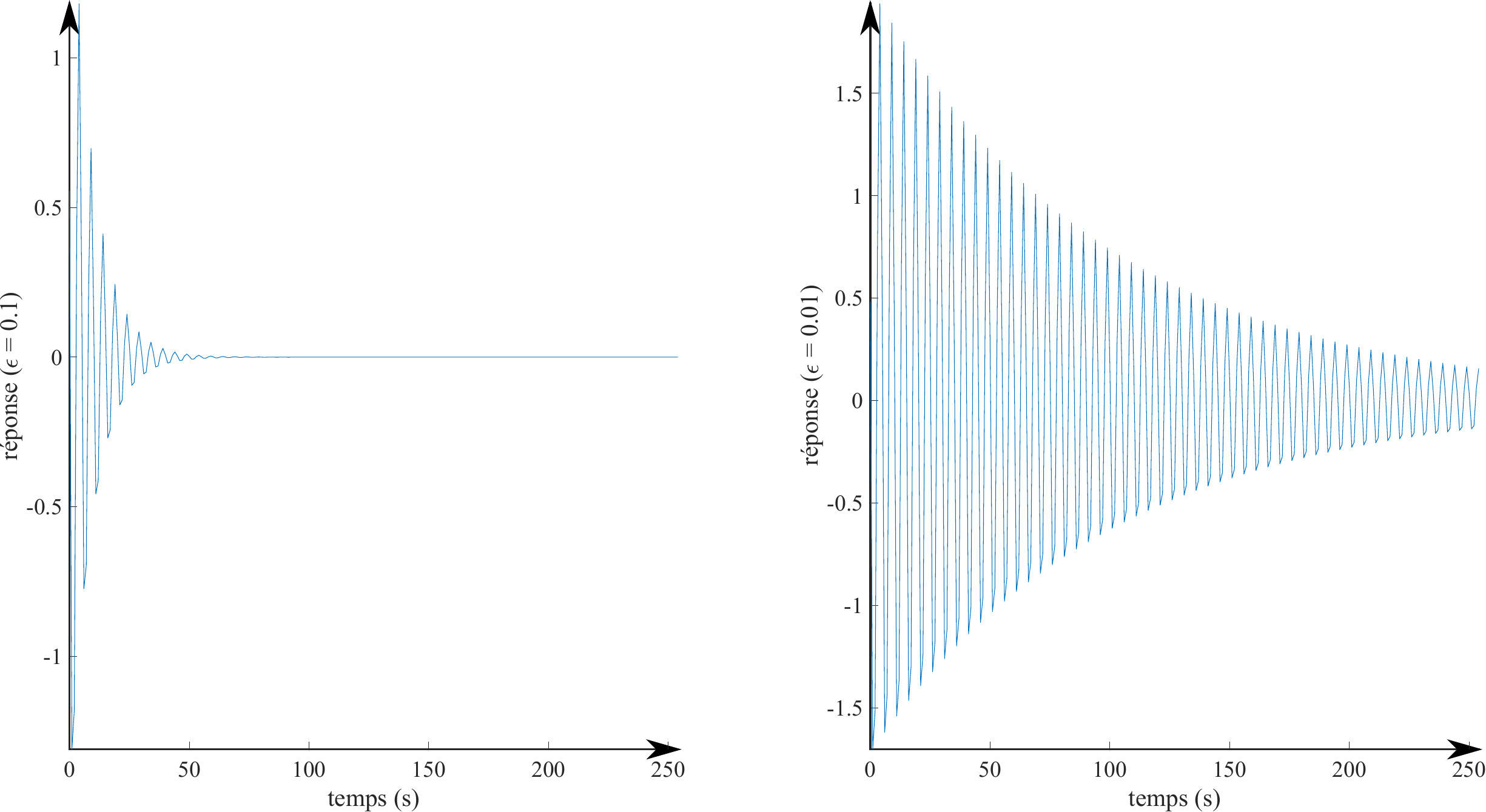}}
	\end{center}
	\caption{Impulse responses of the narrowband filters shown in Figure \ref{bande_etroite}. The filter constructed with $\epsilon=0.1$ has a much shorter impulse response (on the left) compared to the one (on the right) constructed with $\epsilon=0.01$. These differences in duration explain the more or less significant boundary effects that occur when implementing recursive filters.\label{be_repimp}}
\end{figure}

\paragraph{An example}
The filter can be applied using these recursive formulas, one advantage of which is speed: the above operation requires only 7 operations (additions and multiplications), whereas filtering by convolution using the filter in its non-recursive form,
\begin{equation}
	\begin{split}
	F^{^{\prime}}\left(Z\right) & = 	1+ZZ_{0}^{^{\prime}}+\left(ZZ_{0}^{^{\prime}}\right)^{2}+\left(ZZ_{0}^{^{\prime}}\right)^{3}+\cdots\\
 &   +1+ZZ_{0}^{^{\prime\prime}}+\left(ZZ_{0}^{^{\prime\prime}}\right)^{2}+\left(ZZ_{0}^{^{\prime\prime}}\right)^{3}+\cdots
	\end{split}
\end{equation}

requires significantly more operations. For example, if $\epsilon=0.05$, truncating the filter to a coefficient equal to 10\% of $f_{0}$ requires extending to $n=45$, which costs 90 operations! The narrowband filter we constructed allows for isolating a spectral component within a strong noise. This is illustrated in Figure \ref{bande_etoite_eps}, where, in particular, the boundary effects associated with the recursive formula are noticeable. These boundary effects become more pronounced as the filter's impulse response is longer (\cf Figure \ref{be_repimp}).

\section{Filter Stability}
\subsection{Back to narrow band filter}
In the previous section, we modified the ideal filter by setting,
\begin{equation}
	F^{^{\prime}}\left(Z\right)=\frac{1}{1-ZZ_{0}^{^{\prime}}}+\frac{1}{1-ZZ_{0}^{^{\prime\prime}}}.
\end{equation}

By expanding the terms,
\begin{equation}
	\frac{1}{1-ZZ_{0}^{^{\prime}}}=1+\left(1-\epsilon\right)ZZ_{0}+\left(1-\epsilon\right)^{2}\left(ZZ_{0}\right)^{2}+\cdots
\end{equation}

and,
\begin{equation}
	\frac{1}{1-ZZ_{0}^{^{^{\prime\prime}}}}=1+\left(1-\epsilon\right)Z/Z_{0}+\left(1-\epsilon\right)^{2}\left(Z/Z_{0}\right)^{2}+\cdots,
\end{equation}

We observe that convergence is achieved only if $\epsilon>0$, meaning that the poles of the filter must lie outside the unit circle. Otherwise, the series do not converge and the filter is unstable.

\subsection{The general case}
The general problem of filter stability can lead to rather lengthy algebraic developments. However, a sufficient condition to ensure the stability of recursive filters,
\begin{equation}
	F\left(Z\right)=\frac{N\left(Z\right)}{D\left(Z\right)}
\end{equation}

can be easily obtained by expressing the denominator in the form,
\begin{equation}
	D\left(Z\right)=D_{0}\prod_{l=1}^{L}\left(1-ZZ_{l}\right)
\end{equation}

which yields,
\begin{equation}
	\begin{split}
	F\left(Z\right) & =  \frac{N\left(Z\right)}{D_{0}}\prod_{l=1}^{L}\left(1-ZZ_{l}\right)^{-1}\\
 & =  \frac{N\left(Z\right)}{D_{0}}\prod_{l=1}^{L}\left[1+ZZ_{l}+\left(ZZ_{l}\right)^{2}+\cdots\right].
	\end{split}
\end{equation}

\begin{figure}[H]
	\begin{center}
		\tcbox[colback=white]{\includegraphics[width=16cm]{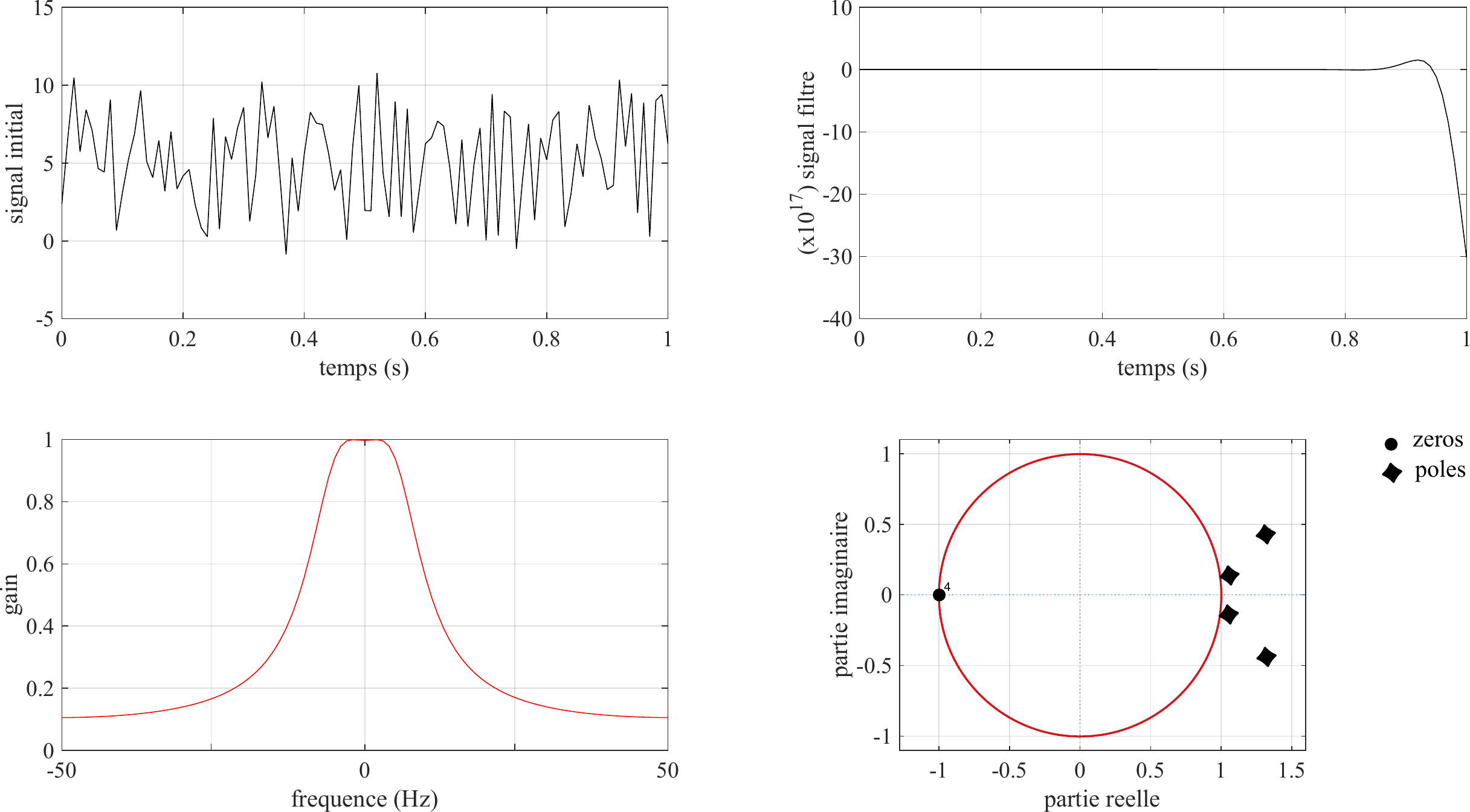}}
	\end{center}
	\caption{An unstable low-pass filter. This filter, designed to retain only the low frequencies of the initial signal (top left), is unstable and produces an unusable filtered signal (top right) showing exponential numerical divergence. Although the filter's gain (bottom left) is as expected, instability occurs because some poles are inside the unit circle (bottom right).\label{filtre_instable}}
\end{figure}

\begin{figure}[H]
	\begin{center}
		\tcbox[colback=white]{\includegraphics[width=16cm]{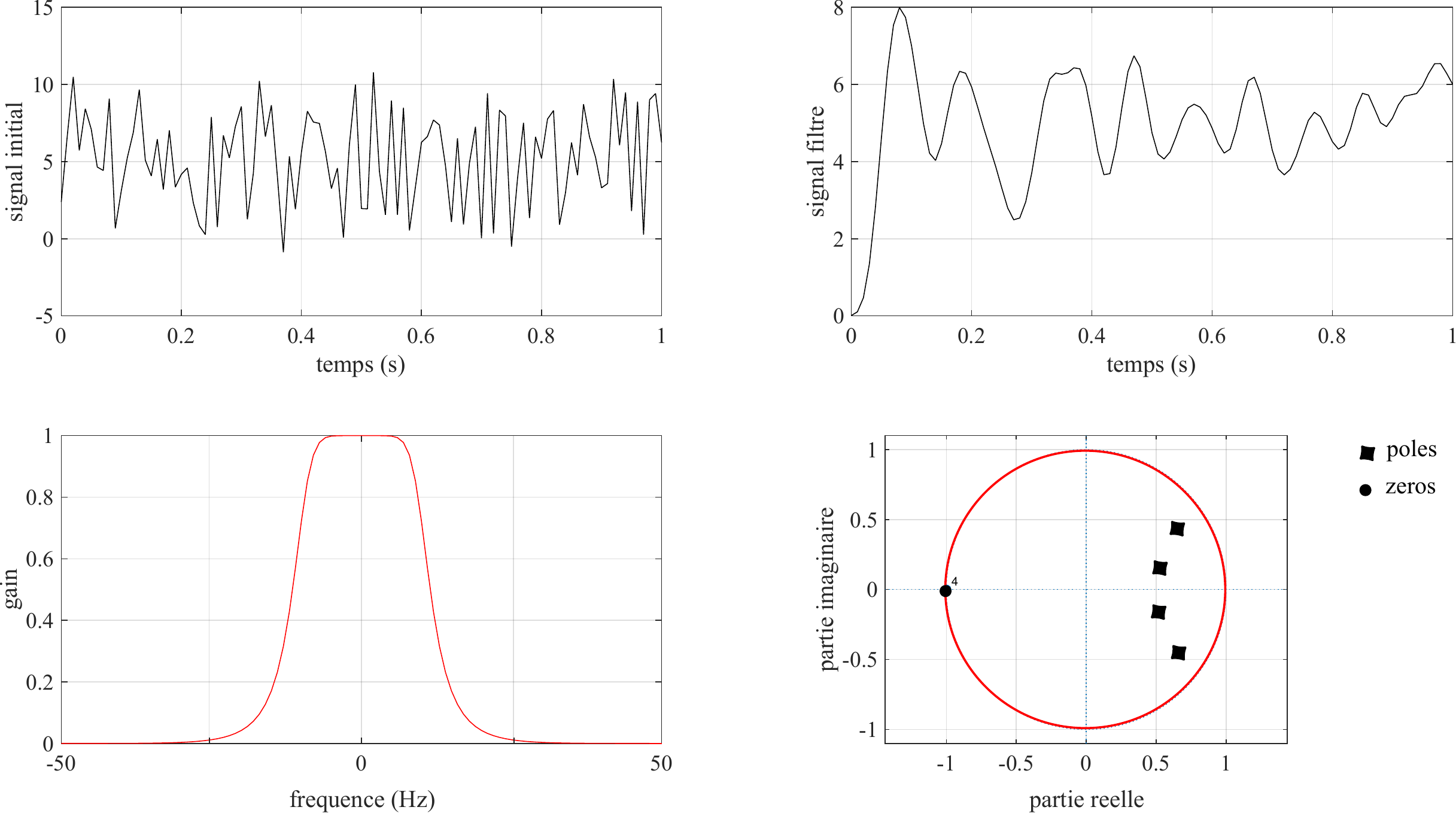}}
	\end{center}
	\caption{A stable low-pass filter. This filter is stable, as shown by the filtered signal (top right). It was obtained from the unstable filter in Figure \ref{filtre_instable} by making $D\left(Z\right)$ a minimum-phase filter, which does not alter the gain (bottom left) but places all the poles outside the unit circle (bottom right) \label{filtre_stable}}
\end{figure}

It is clear that the filter will be stable only if all the series within the product converge, that is, \warning{if all the poles $Z_{l}^{*}$ of the filter are outside the unit circle}. This means that all dipoles $\left\{ 1;-Z_{l} \right\}$ must satisfy $\left| Z_{l} \right| <1$. Such dipoles are said to be of minimum phase, and their convolution is as well. An unstable filter is unusable even though its gain might perfectly meet expectations (Figure \ref{filtre_instable}); however, it can be stabilized by making $D\left(Z\right)$ a minimum-phase filter (Figure \ref{filtre_stable}). Figures obtained using the program \couleur{ex\_filtre\_stable\_instable.m}

\section{\couleur{Butterworth} Filters}
\section{General Overview}
\begin{figure}[H]
	\begin{center}
		\tcbox[colback=white]{\includegraphics[width=16cm]{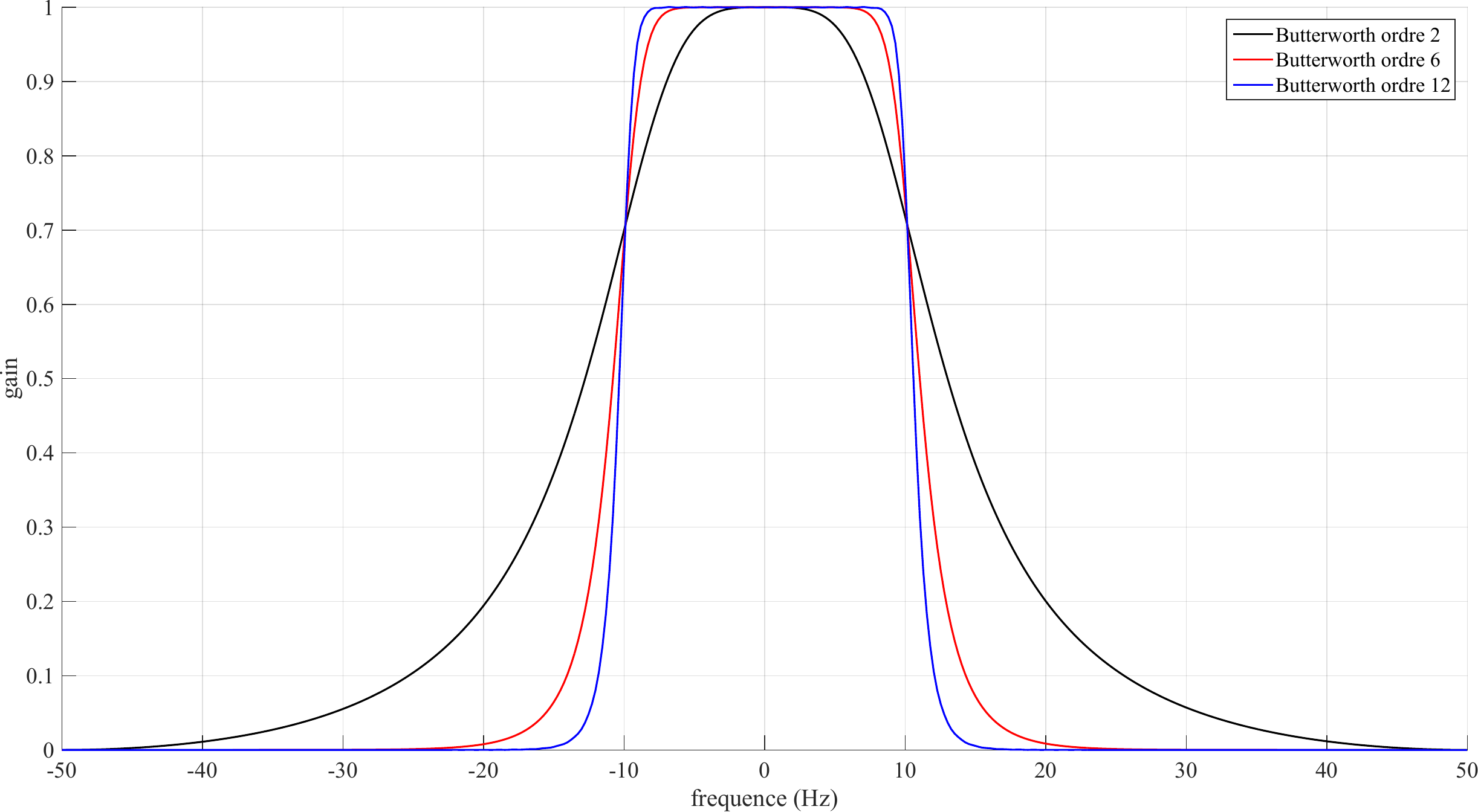}}
	\end{center}
	\caption{Examples of Butterworth low-pass filters (\couleur{ex\_butter\_2\_6\_12.m}) for orders of 2, 6, and 12. The higher the order, the closer the filter is to a window function.\label{ex_filtre_butter}} \end{figure}

The discretization and truncation of signals make it impossible to realize ideal band-pass filters\footnote{That is, filters constructed using window functions.}. A good approximation of these filters can be obtained using Butterworth filters\index{Butterworth}, whose low-pass gain function (Figure \ref{ex_filtre_butter}) is given by the following relation,
\begin{equation}
	\left|F\left(u\right)\right|^{2}=\frac{1}{1+\left(u/u_{c}\right)^{2n}}
\end{equation}

approaches a window function as the order $n \longrightarrow +\infty$. Furthermore
\begin{equation}
	\begin{cases}
		\left|F\left(\pm u_{c}\right)\right|^{2} \quad = \qquad 1/2\\
		\left|F\left(0\right)\right|^{2} \qquad \ = \qquad \ 1
	\end{cases}
\end{equation}

the attenuation at $u=\pm u_{c}$ is $10\log\left(1/2\right)=-3\text{dB}$, which defines the filter's bandwidth as $\left[-u_{c};+u_{c}\right]$. Outside this band, the higher the filter order, the more rapid the roll-off. For example, a roll-off of at least $48\text{ dB}$ per octave in the range $\left[u_{c};2u_{c}\right]$ is achieved for orders such that,
\begin{equation}
	20\log\left(\frac{\left|F\left(u_{c}\right)\right|}{\left|F\left(2u_{c}\right)\right|}\right)\geq48,
\end{equation}

which justifies the choice of $n \geq 9$, which provides a minimum attenuation of $51\text{ dB}$
	
	The construction of a high-pass filter is easily achieved using a low-pass filter and a passthrough filter\footnote{That is, a filter with an impulse response equal to the Dirac delta function.}
\begin{equation}
	\begin{split}
	\left|F\left(u\right)\right|^{2} & =  1-\frac{1}{1+\left(u/u_{c}\right)^{2n}}\\
 & =  \frac{\left(u/u_{c}\right)^{2n}}{1+\left(u/u_{c}\right)^{2n}}
 	\end{split}
\end{equation}
 
Similarly, a band-pass filter is the intersection of a low-pass filter and a high-pass filter,
\begin{equation}
	\left|F\left(u\right)\right|^{2}=\left[\frac{1}{1+\left(u/u_{h}\right)^{2n}}\right]\times\left[\frac{\left(u/u_{b}\right)^{2n}}{1+\left(u/u_{b}\right)^{2n}}\right]
\end{equation}

with the passband being $\left[u_{b};u_{h}\right]$.

\section{The Bilinear Transformation}
Applying filters via a recursive formula is recommended when the volume of data to be processed is large or when real-time filtering is required. This raises the issue of obtaining the recursive formula corresponding to a filter for which we only know \apriori the gain. The problem is as follows: given the magnitude of the filter's \couleur{Fourier} transform, how can we compute the coefficients of the same filter in the physical space to be able to use a recursive formula? If the general form of the recursive formula is,
\begin{equation}
	s_{n}=\sum_{k=0}^{M}\alpha_{k}e_{n-k}+\sum_{l=1}^{L}\beta_{l}s_{n-l}
\end{equation}

then the Fourier transform of the filter can be written as,
\begin{equation}
	F\left(u\right)=\frac{\sum_{k=0}^{M}\alpha_{k}Z^{k}}{1-\sum_{l=1}^{L}\beta_{l}Z^{l}}
\end{equation}

\begin{figure}[H]
	\begin{center}
		\tcbox[colback=white]{\includegraphics[width=16cm]{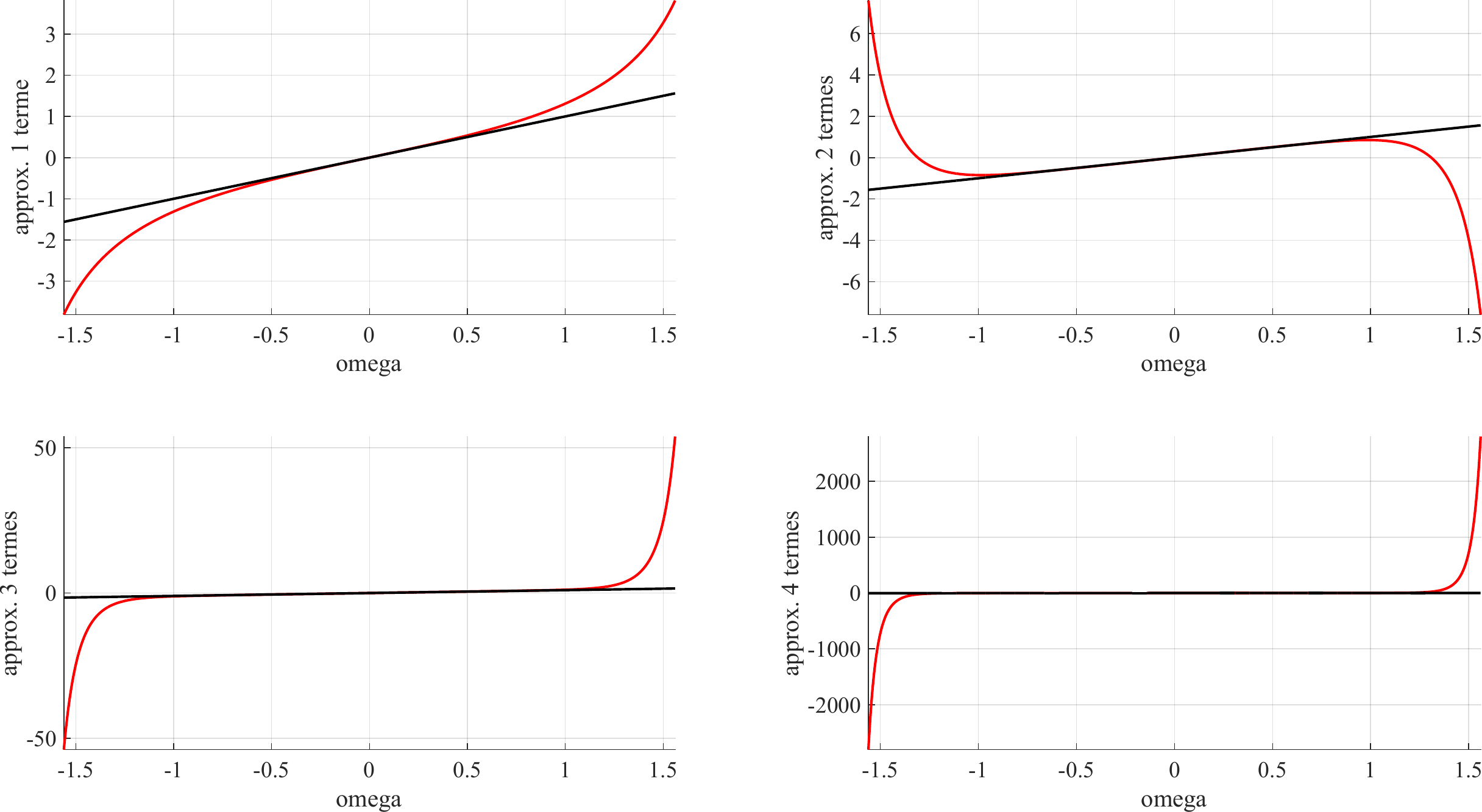}}
	\end{center}
	\caption{The first four terms of the "bilinear approximation". The bilinear transformation, in the strict sense, has a validity range restricted to approximately Nyquist/4 (top left), as shown by the comparison with the line of slope 1 (black). The approximation using only the first two terms of the expansion (top right) has a broader validity range but is practically unusable due to spectral aliasing caused by the function not being bijective. The approximation with three terms (bottom left) is practically usable and has a validity range significantly larger than the classical approximation. \etc \label{bilinear_4}} \end{figure}

Obtaining the coefficients $\alpha_{k}$ and $\beta_{l}$ thus requires that $F\left(u\right)$ be expressed as a ratio of two polynomials in $Z$, which is not always straightforward since the variable change $u \longrightarrow Z$ is rarely exact. It is then necessary to use an approximate correspondence between $u$ and $Z$ via the \index{Bilinear approximation} development
\begin{equation}
	\begin{split}
	-2i\pi u\tau & =  \ln\left(Z\right)\\
 & =  -2\left[\frac{1-Z}{1+Z}+\frac{1}{3}\left(\frac{1-Z}{1+Z}\right)^{3}+\frac{1}{5}\left(\frac{1-Z}{1+Z}\right)^{5}+\cdots\right]
	\end{split}
\end{equation}

of which the first term provides the bilinear approximation,
\begin{equation}
	u\approx\frac{1}{i\pi\tau}\frac{1-Z}{1+Z}
\end{equation}

which is valid (with an error of less than 5\%) only for,
\begin{equation}
	\left|u\right|\leq\frac{1}{10\tau}
\end{equation}

which is a much more restrictive condition than that of \couleur{Shannon} discussed in the chapter on sampling (Figure \ref{bilinear_4}). From a practical standpoint, a filter constructed using the bilinear approximation will only function correctly for frequencies adhering to this constraint; otherwise, the filter will exhibit performance different from what was specified during its design. It is possible to mitigate this limitation by adopting higher-order approximations\footnote{Note that an odd order should always be chosen.}, but this will result in a longer recursive formula; thus, a trade-off must be found (\couleur{ex\_bilinear\_4\_termes.m}).
	
\section{An example}
\begin{figure}[H]
	\begin{center}
		\tcbox[colback=white]{\includegraphics[width=16cm]{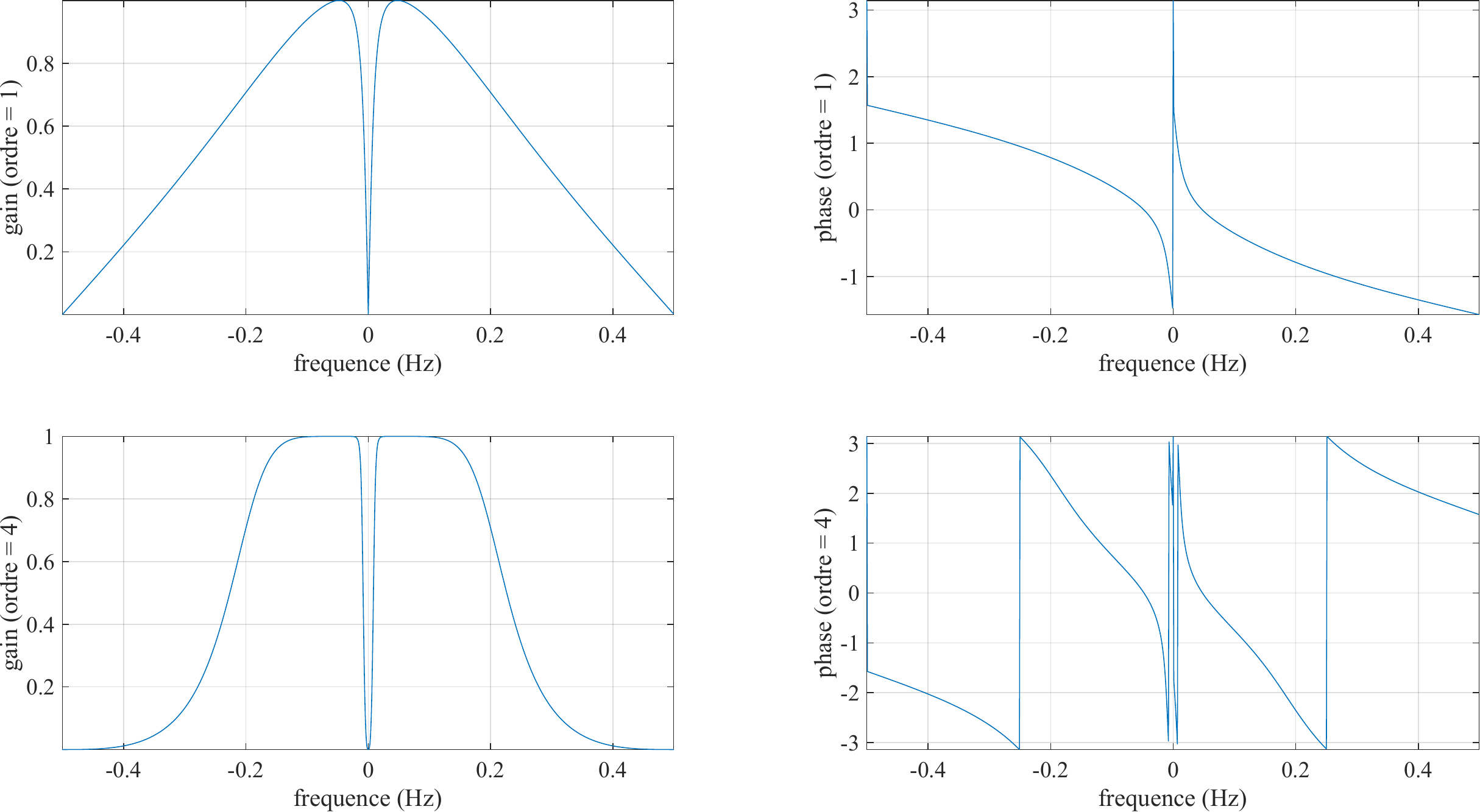}}
	\end{center}
	\caption{Butterworth band-pass filters. The filter at the top is a first-order filter with a passband of 0.04-0.14 Hz, while the filter at the bottom is of the same passband but of fourth order.  \label{book7j}}
\end{figure}

\begin{figure}[H]
	\begin{center}
		\tcbox[colback=white]{\includegraphics[width=16cm]{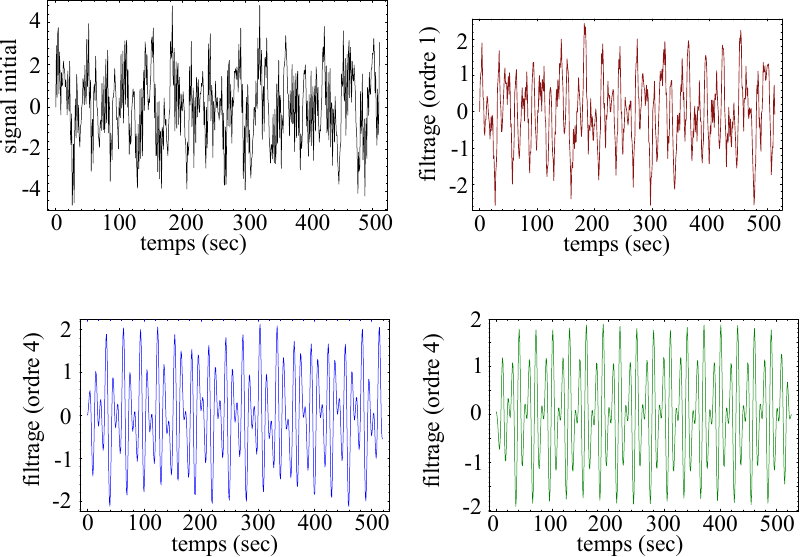}}
	\end{center}
	\caption{Band-pass filtering. The initial signal (top left) contains spectral lines between 0.04 Hz and 0.14 Hz, which we will isolate using the filters shown in Figure \ref{book7j}. The filtering performed with the first-order filter is shown in the top right, and that with the fourth-order filter is shown in the bottom left. Both of these filtrations were performed in a forward-only manner and are therefore phase-shifting. A non-phase-shifting filtration using the fourth-order filter is shown in the bottom right. \label{book7k}}
\end{figure}

The bilinear approximation makes the transformation $u \longrightarrow Z$ straightforward; for example, in the case of a first-order band-pass filter,
\begin{equation}
	\left|F\left(u\right)\right|^{2}=\left[\frac{u_{h}^{2}}{u^{2}+u_{h}^{2}}\right]\times\left[\frac{u^{2}}{u^{2}+u_{b}^{2}}\right]
\end{equation}

pour lequel on peut choisir,

\begin{equation}
	F\left(u\right)=\left[\frac{-iu_{h}}{u-iu_{h}}\right]\times\left[\frac{u}{u-iu_{b}}\right],
\end{equation}

the variable transformation yields,
\begin{equation}
	F\left(Z\right)\approx\frac{\alpha_{0}+\alpha_{2}Z^{2}}{1+\beta_{1}Z+\beta_{2}Z^{2}}
\end{equation}

and the recursive formula,
\begin{equation}
	s_{n}=\alpha_{0}e_{e}+\alpha_{2}e_{n-2}-\beta_{1}s_{n-1}-\beta_{2}s_{n-2}
\end{equation}

which requires only 7 operations. The gain and phase of this filter are shown in Figure \ref{book7j}. As you can see in the same figure, the fourth-order filter has a gain that is evidently closer to the ideal window. This higher-order filter has a more complex $Z$-transform,
\begin{equation}
	F\left(Z\right)=\frac{\alpha_{0}+\alpha_{2}Z^{2}+\alpha_{4}Z^{4}+\alpha_{6}Z^{6}+\alpha_{8}Z^{8}}{1+\beta_{1}Z+\beta_{2}Z^{2}+\beta_{3}Z^{3}+\beta_{4}Z^{4}+\beta_{5}Z^{5}+\beta_{6}Z^{6}+\beta_{7}Z^{7}+\beta_{8}Z^{8}},
\end{equation}

and the corresponding recursive formula,
\begin{equation}
	s_{n}=\sum_{k=0}^{4}\alpha_{2k}e_{n-2k}-\sum_{l=1}^{8}\beta_{l}s_{n-l}
\end{equation}

and the corresponding recursive formula is longer than that of the first-order filter. This is the price to pay for achieving a filter that is closer to the ideal. Note that the filters we have just created were based on the magnitude of their Fourier transform without concern for the phase, which, in this specific example, is not zero (Figure \ref{book7j}). These filters are thus phase-shifting (Figure \ref{book7k}). A non-phase-shifting filter (Figure \ref{book7k}) can be easily realized by applying the phase-shifting filter in a forward and backward manner; the gain of the resulting filter is then equal to the square of that of the initial phase-shifting filter. Note that the ability to perform non-phase-shifting filtering requires a backward filtering operation, which is anti-causal. This aligns with what we observed at the beginning of this chapter, namely that a non-phase-shifting filter is necessarily anti-causal. 	

\section{Wiener Filters}
\section{Wiener Filtering in the Frequency Domain}
Unlike the recursive filtering we have just discussed, where the processed signal $d(t)$ is deterministic, Wiener filtering\index{Wiener filter} accounts for the presence of noise $b(t)$ in the signal to be filtered,
\begin{equation} 
	s\left(t\right)=d\left(t\right)+b\left(t\right)
\end{equation}

	The problem is to construct a linear filter that, when applied to $s(t)$, provides an output as close as possible to $d(t)$. In the case of Wiener filtering, 'as close as possible' means 'in the least squares sense,' and the desired filter $f_{W}$ is such that,

\begin{equation}
	f_{W}*s\stackrel{\mathbf{L}_{2}}{=}d
\end{equation}
	The obtained filter will be optimal in the probabilistic sense if the noise statistics are Gaussian. Otherwise, a different norm would need to be adopted for optimization. The equality above can be expressed as,

\begin{equation}
	\int_{-\infty}^{+\infty}\left|\left[f_{W}*s\right]\left(t\right)-d\left(t\right)\right|^{2}\mathrm{d}t\;\;\mathrm{MINIMUM}
\end{equation}

in the time domain, or
\begin{equation}
	\int_{-\infty}^{+\infty}\left|F_{W}\left(u\right)S\left(u\right)-D\left(u\right)\right|^{2}\mathrm{d}u\;\;\mathrm{MINIMUM}
\end{equation}

in the frequency domain. By expanding this latter expression,
\begin{equation}
	\begin{split}
 &  & \int_{-\infty}^{+\infty}\left[\left|D\left(u\right)\right|^{2}\left|F_{W}\left(u\right)-1\right|^{2}+\left|F_{W}\left(u\right)\right|^{2}\left|B\left(u\right)\right|^{2}\right]\mathrm{d}u\\
 &  & +\int_{-\infty}^{+\infty}D\left(u\right)B^{*}\left(u\right)\left[\left|F_{W}\left(u\right)\right|^{2}-F_{W}^{*}\left(u\right)\right]\mathrm{d}u\\
 &  & +\int_{-\infty}^{+\infty}D^{*}\left(u\right)B\left(u\right)\left[\left|F_{W}\left(u\right)\right|^{2}-F_{W}\left(u\right)\right]\mathrm{d}u\\
 &  & \mathrm{MINIMUM}
	\end{split}
\end{equation}

An important simplification of this expression occurs if we assume that the noise and the deterministic component are uncorrelated. The last two integrals are identically zero, and the filter must be such that,
\begin{equation}
	\int_{-\infty}^{+\infty}\left[\left|D\left(u\right)\right|^{2}\left|F_{W}\left(u\right)-1\right|^{2}+\left|F_{W}\left(u\right)\right|^{2}\left|B\left(u\right)\right|^{2}\right]\mathrm{d}u\;\;\mathrm{MINIMUM}
\end{equation}

By requiring that $F_{w} \in \mathbf{R}$, the minimization condition becomes,
\begin{equation}
	\frac{\mathrm{d}}{\mathrm{d}F_{W}\left(u\right)}\int_{-\infty}^{+\infty}\left[\left|D\left(u\right)\right|^{2}\left(F_{W}\left(u\right)-1\right)^{2}+F_{W}\left(u\right)^{2}\left|B\left(u\right)\right|^{2}\right]\mathrm{d}u
\end{equation}

and the filter,
\begin{equation}
	F_{W}\left(u\right)=\frac{\left|D\left(u\right)\right|^{2}}{\left|D\left(u\right)\right|^{2}+\left|B\left(u\right)\right|^{2}}
\end{equation}

\begin{figure}[!h]
	\begin{center}
		\tcbox[colback=white]{\includegraphics[width=15cm]{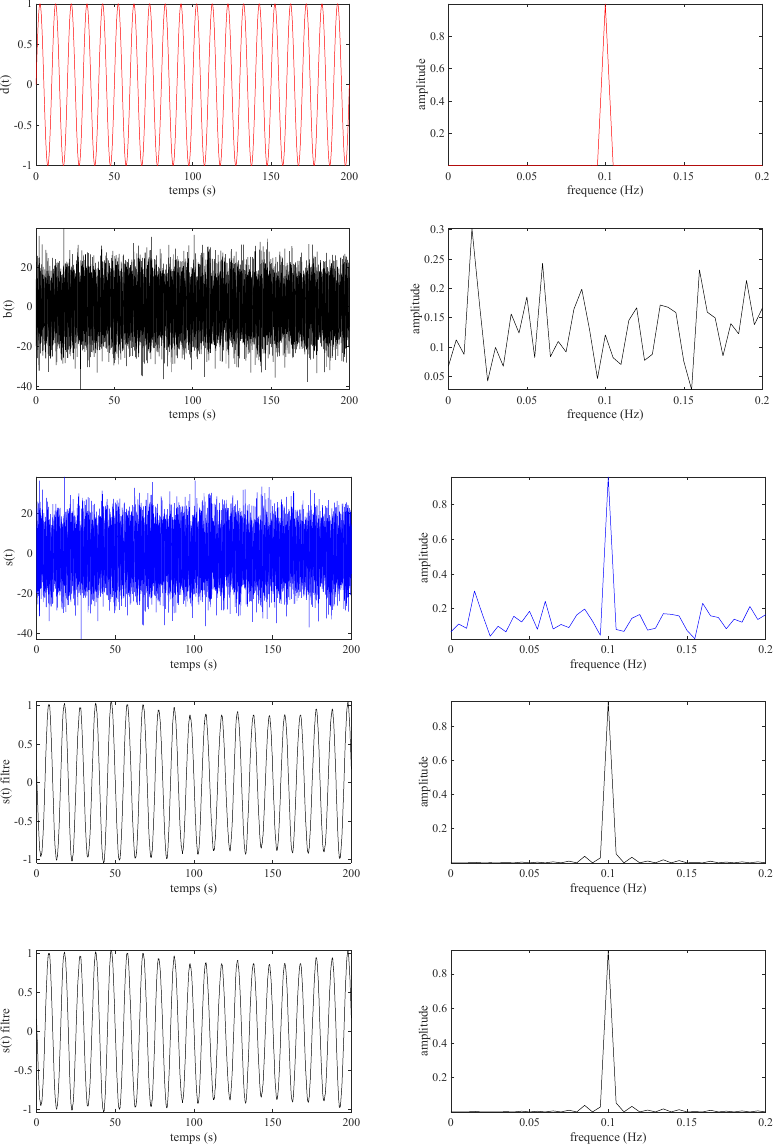}}
	\end{center}
	\caption{Wiener Filtering  \label{book7l}}
\end{figure}

The filter can only be constructed if the energy spectra of the components $b(t)$ and $d(t)$ are known (Figure \ref{book7l}), which is generally possible only through prior information. This reflects the ongoing ambiguity in signal processing that we mentioned in the introduction. While this information may seem difficult to obtain, it is important to remember that the filter was constructed through a minimization process (least squares) that nullifies the first derivative of the cost function. Thus, errors in the filter definition will only start to manifest in the second order, which mitigates their impact (Figure \ref{book7l}). The deterministic signal (top left) is a pure sinusoid (0.1 Hz) and is contaminated by Gaussian noise (second row left, black curve) before filtering (last row left, black curve). The energy spectra of the noise (second row right) and the deterministic signal (top right) are used to construct the gain of the Wiener filter (fourth row right). The filtered signal is shown at the bottom left (fourth row). The robustness of Wiener filtering can be appreciated in this figure, where the filter gain (left) was constructed by replacing the noise energy spectrum with its average value. It is observed that the filtered signal (right) is not significantly affected by this simplification. Figure (\ref{book7l}) was obtained using the function \couleur{ex\_wiener.m}.

\chapter{\titrechap{Spectral analyses}}
\minitoc
\section{The spectral analysis models}
	This chapter deals with the problem of spectral analysis, which is the study of the distribution of the energy of a signal as a function of frequency. This distribution law is known as the energy spectrum, which is defined as the square of the modulus of the \couleur{Fourier} transform of the signal,
\begin{equation}
	E\left(u\right)\equiv\left|S\left(u\right)\right|^{2}
\end{equation}

The simplest method of calculating the energy spectrum of a sampled signal is to use the Discrete \couleur{Fourier} Transform. Although effective, this method has certain drawbacks which have led to the development of alternative techniques whose main advantage is the ability to achieve very fine frequency resolutions. This is not a miraculous violation of the Uncertainty Principle discussed earlier, but rather a consequence of the fact that these techniques are autoregressive and implicitly extrapolate the analysed signal beyond the observation interval, thereby increasing the frequency resolution. However, this extrapolation comes with restrictive assumptions\footnote{which many authors conveniently overlook!} that limit the applicability of these methods to certain categories of signals. We refer the interested reader to the excellent article by \shortciteN{kay1981spectrum} for a critical review of these methods. In general, any spectral analysis method is based on fitting a model to the data and calculating a spectrum from the parameters of that model. Seen in this light, it is clear that spectral analysis falls within the scope of inverse problem theory.

\subsection{\textit{Prony, Hildebrand, Pisarenko and Schuster} : Trigonometric series}
The oldest model was proposed in 1795 by Baron de \shortciteN{prony1795essai} \index{Prony}, not for spectral analysis, but to describe the behaviour of certain gases. This model,
\begin{equation}
	s_{\mathrm{PRONY}}\left(t\right)=\sum_{m=1}^{M}S_{m}\exp\left(\alpha_{m}t\right)\exp\left(2i\pi u_{m}t\right)
\end{equation}

is composed of damped sinusoids and has strong links with \couleur{Fourier} analysis. This model is very general, with adjustable parameters
\begin{equation}
	\left\{ M,S_{m},\alpha_{m},u_{m}\right\} \;\;\left(m=1,\cdots,M\right)
\end{equation}

make the inverse problem highly non-linear. Note that even the number\footnote{Called the order of the model.} $M$ of elements in the sum is \apriori unknown. This is an inverse problem where the exact number of parameters is not known. The \couleur{Prony} model is often used in signal processing, and the inverse problem is generally simplified and not treated in a non-linear way. The solutions obtained are approximate and have biases that become more significant as the signal-to-noise ratio deteriorates. Other models used in spectral analysis can be considered as simplified versions of the \couleur{Prony} model. For example, the model of \shortciteN{hildebrand1956introduction},
\begin{equation}
	s_{\mathrm{HILDEBRANT}}\left(t\right)=\sum_{m=1}^{M}S_{m}\exp\left(2i\pi u_{m}t\right)
\end{equation}

is obtained by setting $\alpha_{m}=0$ in the \couleur{Prony} model. The adjustable parameters are,
\begin{equation}
	\left\{ M,S_{m},u_{m}\right\} \;\;\left(m=1,\cdots,M\right)
\end{equation}

The model used in the method of \shortciteN{pisarenko1973retrieval},
\begin{equation}
	s_{\mathrm{PISARENKO}}\left(t\right)=\sum_{m=1}^{M}S_{m}\exp\left(2i\pi u_{m}t\right)+\sigma_{b}b\left(t\right)
\end{equation}

is very similar to \couleur{Hildebrand}, but explicitly takes into account that the data are contaminated by white noise, $b\left(t\right)$, whose variance, $\sigma_{b}^{2}$, is part of the set of adjustable parameters,
\begin{equation}
	\left\{ M,\sigma_{b},S_{m},u_{m}\right\} \;\;\left(m=1,\cdots,M\right)
\end{equation}

	The solution provided by \couleur{Pisarenko} involves working from the autocorrelation function of the data and does not allow for the recovery of phases. One only has access to the energy spectrum. The models of \couleur{Hildebrand} and \couleur{Pisarenko} are suited for representing data with a line spectrum. In contrast, due to the presence of the damping coefficients, $\alpha_{m} \neq 0$, the \couleur{Prony} model allows for the analysis of continuous spectra, which may also contain lines.

	All of these models are highly non-linear, and estimating their parameters poses significant challenges. The suboptimal solutions typically computed are often unsatisfactory when the data are noisy. Estimating the order, $M$, of these models can be done more or less accurately and is undoubtedly a critical stage of these techniques. This probably explains the popularity of the \couleur{Schuster} model,
\begin{equation}
	s_{\mathrm{SCHUSTER}}\left(n\tau\right)=\sum_{k=0}^{N-1}S_{k}\exp\left(2i\pi kn/N\right)\;\;\left(n=0,1,\cdots,N-1\right)
\end{equation}

whose frequencies, fixed \apriori, correspond to the number $N$ of available data\footnote{Assumed to be sampled at a constant interval $\tau$.}. The set of parameters
\begin{equation}
	\left\{ S_{k}\right\} \;\;\left(k=0,\cdots,N-1\right)
\end{equation}

 is reduced to those that appear linearly in the \couleur{Prony} model. We will see that this model, fitted to the data by least squares, gives a spectral analysis by discrete \couleur{Fourier} transform.

\subsection{\textit{Burg, Pisarenko},\ldots : autoregressive models}
	We will now delve into the realm of spectral analysis using autoregressive models. Many methods employ such models, with one of the most popular being the maximum entropy analysis method. The simplest way to understand the role of autoregressive models is to start with the \couleur{Fourier} transform,
\begin{equation}
	S_{\tau}\left(u\right)=\tau\sum_{n=-\infty}^{+\infty}s\left(n\tau\right)Z^{n}
\end{equation}

	of the discrete signal $s_{n} \equiv s(n\tau)$. When the signal is truncated, we have seen that the sum in the above equation is bounded,
\begin{equation}
	S_{\tau,T}\left(u\right)=\tau\sum_{n=0}^{N-1}s_{n}Z^{n}
\end{equation}

where $T = N\tau$. The associated energy spectrum,
\begin{equation}
	E_{\tau,T}\left(u\right)=\tau\left|\sum_{n=0}^{N-1}s_{n}Z^{n}\right|^{2},
\end{equation}

is represented by a finite number of terms, which poses problems for analytical representation if the true spectrum contains lines. A better representation of such a spectrum can be achieved by using an autoregressive model of the type,
\begin{equation}
	s_{n}=-\sum_{m=1}^{M}b_{m}s_{n-m}
\end{equation}

	where $M$ is the model order\footnote{Such models are often referred to in the technical literature by the notation AR(M)}.

	This justification for autoregressive models can be further supported by noting that a discrete monochromatic signal leads directly to an autoregressive expression AR(2),
\begin{equation}
	\begin{split}
s_{n} & =  \sin\left(2\pi u_{0}\tau n\right)\\
 & =  2\cos\left(2\pi u_{0}\tau\right)\sin\left(2\pi u_{0}\tau\left(n-1\right)\right)-\sin\left(2\pi u_{0}\tau\left(n-2\right)\right)\\
 & =  2\cos\left(2\pi u_{0}\tau\right)s_{n-1}-s_{n-2}
	\end{split}
\end{equation}
  
The initialisation of this recursive formula is necessary when the signal is truncated, and two initial values, $e_{0}$ and $e_{1}$, must be provided. These values determine the amplitude and phase of the sine wave to be generated. The recursive formula is,
\begin{equation}
	\begin{cases}
s_{0} \qquad = \qquad e_{0}\\
s_{1} \qquad = \qquad 2\cos\left(2\pi u_{0}\tau\right)s_{0}+e_{1}\\
s_{n} \qquad = \qquad 2\cos\left(2\pi u_{0}\tau\right)s_{n-1}-s_{n-2}\;\;\left(n>1\right)
	\end{cases}
\end{equation}

	and its $Z$-transform provides,
\begin{equation}
	S\left(Z\right)=\frac{e_{0}+e_{1}Z}{1-2\cos\left(2\pi u_{0}\tau\right)Z+Z^{2}}
\end{equation}

	which is none other than the $Z$-transform of the narrowband filter studied in the chapter on filtering. Extending this to a signal composed of $M$ frequencies,
\begin{equation}
	s_{n}=\sum_{l=1}^{M}S_{l}\exp\left(2i\pi u_{l}\tau n\right)\;\;\left(n\geq0\right)
\end{equation}

	it's not much more difficult. Indeed,
\begin{equation}
	s_{n-m}=\sum_{l=1}^{M}S_{l}\exp\left[2i\pi u_{l}\tau\left(n-m\right)\right]\;\;\left(n\geq0\right),\end{equation}
	
	and by multiplying both sides of this equation by a coefficient $b_{m}$ and summing $M+1$ such equations,
\begin{equation}
	\begin{split}
	\sum_{m=0}^{M}b_{m}s_{n-m} & =  \sum_{m=0}^{M}b_{m}\sum_{l=1}^{M}S_{l}\left[\exp\left(2i\pi u_{l}\tau\right)\right]^{n-m}\\
 & =  \sum_{l=1}^{M}S_{l}\left[\exp\left(2i\pi u_{l}\tau\right)\right]^{n-M}\sum_{m=0}^{M}b_{m}\left[\exp\left(2i\pi u_{l}\tau\right)\right]^{M-m}
	\end{split}
\end{equation}

	valid for $n \geq M$. Let's choose the coefficients $b_{m}$ such that,
\begin{equation}
	b_{0}=1
\end{equation}

	and,
\begin{equation}
	\sum_{m=0}^{M}b_{m}\left[\exp\left(2i\pi u_{l}\tau\right)\right]^{M-m}=0,
\end{equation}

	we obtain the recursive formula directly,
\begin{equation}
	s_{n}=-\sum_{m=1}^{M}b_{m}s_{n-m}\;\;\left(n\geq M\right)
\end{equation}
	
Line spectra can thus be modelled by autoregressive models, for which the task now is to determine the parameters $b_{m}$.

\section{Discrete \couleur{Fourier} Transform Analysis}
\subsection{\couleur{Schuster}'s periodogram}

	This technique involves the representation of the observed signal,
\begin{equation}
	s_{n}^{obs}\equiv s\left(n\tau\right)\;\;\left(n=0,1,\cdots,N-1\right)
\end{equation}

	using the model,
\begin{equation}
	s_{n}^{mod}=\sum_{k=0}^{N-1}S_{k}\exp\left(2i\pi kn/N\right)\;\;\left(n=0,1,\cdots,N-1\right),
\end{equation}

	consisting of $N$ sinusoids with frequencies that are multiples of $\nu = 1/N\tau$. Note that this model is highly constrained: the frequencies are fixed \apriori in both value and number, and the nature of the functions is also predetermined; they are sinusoids and nothing else. The only adjustable parameters are the $S_{k}$, which allow  the amplitudes and phases of each sinusoid in the model to be adjusted. Several generalisations have been proposed to also adjust the number of frequencies and their values. Although these generalisations are quite legitimate, their main drawback is that they render the problem highly non-linear and practically very difficult to solve. Let us rewrite our initial model in its expanded form,
\begin{equation}
	\left(
	\begin{array}{c}
		s_{0}^{mod}\\
		\vdots\\
		s_{n}^{mod}\\
		\vdots\\
		s_{N-1}^{mod}
	\end{array}\right)=\left[
	\begin{array}{ccccc}
		1 & \cdots & 1 & \cdots & 1\\
		\vdots &  & \vdots &  & \vdots\\
		1 & \cdots & \exp\left[\frac{2i\pi nk}{N}\right] & \cdots & \exp\left[\frac{2i\pi n\left(N-1\right)}			{N}\right]\\
		\vdots &  & \vdots &  & \vdots\\
		1 & \cdots & \exp\left[\frac{2i\pi k\left(N-1\right)}{N}\right] & \cdots & 			\exp\left[\frac{2i\pi\left(N-1\right)^{2}}{N}\right]
	\end{array}\right]\times\left(
	\begin{array}{c}
		S_{0}\\
		\vdots\\
		S_{k}\\
		\vdots\\
		S_{N-1}
	\end{array}\right);
\end{equation}

	or, in a more compact form,
\begin{equation}
	\overrightarrow{s}_{mod}=\mathbf{W}\overrightarrow{S}
\end{equation}

	The problem now is the computation of the components of the vector $\overrightarrow{S}$, so that
\begin{equation}
	\overrightarrow{s}_{mod}\approx\overrightarrow{s}_{obs}
\end{equation}

This fitting is not unique and of course depends on the criterion chosen to determine whether the model predictions are close to the observed data: a norm must be chosen. The classical choice of the $\Bbb{L}_{2}$ norm leads to the optimal least squares fitting criterion, for which the best model is such that
\begin{equation}
	\left\Vert \overrightarrow{s}_{mod}-\overrightarrow{s}_{obs}\right\Vert ^{2}\;\;\mathrm{MIMIMUM}.
\end{equation}
	
The solution obtained by applying this criterion is
\begin{equation}
	\begin{split}
\overrightarrow{S} & =  \left[\mathbf{W}^{H}\mathbf{W}\right]^{-1}\mathbf{W}^{H}\overrightarrow{s}_{obs}\\
 & =  \frac{1}{N}\mathbf{W}^{*}\overrightarrow{s}_{obs}
	\end{split}
\end{equation}

	where we have used the fact that,
\begin{equation}
	\left[\mathbf{W}^{H}\mathbf{W}\right]^{-1}=\frac{1}{N}\mathbf{I}
\end{equation}

and,
\begin{equation}
	\mathbf{W}^{H}=\mathbf{W}^{*}
\end{equation}

	Let us rewrite this solution in its extended form,
\begin{equation}
	S_{k}=N^{-1}\sum_{n=0}^{N-1}s_{n}^{obs}\exp\left(-2i\pi kn/N\right)\;\;\left(k=0,1,\cdots,N-1\right).
	\label{schusterf}
\end{equation}

	(\ref{schusterf}) is a slightly modified form of the discrete \couleur{Fourier} transform. The least-squares fitting of the \couleur{Schuster} model presented at the beginning of this section is thus equivalent to the spectral analysis method based on the discrete \couleur{Fourier} transform of the observed signal. This method, which is by far the most commonly used, is therefore very precise; in particular, it only provides optimal solutions when the noise contaminating the data is \couleur{Gaussian} and white. Otherwise\footnote{for example, in the presence of outliers in the signal.}, the solution obtained can be significantly biased, as indicated by the notable lack of robustness of the least squares criterion.

\subsection{Signal truncation effects}
	The equivalence between \couleur{Schuster}'s method and the discrete \couleur{Fourier} transform allows us to make direct use of some previously established results. For example, the fact that the observed signal is a truncated version of the real signal.
\begin{equation}
	s_{T}\left(t\right)=s\left(t\right)\Pi\left(t/T\right)
\end{equation}

means that the computed \couleur{Fourier} transform is a degraded version of the real signal,
\begin{equation}
	S_{T}\left(u\right)=TS\left(u\right)*\textrm{sinc}\left(uT\right)
\end{equation}
	
	From a practical point of view, this degradation manifests itself in two effects: the limitation of the frequency resolution and the phenomenon of \textit{leakage}.\index{leakage}.

\begin{figure}[H]
	\begin{center}
		\tcbox[colback=white]{\includegraphics[width=16cm]{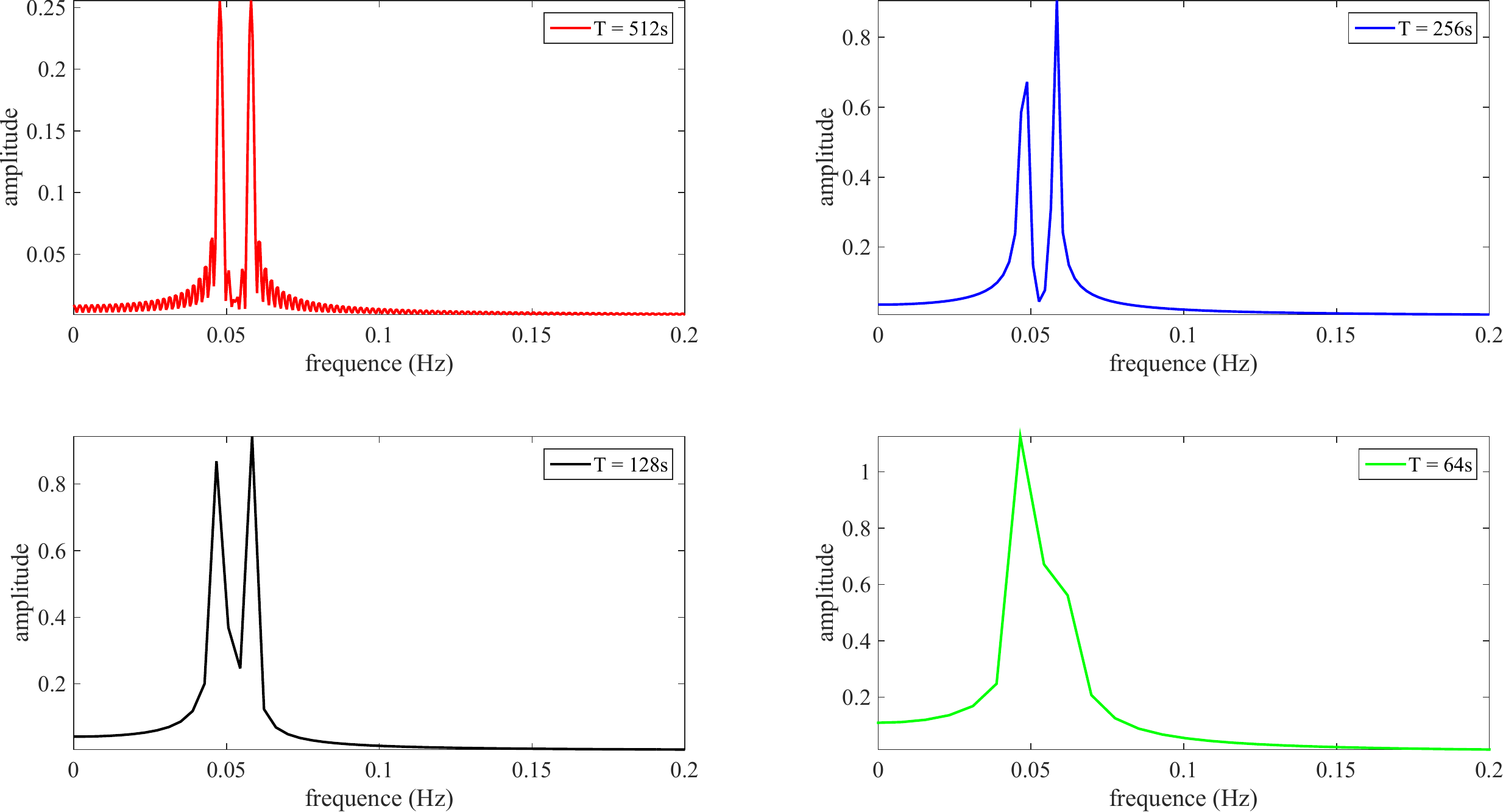}}
	\end{center}
	\caption{Signal truncation limits the frequency resolution. The signal analysed in this example consists of two sinusoids (0.048 Hz and 0.058 Hz) sampled with $\tau = 1; s$. The frequency resolution, approximately equal to $1/T$, is sufficient to resolve the spectral lines at $T = 512; s$ (top left), $T = 256; s$ (top right), and $T = 128; s$ (bottom left). However, a stronger truncation, $T = 64; s$ (bottom right), no longer allows the resolution of the two spectral lines. Note that the amplitude of the peaks in these spectra decreases as they widen. Note also the increasing prominence of the secondary lobes associated with the main peaks as the duration of the signal analysed decreases. This phenomenon, known as \textit{leakage}, can be reduced by using apodization windows.	
	  \label{leakage}}
\end{figure}

Frequency resolution, as we have seen, can be defined as the width of the main lobe of the sinc function,
\begin{equation}
	\delta u\approx\frac{1}{T}
\end{equation}

	and it is clear that severe truncation can prevent the resolution of closely spaced spectral lines (Figure \ref{leakage}). The remedy is to increase the length of the analysed signal or to use a spectral analysis method other than \couleur{Schuster}'s. If you choose the latter solution, make sure that the 'miracle' method you intend to use is suitable for your signal.

	\textit{"leakage"} is a phenomenon caused by the secondary lobes of the sinc function that appear in the convolution described above. If the spectrum of the signal being analysed contains a mixture of large and small energy peaks, these secondary lobes can completely obscure the smaller energy peaks near the larger ones. This effect results in a transfer of energy from the original frequency to neighbouring frequencies, hence the term \textit{"leakage"}. If the original peak is very intense, this transfer can affect a significant portion, or even all, of the calculated spectrum, so it is sometimes necessary to reduce this effect by using apodization windows.
	\newpage

\subsection{Apodization windows}
\begin{figure}[H]
	\begin{center}
		\tcbox[colback=white]{\includegraphics[width=16cm]{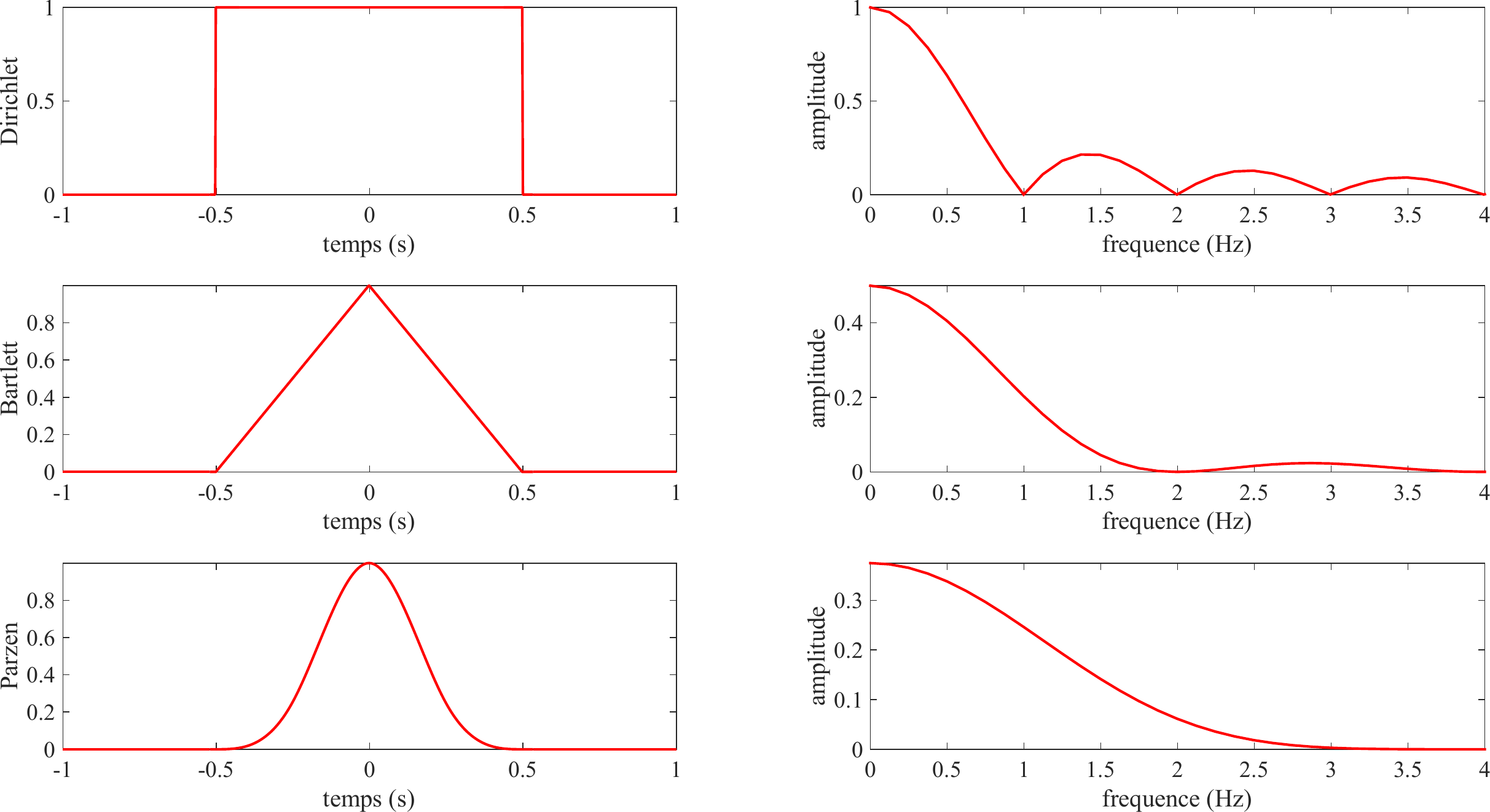}}
	\end{center}
	\caption{Apodization windows. These three windows are the \couleur{Dirichlet} (top left), \couleur{Bartlett} (middle left) and \couleur{Parzen} (bottom left) windows. They are obtained by successive auto-convolutions of the function $\Pi(t)$. As the number of auto-convolutions increases (from top to bottom), the window becomes smoother, and its Fourier transform (right) has attenuated secondary lobes and a wider central lobe, corresponding to a degradation in frequency resolution. From the central limit theorem illustrated in a previous chapter, you know that the limiting window obtained by this auto-convolution process is the Gaussian window, which is not very different from the \couleur{Parzen} window
 \label{appodisation_a}} 
\end{figure}

\begin{figure}[H]
	\begin{center}
		\tcbox[colback=white]{\includegraphics[width=16cm]{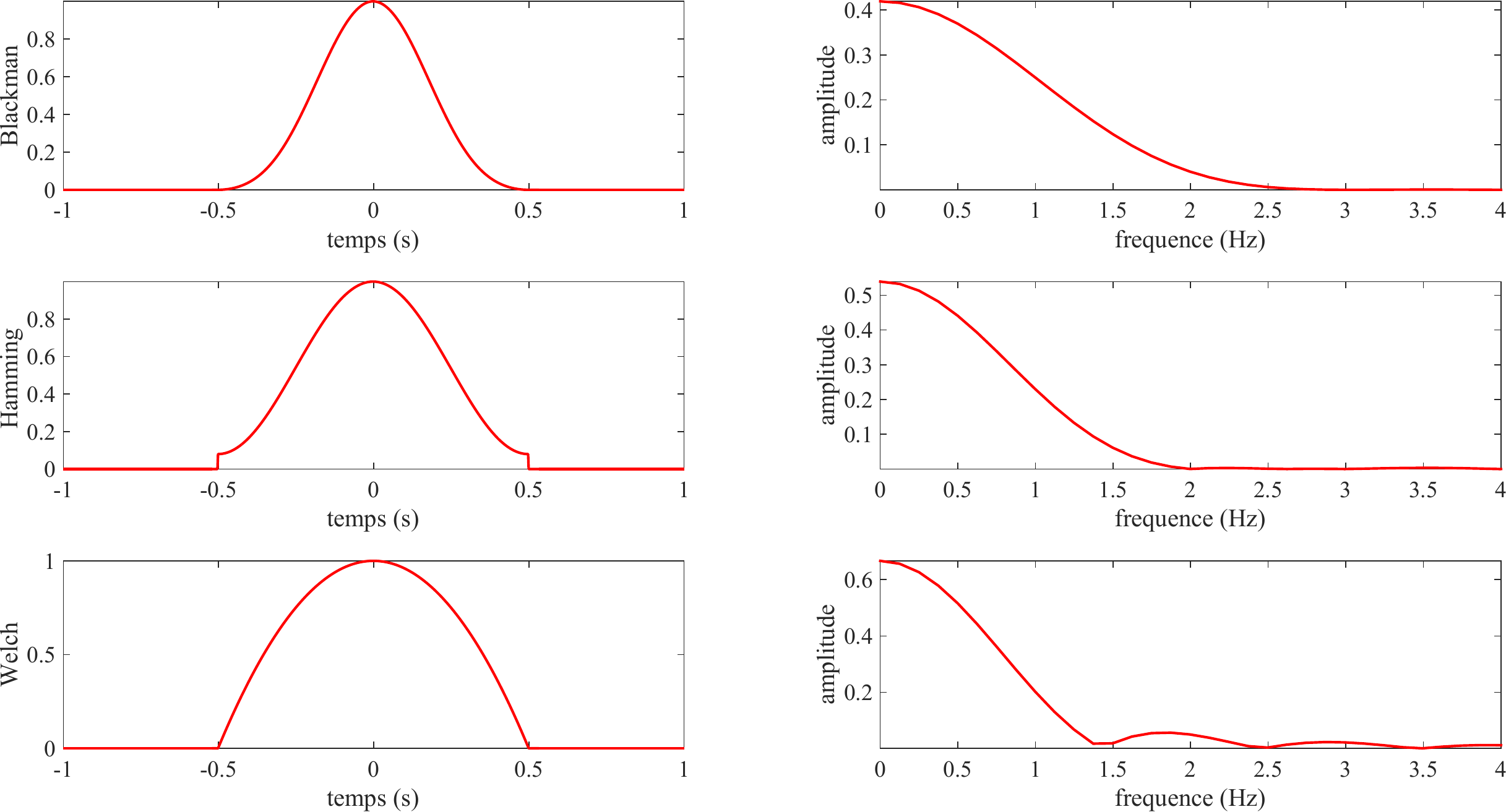}}
	\end{center}
	\caption{Apodization windows. These three windows are the \couleur{Blackman} (top left), \couleur{Hamming} (middle left) and \couleur{Welch} (bottom left) windows. Note that the \couleur{Blackman} window is very similar to the  \couleur{Parzen} window. \label{appodisation_b}}
\end{figure}

\begin{figure}[H]
	\begin{center}
		\tcbox[colback=white]{\includegraphics[width=16cm]{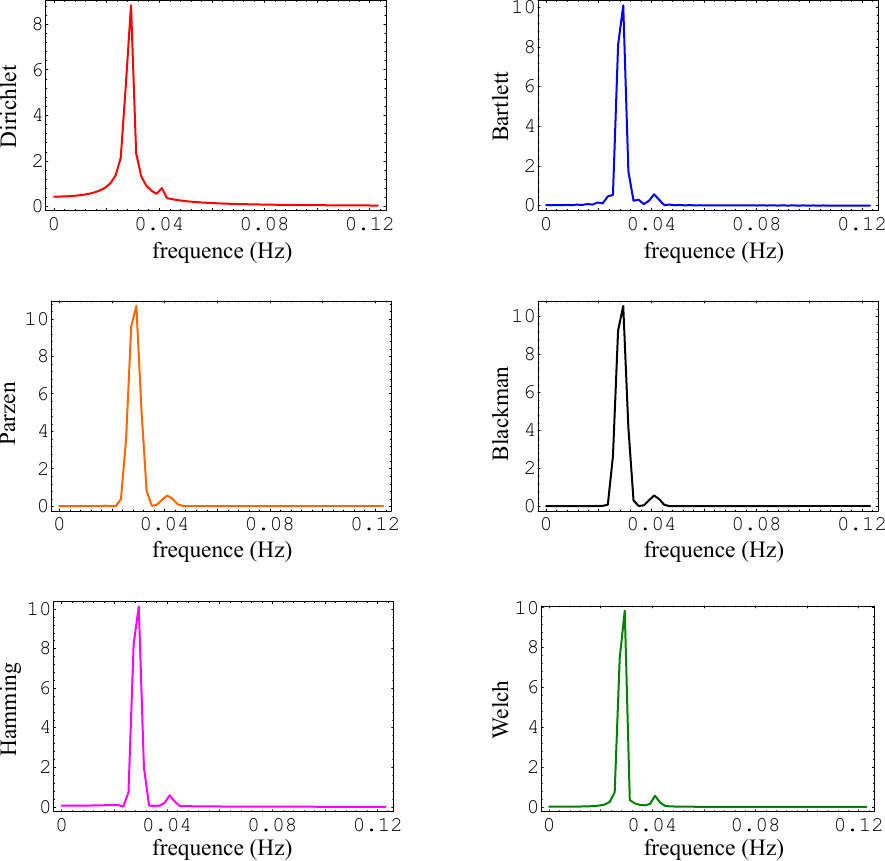}}
	\end{center}
	\caption{Effects of the apodisation window. The analysed signal consists of two sinusoids with different amplitudes (1 and 0.05), sampled with $\tau = 1; s$ and $T = 512; s$. The amplitude spectra obtained after apodising the signal with windows smoother than the \couleur{Dirichlet} window (top left) allow a better resolution of the low amplitude spectral line.  \label{appodisation_c}}
\end{figure}

	Apodisation windows are used to reduce the leakage phenomenon. These are functions $f(t)$ whose Fourier transform has smaller secondary lobes than those of $\textrm{sinc}(u)$ (Figures \ref{appodisation_a} and \ref{appodisation_b}, \cf \couleur{ex\_appodisation.m}). In this case, the apodised sample,
\begin{equation}
	s_{T}\left(t\right)=\frac{1}{A_{f}}f\left(t/T\right)s\left(t\right)
\end{equation}

where the normalization factor,
\begin{equation}
	A_{f}=\int_{-T/2}^{+T/2}f\left(t/T\right)dt
\end{equation}

corrects for the artificial attenuation introduced by the window. The resulting Fourier transform (Figure \ref{appodisation_c}),
\begin{equation}
	S_{T}\left(u\right)=\frac{T}{A_{f}}S\left(u\right)*F\left(uT\right).
\end{equation}

	Among the many possible apodisation windows (Figures \ref{appodisation_a} and \ref{appodisation_b}), all of which are zero outside the interval $\left[-T/2;+T/2\right]$, we can mention the \couleur{Dirichlet} window,
\begin{equation}
	\Pi\left(t/T\right)
\end{equation}

	which is none other than the window discussed in the chapter on signal truncation. It is important to note that, as in this case, there is no magic solution: the reduction in \textit{"leakage"} comes at the cost of a reduction in frequency resolution.

\subsection{Impact of a trend}
\begin{figure}[H]
	\begin{center}
		\tcbox[colback=white]{\includegraphics[width=16cm]{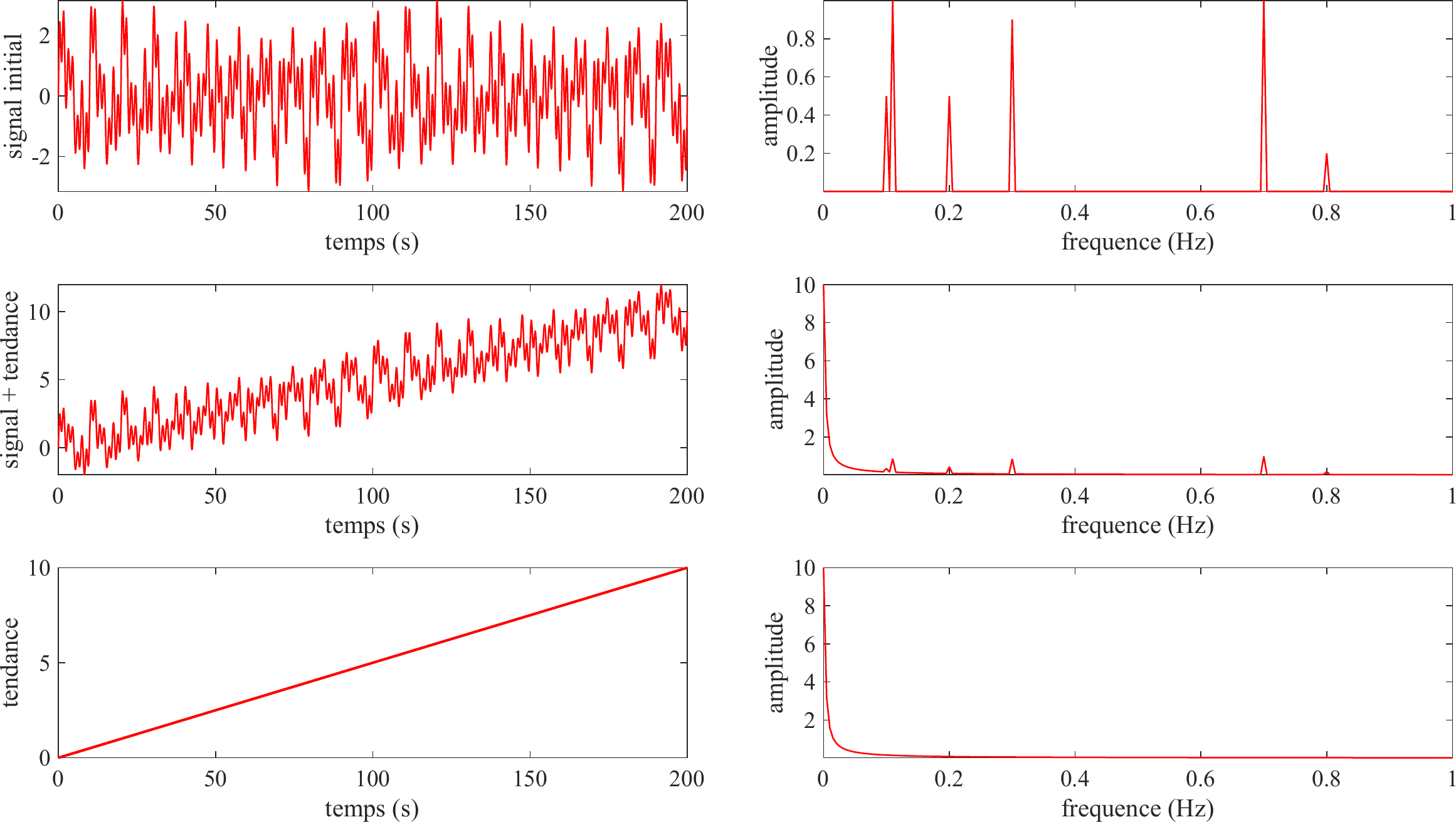}}
	\end{center}
	\caption{Effects of the presence of a trend in the analysed signal. When the signal contains a significant trend (middle left), its spectrum (middle right) is primarily representative of that of the trend alone (bottom left). Some details that are visible in the spectrum (top right) of the signal without the trend (top left) may then be obscured.  \label{tendance}}
\end{figure}

	We will refer to a trend as the component of the sampled signal characterised by oscillations with periods longer than the duration of the sample itself. Ideally, the energy of this trend should be entirely contained within the spectral coefficient corresponding to the zero frequency; in practice, as we have seen, \textit{"leakage"} causes some of this energy to spill over to neighbouring frequencies. If the trend is significant, and therefore energetic, this leakage will cause significant distortion in the spectral coefficients corresponding to the lower frequencies of the spectrum (Figure \ref{tendance}). There will also be additional effects due to the nature of the \couleur{Schuster} model, which can only produce signals of period $T$. Adopting this model implicitly assumes that the signal being analysed is itself periodic, and the presence of a trend means that this periodic signal will essentially exhibit a sawtooth pattern, with its spectrum dominating the rest. As a result, a large portion of the spectrum obtained can become contaminated and difficult to interpret. The presence of a trend in a signal is therefore an unfortunate event; its removal is necessary to obtain a usable spectrum. However, this removal is generally not straightforward and requires a good understanding of the physics of the signal to develop an appropriate model for the trend to be removed.

\subsection{Statistical issues}
	We will consider the case where the signal contains white Gaussian noise. Due to the linearity of the discrete Fourier transform, the real and imaginary parts of the spectral estimates\footnote{The presence of a tilde indicates that we have an estimate of the parameter in question.} $\widetilde{S}_{k}$ will be Gaussian variables, and the coefficients of the power spectrum,
\begin{equation}
	\widetilde{E}_{k}=\left|\widetilde{S}_{k}\right|^{2}
\end{equation}

 follow a $\chi_{2}$ distribution
 \begin{equation}
	\frac{\widetilde{E}_{k}}{E_{k}}=\chi_{2}^{2}
\end{equation}

where the values, 
\begin{equation}
	E_{k}=\left|S_{k}\right|^{2}
\end{equation}

	are the true (but unknown) values. There are two degrees of freedom because the coefficients of the power spectrum are the sum of two squared Gaussian variables (the imaginary and real parts). The variance of the reduced variable $\widetilde{E}{k}/E{k}$ is 4 and does not decrease as the signal length increases because the number of spectral estimates increases in the same proportion. The only way to reduce the variance is to average $M$ independent spectra.
\begin{equation}
	\widetilde{E}_{k}=\frac{1}{M}\sum_{m=1}^{M}\widetilde{E}_{m,k}
\end{equation}

so that,
\begin{equation}
	\frac{\widetilde{E}_{k}}{E_{k}}=\chi_{2M}^{2}.
\end{equation}

	The variance of the estimator is now reduced to $4/M$. At this point, it is important to note that if the signal being analysed is real, the Fourier coefficients corresponding to negative frequencies do not provide any information beyond that already contained in the positive frequencies. It is therefore illusory to hope for a further reduction in variance by extending the above sum to include negative frequencies.

	If only a single signal is available, it is possible to divide it into segments to perform the averaging recommended earlier. However, in accordance with the uncertainty principle, improving the statistical resolution of the estimates $\widetilde{E}_{k}$ will result in a degradation of the frequency resolution. More specifically, the frequency resolution is such that,
\begin{equation}
	\delta u\approx\frac{M}{T},
\end{equation}

and the standard deviation of the estimator is,
\begin{equation}
	\sigma_{E}=\frac{2}{\sqrt{M}}.
\end{equation}

The uncertainty relation is derived from these results,
\begin{equation}
	\sigma_{E}\times\delta u\approx\frac{2\sqrt{M}}{T}.
\end{equation}

	Assuming that the noise contaminating the data is white and Gaussian, it is possible to use the previous results to calculate the bounds of the confidence intervals associated with the $\widetilde{E}_{k}$ estimates,
\begin{equation}
	\frac{2M\widetilde{E}_{k}}{\chi_{2M}^{2}\left(\alpha/2\right)}\leq E_{k}\leq\frac{2M\widetilde{E}_{k}}{\chi_{2M}^{2}\left(1-\alpha/2\right)}.
\end{equation}

	where $\alpha$ is the probability that the true value is not within the interval. The use of a window function $f\left(t\right)$ results in a reduction in the number of degrees of freedom, which must be taken into account in the previous calculations. In this case, $M$ should be replaced by,
\begin{equation}
	M_{a}\approx\frac{M}{T}\int_{-T/2}^{+T/2}f\left(t/T\right)\mathrm{d}t
\end{equation}

	In the case of the \couleur{Hamming} window, this reduction is approximately 50\%\footnote{Some authors suggest overlapping the signal segments by the same proportion to preserve all the initial information.}.

\section{Autoregressive model analysis}
	We have seen that a signal consisting of a sum of harmonic functions satisfies a recursive formula where the coefficients $b_{m}$ determine the spectrum. We will now examine some of the ways to estimate the autoregressive coefficients for spectral analysis. There are several possible approaches, generally named after their developers. For example, the \couleur{Pisarenko} model, which is a sum of sinusoids, can be considered an autoregressive model. This is what Pisarenko chose to do, using the method of least squares to determine the model parameters. \couleur{Burg}, on the other hand, takes a different approach and chooses to fit the autoregressive parameters by maximising the entropy of the discrepancy between the data and the signal reconstructed by the autoregressive model.

\subsection{The prediction error filter}
	In practice, the estimation of the parameters $b_{m}$ of the autoregressive model involves the use of a quality criterion for the fit, which may involve a number of \apriori constraints on the nature of the signal being analysed. The criterion used by \couleur{Burg} involves minimising the total energy of the prediction error, defined by
\begin{equation}
	e_{n}\equiv s_{n}+\sum_{m=1}^{M}b_{m}s_{n-m}
\end{equation}

that is, to make,
\begin{equation}
	\sum_{n}e_{n}^{2}\;\;\;\mathrm{MINIMUM}.
\end{equation}

	The expression for the prediction error can be rewritten in the form of a convolution,
\begin{equation}
	\left\{ e_{n}\right\} =\left\{ s_{n}\right\} *\left\{ 1,b_{1},b_{2},\cdots,b_{M}\right\} 
\end{equation}

where the causal filter appears,
\begin{equation}
	fep\equiv\left\{ 1,b_{1},b_{2},\cdots,b_{M}\right\} 
\end{equation}

is called the prediction error filter. The coefficients $b_{m}$ that minimise the energy of the prediction error are such that,
\begin{equation}
	\begin{split}
0 & = & \frac{1}{2}\frac{\partial}{\partial b_{m}}\sum_{n}e_{n}^{2}\\
 & = & \sum_{n}e_{n}\frac{\partial e_{n}}{\partial b_{m}}\\
 & = & \sum_{n}e_{n}s_{n-m}
	\end{split}
\end{equation}

	If the number $M$ of autoregressive coefficients is unlimited, a simple change of variable allows us to rewrite the last line in the form,
\begin{equation}
	\sum_{n}e_{n+k}s_{n-l}=0\;\;\;\left(k>0,\, l\geq0\right),
\end{equation}

	which is still true after multiplication by a constant,
\begin{equation}
	\sum_{n}e_{n+k}b_{l}s_{n-l}=0\;\;\;\left(k>0,\, l\geq0\right).
\end{equation}

	Of course, the sum of such expressions remains equal to zero, and in particular, we have the following,
\begin{equation}
	\sum_{n}e_{n+k}\sum_{l\in\Bbb{N}}b_{l}s_{n-l}=0\;\;\;\left(k>0\right)
\end{equation}

	which can be simplified using the definition of the prediction error itself, to find that the autocorrelation
\begin{equation}
	\sum_{n}e_{n+k}e_{n}=r_{e,e}\left(k>0\right)=0.
\end{equation}

	Since the autocorrelation is a symmetric function, we can modify the condition on $k$ to obtain,
\begin{equation}
	r_{e,e}\left(k\neq0\right)=0
\end{equation}

Cette expression montre que, 
\begin{center}
\warning{The autocorrelation function of the prediction error produced by an infinite duration prediction error filter is that of white noise.}
\end{center}

\subsection{Prediction error filter utility}
	The prediction error filter has the ability to transform a signal, $s_{n}$ into white noise, $e_{n}$. In \couleur{Fourier} space, this is expressed by the relation,
\begin{equation}
	\begin{split}
	S\left(Z\right)\times FEP\left(Z\right) & =  E\left(Z\right)\\
 & =  \sigma_{e}
 	\end{split}
\end{equation}

	where $\sigma_{e}^{2}$ is the energy of the white noise $e_{n}$. This relationship allows us to obtain the \couleur{Fourier} transform of the signal $s_{n}$.
\begin{equation}
	\begin{split}
	S\left(Z\right) & =  \frac{\sigma_{e}}{FEP\left(Z\right)}\\
 & =  \frac{\sigma_{e}}{1+b_{1}Z+b_{2}Z^{2}+\cdots}
 	\end{split}
\end{equation}

In practice, the spectral division above is very unstable and generally yields poor results. Stabilization can be achieved by replacing the filter $\left\{ 1, b_{1}, b_{2}, \cdots \right\}$ with its associated minimum-phase filter. By doing so, the phases are destroyed, and it is only possible to recover the amplitude spectrum of the signal,
\begin{equation}
	\left|S\left(Z\right)\right|=\frac{\sigma_{e}}{\left|\mathcal{DM}\left\{ 1+b_{1}Z+b_{2}Z^{2}+\cdots\right\} \right|}.
\end{equation}
\chapter{\titrechap{Wavelet transform analysis}}
\minitoc
\section{Wavelets: A brief history}
\subsection{Recent history}
	Wavelet analysis emerged in the early 1980s and was the subject of significant mathematical research for about a decade. Following this period of emergence, wavelet analysis methods have been fundamental to numerous applications in fields as diverse as geophysics, medical imaging, astrophysics, data compression, etc. Today, theoretical work continues and is published in particular in the journal \href{https://www.journals.elsevier.com/applied-and-computational-harmonic-analysis}{\textit{Applied and Computational Harmonic Analysis}}. A common feature of various wavelet techniques is the analysis of signals with fluctuations over a wide range of spatial or temporal scales. This analysis is performed via decompositions based on families of functions, which have the remarkable property of being derived by dilating a base function - the analysing wavelet - in such a way that all functions in a given family have the same shape. Depending on the analysing wavelet chosen, the resulting wavelet family may be orthogonal or non-orthogonal, with mathematical properties more or less appropriate to the signals being analysed.

	In retrospect, it has been recognised that the concept of wavelets with a constant shape was introduced by the Hungarian mathematician \couleur{Alfréd Haar}\index{Haar wavelets} in the early 20th century \shortcite{haar1909theorie}. However, \couleur{Haar}'s orthogonal wavelets were not the starting point for wavelet theory in its current form. It was the work of \couleur{Jean Morlet} in the early 1980s that really launched the field. The wavelets proposed by \couleur{Jean Morlet} are non-orthogonal and are a fairly direct adaptation of Fourier analysis by segments \shortcite{morlet1982wave}. However, it was in fact the concept of constant-shape wavelets, introduced somewhat empirically by Morlet, that served as the basis, in particular thanks to a highly theoretical paper by \couleur{Alex Grossmann} and \couleur{Jean Morlet} entitled "Decomposition of Hardy function into square-integrable wavelets of constant shape", published in an applied mathematics journal \shortcite{grossmann1984decomposition}. The generalisation of Morlet's wavelet transform laid the foundation for continuous wavelet transform.

A little later, the orthogonal wavelet transform was developed under the direction of \couleur{Yves Meyer}, who was then a professor at the Centre de Recherche en Mathématiques de la Décision (\href{https://www.ceremade.dauphine.fr/}{CEREMADE}) at the University of Paris Dauphine. The collective volume \couleur{"Fundamental Papers in Wavelet Theory"}, published in 2006, provides an insight into the emergence of wavelet theory and shows that several fundamental foundations had already been established for some time, although they had not yet triggered the synthesis work of the 1980s \shortcite{daubechies2006fundamental}. As is often the case in research, serendipity played a role in the history of wavelets when \couleur{Yves Meyer} discovered the paper by \couleur{Grossmann} and \couleur{Morlet} while waiting his turn at the photocopier in his laboratory, leafing through journals brought in by a colleague.

\subsection{From \couleur{Joseph Fourier} to \couleur{Dennis Gabor}}
\begin{figure}[H]
	\begin{center}
		\tcbox[colback=white]{\includegraphics[width=16cm]{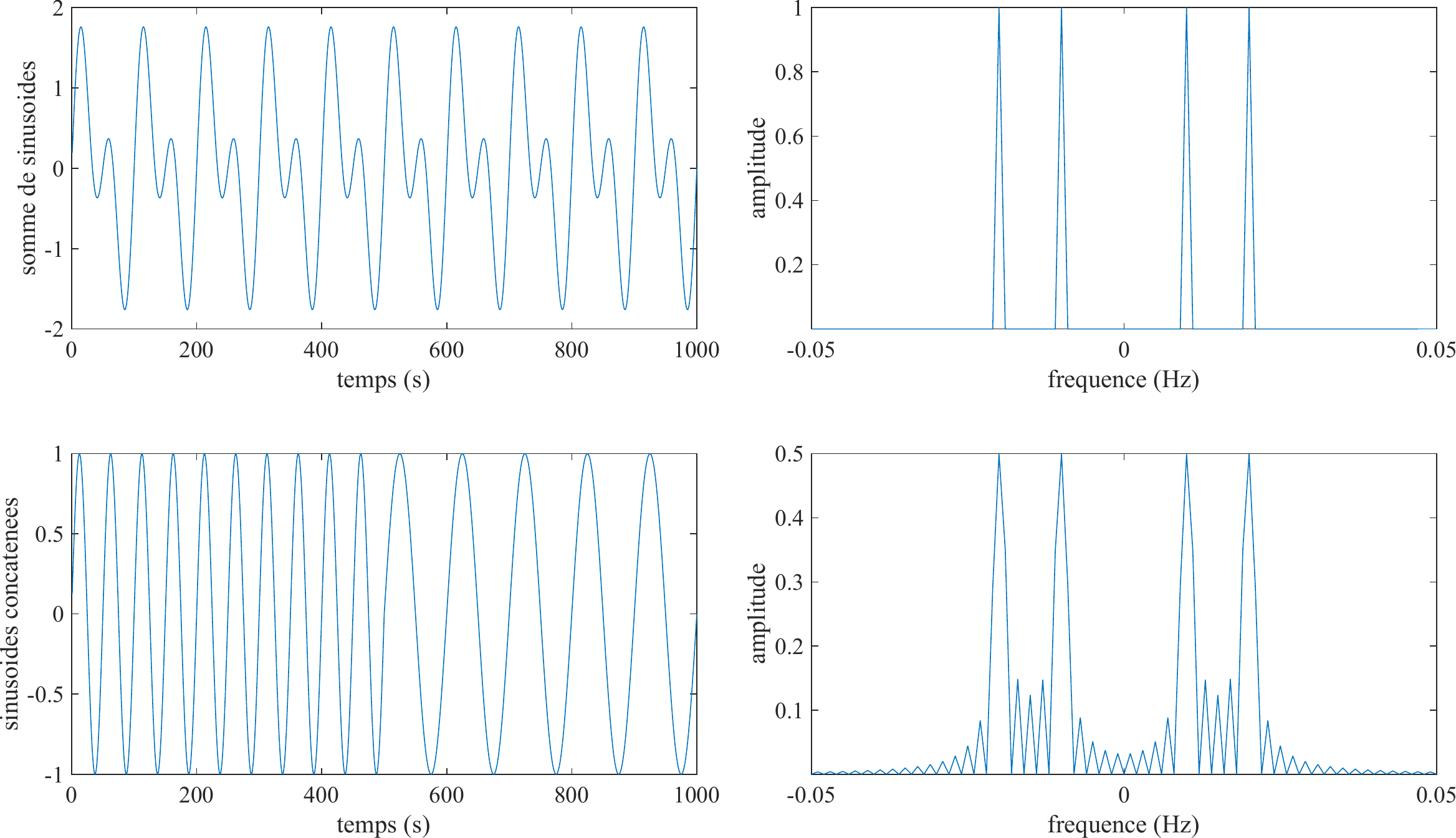}}
	\end{center}
	\caption{Illustration of the inadequacy of the \couleur{Fourier} transform for non-stationary signals: a signal consisting of two successive sinusoids (bottom left) has an amplitude spectrum (bottom right) very similar to that of a signal consisting of the superposition of the two sinusoids (top left), despite their different temporal structures (top right).}
	\label{ondelettes01}
\end{figure}

	We have already noted that non-stationary signals are very common in geophysics, and that a significant part of the information they contain is embedded precisely in this non-stationarity. The \couleur{Fourier} transform, by completely neglecting the time domain, is poorly suited to the analysis of non-stationary signals. This is illustrated in the figure (\ref{ondelettes01}) obtained with the code \couleur{ondelette01.m}.

This figure (\ref{ondelettes01}) shows that the amplitude spectrum of a non-stationary signal composed of two successive sinusoids is little different from that of two superimposed sinusoids. In both cases the amplitude spectrum shows peaks at the frequencies of the sinusoids. The information about the transition from one sinusoid to another in the non-stationary signal is contained in the low amplitude peaks of the spectrum and in the phase of the \couleur{Fourier} transform. Therefore, information initially localised at a specific point on the time axis is dispersed in the frequency domain, making it difficult to retrieve. A simple solution to preserve, at least partially, the information about the transition from one sinusoid to another is to perform a \couleur{Fourier} analysis on successive segments of the signal. This was the idea of \couleur{Gabor}\index{Gabor} when calculating the spectrogram,
\begin{equation}
\mathcal{G}\left[w,f\right]\left(u,t\right)=\int_{-\infty}^{+\infty}f\left(\tau\right)w\left(t-\tau\right)\exp\left[-2i\pi u\left(t-\tau\right)\right]\mathrm{d}\tau.
\end{equation}

\begin{figure}[H]
	\begin{center}
		\tcbox[colback=white]{\includegraphics[width=16cm]{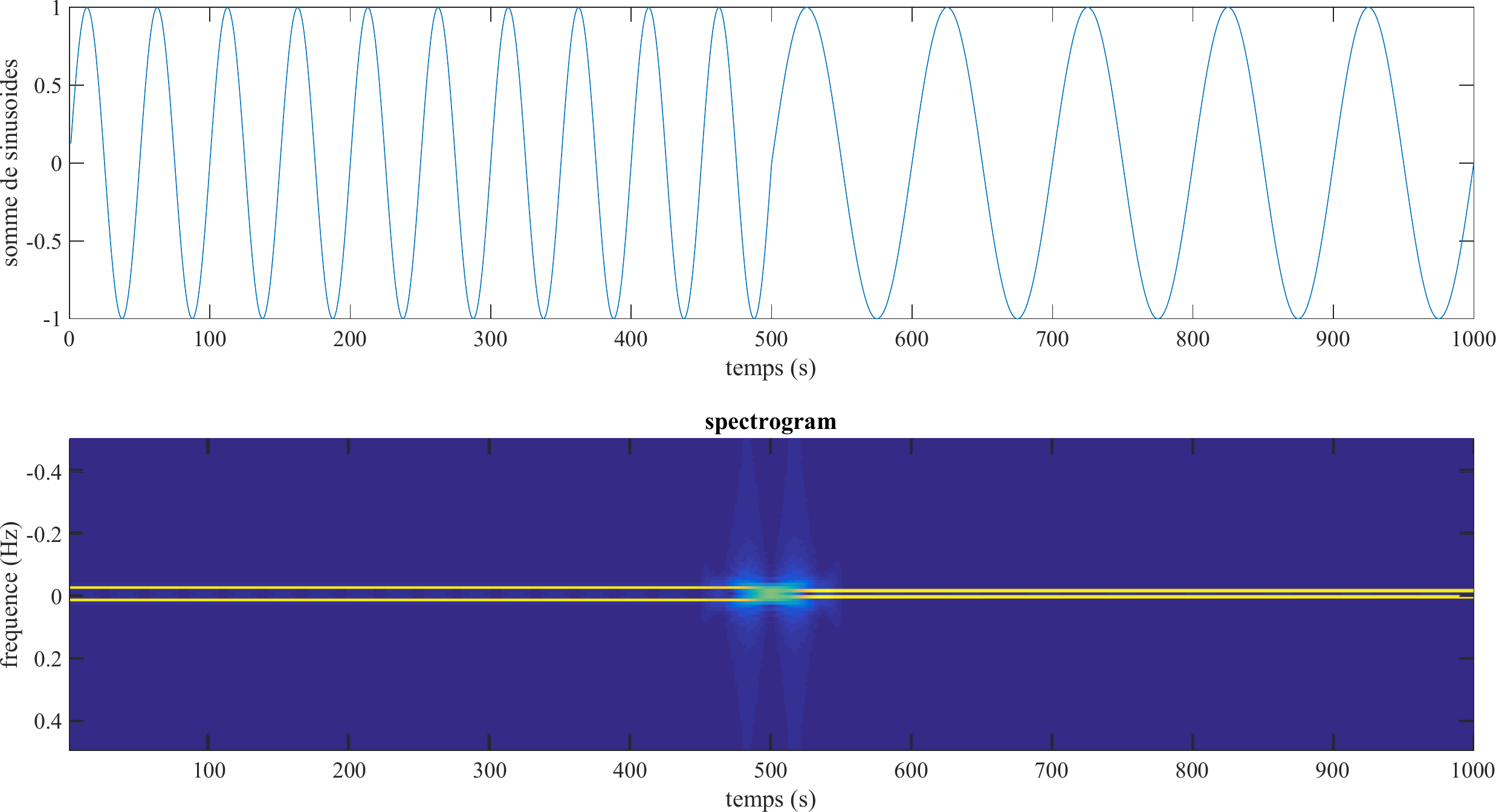}}
	\end{center}
	\caption{Example of the calculation of a simple spectrogram using a \couleur{Dirichlet} window to extract segments of the signal. The amplitude of the spectrogram is plotted in the time-frequency plane.}
	\label{ondelettes02}
\end{figure}

The function $w(t)$ is a window used to extract a segment of the signal. The code \couleur{ondelettes02.m} computes a simple spectrogram using a \couleur{Dirichlet} window to extract successive segments of the signal. The result is shown in figure (\ref{ondelettes02}), where it can be observed that the frequency and time information carried by the signal is partially recovered in the time-frequency plane representing the spectrogram

	The spectrogram allows the time-frequency analysis of a signal, for example by displaying its energy $\left|F\left(u,t\right)\right|^{2}$. The choice of the window function $w\left(t\right)$ is, \apriori, quite flexible, but it is advantageous for this window to be optimal with respect to \couleur{Heisenberg}'s uncertainty principle\footnote{See the chapter on \couleur{Time-Frequency Duality} for more details on the uncertainty principle.}. For this reason, \couleur{Gabor} chose the Gaussian window, which leads to the following expression for the spectrogram,
\begin{equation}
\mathcal{G}\left[\exp\left(-\frac{\pi t^{2}}{T^{2}}\right),f\right]\left(u,t\right)=\int_{-\infty}^{+\infty}f\left(\tau\right)\exp\left(-\frac{\pi (t-\tau)^{2}}{T^{2}}\right)\exp\left[-2i\pi u\left(t-\tau\right)\right]\mathrm{d}\tau.
\end{equation}

	By defining the analysis function $g_{T}$ as a\footnote{sometimes referred to as the "\couleur{gaborette}" in french},\begin{equation}
	g_{T}\left(u,t\right)\equiv\exp\left(-\frac{\pi t^{2}}{T^{2}}\right)\exp\left(-2i\pi ut\right),
\end{equation}

	it can be seen that the spectrogram can be rewritten in the form of a convolution product,
\begin{equation}
	\mathcal{G}\left[g_{T},f\right]\left(u,t\right)=\left[g_{T}\left(u,\cdot\right)*f\left(\cdot\right)\right]\left(t\right).
\end{equation}

\begin{figure}[H]
	\begin{center}
		\tcbox[colback=white]{\includegraphics[width=16cm]{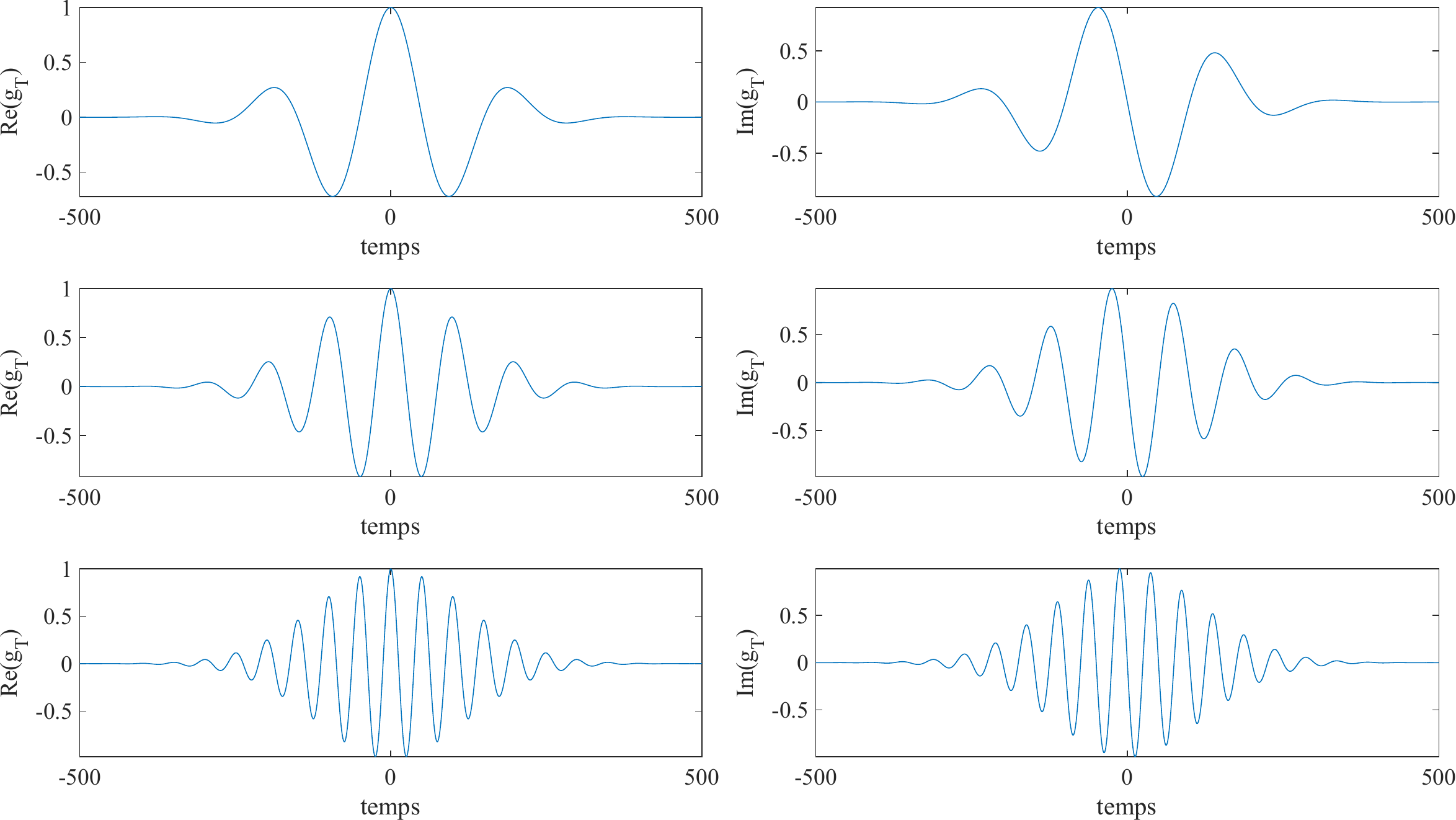}}
	\end{center}
	\caption{Analysis function $g_{T}$ for three different frequencies.}
	\label{ondelettes03}
\end{figure}

	The code \couleur{ondelettes03.m} allows you to calculate the function $g_{T}$, with some examples shown in the figure (\ref{ondelettes03})

	The analysis function is parameterised by the duration $T$, which defines the width of the window. The Gaussian in the time domain corresponds to another Gaussian in the frequency domain, and these two functions determine the time and frequency resolutions, $\delta t$ and $\delta u$, that satisfy,
\begin{equation}
	\delta t\times\delta u=\frac{1}{4\pi},
\end{equation}

	and remain constant over the whole of the $\left(u,t\right)$ plane:
\begin{equation}
	\delta t=\frac{T}{2\sqrt{\pi}}\;\;\mathrm{et}\;\;\delta u=\frac{1}{2T\sqrt{\pi}}.
\end{equation}

\subsection{From \couleur{Dennis Gabor} to \couleur{Jean Morlet}}
	It was in the early 1980s that a significant modification of \couleur{Gabor}'s spectrogram was proposed by \couleur{Jean Morlet}, leading to the development of the wavelet transform. The modification consisted in adjusting the duration $T$ of the window according to the frequency $u$. \couleur{Jean Morlet} chose the following setting,
\begin{equation}
	T=\frac{\sqrt{\alpha}}{u},
\end{equation}

	where $\alpha$ is a parameter whose meaning will be discussed later. Using this new definition of the window duration, the analysis function of \couleur{Gabor} becomes,
\begin{equation}
	m_{\alpha}\left(u,t\right)=\exp\left[-\frac{\pi}{\alpha}\left(ut\right)^{2}\right]\exp\left(-2i\pi ut\right).
\end{equation}

	By performing the variable change $u \longmapsto a^{-1}$, which introduces the dilation $a$, we obtain the classical expression of the normalised \couleur{Morlet} wavelet,
\begin{equation}
	m_{\alpha}\left(\frac{t}{a}\right)=\frac{1}{a}\exp\left[-\frac{\pi}{\alpha}\left(\frac{t}{a}\right)^{2}\right]\exp\left(-\frac{2i\pi t}{a}\right).
	\label{WaveletMorlet} 
\end{equation}

	The $\alpha$ parameter allows you to adjust the ratio between the width of the Gaussian envelope and the dominant period of the wavelet, which in the case of the \couleur{Morlet} wavelet is $1/a$. For reasons we will discuss later, it is necessary that $\alpha > 2$ for the wavelet to be considered as having zero mean. The larger $\alpha$, the better the frequency resolution $\delta u$, but at the expense of the time resolution $\delta t$.

\begin{figure}[H]
	\begin{center}
		\tcbox[colback=white]{\includegraphics[width=16cm]{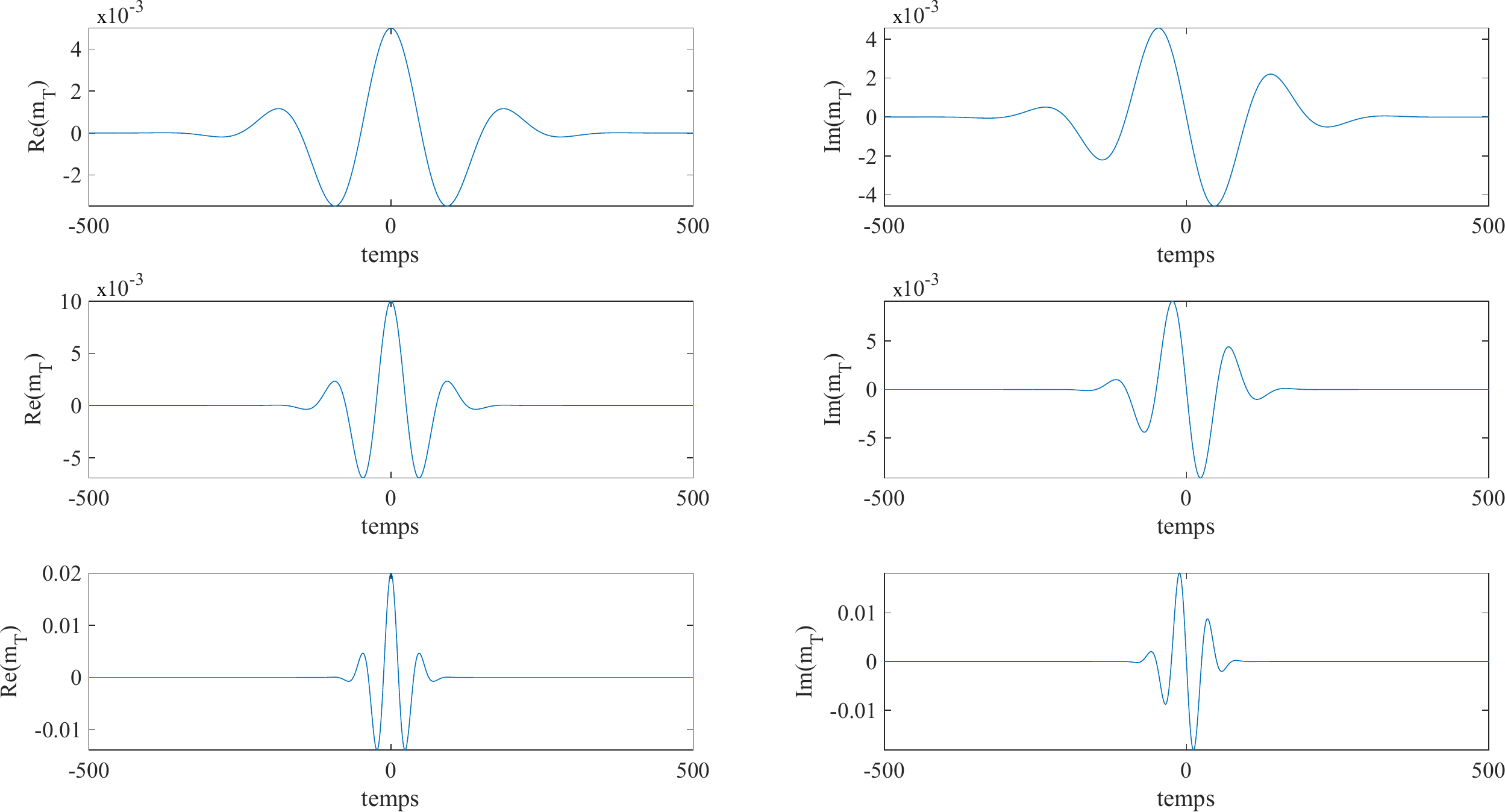}}
	\end{center}
	\caption{\couleur{Morlet} wavelet for three different dilations $a$}
	\label{ondelettes04}
\end{figure}

	The code \couleur{ondelettes04.m} calculates the function $m_{\alpha}(t/a)$, with some examples shown in figure (\ref{ondelettes04}). A comparison with figure (\ref{ondelettes03}) clearly illustrates the fundamental property of wavelets: their constant shape. All wavelets in the same family are obtained by dilating a single analysing wavelet. This property is the basis of all wavelet transforms: continuous, discrete, orthogonal, \etc 

	The \couleur{Gabor} spectrogram thus becomes the \couleur{Morlet} wavelet transform,
\begin{equation}
	\mathcal{W}\left[m_{\alpha},f\right]\left(a,t\right)=\frac{1}{a}\left[m_{\alpha}\left(\frac{t}{a}\right)*f\left(t\right)\right]\left(t\right).
	\label{WTMorlet} 
\end{equation}

	Since we are generally interested in real signals, the symmetry properties of the \couleur{Fourier} transform imply that it is sufficient to compute the \couleur{Gabor} spectrogram for $u \geq 0$, \ie $a > 0$.

	The introduction of the dilation parameter $a$ significantly changes the properties of the wavelet transform compared to those of the \couleur{Gabor} spectrogram. In particular, the wavelet transform adapts well to non-stationarities because, whatever the time constant of a sudden change in the signal, there will always be wavelets of appropriate size to localise this change. This is due to the fact that the time resolution of the \couleur{Morlet} wavelet transform is given by,
\begin{equation}
	\delta t=\frac{a}{2\sqrt{\alpha\pi}},
\end{equation}

	and is therefore not constant in the half-plane\footnote{This is called the \couleur{Poincaré} half-plane.} $\left(a > 0, t\right)$. Of course, in accordance with the uncertainty principle mentioned earlier, the frequency resolution varies inversely with dilation.
\begin{equation}
	\delta u=\frac{1}{2a}\sqrt{\frac{\alpha}{\pi}}.
\end{equation}

\begin{figure}[H]
	\begin{center}
		\tcbox[colback=white]{\includegraphics[width=16cm]{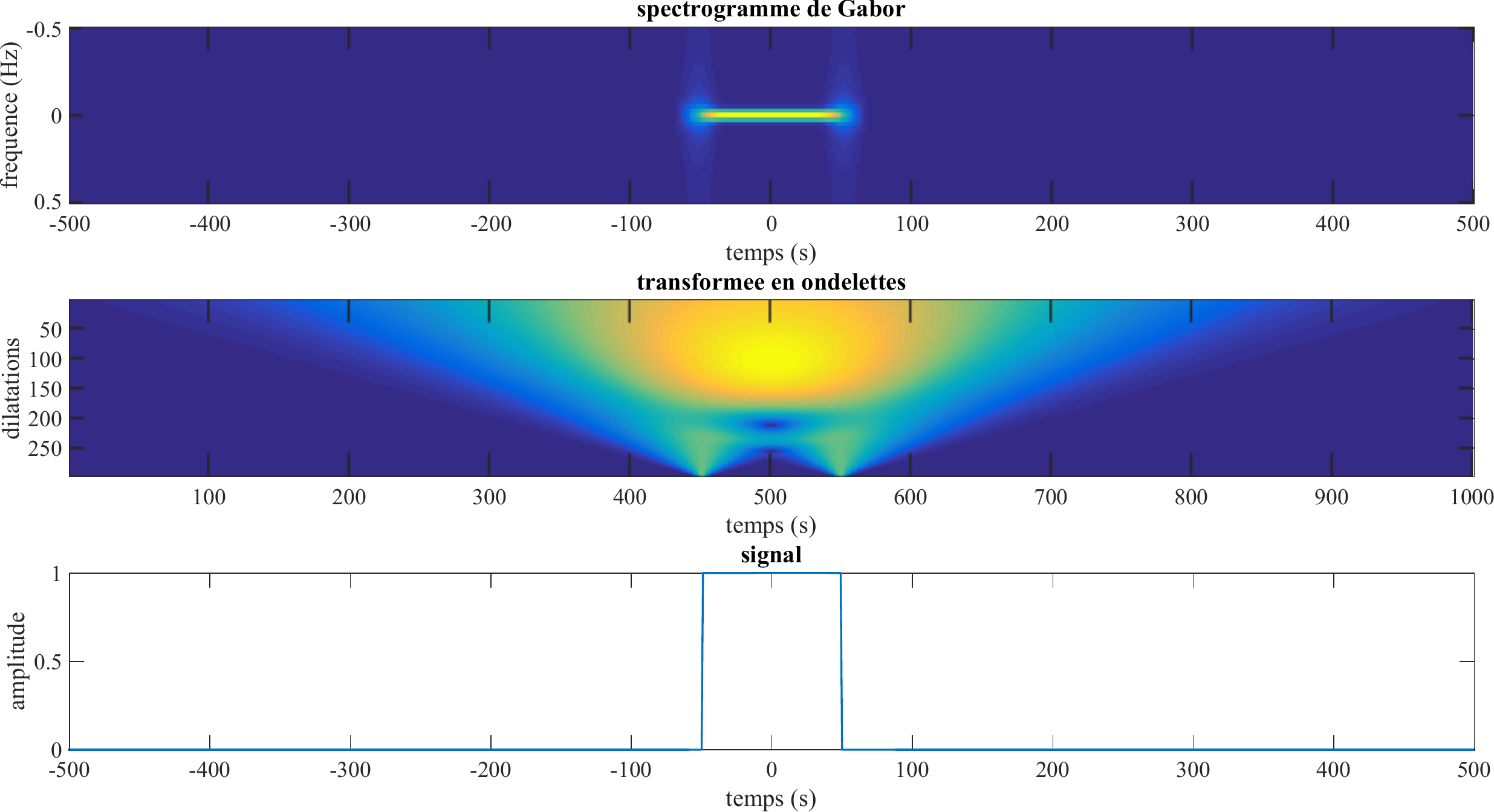}}
	\end{center}
	\caption{The magnitude of the spectrogram (top) and the \couleur{Morlet} wavelet transform (middle) of a rectangular pulse.}
	\label{ondelettes05}
\end{figure}

This ability of the wavelet transform to adapt to the finest details of a signal has earned it the nickname 'the mathematical microscope'. This property is illustrated by the following code, which computes both the spectrogram and the wavelet transform of a Dirichlet window. The result is shown in figure (\ref{ondelettes05}), obtained using \couleur{ondelettes05.m}.

	Figure (\ref{ondelettes05}) effectively illustrates the multi-scale analysis capabilities of the wavelet transform. Wavelets with small dilation focus on the discontinuities in the signal, while wavelets with dilation matched to the width of the window correspond to a maximum amplitude in the wavelet transform.

\subsection{Questions addressed in this chapter}
	Wavelet analysis has become an important field in mathematical analysis as well as in signal and image processing. Based on a strong theoretical framework, wavelet methods are used in numerous applications thanks to readily available algorithms, the most famous of which are those developed by \href{http://perso.ens-lyon.fr/patrick.flandrin/software2.html}{\couleur{Patrick Flandrin}} and his colleagues\footnote{\texttt{http://perso.ens-lyon.fr/patrick.flandrin/software2.html}} and those from the Statistics Department of Stanford University\footnote{\texttt{http://www-stat.stanford.edu/~wavelab/}}. These software tools will be very useful complements to the functions developed in this course.

	In the remainder of this extensive chapter, we will focus specifically on the use of wavelets for signal analysis. We will explore how it is possible to \warning{teach physics to wavelets}, so that they allow us to extract certain information about physical systems or phenomena. For reasons that will become clear later, it is primarily the continuous wavelet transform, obtained by generalising the equation \ref{WTMorlet}, that will enable us to achieve our goals. Therefore, in contrast to most texts, we will only moderately cover the topic of orthogonal wavelets. Due to space limitations in this short introduction, our discussion will primarily be of one-dimensional (1D) wavelets,

\begin{itemize}[label=$\ast$]
	\item Non-orthogonal wavelets are functions that can be chosen with considerable flexibility, allowing them to be tailored to the physical characteristics of the signals being analysed.
	\item The continuous wavelet transform allows the wavelets to be precisely localised on the events that make up the signals being analysed.
	\item The theory of the continuous wavelet transform is straightforward, and its integration into physical theories such as potential theory, wave phenomena, \etc is more feasible than with orthogonal wavelets.
\end{itemize}

\section{Continuous Wavelets — Discrete Wavelets — Orthogonal Wavelets}
	Before looking at specific aspects of wavelet analysis, we will first establish some basic principles that characterise the two main families of wavelets: continuous wavelets and orthogonal wavelets.

\subsection{Continuous Wavelet Transform}
	The continuous wavelet transform is easily obtained by generalizing the \couleur{Morlet} wavelet transform. For reasons that will become clearer later, we choose to define the continuous wavelet transform as a convolution product,
\begin{equation}
\mathcal{W}\left[\psi,f\right]\left(a,t\right)\equiv\left[f\left(\cdot\right)*\psi_{a}\left(\cdot\right)\right]\left(t\right),	
\end{equation}

	where the wavelet is such that,
\begin{equation}
	\psi_{a}\left(t\right)\equiv\frac{1}{a}\psi\left(\frac{t}{a}\right).
\end{equation}

	The scale parameter $a > 0$, also known as the dilation, affects the analysing wavelet $\psi(t)$ by stretching if $a > 1$ or compressing if $a < 1$.

	As defined above, the continuous wavelet transform is a bank of filters applied to the signal $f$. Since wavelets are obtained by dilation, their Fourier transforms, which are the corresponding filters, are also a family of functions generated by dilation. We will see later that the choice of wavelet is quite flexible, which allows us to give the wavelet transform special properties, including giving it physical meaning. In fact, the primary condition that a wavelet must satisfy is the admissibility condition,
\begin{equation}
	\int_{0}^{+\infty}\left|\Psi\left(u\right)\right|^{2}\frac{\mathrm{d}u}{u}<\infty,
\end{equation}

	which requires the wavelet to have a zero mean. We will discuss later that this condition is necessary to establish the reconstruction formula corresponding to the inverse wavelet transform

\subsection{Orthogonal Wavelets}
	In the modern history of wavelets, orthogonal wavelets were not discovered immediately after the introduction of the continuous wavelet transform \etc although \couleur{Haar} wavelets, discovered in the early 20th century, are indeed orthogonal! It is also interesting to note that \couleur{Haar} wavelets were used to filter signals in the 1970s\shortcite{gubbins1971two}\footnote{Gubbins, D., 'Two dimensional digital filtering with Haar and Walsh transforms', \textit{Annales de Géophysique}, \textbf{27}, 85-104, 1971.}, well before the advent of wavelet theory. It is these wavelets that we will use as an example to introduce orthogonal wavelets and their main properties.

	Orthogonality requires a scalar product, which we will define here as,
\begin{equation}
	\mathcal{W}\left[\psi,f\right]\left(a,t\right)\equiv\left[f\left(\cdot\right)*\psi_{a}\left(\cdot\right)\right]\left(t\right).
\end{equation}

	The \couleur{Haar} wavelets are constructed from the function consisting of a positive rectangular window followed by a negative one,
\begin{equation}
	\psi_{H}\left(t\right) = \Pi\left(t+\frac{1}{2}\right) - \Pi\left(t-\frac{1}{2}\right).
\end{equation}

	These functions have compact support, and an initial subset of orthogonal functions is trivially obtained by keeping only those functions whose supports are disjoint while densely covering $\mathbb{R}$.
\begin{equation}
	\mathcal{H}_{1} = \{ \psi_{H}\left(t - 2m\right) \; m \in \mathbb{Z} \}.
\end{equation}

	A second subset of functions which are orthogonal to each other and also orthogonal to the family $\mathcal{H}{1}$ is formed by dilating the functions in $\mathcal{H}{1}$ by a factor of $a = 2$,
\begin{equation}
\mathcal{H}_{2} = \{ \psi_{H}\left(\frac{t}{2} - 2m\right) \; m \in \mathbb{Z} \}.
\end{equation}

	The \couleur{Haar} basis is obtained by iterating this process,
\begin{equation}
\mathcal{H} = \bigcup_{n \in \mathcal{Z}} \mathcal{H}_{2^n}.
\end{equation}

	The example of \couleur{Haar} wavelets shows that orthogonality is achieved if the dilations are powers of $2$. This is why the term 'octave' is often used in wavelet theory terminology. The fact that the allowed dilations are powers of $2$ is a rather fundamental property that holds for most orthogonal wavelets in use. However, it is not an absolutely necessary property, since in general orthogonality can be satisfied if the dilation is given by,
\begin{equation}
a = q^n \; \mathrm{avec} \; q \in \mathbb{P}.
\end{equation}

	Another important property highlighted by the example of Haar wavelets is that orthogonality requires the wavelets to be translated according to a dyadic tiling when $a = 2^n$, triadic for $a = 3^n$, and so on. This constraint is of particular practical importance because it implies that the orthogonal wavelet transform is not invariant under translation. This can cause serious problems in signal analysis, since adding or removing a few values at the beginning of a signal can significantly alter its orthogonal wavelet transform.

\begin{figure}[H]
	\begin{center}
		\tcbox[colback=white]{\includegraphics[width=16cm]{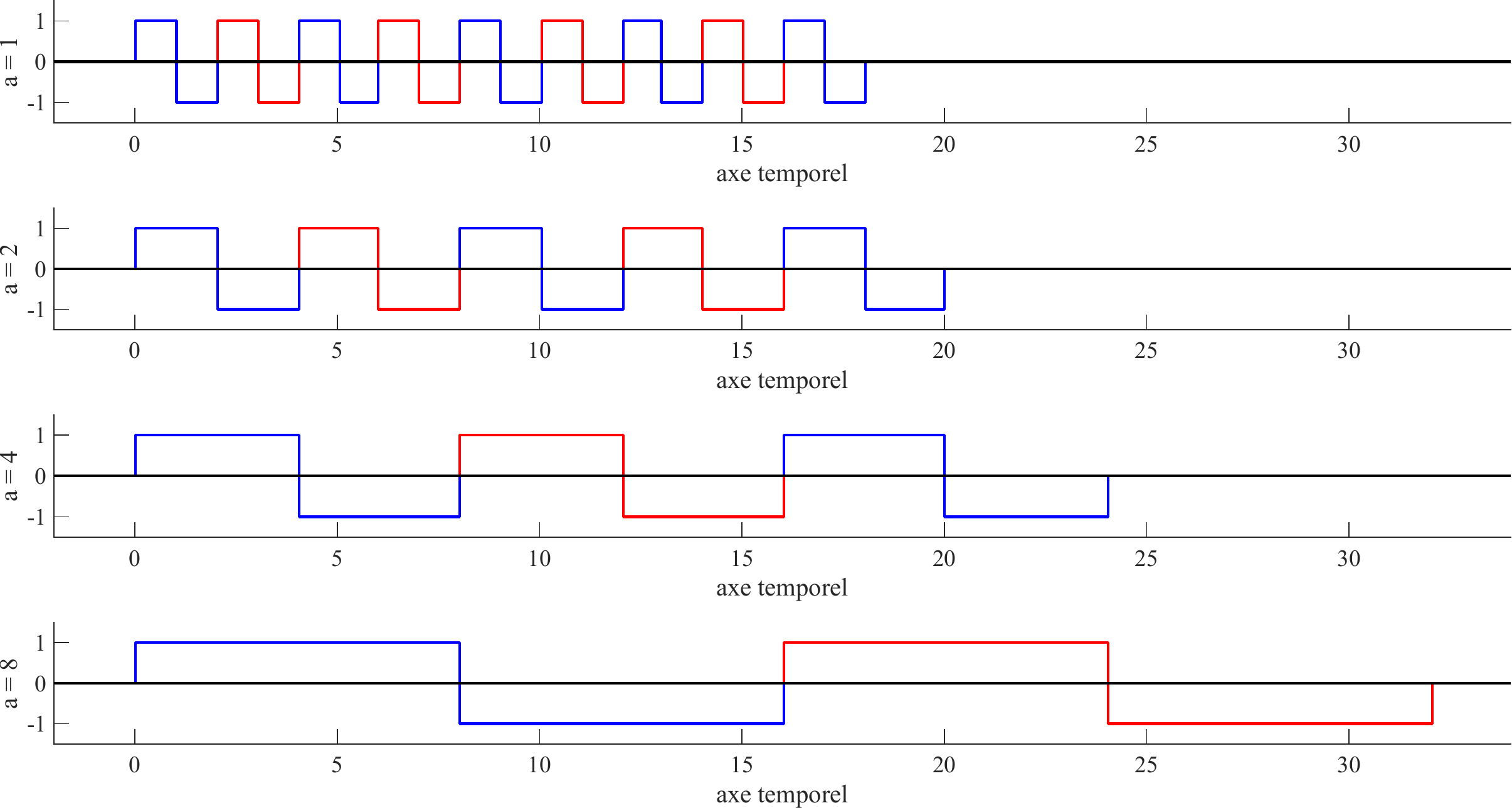}}
	\end{center}
	\caption{Some functions of the Haar basis for $a = 1$, $2$, $4$, and $8$}
	\label{ondelettes06}
\end{figure}

	The code \couleur{ondelettes06.m} calculates the functions of the \couleur{Haar} basis, as shown in the figure (\ref{ondelettes06}).

\section{How is the Orthogonal Wavelet Transform computed?}
\subsection{The pyramid algorithm}
	The very particular construction of the orthogonal wavelet transform, namely the dyadic sampling and octave discretisation of the dilations, allows a fast computation of the wavelet coefficients thanks to an algorithm proposed by \shortciteN{mallat1989theory} and inspired by the pyramid algorithm developed in the 1970s. To understand the principle of this algorithm, let us start with the discrete version of the wavelet coefficients for the minimum dilation $a_0 = 2^0$,
\begin{equation}
	\begin{split}
	& \mathbf{W}_0 = \{ f_{1} - f_{0} \; , \;  f_{3} - f_{2} \; , \; f_{5} -  f_{4} \; \cdots \} \\
	& = \curvearrowright^2_1 \; \{f_{1}-f_{0} \; , \;  f_{2}-f_{1} \; , \; f_{3}-f_{2} \; , \; f_{4}-f_{3} \; , \; f_{5}-f_{4} \; \cdots \} \\
	& = \curvearrowright^2_1 \; \{ +1, \; -1 \} * \{ f_{0} \; , \; f_{1} \; , \;  f_{2} \; , \; f_{3} \; , \; f_{4} \; , \; f_{5} \; \cdots \} \\
	& = \curvearrowright^2_1 \; \{ +1, \; -1 \} * \mathbf{f}_0,
	\end{split}
	\label{pyramidal01}
\end{equation}

	where $\mathbf{f}_0$ represents the initial signal and where the operator $\curvearrowright^2_1$ denotes subsampling by $2$ such that,
\begin{equation}
\curvearrowright^2_1 \; \{ 0,1,2,3,4,5, \cdots \} = \{ 0,2,4, \cdots \}.
\end{equation}

	Equation (\ref{pyramidal01}) shows that wavelet coefficients can be obtained by applying a high-pass filter to the signal and then removing every other value from the filtered signal. The high-pass filter is nothing other than the dilation wavelet $a_0$. Let us now consider the wavelet coefficients for the dilation $a_1 = 2^1$,
\begin{equation}
	\begin{split}
	& \mathbf{W}_1 = \{ (f_{3}+f_{2}) - (f_{1}+f_{0}) \; , \; (f_{7}+f_{6}) - (f_{5}+f_{4}) \; , \; (f_{11}+f_{10}) - (f_{9}+f_{8}) \; \cdots \} \\
	& = \curvearrowright^2_1 \; \{ 1, \; -1 \} * \{ f_{1}+f_{0} \; , \; f_{3}+f_{2} \; , \; f_{5}+f_{4} \; , \; f_{7}+f_{6} \; , \; f_{9}+f_{8} \; , \; f_{11}+f_{10} \; \cdots \} \\
	& = \curvearrowright^2_1 \; \{ 1, \; -1 \} * \big{[} \; \curvearrowright^2_1 \; \{ 1, \; 1 \} * \{ f_{0} \; , \; f_{1} \; , \;  f_{2} \; , \; f_{3} \; , \; f_{4} \; , \; f_{5} \; , \; f_{6}, \cdots \} \big{]} \\
	& = \curvearrowright^2_1 \; \{ 1, \; -1 \} * \mathbf{f}_1.
	\end{split}
	\label{pyramidal02}
\end{equation}

	Equation (\ref{pyramidal02}) shows that the wavelet coefficients for the dilation $a_1$ are obtained by applying the low pass filter ${ 1, ; 1 }$ and a decimation by two to obtain the signal $\mathbf{f}_1$, followed by a high pass filter and another decimation by two to obtain the coefficients $\mathbf{W}_1$. It is easy to show that the subsequent wavelet coefficients are obtained in the same way. This results in the following cascade,
\begin{equation}
	\begin{split}
	&\curvearrowright^2_1 \; \{ +1, \; -1 \} * \mathbf{f}_0 \mapsto \mathbf{W}_0 \\
	&\curvearrowright^2_1 \; \{ +1, \; +1 \} * \mathbf{f}_0 \mapsto \mathbf{f}_1 \\
	&\hspace*{15mm}\curvearrowright^2_1 \; \{ +1, \; -1 \} * \mathbf{f}_1 \mapsto \mathbf{W}_1 \\
	&\hspace*{15mm}\curvearrowright^2_1 \; \{ +1, \; +1 \} * \mathbf{f}_1 \mapsto \mathbf{f}_2 \\
	&\hspace*{30mm}\curvearrowright^2_1 \; \{ +1, \; -1 \} * \mathbf{f}_2 \mapsto \mathbf{W}_2 \\
	&\hspace*{30mm}\curvearrowright^2_1 \; \{ +1, \; +1 \} * \mathbf{f}_2 \mapsto \mathbf{f}_3 \\
	&\hspace*{45mm}\cdots
	\end{split}
	\label{pyramidal03}
\end{equation}

	This cascade represents the Pyramid algorithm. Note that it assumes that the initial signal contains $2^N$ values. For the decomposition to be complete, i.e. for all the information contained in the signal to be represented in the coefficients $\mathbf{W}$, it is necessary to include $\mathbf{f}_N$, which is simply the sum of the signal values. The following code calculates the wavelet coefficients of a signal in the \couleur{Haar} basis,
\newpage

\begin{lstlisting}[commentstyle=\footnotesize\textit]
function w = DirectHaar(f)
    hf = [1 -1]/sqrt(2);       % Haar wavelet = high-pass filter
    lf = [1 1]/sqrt(2);        % low-pass miror filter
    f = f(:); nf = length(f); meanf = mean(f); w = [];
    while nf > 1
        wa = conv(f,hf);       % convolve signal with high-pass filter
        wa = wa(2:2:end);      % decimate to get wavelet coefficients
        f = conv(f,lf);        % convolve signal with low-pass filter
        f = f(2:2:end);        % decimate
        w = [w wa'];           % merge wavelet coefficients 
        nf = nf/2;            %(slow manner !)
    end  
    w = [w meanf];
end
\end{lstlisting}

	An example application is shown below,
\begin{lstlisting}[commentstyle=\footnotesize\textit]
function ondelettes07()
    close all; clc; home;
    nf = 16;
    f = randi(10,1,nf);     
    disp([input signal:  num2str(f)]);
    w = DirectHaar(f);
    nw = length(w); 
    a = 1;                      % dilatation
    disp(----------------------);

    while nw > 1
        nw = nw/2;
        disp([dilatation a =  num2str(a)]);
        disp([Haar coef:  num2str(w(1:nw))]);
        disp(----------------------);
        w = w(nw+1:end);
        a = 2*a;
    end
    disp([last coefficient (mean of signal) =  num2str(w(end))]);
end
\end{lstlisting}

\subsection{Quadrature Mirror Filters}
	The example of the Haar wavelet decomposition illustrates a property that holds for all orthogonal wavelet bases, namely that the wavelet coefficients are obtained by the iterative application of two filters, a high-pass and a low-pass. These two filters completely define the wavelet basis and are clearly not arbitrary with respect to each other. In fact, it is necessary for the information filtered by the high-pass filter to be exactly complementary to the information filtered by the low-pass filter. Two filters with this property are called \couleur{quadrature mirror filters}.

\subsection{The inverse transform}
	Let us now see how to reconstruct the signal $\mathbf{f}_0$ from its Haar coefficients $\mathbf{W}$. The following function reconstructs a signal from its Haar coefficients
\begin{lstlisting}[commentstyle=\footnotesize\textit]
function f = InverseHaar(w)

    hf = [-1 1]/sqrt(2);           
    w = w(:);
    f = w(end)*ones(size(w));   
    nw = length(w)/2;
    H = repmat(hf,nw,length(f)/2/nw);
    while nw >= 1
            W = repmat(w(1:nw),length(w)/2/nw,length(f)/nw);
            s = H'.*W';         
            f = f + s(:); 
            w = w(nw+1:end);          
            nw = nw/2;
            if nw >= 1
               H = reshape(H,nw,length(f)/nw)/sqrt(2);
            end
    end
end
\end{lstlisting}

\section{Filter, denoise and compress signals using orthogonal wavelets}
	Filtering signals using orthogonal wavelet bases is done in the same way as other decompositions: you change the values of the wavelet coefficients and then calculate the inverse transform to reconstruct the filtered signal. It is interesting to note that the 'brutal' zeroing of certain wavelet coefficients does not produce \couleur{Gibbs} oscillations, unlike filtering in the \couleur{Fourier} basis.

\begin{figure}[H]
	\begin{center}
		\tcbox[colback=white]{\includegraphics[width=16cm]{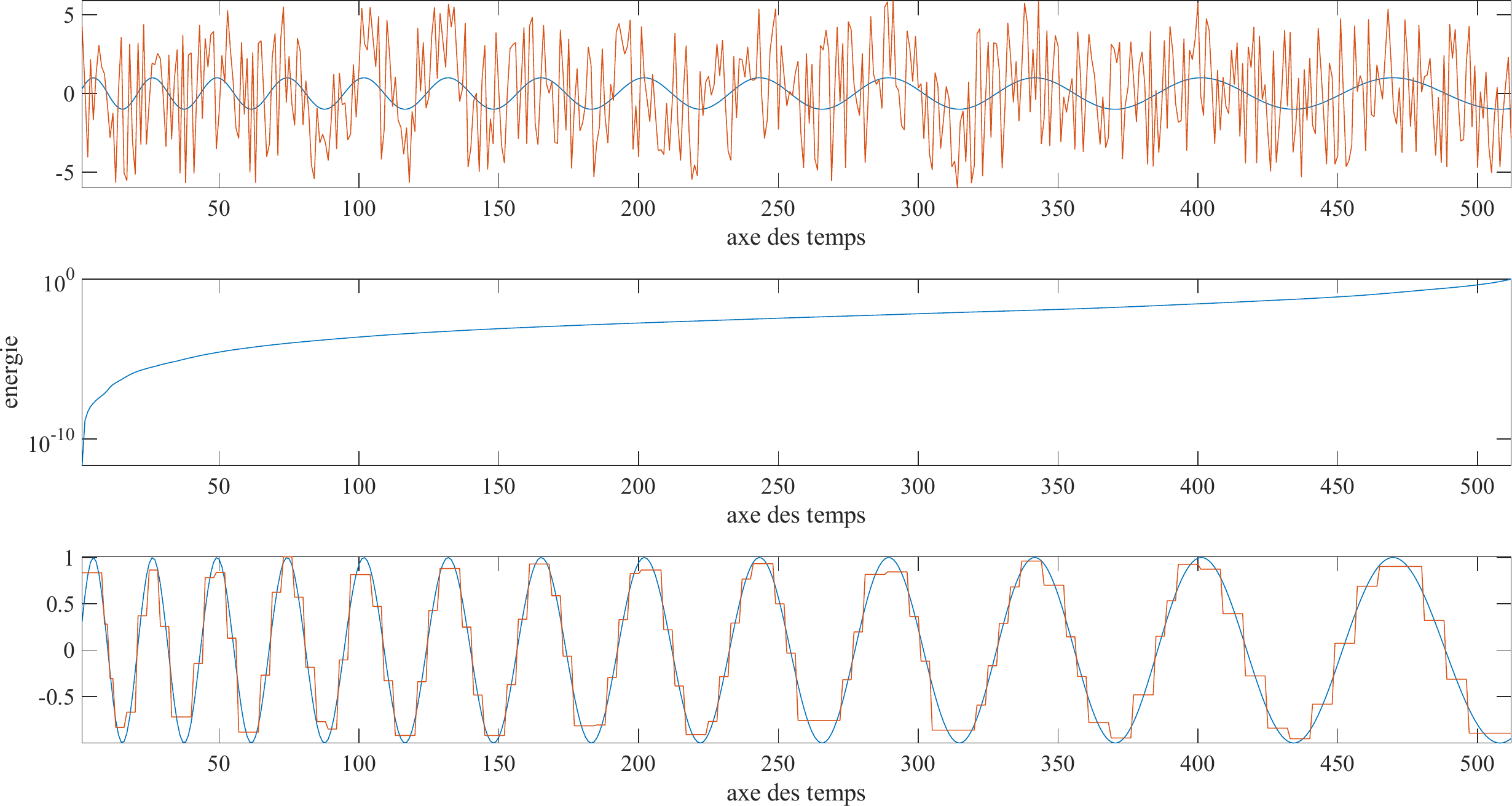}}
	\end{center}
	\caption{Example of filtering in the \couleur{Haar} basis. At the top, the desired signal (in blue) and its noisy version (in green). In the middle, the cumulative energy of the wavelet coefficients. At the bottom, the reconstructed signal retaining the 15\% most energetic coefficients}
	\label{ondelettes08}
\end{figure}

	The filtering performed in the following code shows an example of denoising a sinusoidal signal with a variable period. Figure \ref{ondelettes08} shows the result. In this example, the filtering is performed by calculating the cumulative energy of the wavelet coefficients and removing those whose cumulative energy contributes less than 1\% of the total energy. It is interesting to note that this filtering removes about 85\% of the coefficients, which allows a significant compression of the information.

\begin{lstlisting}[commentstyle=\footnotesize\textit]
function ondelettes08()

        t = 1:512;                               
        T = linspace(20,40,length(t));   
        f = sin(2*pi*t./T);                   
        fn = f + 10*(rand(size(f))-0.5); 
         w = DirectHaar(f);                  
         [ws,iw] = sort(w.^2);
                  
         nwi = length(find(w));
         cutlimit = 0.01;                        
         iwzero = iw(ws <= ws(end)*cutlimit);
         w(iwzero) = 0;
         nwf = length(find(w));
         disp(['ratio of compression = ' num2str(nwi/nwf)])
         ff = InverseHaar(w);
        
        figure
        subplot(311)                           
            plot(t,f,t,fn);axis tight;xlabel('time axis')
            set(gca,'Fontname','Times New Roman','FontSize',18) ;
        subplot(312)
            semilogy(cumsum(ws)/max(cumsum(ws))) ;
            axis tight;xlabel('time axis')
            ylabel('energy');
            set(gca,'Fontname','Times New Roman','FontSize',18) ;             
        subplot(313)
            plot(t,f,t,ff);axis tight;xlabel('time axis')
            set(gca,'Fontname','Times New Roman','FontSize',18) ;
end
\end{lstlisting}

\section{How do you filter with the continuous wavelet transform ?}
\subsection{The Reconstruction Formula}
	We will look for a reconstruction wavelet, $\chi\left(t\right)$, that allows us to reconstruct the signal $f\left(t\right)$ from its transform $\mathcal{W}\left[\psi,f\right]\left(a,t\right)$. Using a reconstruction formula of the form
\begin{equation}
	\begin{split}
	f\left(t\right) & =  \int_{0}^{+\infty}\mathcal{W}\left[\psi,f\right]\left(a,t\right)*\chi_{a}\left(t\right)\mathrm{d}a\\
	 & =  \int_{0}^{+\infty}f\left(t\right)*\psi_{a}\left(t\right)*\chi_{a}\left(t\right)\mathrm{d}a,
	\end{split}
\end{equation}

	which, after Fourier transformation, becomes,
\begin{equation}
	F\left(u\right)=\int_{0}^{+\infty}F\left(u\right)\Psi\left(au\right)Q\left(au\right)\mathrm{d}a,
\end{equation}

	we obtain the following condition,
\begin{equation}
	\int_{0}^{+\infty}\Psi\left(au\right)Q\left(au\right)\mathrm{d}a=1.
\end{equation}

	This equation has a solution,
\begin{equation}
Q\left(bu\right)=\frac{\Psi^{*}\left(bu\right)}{\int_{0}^{+\infty}\left|\Psi\left(au\right)\right|^{2}\mathrm{d}a}\;\;\;\forall b>0
\end{equation}

which, to be acceptable, requires,
\begin{equation}
	0<\int_{0}^{+\infty}\left|\Psi\left(au\right)\right|^{2}\mathrm{d}a<\infty.
\end{equation}

By setting $v=au$, this expression becomes
\begin{equation}
	0<a\int_{0}^{+\infty}\left|\Psi\left(v\right)\right|^{2}\frac{\mathrm{d}v}{v}<\infty.
\end{equation}

	Since $a>0$, it can be eliminated from the above inequalities without changing the direction of the inequalities, giving the admissibility condition in its standard form,
\begin{equation}
	0<C_{\psi}\equiv\int_{0}^{+\infty}\left|\Psi\left(v\right)\right|^{2}\frac{\mathrm{d}v}{v}<\infty.
\end{equation}

	Taking advantage of the fact that,
\begin{equation}
	\mathcal{F}\left[f^{*}\left(-t\right)\right]\left(u\right)=F^{*}\left(u\right)
\end{equation}

	we obtain the expression for the reconstruction wavelet
\begin{equation}
	\chi_{a}\left(t\right)=\frac{\psi_{a}^{*}\left(-t\right)}{aC_{\psi}},
\end{equation}

	and the continuous reconstruction formula,
\begin{equation}
	f\left(t\right)=\frac{1}{C_{\psi}}\int_{0}^{+\infty}\frac{\mathrm{d}a}{a}\int_{-\infty}^{+\infty}\mathcal{W}\left[\psi,f\right]\left(a,\tau\right)\psi_{a}^{*}\left(t-\tau\right)\mathrm{d}\tau
\end{equation}

\subsection{The reproducing kernel}
	The continuous wavelet transform is complete when the entire frequency axis, $u \in \mathbb{R}$, is covered, \ie, when
\begin{equation}
	0<\int_{0}^{+\infty}\left|\Psi\left(au\right)\right|^{2}\mathrm{d}a<+\infty\;\;\forall u\in\Bbb{R},
\end{equation}

	which is automatically satisfied if the wavelet satisfies the admissibility condition discussed in the previous section. The family of wavelets,
\begin{equation}
	\left\{ \psi_{a}\left(t-\tau\right),a\in\Bbb{R}^{+*},\tau\in\Bbb{R}\right\} 
\end{equation}

	is actually redundant, meaning that decomposing a signal over this family is redundant. As a result, the wavelet coefficients, $\mathcal{W}\left[\psi,f\right]\left(a,t\right)$, are correlated, which can be observed by reflexively using the reconstruction formula,
\begin{equation}
	\begin{split}
	\mathcal{W}\left[\psi,f\right]\left(b,t\right) & =  \mathcal{W}\left[\psi,\int_{0}^{+\infty}\mathcal{W}\left[\psi,f\right]\left(a,\cdot\right)*\chi_{a}\left(\cdot\right)\mathrm{d}a\right]\left(b,t\right)\\
 & =  \int_{0}^{+\infty}\psi_{b}\left(\cdot\right)*\psi_{a}\left(\cdot\right)*f\left(\cdot\right)*\chi_{a}\left(\cdot\right)\mathrm{d}a\\
 & =  \int_{0}^{+\infty}\mathcal{W}\left[\psi,f\right]\left(a,\cdot\right)*\left[\psi_{b}\left(\cdot\right)*\chi_{a}\left(\cdot\right)\right]\mathrm{d}a\\
 & =  \int_{0}^{+\infty}\mathcal{W}\left[\psi,f\right]\left(a,t\right)*K_{b,a}\left(t\right)\frac{\mathrm{d}a}{a},
	\end{split}
\end{equation}

	where the reproducing kernel, $K_{b,a}\left(t\right)\equiv a\psi_{b}\left(t\right)*\chi_{a}\left(t\right)$, quantifies the redundancy of the wavelet transform. This kernel is fully defined by the analysing wavelet,
\begin{equation}
	\begin{split}
	K_{b,a}\left(t\right) & =  a\psi_{b}\left(t\right)*\chi_{a}\left(t\right)\\
 & =  \frac{1}{C_{\psi}}\int_{-\infty}^{+\infty}\psi_{b}\left(\tau\right)\psi_{a}^{*}\left(\tau-t\right)\mathrm{d}\tau 
	\end{split}
\end{equation}

\section{Asymptotic signal analysis}
	We have just seen that the continuous wavelet transform is redundant because the information originally contained in the analysed signal is projected onto the \couleur{Poincaré} half-plane. We are moving from a one-dimensional space to a two-dimensional space, and it is interesting to investigate whether the projected information is 'uniformly' distributed or, conversely, 'concentrated' in preferred regions of the half-plane. This investigation can be done by noting that the wavelet transform is similar to a Fresnel-type oscillatory integral to which asymptotic approximations can be applied.

\subsection{Signaux asymptotiques}
	A real signal, $f\left (t\right)$, can always be represented in terms of instantaneous amplitude and phase,
\begin{equation}
	f\left(t\right)=A\left(t\right)\cos\left[\phi\left(t\right)\right].
\end{equation}

	This representation admits an infinite number of solutions $\left(A;\phi\right)$, including the so-called canonical solution,
\begin{equation}
	\begin{split}
	A_{f}\left(t\right) & =  \left|Z_{f}\left(t\right)\right|\\
\phi_{f}\left(t\right) & =  \arg\left[Z_{f}\left(t\right)\right]
	\end{split}
\end{equation}

	where the analytic signal associated with $f\left(t\right)$ is defined by,
\begin{equation}
	Z_{f}\left(t\right)\equiv f\left(t\right)+iF_{Hi}\left(t\right)
\end{equation}

	The canonical solution allows for the definition of the instantaneous frequency,
\begin{equation}
	u_{f}\left(t\right)=\frac{1}{2\pi}\frac{\mathrm{d}\phi_{f}\left(t\right)}{\mathrm{d}t},
\end{equation}
	
	whose physical meaning can sometimes be confusing. This is particularly true when the signal is too slow or when the signal being analysed is the sum of two sinusoids.

	We will say that a signal,
\begin{equation}
	f\left(t\right)=A_{f}\left(t\right)\cos\left[\phi_{f}\left(t\right)\right]
\end{equation}

	is asymptotic if,
\begin{equation}
	\left|\frac{\mathrm{d}\phi_{f}}{\mathrm{d}t}\right|\gg\left|\frac{1}{A_{f}}\frac{\mathrm{d}A_{f}}{\mathrm{d}t}.\right|
\end{equation}

	that is, the signal's oscillations are much faster than its envelope.

\subsection{Asymptotic wavelet analysis}
	We will now focus on the wavelet transform of asymptotic signals when the analysing wavelet, $\psi\left(t\right)$, is itself asymptotic. We will see that the stationary phase method allows us to identify specific sets of points in the half-plane $\left(a>0,t\right)$ from which we can obtain estimates of the wavelet coefficients and recover the modulation laws of the signals. Let $f\left(t\right)$ be a locally monochromatic asymptotic real signal, with the associated analytic signal given by, 
\begin{equation}
	z_{f}\left(t\right)=A_{f}\left(t\right)\exp\left[i\phi_{f}\left(t\right)\right].
\end{equation}

	Naturally,
\begin{equation}
	f\left(t\right)=\textrm{Re}\left[z_{f}\left(t\right)\right].
\end{equation}

The wavelet transform of this signal is,
\begin{equation}
	\begin{split}
	\mathcal{W}\left[\psi,f\right]\left(a,t\right) & =  \psi_{a}\left(t\right)*f\left(t\right)\\
 & =  \mathcal{F}^{-1}\left[F\left(u\right)\Psi\left(au\right)\right]\left(t\right)
	\end{split}
\end{equation}

	If $\psi\left(t\right)$ is an analytic wavelet, meaning that $\Psi\left(u<0\right)=0$, then, noting that,
\begin{equation}
	\frac{1}{2}Z_{f}\left(u\right)=F\left(u\right)\;\textrm{lorsque}\; u\geq0,
\end{equation}

	one finds that,
\begin{equation}
	F\left(u\right)\Psi\left(au\right)=\frac{1}{2}Z_{f}\left(u\right)\Psi\left(au\right)
\end{equation}

	The wavelet transform of the signal can therefore be expressed in terms of the wavelet transform of the corresponding analytical signal,
\begin{equation}
	\begin{split}
	\mathcal{W}\left[\psi,f\right]\left(a,t\right) & =  \frac{1}{2}\mathcal{F}^{-1}\left[Z_{f}\left(u\right)\Psi\left(au\right)\right]\left(t\right)\\
 & =  \frac{1}{2}\mathcal{W}\left[\psi,z_{f}\right]\left(a,t\right)
	\end{split}
\end{equation}

	If the wavelet $\psi\left(t\right)$ is an asymptotic wavelet, it can be written in the following form,
\begin{equation}
	\psi\left(t\right)=A_{\psi}\left(t\right)\exp\left[i\phi_{\psi}\left(t\right)\right]
\end{equation}

and $\psi_{a}\left(t\right)$,
\begin{equation}
	\psi_{a}\left(t\right)=\frac{1}{a}A_{\psi}\left(\frac{t}{a}\right)\exp\left[i\phi_{\psi}\left(\frac{t}{a}\right).\right]
\end{equation}

	Using this expression in the expression for the wavelet transform of the signal, we obtain,
\begin{equation}
	\mathcal{W}\left[\psi,f\right]\left(a,t\right)=\frac{1}{2a}\int_{-\infty}^{+\infty}A_{f}\left(\tau\right)A_{\psi}\left(\frac{t-\tau}{a}\right)\exp\left[i\phi_{f}\left(\tau\right)+i\phi_{\psi}\left(\frac{t-\tau}{a}\right)\right]\mathrm{d}\tau 
\end{equation}

	which is an oscillatory integral that can be approximated using the stationary phase method. This method exploits the fact that the integral takes most of its value around the points $t_{f}\left(a,t\right)$ where,
\begin{equation}
	\frac{\mathrm{d}}{\mathrm{d}\tau}\left.\left[\phi_{f}\left(\tau\right)+\phi_{\psi}\left(\frac{t-\tau}{a}\right)\right]\right|_{\tau=t_{f}}=0,
\end{equation}

	that is, where,
\begin{equation}
	\phi_{f}^{^{\prime}}\left(t_{f}\right)=\frac{1}{a}\phi_{\psi}^{^{\prime}}\left(\frac{t-t_{f}}{a}\right).\end{equation}

	The stationary phase approximation yields,
\begin{equation}
	\begin{split}
\mathcal{W}\left[\psi,f\right]\left(a,t\right) & \simeq  \frac{\sqrt{\pi}}{a\sqrt{2}}\frac{\exp\left\{ i\frac{\pi}{4}\textrm{sgn}\left[\phi_{f}^{^{\prime\prime}}\left(t_{f}\right)+a^{-2}\phi_{\psi}^{^{\prime\prime}}\left(\frac{t-t_{f}}{a}\right)\right]\right\} }{\sqrt{\left|\phi_{f}^{^{\prime\prime}}\left(t_{f}\right)+a^{-2}\phi_{\psi}^{^{\prime\prime}}\left(\frac{t-t_{f}}{a}\right)\right|}}\\
 &   \times A_{f}\left(t_{f}\right)A_{\psi}\left(\frac{t-t_{f}}{a}\right)\exp\left\{ i\left[\phi_{f}\left(t_{f}\right)+\phi_{\psi}\left(\frac{t-t_{f}}{a}\right)\right]\right\} \\
 & =  \sqrt{\frac{\pi}{2}}\frac{\exp\left\{ i\frac{\pi}{4}\textrm{sgn}\left[\phi_{f}^{^{\prime\prime}}\left(t_{f}\right)+a^{-2}\phi_{\psi}^{^{\prime\prime}}\left(\frac{t-t_{f}}{a}\right)\right]\right\} }{\sqrt{\left|\phi_{f}^{^{\prime\prime}}\left(t_{f}\right)+a^{-2}\phi_{\psi}^{^{\prime\prime}}\left(\frac{t-t_{f}}{a}\right)\right|}}Z_{f}\left(t_{f}\right)\psi_{a}\left(t-t_{f}\right)
	\end{split}
\end{equation}

\subsection{The stationary phase method}
	This method was used by \couleur{Lord Kelvin} in 1887 to study integrals of the form,
\begin{equation}
	I=\int_{-\infty}^{+\infty}A\left(t\right)\exp\left[i\phi\left(t\right)\right]\mathrm{d}t
\end{equation}

	where $A\left(t\right)$ and $\phi\left(t\right)$ are regular functions. The idea behind this method is to exploit the fact that the integral takes most of its value near points where the phase $\phi\left(t\right)$ is stationary. Suppose, without limiting the generality of our discussion, that there is only one point, $t_{0}$, for which this is true,
\begin{equation}
	\phi^{^{\prime}}\left(t_{0}\right)=0.
\end{equation}

We have,
\begin{equation}
	\begin{split}
	I & =   \int_{-\infty}^{+\infty}A\left(t\right)\exp\left[i\phi\left(t\right)\right]\mathrm{d}t\\
 & \simeq  \int_{-\infty}^{+\infty}A\left(t\right)\exp\left[i\left(\phi\left(t_{0}\right)+\frac{1}{2}\phi^{^{\prime\prime}}\left(t_{0}\right)\left(t-t_{0}\right)^{2}\right)\right]\mathrm{d}t\\
 & \simeq  A\left(t_{0}\right)\exp\left[i\phi\left(t_{0}\right)\right]\int_{-\infty}^{+\infty}\exp\left[\frac{i}{2}\phi^{^{\prime\prime}}\left(t_{0}\right)\left(t-t_{0}\right)^{2}\right]\mathrm{d}t\\
 & =  A\left(t_{0}\right)\exp\left[i\phi\left(t_{0}\right)\right]\int_{-\infty}^{+\infty}\exp\left[\frac{i}{2}\phi^{^{\prime\prime}}\left(t_{0}\right)\xi^{2}\right]\mathrm{d}\xi
	\end{split}
\end{equation}

	Performing the change of variable,
\begin{equation}
	\chi=\left|\frac{\phi^{^{\prime\prime}}\left(t_{0}\right)}{2}\right|^{1/2}\xi
\end{equation}

	one deduces,
\begin{equation}
	\begin{split}
I & \simeq & \frac{A\left(t_{0}\right)\exp\left[i\phi\left(t_{0}\right)\right]}{\sqrt{\left|\phi^{^{\prime\prime}}\left(t_{0}\right)/2\right|}}\int_{-\infty}^{+\infty}\exp\left[i\textrm{sgn}\left(\phi^{^{\prime\prime}}\left(t_{0}\right)\right)\chi^{2}\right]\mathrm{d}\chi\\
 & = & \frac{2A\left(t_{0}\right)\exp\left[i\phi\left(t_{0}\right)\right]}{\sqrt{\left|\phi^{^{\prime\prime}}\left(t_{0}\right)/2\right|}}\left\{ \int_{0}^{+\infty}\cos\chi^{2}\mathrm{d}\chi+i\textrm{sgn}\left[\phi^{^{\prime\prime}}\left(t_{0}\right)\right]\int_{0}^{+\infty}\sin\chi^{2}\mathrm{d}\chi\right\} \\
 & = & \frac{2A\left(t_{0}\right)\exp\left[i\phi\left(t_{0}\right)\right]}{\sqrt{\left|\phi^{^{\prime\prime}}\left(t_{0}\right)/2\right|}}\left\{ \sqrt{\frac{\pi}{8}}+i\textrm{sgn}\left[\phi^{^{\prime\prime}}\left(t_{0}\right)\right]\sqrt{\frac{\pi}{8}}\right\} \\
 & = & \frac{\sqrt{2\pi}A\left(t_{0}\right)\exp\left[i\phi\left(t_{0}\right)\right]}{\sqrt{\left|\phi^{^{\prime\prime}}\left(t_{0}\right)\right|}}\exp\left\{ i\frac{\pi}{4}\textrm{sgn}\left[\phi^{^{\prime\prime}}\left(t_{0}\right)\right]\right\} 	
	\end{split}
\end{equation}

\subsection{The Wavelet Transform Ridge}
	We will define the edge, $a_{r}\left(t\right)$, of the continuous wavelet transform as the set of points $\left(a,t\right)$ such that,
\begin{equation}
	t_{f}\left(a_{r},t\right)=t
\end{equation}

Since,
\begin{equation}
	\phi_{f}^{^{\prime}}\left(t_{f}\right)=\frac{1}{a}\phi_{\psi}^{^{\prime}}\left(\frac{t-t_{f}}{a}\right),\end{equation}

on the ridge,
\begin{equation}
	a_{r}\left(t\right)=\frac{\phi_{\psi}^{^{\prime}}\left(0\right)}{\phi_{f}^{^{\prime}}\left(t\right)}.
\end{equation}

	It is therefore possible to recover the modulation law of the signal, $\phi_{f}^{^{\prime}}\left(t\right)$, from the edge of its wavelet transform. The problem now is to calculate the edge; we will use the phase to do this,
\begin{equation}
	\Phi_{\psi}f\left(a,t\right)\equiv\arg\left[\mathcal{W}\left[\psi,f\right]\left(a,t\right)\right]
\end{equation}

	whose estimator is unbiased, unlike the magnitude\footnote{It is this stochastic behaviour that led us to choose the phase rather than the magnitude. It should be noted, however, that calculations equivalent to those we will develop for the phase can be made for the magnitude.}. The expression resulting from the stationary phase approximation is,
\begin{equation}
	\Phi_{\psi}f\left(a,t\right)=\phi_{f}\left(t_{f}\right)+\phi_{\psi}\left(\frac{t-t_{f}}{a}\right)+Cte
\end{equation}

and on the ridge, 
\begin{equation}
	\begin{split}
\left.\frac{\partial}{\partial t}\Phi_{\psi}f\left(a,t\right)\right|_{t=t_{f}} & =  \left[\phi_{f}^{^{\prime}}\left(t\right)+\frac{\partial a}{\partial t}\times\frac{\partial\left(\frac{t-t_{f}}{a}\right)}{\partial a}\times\frac{\partial}{\partial\left(\frac{t-t_{f}}{a}\right)}\phi_{\psi}\left(\frac{t-t_{f}}{a}\right)\right]_{t=t_{f}}\\
 & =  \left[\frac{1}{a}\phi_{\psi}^{^{\prime}}\left(\frac{t-t_{f}}{a}\right)-\left(\frac{\partial a}{\partial t}\right)\frac{t-t_{f}}{a^{2}}\phi_{\psi}^{^{\prime}}\left(\frac{t-t_{f}}{a}\right)\right]_{t=t_{f}}\\
 & =  \frac{1}{a_{r}}\phi_{\psi}^{^{\prime}}\left(0\right),
	\end{split}
\end{equation}

which is the property we will use to extract the edge from the wavelet transforms. We also have,
\begin{equation}
	\begin{split}
\left.\frac{\partial}{\partial a}\Phi_{\psi}f\left(a,t\right)\right|_{t=t_{f}} & =  \left.-\frac{t-t_{f}}{a^{2}}\phi_{\psi}^{^{\prime}}\left(\frac{t-t_{f}}{a}\right)\right|_{t=t_{f}}\\
 & =  0.
 	\end{split}
\end{equation}

\subsection{Use of non-asymptotic wavelets}
	The previous calculations were made assuming an asymptotic wavelet; let us see how they change when this is not the case. The main difference arises from the fact that the instantaneous amplitude of the wavelet varies too rapidly to be taken out of the integral in the stationary phase approximation. Therefore we have,
\begin{equation}
	\begin{split}
\mathcal{W}\left[\psi,f\right]\left(a,t\right) & \simeq  \frac{1}{2a}A_{f}\left(t_{f}\right)\exp\left\{ i\left[\phi_{f}\left(t_{f}\right)+\phi_{\psi}^{^{\prime\prime}}\left(\frac{t-t_{f}}{a}\right)\right]\right\} \\
 &   \times\int_{-\infty}^{+\infty}A_{\psi}\left(\frac{t-\tau}{a}\right)\exp\left\{ \frac{i}{2}\left(\tau-t_{f}\right)^{2}\left[\phi_{f}^{^{\prime\prime}}\left(t_{f}\right)+\frac{1}{a^{2}}\phi_{\psi}^{^{\prime\prime}}\left(\frac{t-t_{f}}{a}\right)\right]\right\} \mathrm{d}\tau. 	
 	\end{split}
\end{equation}

	The calculations can be carried out in the case of the \couleur{Morlet} wavelet,
\begin{equation}
	\psi_{a}\left(t\right)=\frac{1}{a}\exp\left(i\frac{\pi t}{a}\right)\exp\left[-\frac{1}{2}\left(\frac{t}{2\sigma a}\right)^{2}\right]
\end{equation}

	for which,
\begin{equation}
	\begin{split}
A_{\psi}\left(t\right) & =  \exp\left[-\frac{1}{2}\left(\frac{t}{2\sigma}\right)^{2}\right]\\
\phi_{\psi}\left(t\right) & =  \pi t
 	\end{split}
\end{equation}

	Direct but rather lengthy calculations yield,
\begin{equation}
\mathcal{W}\left[\psi,f\right]\left(a,t\right)\simeq\frac{\sigma\sqrt{2\pi}}{\left[1+\left(4\pi\sigma^{2}a_{r}^{^{\prime}}\right)^{2}\right]^{1/4}}\exp\left[\frac{i}{2}\arctan\left(-4\pi\sigma^{2}a_{r}^{^{\prime}}\right)\right]Z_{f}\left(t_{f}\right)
\end{equation}

where we have used the result\footnote{Found in the tables of \couleur{Gradshteyn and Ryzhik}, page 485.},
\begin{equation}
	\begin{split}
 &   \int_{-\infty}^{+\infty}\exp\left[-\left(\alpha x^{2}+2\beta x+\gamma\right)\right]\exp\left[i\left(px^{2}+2qx+r\right)\right]\mathrm{d}x\\
 & =  \frac{\sqrt{\pi}}{\left(\alpha^{2}+p^{2}\right)^{1/4}}\exp\left[\frac{\alpha\left(\beta^{2}-\alpha\gamma\right)-\left(\alpha q^{2}-2\beta pq+\gamma p^{2}\right)}{\alpha^{2}+p^{2}}\right]\\
 &   \times\exp\left\{ i\left[\frac{1}{2}\arctan\left(\frac{p}{\alpha}\right)-\frac{p\left(q^{2}-pr\right)-\left(p\beta^{2}-2q\alpha\beta+r\alpha^{2}\right)}{\alpha^{2}+p^{2}}\right]\right\} . 	\end{split}
\end{equation}

\chapter{\titrechap{Singular Spectrum Analysis}}
\minitoc
	As we have seen so far, in the series decompositions of \couleur{Fourier} or in wavelets, the orthogonal basis on which the signal is projected is imposed by the method. In the context of \couleur{Fourier} analysis, these are complex exponentials, not to mention infinite sines; for wavelets, they are specific functions that can be expanded or contracted at will, or almost. In the pragmatic approach followed in our field, we tend to lean towards \couleur{Fourier} analysis primarily to filter our geophysical signals; as for wavelets, they appear surprisingly in their continuous form, with the underlying notion of our contemporaries being to represent the evolution of the frequency support contained within a time series of \ldots over time. Although this perspective is somewhat reductive for each of the two approaches, the question arises: can we decompose our signal, for filtering, analysis, compression, \etc, on a basis that is the most optimal and intrinsic to the original signal? As always, geophysicists have pondered this question, their intention at the time being to "fill in" gaps in a palaeoclimatic series (\shortciteN{vautard1989singular}, \shortciteN{vautard1992singular}).

\section{Singular Spectum Analysis (SSA)}
\subsection{Simple algorithm presentation}
	Consider a discrete time series ($\mathcal{X}$) of length N (with N>2) and, of course, non-zero,
\begin{equation}
	\mathcal{X}_N =(x_1,\ldots, x_N)
\end{equation}

\paragraph{Step 1: the trajectory matrix} $\mathcal{X}$ is segmented into $K$ sections of length $L$ to form a matrix \textbf{X} of dimension $K \times N$, where $K = N - L + 1$. This length $L$ will henceforth be referred to as the analysis window $\mathcal{L}$, and as we will see later, after describing and discussing \textbf{X}, it will become clear that the choice of the dimension of $\mathcal{L}$ will dictate our decomposition. This is the first tuning parameter. This phase of embedding $\mathcal{X}$ in \textbf{X} is the first step of the SSA algorithm, which the Anglo-Saxons call the \textit{embedding step}. The expression for \textbf{X} is
\begin{equation}
\textbf{X} = 
\begin{pmatrix} 
x_{1}   & x_{2}  & x_{3} \cdots  & x_{K}\\
x_{2}   & x_{3}  & x_{4} \cdots  & x_{K+1}\\
x_{3}   & x_{4}  & x_{5} \cdots  & x_{K+2}\\
\vdots & \vdots & \vdots \ddots & \vdots\\ 
x_{L}   & x_{L+1}  & x_{L+2} \cdots  & x_{N}
\end{pmatrix}
\label{hankel1}
\end{equation}

	As we can see, each column of \textbf{X} is a segment of the realisation of $\mathcal{X}$, shifted or delayed by one sample. In fact, the regularity of the shift is not important; we do not have the constraint of "dt" as in the expression of the \couleur{Fourier} transform. Therefore, the column vectors of \textbf{X} are called $\mathcal{L}$-\textit{lagged vectors} and \textbf{X} is called the $\mathcal{L}$-\textit{trajectory matrix} or trajectory matrix. By construction, for any element $(i,j)$ of \textbf{X} we have $x_{i,j} = x_{i+-1,j+1}$, which makes it an antidiagonal matrix defined by $i+j = \text{constant}$. It is a \couleur{Hankel} matrix\index{Hankel matrix} provided it is square; otherwise it is quite easy to make it square. The values of \textbf{X} are constant along the ascending diagonals. \textbf{X} would be a \couleur{Toeplitz} matrix if they were constant along the descending diagonals. This \couleur{Hankel} matrix is very useful in the context of non-stationary signal decomposition, which will make it quite attractive to us later; it is also similar to an autocorrelation matrix, hence our earlier remark about the size of $\mathcal{L}$.
	
\paragraph{Step 2: Singular Value Decomposition (SVD)} 
	At this stage we are going to perform the Singular Value Decomposition, or SVD (\shortciteNP{golub1971singular}), of the matrix \textbf{X}; this step is a bit like going from the data space to the dual space. Let us construct $\mathcal{S} = \textbf{X}^t\textbf{X}$, the product of the transpose of \textbf{X} with itself, to obtain a square matrix just for the purpose of using the terminology below (in fact, we could decompose \textbf{X} directly): let $\lambda_1, \lambda_2, \ldots, \lambda_L$ be the eigenvalues of $\mathcal{S}$, in decreasing order of magnitude (\eg $\lambda_1 \geqslant \lambda_2 \geqslant \ldots \geqslant \lambda_L \geqslant 0$), and $U_1, U_2, \ldots, U_L$ the orthonormal basis of the associated eigenvectors.

	The rank d of \textbf{X}, defined by d = rank \textbf{X} = max$\{i \vert \lambda_i > 0\}$, allows us to express \textbf{X} as a sum of d unitary matrices using SVD,
\begin{equation}
\textbf{X} = \textbf{X}_1 +\textbf{X}_2 + \ldots + \textbf{X}_d
\label{ssa_unit}
\end{equation}

	In real life, \ie for real signals, the rank d of \textbf{X} is often simply the minimum of L and K (d=min${L,K}$). The relationship (\ref{ssa_unit}) is analogous to that of the discrete \couleur{Fourier} transform; it is always possible to consider a signal as the sum of orthogonal sub-signals. Orthogonality ensures the linearity and uniqueness of the decomposition basis; in other words, energy is normally conserved from one space to another. However, it is important to note an important difference here: we are summing real numbers. Each of these unitary matrices \textbf{X}$_i$, which are rank-1 matrices, is computed from the transpose of the original matrix \textbf{X} and its eigenvalues and eigenvectors. The matrix i$^{th}$ (i=1, \ldots, d) is defined, 
\begin{equation}
	\textbf{X}_i = \sqrt{\lambda_i} U_i V^t_i \qquad avec, \qquad V^t_i = \textbf{X}^t U_i/\sqrt{\lambda_i}.
\end{equation}

\paragraph{Step 3: Reconstruction}
	As we have just seen, the matrices \textbf{X}$_i$ are unitary matrices, and indeed, with the same philosophy as in the classical approach, it is possible to "group" these matrices into a physically homogeneous set, energetically homogeneous, \etc. This is the second tuning parameter of the SSA algorithm: how to group the unitary matrices. For this purpose, the index set i $\{1, \ldots, d\}$ is divided into m disjoint index subsets $\{I_1, \ldots, I_m\}$.
	
	Let $I$ be the set of $p$ indices of $i$, $I={i_1,i_2, \ldots, i_p}$. Since the relation (\ref{ssa_unit}) is linear, the resulting matrix \textbf{X}$_I$, which groups the indices $I$, is expressed as follows,
\begin{equation}
\textbf{X}_I = \textbf{X}_{I1} +  \textbf{X}_{I2}  + \ldots + \textbf{X}_{Im} 
\label{ssa_rel2}
\end{equation}

	We call this step the grouping of the eigentriplets ($\lambda$s, $U$ and $V$). Obviously, in the limiting case where 
$m=d$, the relation (\ref{ssa_rel2}) reduces rigorously to the relation (\ref{ssa_unit}), and we obtain our unitary matrices.

\paragraph{Step 4: Diagonale average or Hankelization} This is the final step. Once the submatrices \textbf{X}$_I$ have been constructed, the task is to return to the data space, that is, to calculate the time series of length N associated with these matrices. Let \textbf{Y} be a matrix of dimension $L\times K$, where for each element $y_{i,j}$, we have $1 \leqslant i \leqslant L$ and $1 \leqslant j \leqslant K$. Let $L^\ast$  be the minimum between $L$ and $K$ ($min\{L,K\}$), and let $K^\times$  be the maximum between $L$ and $K$ ($max\{L,K\}$). We always have $N=L+K-1$. Finally, let $y^{\ast}_{ij}$ = $y_{ij}$ if $L<K$, and $y^{\ast}_{ij}$ = $y_{ji}$ otherwise. The diagonal average, applied to the $k^{th}$  index of the time series y associated with the matrix \textbf{Y}, yields,
\begin{equation}
y_k= \left\{ 
\begin{array}{cccc} 
\frac{1}{k} &\sum\limits_{m=1}^k y^{\ast}_{m,k-m+1} &\qquad 1 \leqslant k \leqslant L^\ast \\ 
\frac{1}{L^\ast} &\sum\limits_{m=1}^{L^\ast} y^{\ast}_{m,k-m+1} &\qquad L^\ast \leqslant k \leqslant K^\ast \\ 
\frac{1}{N-K+1} &\sum\limits_{m=k-K\ast +1}^{N-K\ast +1} y^{\ast}_{m,k-m+1} &\qquad K^\ast \leqslant k \leqslant N^\ast 
 \end{array}
 \right.
 \label{diag_average}
\end{equation}

	The relation (\ref{diag_average}) corresponds to the average of the $k^{th}$ element along the anti-diagonal where $i+j=k+1$. For $k=1$,  y$_1$=y$_{1,1}$, for  k=2, y$_2$=(y$_{1,2}$+y$_{2,1}$)/2, \etc.  Thus, from the matrices of step n\num 3, we reconstruct the corresponding time series of length $N$. A note on terminology: when the diagonal mean is applied to the unitary matrices, the resulting series are called elementary series.

\warning{We note that nothing prevents us from extending SSA naturally from real signals to complex signals. It is sufficient to replace all transposes (symbols $^t$ in our demonstration) by complex conjugates.}

\subsection{To see how this works in practice}
\begin{figure}[H]
	\begin{center}
		\tcbox[colback=white]{\includegraphics[width=16cm]{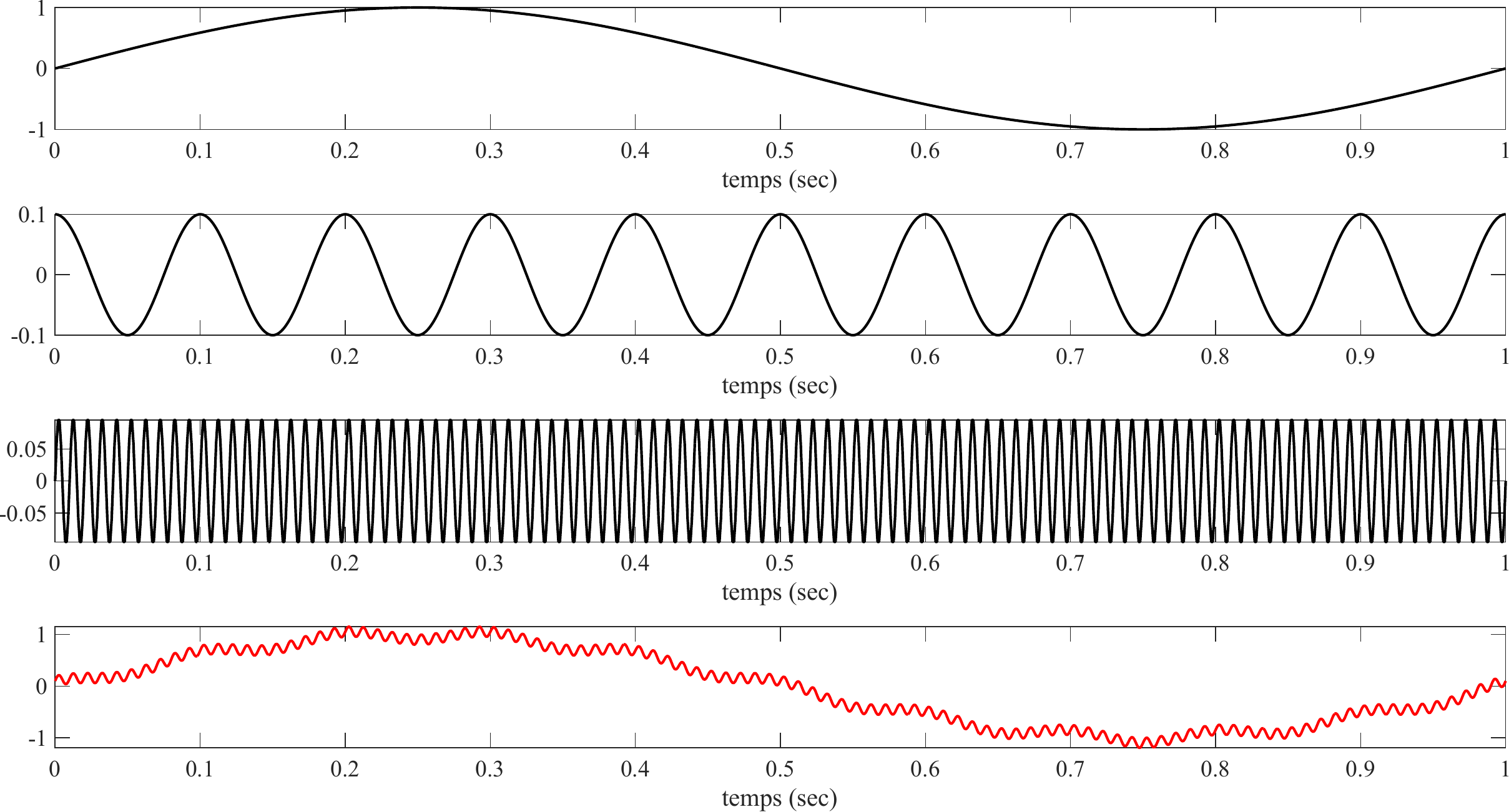}}
	\end{center}
	\caption{At the top is the 1 Hz sine wave, followed in descending order by the 10 Hz and 100 Hz sine waves. These three sinusoids are shown in black. The last one at the bottom, in red, is the sum of these sinusoids.}
	\label{ssa_01}
\end{figure}

	For this example, obtained using the MATLAB program \couleur{ex\_ssa01.m}, we consider three sinusoids with increasing frequencies of 1 Hz, 10 Hz, and 100 Hz, all sampled at 1 kHz and with different amplitudes. We sum them, and Figure (\ref{ssa_01}) shows the initial situation.

\begin{figure}[H]
	\begin{center}
		\tcbox[colback=white]{\includegraphics[width=16cm]{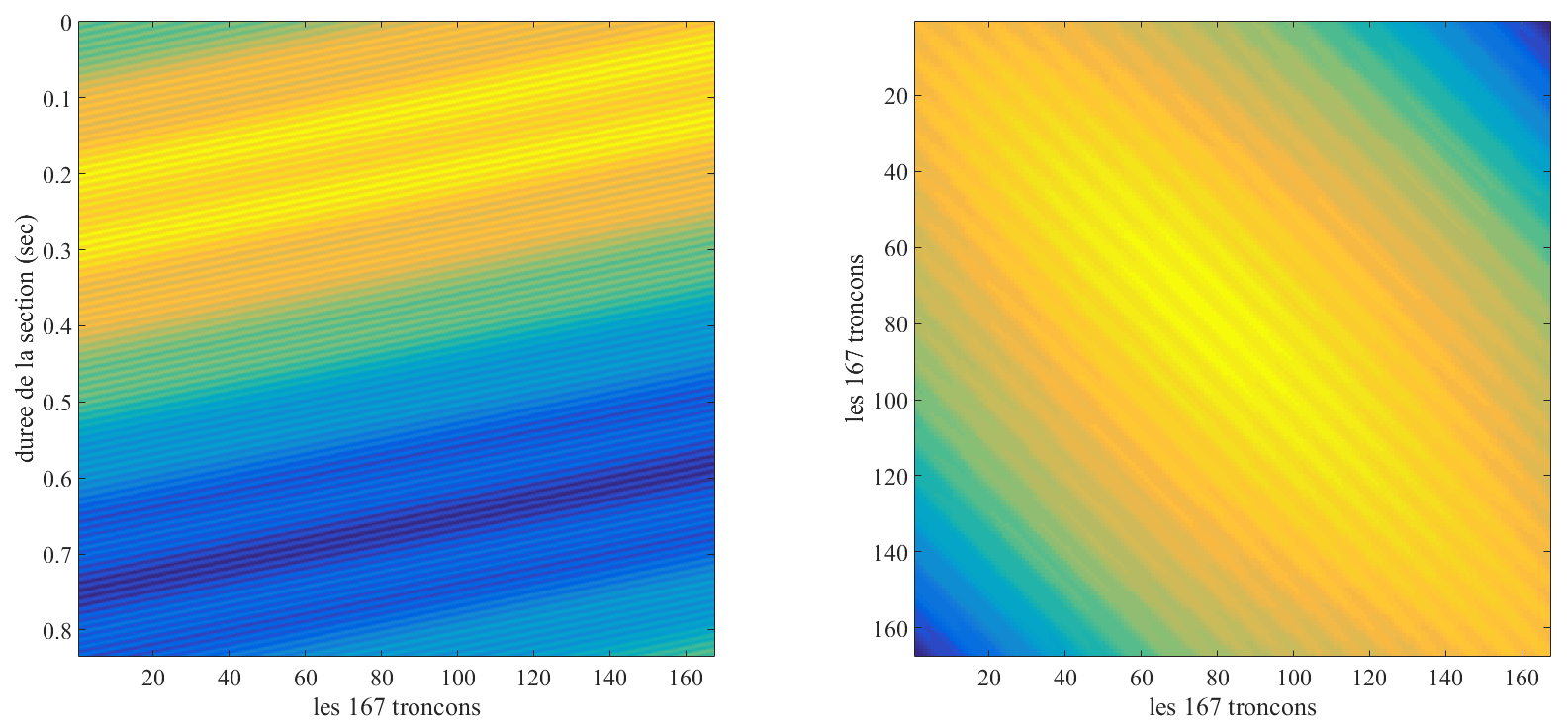}}
	\end{center}
	\caption{On the left is the rectangular matrix \textbf{X}, obtained using the \couleur{hankel.m} function in \Matlab. On the right is the product of \textbf{X} with its transpose, giving a square matrix.}
	\label{ssa_02}
\end{figure}

	We will present (\cf Figure \ref{ssa_02}) the \couleur{Hankel} matrices, or close to it, of the red signal by using the expression (\ref{hankel1}) and rigorously computing the expression of the trajectory matrix $\mathcal{S}$ in order to adhere to the framework and be able to discuss eigenvectors and eigenvalues. The size of the analysis window is 5/6 of the length of the red signal.
\newpage
	
\begin{figure}[H]
	\begin{center}
		\tcbox[colback=white]{\includegraphics[width=16cm]{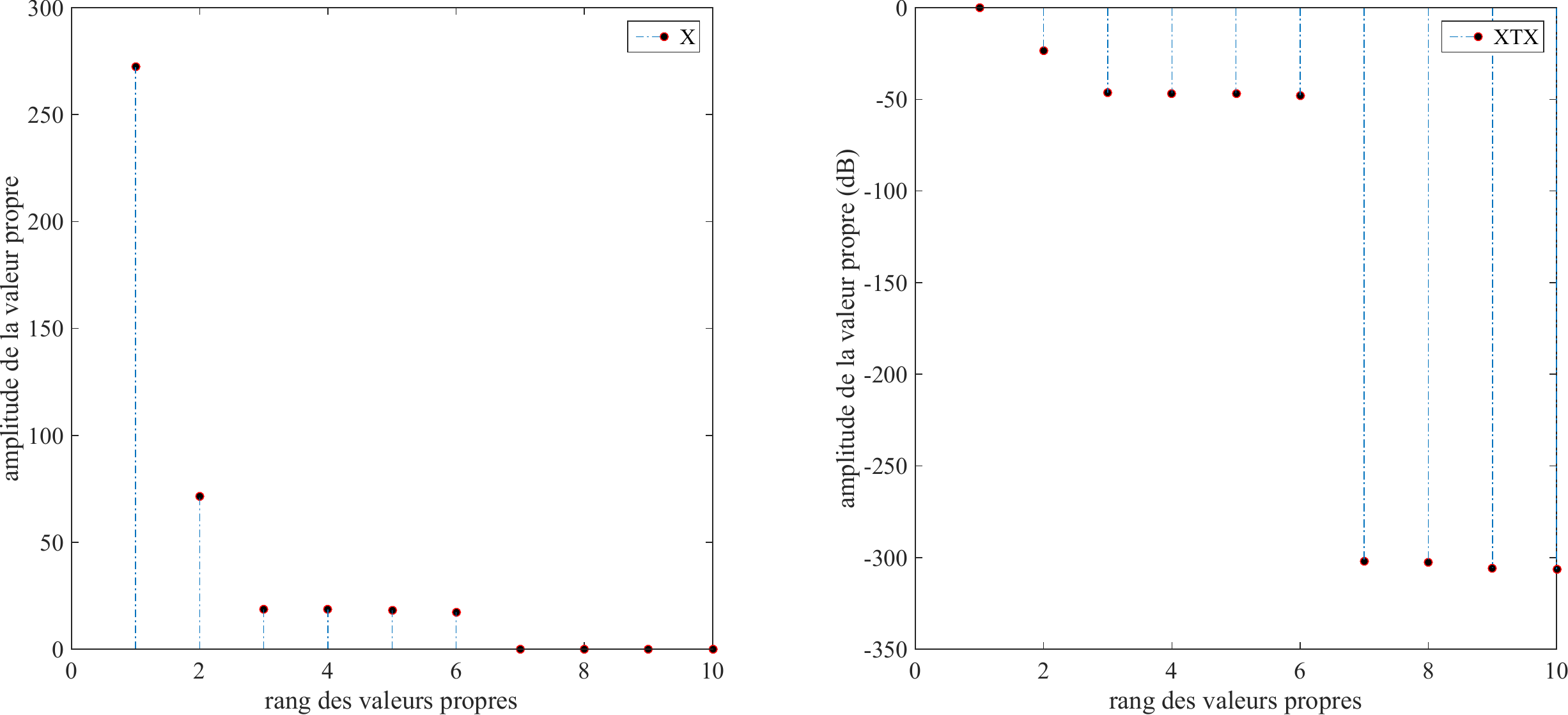}}
	\end{center}
	\caption{On the left, the first 10 singular values of \textbf{X}. On the right, the logarithm (in dB) of the first 10 eigenvalues of \textbf{X}$^t$\textbf{X}.}
	\label{ssa_03}
\end{figure}

	We can proceed to step n\num 2 and apply SVD processing to these two matrices. Only the first 10 singularities and eigenvalues are shown here (Figure \ref{ssa_03}).

	The logarithm was necessary because the square of an eigenvalue can be quite large. Nevertheless, we observe that in both cases there seems to be no significant energy above the 7$^{\text{th}}$ eigenvalue (or singular value). For this example we have chosen the limiting case where $m=d$, i.e. we will reconstruct the 6 first unitary matrices and thus the 6 first elementary signals. One last point of clarification: since it seems that using \textbf{X} instead of \textbf{X}$^t$\textbf{X} does not change the result, except for having to adjust the square of the $\lambda$s in the reconstruction formula and the sign of the original signal, we will use \textbf{X} exclusively from now on.

\begin{figure}[H]
	\begin{center}
		\tcbox[colback=white]{\includegraphics[width=16cm]{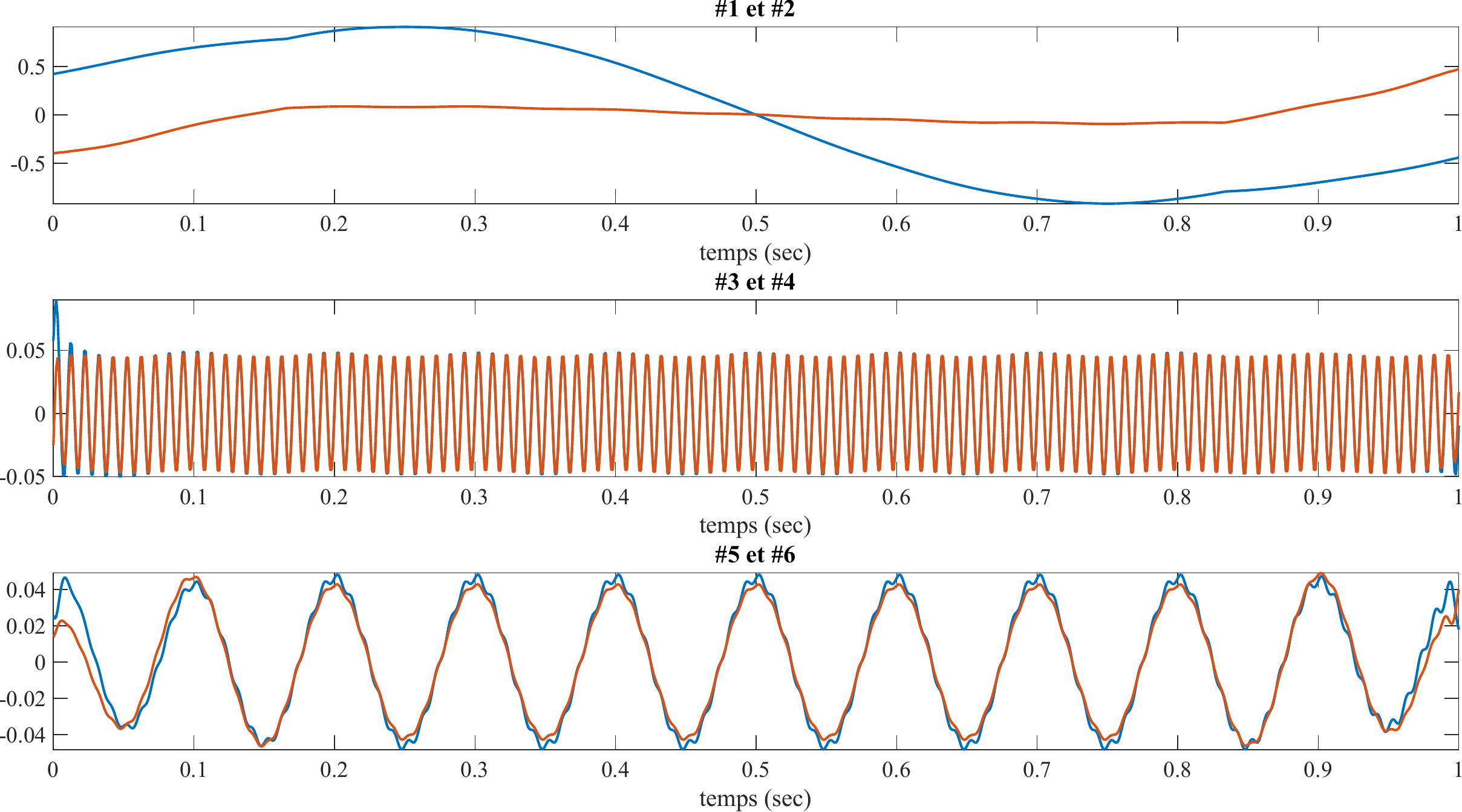}}
	\end{center}
	\caption{From top to bottom, the first 2 elementary signals appear to represent the 1 Hz sinusoid, followed by signals number 3 and 4, which correspond to the 10 Hz sinusoid, and finally, the last 2 signals can be attributed to the 100 Hz oscillation.}
	\label{ssa_04}
\end{figure}

	We are left with the final step, sometimes referred to in the literature as "\textit{hankelization}", which is diagonal averaging. Figure (\ref{ssa_04}) shows the 6 elementary signals reconstructed by SSA. We have paired them for an obvious reason: the similarity of the patterns.

\begin{figure}[H]
	\begin{center}
		\tcbox[colback=white]{\includegraphics[width=16cm]{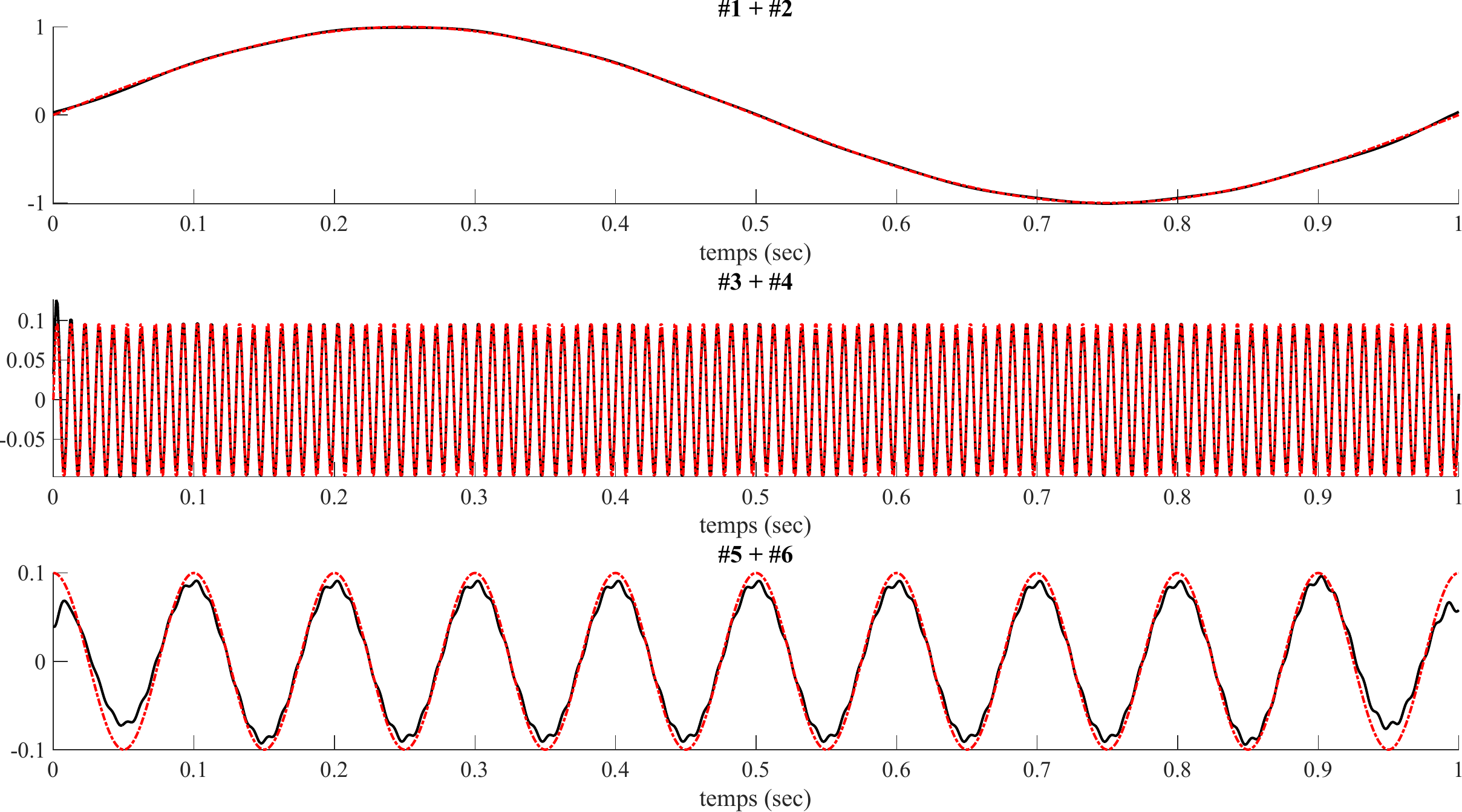}}
	\end{center}
	\caption{The red and blue curves for each pair in Figure (\ref{ssa_04}) have been simply summed (black curves). They are compared to their respective original signals (red curves).}
	\label{ssa_05}
\end{figure}

	As noted at the beginning of this chapter, unlike \couleur{Fourier} analysis, we are working in real space ($\Re$) for both the original signal and the grouping matrices. This allows us to sum the contributions of interest at each step of the operation, focusing here on shape similarity. Figure (\ref{ssa_05}) shows these sums.

\begin{figure}[H]
	\begin{center}
		\tcbox[colback=white]{\includegraphics[width=16cm]{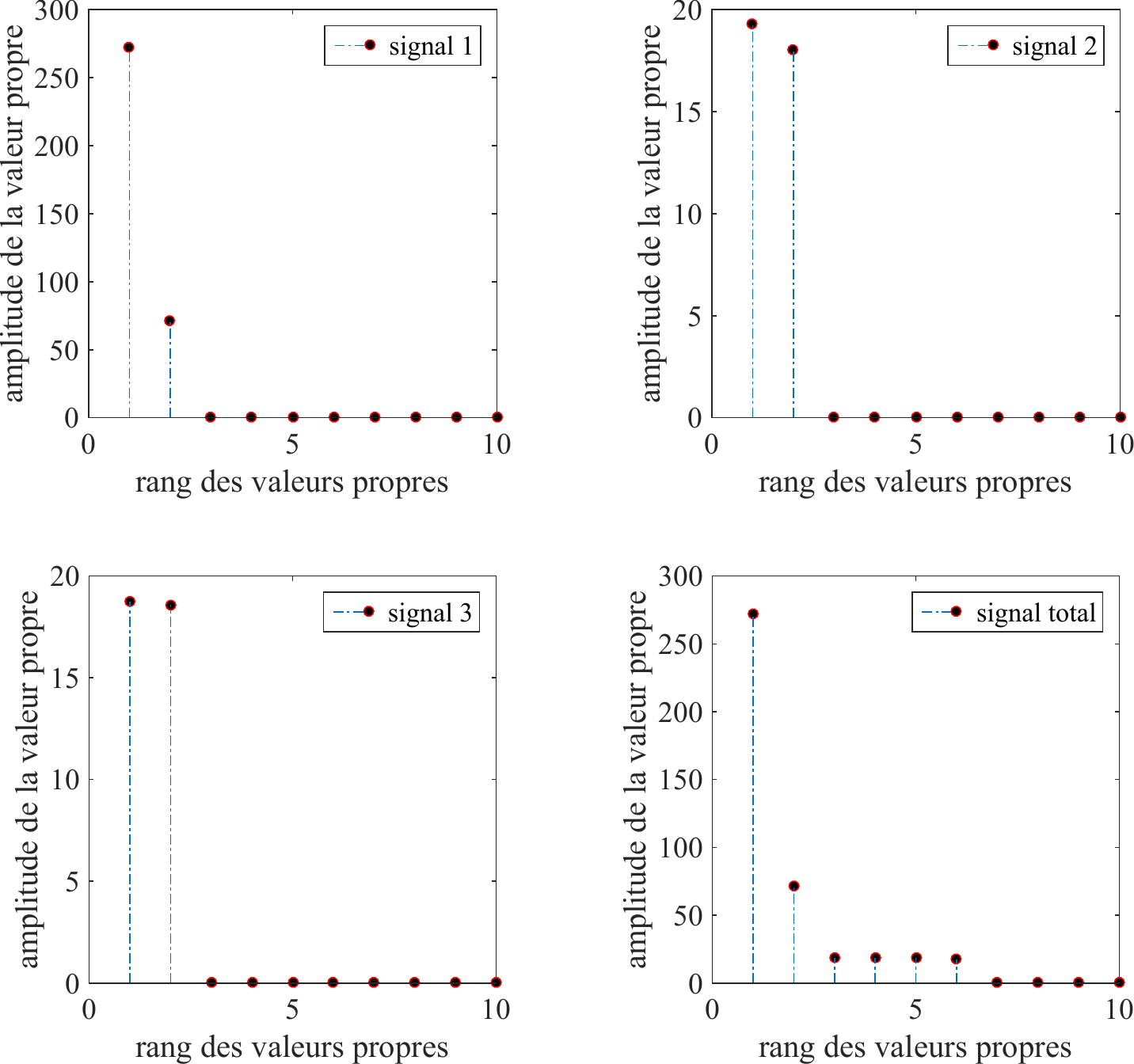}}
	\end{center}
	\caption{The top left shows the signal with the lowest period (1 Hz), the top right shows the singular values associated with the 10 Hz oscillation, the bottom left shows those associated with the 100 Hz oscillation, and finally the bottom right shows the singular values we have already presented for the total signal (Figure \ref{ssa_03}).}
	\label{ssa_06}
\end{figure}

	The result is quite remarkable; the SSA analysis has successfully detected and separated each contribution in terms of both phase and amplitude. However, the reconstruction is not perfect for several reasons, the most important of which is the size of the analysis window $L$. Here we have chosen it somewhat arbitrarily, but as with wavelets, its detection capability depends significantly on its length. Figure (\ref{ssa_06}) shows the first 10 singular values computed by SSA for the three individual sinusoids and for the combined signal.

Two things become clear: first, it seems that two singular values, and thus two singular vectors, are needed to reconstruct a pure oscillation. These are called \couleur{Hilbert} pairs. The second observation is that theoretically these two pairs should have equal amplitudes, which is clearly not the case here. The \couleur{Hilbert} pairs for the 1 Hz oscillation are quite problematic, while those for the fastest oscillation are almost perfect. It is time to analyse these problems.
\newpage	
	
\section{Analysis of the 4 stages of SSA}
\subsection{Embedding}
	The first stage of SSA analysis, embedding, involves projecting the one-dimensional time series $\mathcal{X}_N = (x_1, \ldots, x_N)$ into a multidimensional series space $(\text{X}_1, \ldots, \text{X}_k)$ such that the vectors $\text{X}i = (x_i, \ldots, x{i+L-1})^t$ belong to the space $\mathcal{R}^L$, where $K = N - L + 1$. This somewhat succinct definition was proposed and demonstrated in the early 1980s by \shortciteN{mane1981dimension} and \shortciteN{takens1981detecting}, with the aim of constructing a space that accurately describes strange attractors, often a Banach space. A strange attractor is an object whose dynamical properties can evolve into chaos and are therefore non-linear in nature. The parameter controlling the embedding is $L$, the size of the analysis window; $L$ is an integer between 2 and $N-1$. The Hankel matrix (\eg \ref{hankel1}) has symmetry properties; its transpose $\textbf{X}^t$, known as the trajectory matrix, has dimension $K$. Embedding is a mandatory step in the analysis of nonlinear series; formally, it involves empirically evaluating all pairs of distances between two shifted, lagged vectors to calculate the correlation dimension of the series under analysis. This dimension is quite close to the fractal dimension of the strange attractors that could generate such series, and in this particular case it is advisable to choose very small window sizes $L$ (i.e. very large $K$). \textit{A contrario} for SSA, $L$ must be sufficiently large so that each vector contains a significant part of the information contained in the original time series ($\mathcal{X}_N$); from a mathematical point of view, one must consider the framework of \textit{Structural Total Least Squares} (STLS) for a Hankel matrix \shortcite{lemmerling2001analysis}, which contrasts with the fractal dimension discussed above. A second advantage of using very large values for $L$ is the ability to consider the sub-vectors ($\text{X}_i$) as independent sub-series with different dynamics, thus allowing the identification of common features within collections of these sub-series.

\subsection{Singular Value Decomposition\index{SVD}}
	The SVD of the non-zero trajectory matrix (\textbf{X}), which has dimensions $L \times K$, is a decomposition of the form,
\begin{equation}
	\textbf{X} = \sum_{i=1}^d \sqrt{\lambda_i}U_i V_i^t
\label{decomp_svd}
\end{equation}
	relation (\ref{decomp_svd}), in which we find the eigenvalues $\lambda_i$ ($i=1, \ldots, L$) of the matrix \textbf{S} = \textbf{XX}$^T$, arranged in descending order of magnitude, the corresponding (left) eigenvectors $U_i$, and finally the (right) eigenvectors $V_i$ given by the following relation,
\begin{equation}
	V_i = \textbf{X}^T U_i/\sqrt{\lambda_i}.
\label{decomp_svd_Vi}
\end{equation}

	The equality (\ref{decomp_svd}) shows that the SVD has special symmetry properties, which leads to the fact that the (right) eigenvectors $V_1$, \ldots, $V_2$, which also form an orthonormal basis, are arranged in the same order as the eigenvalues ($\lambda_i$). Let $\textbf{X}_i$ be a submatrix of $\textbf{X}$,
\begin{equation}
 \textbf{X}_i=\sqrt{\lambda_i}U_i V_i^t,
 \label{decomp_svd_Vi}
\end{equation}

then the embedding matrix \textbf{X} can be represented as a simple linear sum of elementary matrices \textbf{X}$_i$. If all the eigenvalues are equal to one, then (\ref{ssa_unit}) is uniquely defined.

	Now to the nature and characteristics of the embedding matrix: Note that its rows and columns are subsets of the original time signal. Consequently, the eigenvectors $(U_i, V_i)$ have a temporal structure and can therefore be considered as a representation of the time series data. Let \textbf{X} be a sequence of $L$ delayed parts of $\mathcal{X}$ and $(\text{X}_1, \ldots, \text{X}_K)$ the linear basis of these eigenvectors. If we set,
\begin{equation}
	 Z_i = \sum_{i=1}^d  \sqrt{\lambda_i}V_i,
\end{equation}

	with $i = 1, \ldots, d$, then (\ref{decomp_svd}) can be expressed in the form,
\begin{equation}
	 \textbf{X} = \sum_{i=1}^d U_i Z_i^t
\end{equation}

	\ie for the elementary matrix $j^{\text{th}}$,
\begin{equation}
	 X_j = \sum_{i=1}^d z_{ji} U_i
\end{equation}

	where $z_{ji}$ is a component of the vector $Z_i$. This means that the vector $Z_i$ is composed of the $i^{\text{th}}$ components of the vector $X_j$. In the same way, if we introduce,
\begin{equation}
	 Y_i = \sum_{i=1}^d  \sqrt{\lambda_i}U_i
\end{equation}

	we obtain for the transposed trajectory matrix,
\begin{equation}
	 X_j^t = \sum_{i=1}^d  U_i Y_i^t
\end{equation}

	which corresponds to a representation of the $K$ lagged vectors in the orthogonal basis $(V_1, \ldots, V_d)$. This illustrates why the SVD is an excellent choice for analyzing the embedding matrix, as it provides us with two geometric descriptions.

\warning{\paragraph{Please note 1} There are strong similarities between performing an SVD of the trajectory matrix, as in the case of SSA, and multivariate analyses such as Principal Component Analysis (PCA) or Karhunen-Loève (KL) decompositions commonly used in time series analysis. However, SSA differs in the nature of its trajectory matrix; it is a Hankel matrix with a particular structure, where its rows and columns are subsets of the signal being analysed and thus have a meaningful temporal and physical sense relative to each other. This is not the case for PCA and KL.}

\warning{\paragraph{Please note 2} In general, the orthonormal basis $(U_i)$ associated with the trajectory matrix and obtained by SVD can be replaced by any orthonormal basis $(P_i)$. In this case, the relation (\ref{ssa_unit}) becomes $X_i = P_i Q_i^t$ with $Q_i = X^t P_i$. A classic example of an alternative basis are the eigenvectors of an autocovariance matrix (Toeplitz SSA).}

\subsection{Grouping of SVD components}
	The topic here is the separation of additive components of a time series, which involves addressing the critically important question: the concept of "separability.

Let $\mathcal{X}$ be the sum of two time series $\mathcal{X}^{(1)}$ and $\mathcal{X}^{(2)}$ such that $x_i = x_i^{(1)} + x_i^{(2)}$ for all $i \in [1, N]$. Let $L$ be the fixed-length analysis window, and let $X$, $X^{(1)}$ and $X^{(2)}$ be the embedding matrices for the series $\mathcal{X}$, $\mathcal{X}^{(1)}$ and $\mathcal{X}^{(2)}$. These two subsets are separable (even weakly) in relation (\ref{ssa_unit}) if there exists a collection of indices $\mathcal{I} \subset {1, \ldots, d}$ such that $\textbf{X}^{(1)} = \sum_{i \in \mathcal{I}} \textbf{X}i$, or if there exists a collection of indices such that $\textbf{X}^{(2)} = \sum{i \not \in \mathcal{I}} \textbf{X}_i$. 
	
	For example, in the case of separability, the contribution of $\textbf{X}^{(1)}$ corresponds to the simple ratio of its eigenvalues ($\sum_{i \in \mathcal{I}} \lambda_i$) to the total eigenvalues ($\sum_{i=1}^d \lambda_i$). We have illustrated this case with figure (\ref{ssa_06}).
	
	Still in the context of the relation (\ref{ssa_unit}), let $\mathcal{I} = \mathcal{I}1$ be the set of indices corresponding to the first signal, with the corresponding matrix denoted by $X{\mathcal{I}1}$. If this matrix, as well as the matrix corresponding to the second signal ($\textbf{X}{\mathcal{I}2} = \textbf{X} - \textbf{X}{\mathcal{I}_1}$), are close to a Hankel matrix, or are Hankel matrices themselves, then the signals are separable or approximately separable. It is therefore clear that the concept of grouping SVD components can be summarised\footnote{only theoretically, as the actual problem is much more complex} as the decomposition of the initial trajectory matrix into several elementary matrices\footnote{whose structures should be as close as possible to that of a Hankel matrix}.
	
	\warning{We will stop here, as the problem is very complex. Although the idea is simple, several procedures are available to us; these will be discussed later in this chapter.}.
	
\section{What SSA can do}
\subsection{Trend extraction}
\begin{figure}[H]
	\begin{center}
		\tcbox[colback=white]{\includegraphics[width=16cm]{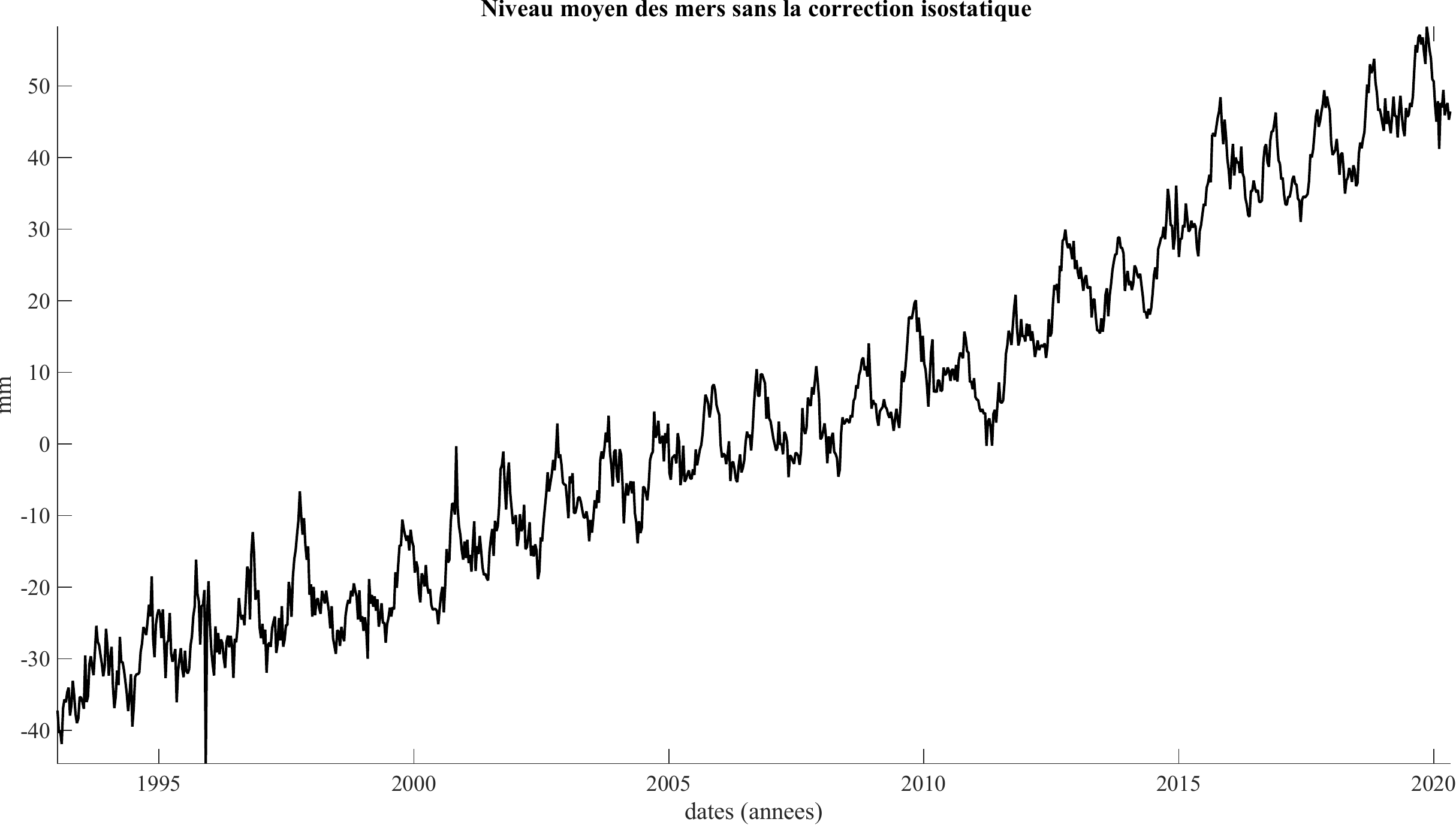}}
	\end{center}
	\caption{Mean sea level from 1993 up to the present day}
	\label{ssa_trend}
\end{figure}

To illustrate our point, we will apply SSA to physical data. Since the early 1990s, NASA has been measuring and providing\footnote{\href{https://climate.nasa.gov/vital-signs/sea-level/}{https://climate.nasa.gov/vital-signs/sea-level/}} mean sea level (\cf Figure (\ref{ssa_trend})). One of the questions for geodesists is the effect of isostasy on tectonics in general and on the axis of rotation of the poles in particular (\eg \shortciteNP{courtillot2022sea}). For example, how does the melting of ice and the redistribution of surface masses affect the Earth's axis of rotation? One way to understand this phenomenon is to study the evolution of global mean sea level from satellite measurements (Poseidon/Topex, Jason I, II and III).

\begin{figure}[H]
	\begin{center}
		\tcbox[colback=white]{\includegraphics[width=16cm]{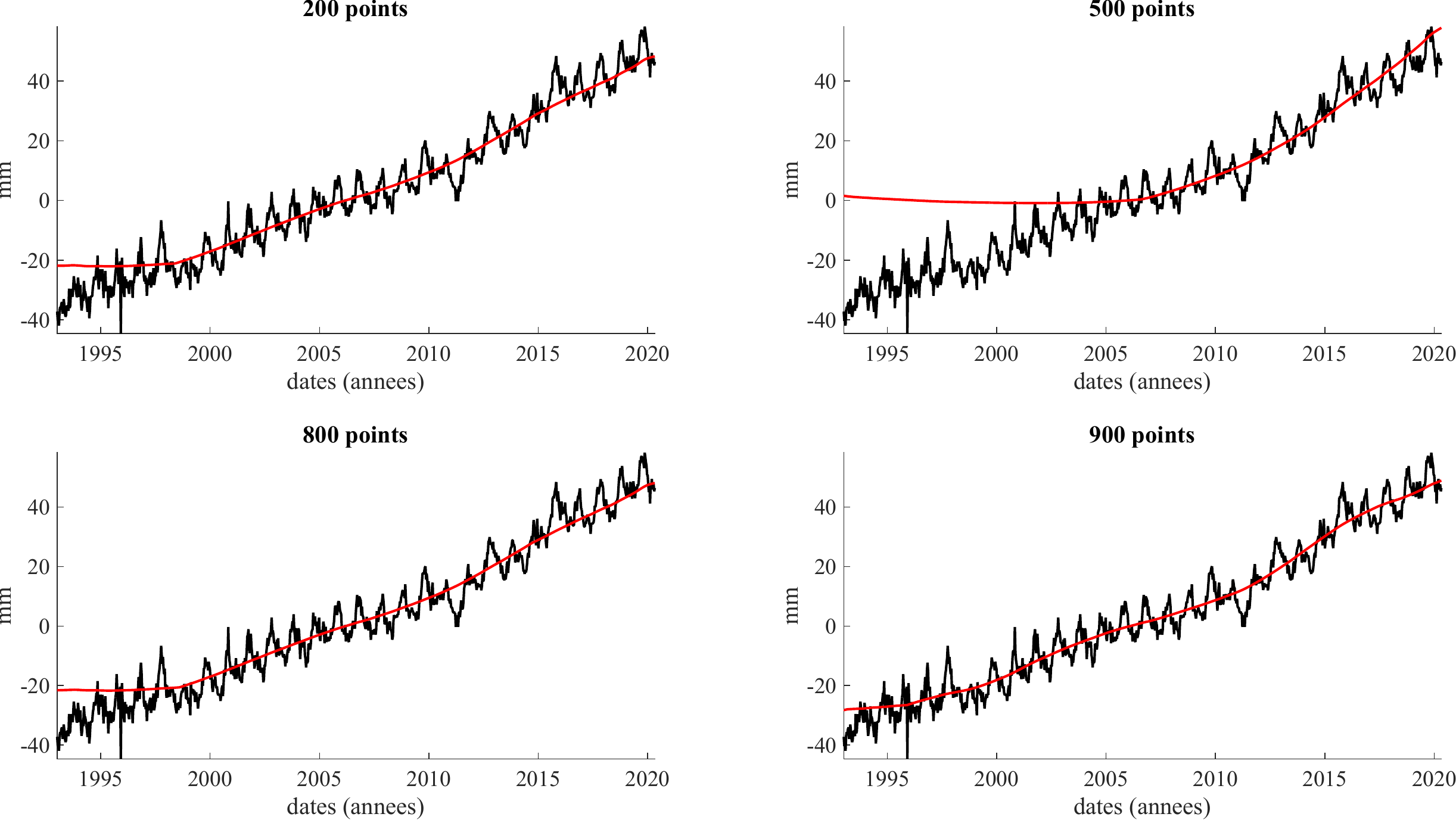}}
	\end{center}
	\caption{Superposées sur la courbe du niveau moyen des océans brute, les tendances extraites par SSA pour différentes valeurs de L (courbes rouges). }
	\label{ssa_trend_02}
\end{figure}

	In figure (\ref{ssa_trend}) we have plotted this mean sea level curve, which obviously shows a superposition of a more or less non-linear trend and an annual oscillation due to the Earth's rotation. The data start in 1993 and extend to September 2020, with a temporal sampling of about one point every 10 days. We will perform the SSA without any precautions and represent the first computed component, the trend, using the first elementary matrix $\textbf{X}_{\mathcal{I}_1}$ with $\mathcal{I}_1 = 1$. We have chosen different values of $L$: 200 points ($\approx$ 5.5 years), 500 points ($\approx$ 13.7 years), 800 points ($\approx$ 21.9 years) and 900 points ($\approx$ 24.6 years). The shift between two consecutive vectors in $\textbf{X}$ is a sample point. The trends obtained with the script \couleur{ex\_ssa02.m} are shown in figure (\ref{ssa_trend_02}).

	First important observation: SSA has a significant smoothing power; the trends obtained (red curves), although non-linear, are all smooth. Next, we observe quite different behaviour, especially for a value of $L$ of 500 points. Formally, for the first eigentriplet\index{eigentriplet}\footnote{a single and unique triplet, \eg for $i=1$ ($\lambda_1$, $U_1$, $V_1^t$)}, there are only minor differences between a moving average and the first component extracted by SSA. A priori, for a window length of 13.7 years (500 points), shifted one sample point at a time, the conditions seem to be met to characterise two behaviours in the data: a plateau from 1993 to around 2007, followed by an affine trend from 2007 to the present. As mentioned in the previous section, the length of the analysis window should be as large as possible; this is not a mathematical criterion. Figure (\ref{ssa_trend_02}) shows that the trends obtained for 200, 800 and 900 points belong to the same family. SSA, like other tools presented in this paper, is by no means a magic tool. Depending on the question, only the geophysicist should have the final say.

\subsection{Pseudo cycle separation}
\begin{figure}[H]
	\begin{center}
		\tcbox[colback=white]{\includegraphics[width=16cm]{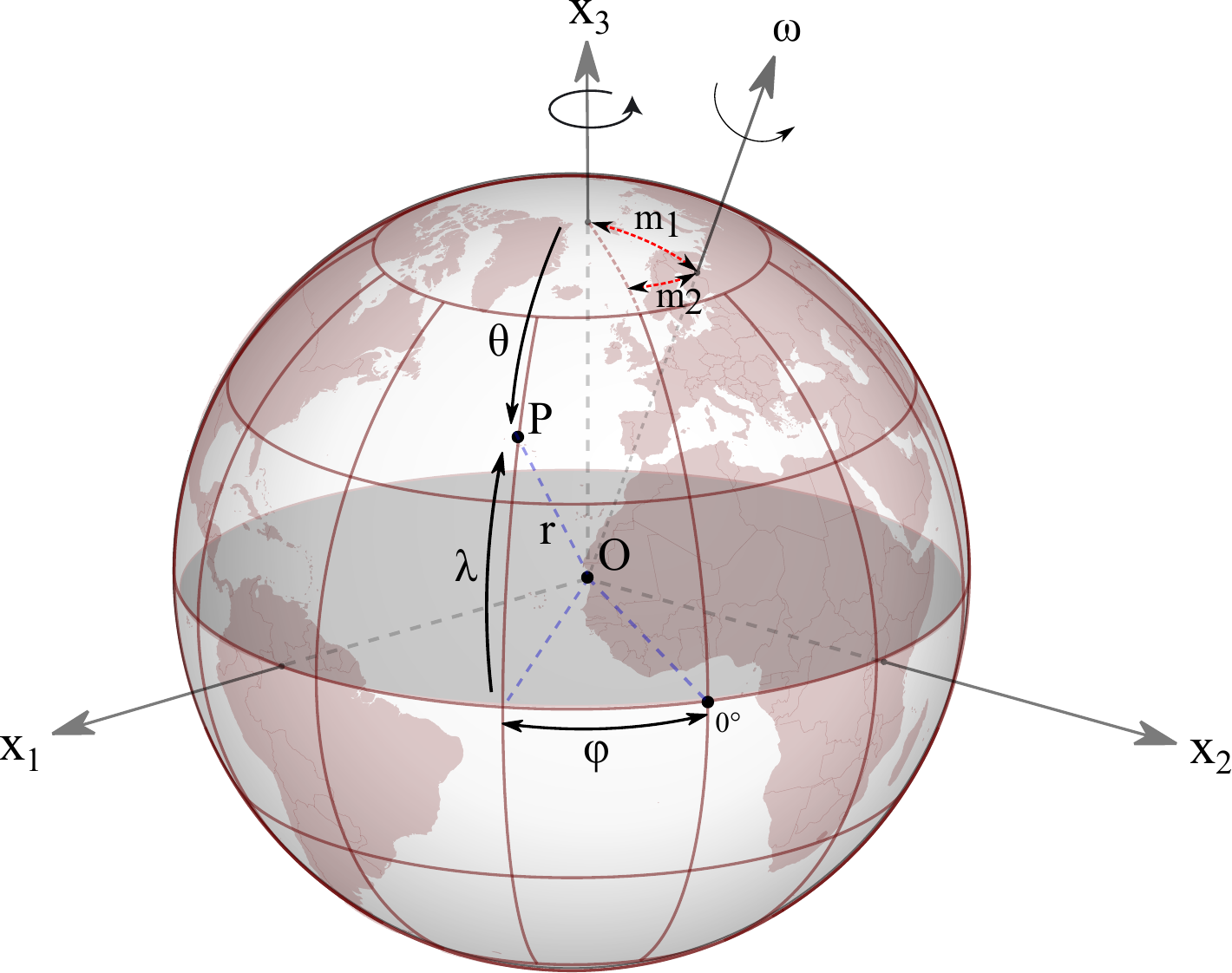}}
	\end{center}
	\caption{Geodetic Reference for the Movement of the Pole. m$_1$  is the North-South distance from the geographic North Pole, and m$_2$  is the East-West distance, with reference to the Greenwich Meridian}
	\label{m1m2_repere}
\end{figure}

\begin{figure}[H]
	\begin{center}
		\tcbox[colback=white]{\includegraphics[width=16cm]{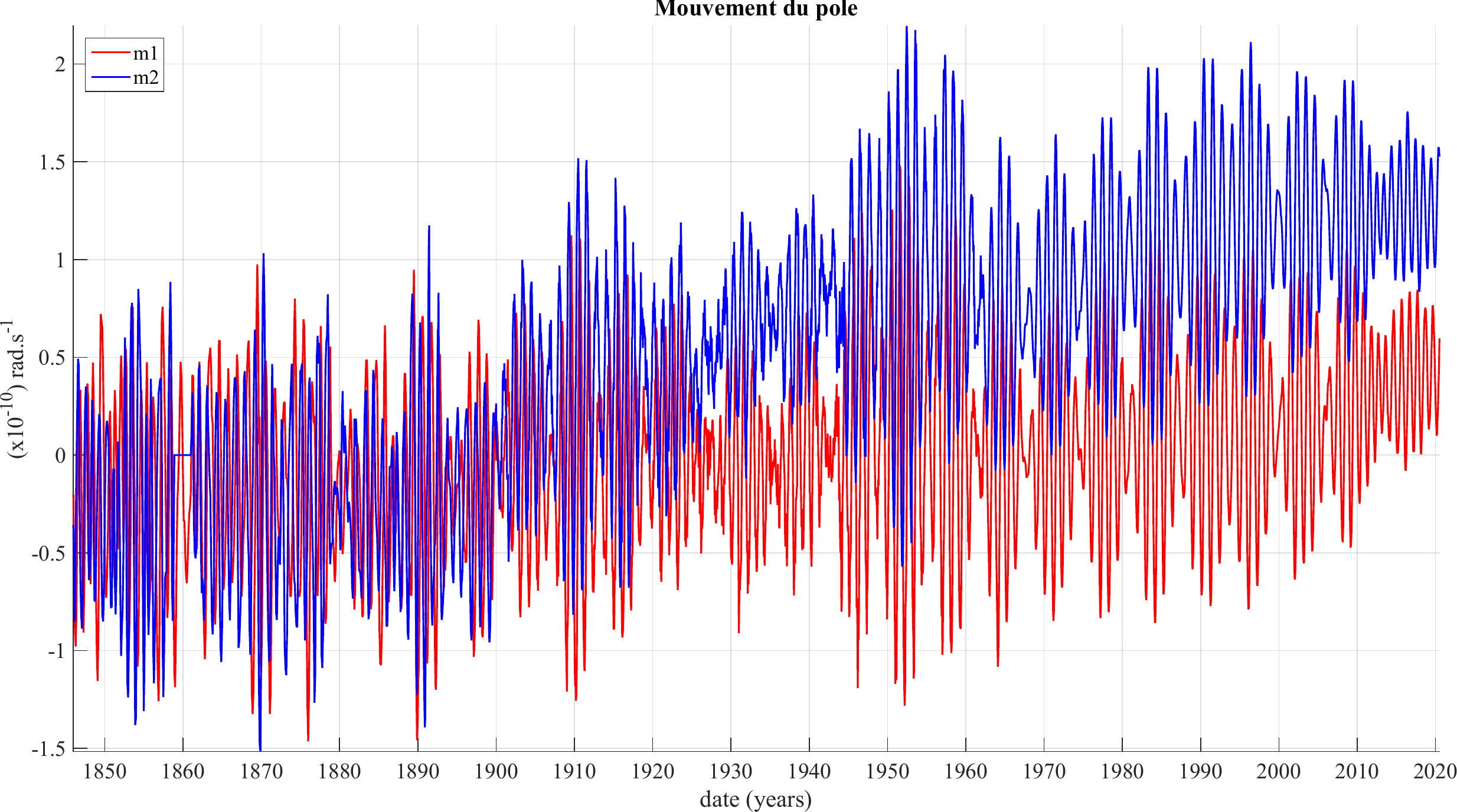}}
	\end{center}
	\caption{Temporal evolution of the components (m$_1$, m$_2$) from 1846 to the present.}
	\label{m1m2_raw}
\end{figure}

	Continuing with the trend extraction just discussed, we will now analyse a new real signal to extract its main pseudo-cycles\footnote{cycles whose periods and amplitudes vary over time}, namely the movement of the Earth's rotation pole (\eg \shortciteNP{lopes2021shoulders}). The movement of the Earth's rotation pole has been measured since 1846, initially using stars and now using laser measurements from satellites. The \textit{International Earth Rotation and Reference Systems Service}\footnote{\href{https://www.iers.org/IERS/EN/DataProducts/EarthOrientationData/eop.html}{https://www.iers.org/IERS/EN/DataProducts/EarthOrientationData/eop.html}} provides us with the time series of the pair (m$_1$, m$_2$), the coordinates of the rotation pole (\cf Figure (\ref{m1m2_repere})). The time series of this pair is shown in Figure (\ref{m1m2_raw}).

\begin{figure}[H]
	\begin{center}
		\tcbox[colback=white]{\includegraphics[width=16cm]{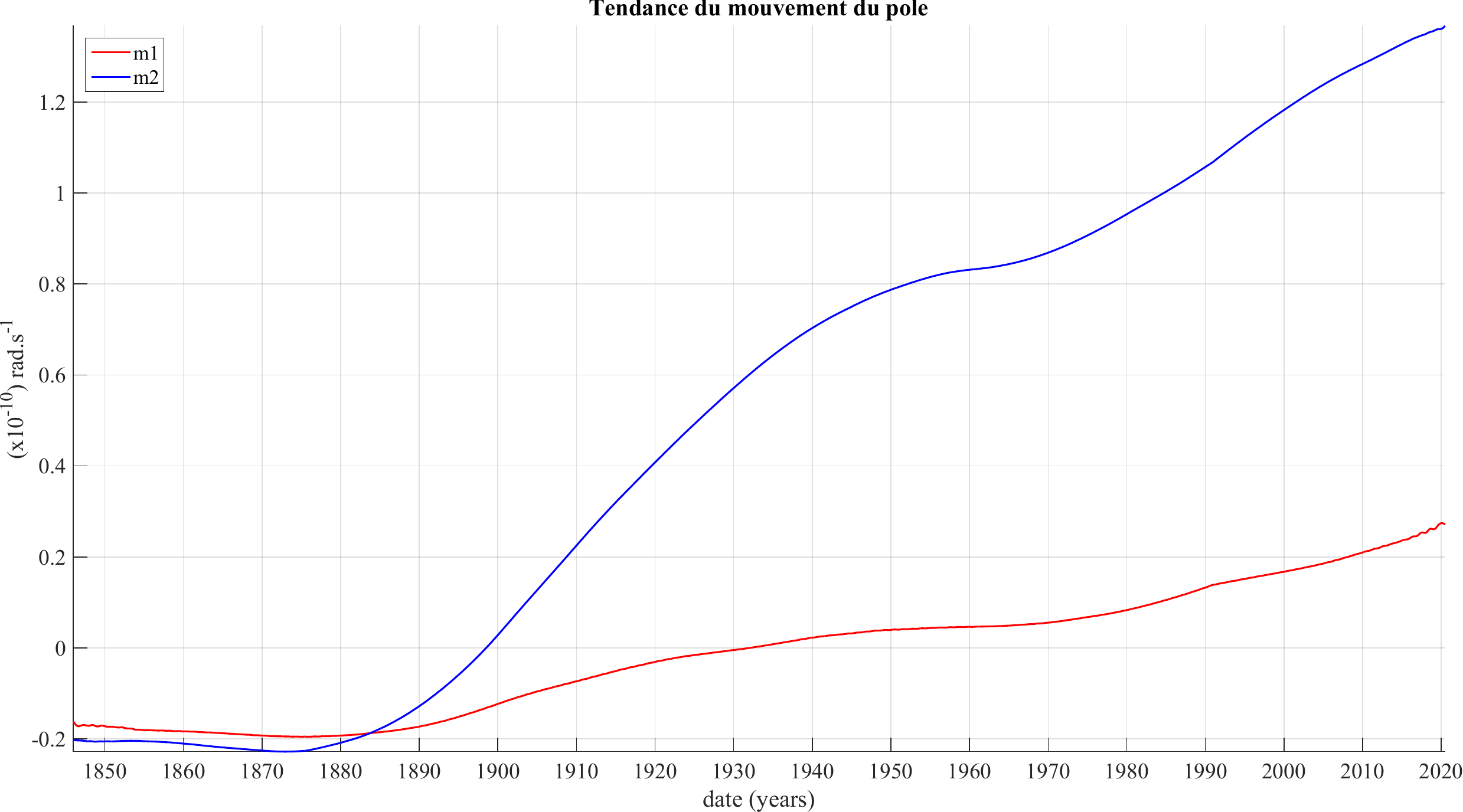}}
	\end{center}
	\caption{Trends of the Pair (m$_1$, m$_2$) Extracted by SSA}
	\label{m1m2_trends}
\end{figure}

	It has been known since the late 18th century that the movement of the pole follows the first-order linear partial differential equations of \couleur{Liouville-Euler}. This system exhibits a forced oscillation, traditionally called the annual oscillation, resulting from the Earth's revolution around the Sun; and a free oscillation, known as the Chandler oscillation (\shortciteN{chandler1891variationI}; \shortciteN{chandler1891variationII}), characterised by a dramatic phase jump during the 1920s and 1940s. These two pseudo-oscillations are superimposed on a pole drift discussed in the previous section, with time constants corresponding to the drift of the plates ($\approx$ 10 cm/year). This drift was first identified in the 1960s by \couleur{Markowitz} \shortcite{Markowitz1968continental}. The following figures have been produced using the script \couleur{ssa\_03.m}. Figure (\ref{m1m2_trends}) shows the SSA analysis components 1 (m$_1$) and 5 (m$_2$). As can be seen in figure (\ref{m1m2_raw}), since component m$_2$ drifts more than its longitudinal counterpart, it is normal for its trend to appear before that of m$_1$.
\newpage 

\begin{figure}[H]
	\begin{center}
		\tcbox[colback=white]{\includegraphics[width=16cm]{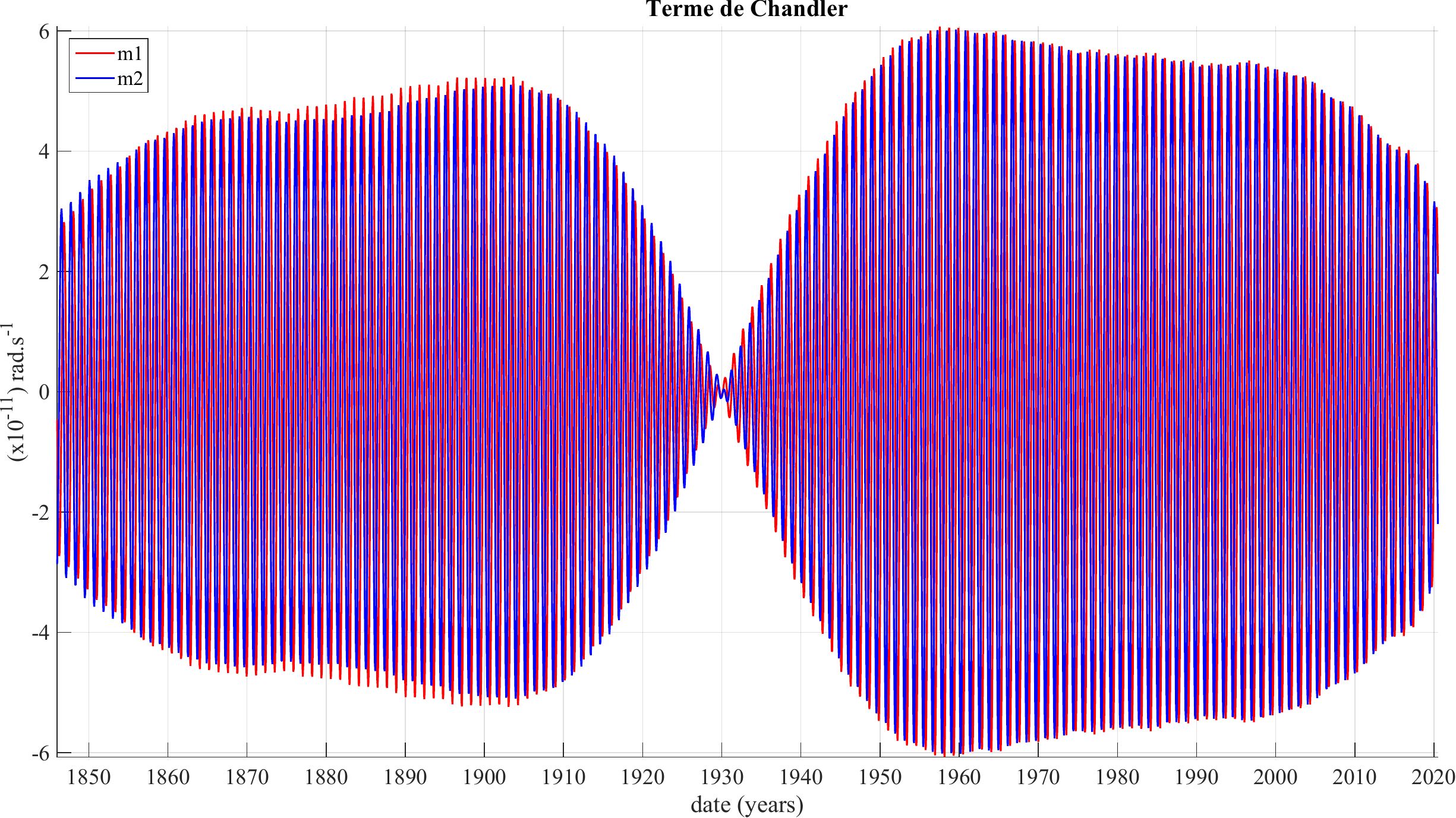}}
	\end{center}
	\caption{Chandler pseudo-cycles of the pair (m$_1$, m$_2$) extracted by SSA.}
	\label{m1m2_chandler}
\end{figure}

	Next, in the same order, are 1) Chandler oscillations (components 1 and 2 for m$_1$, 2 and 3 for m$_2$; \cf Figure (\ref{m1m2_chandler})), and 2) forced oscillations (components 3 and 4 for m$_1$, 4 and 5 for m$_2$; \cf Figure (\ref{m1m2_forced})).
	
\begin{figure}[H]
	\begin{center}
		\tcbox[colback=white]{\includegraphics[width=16cm]{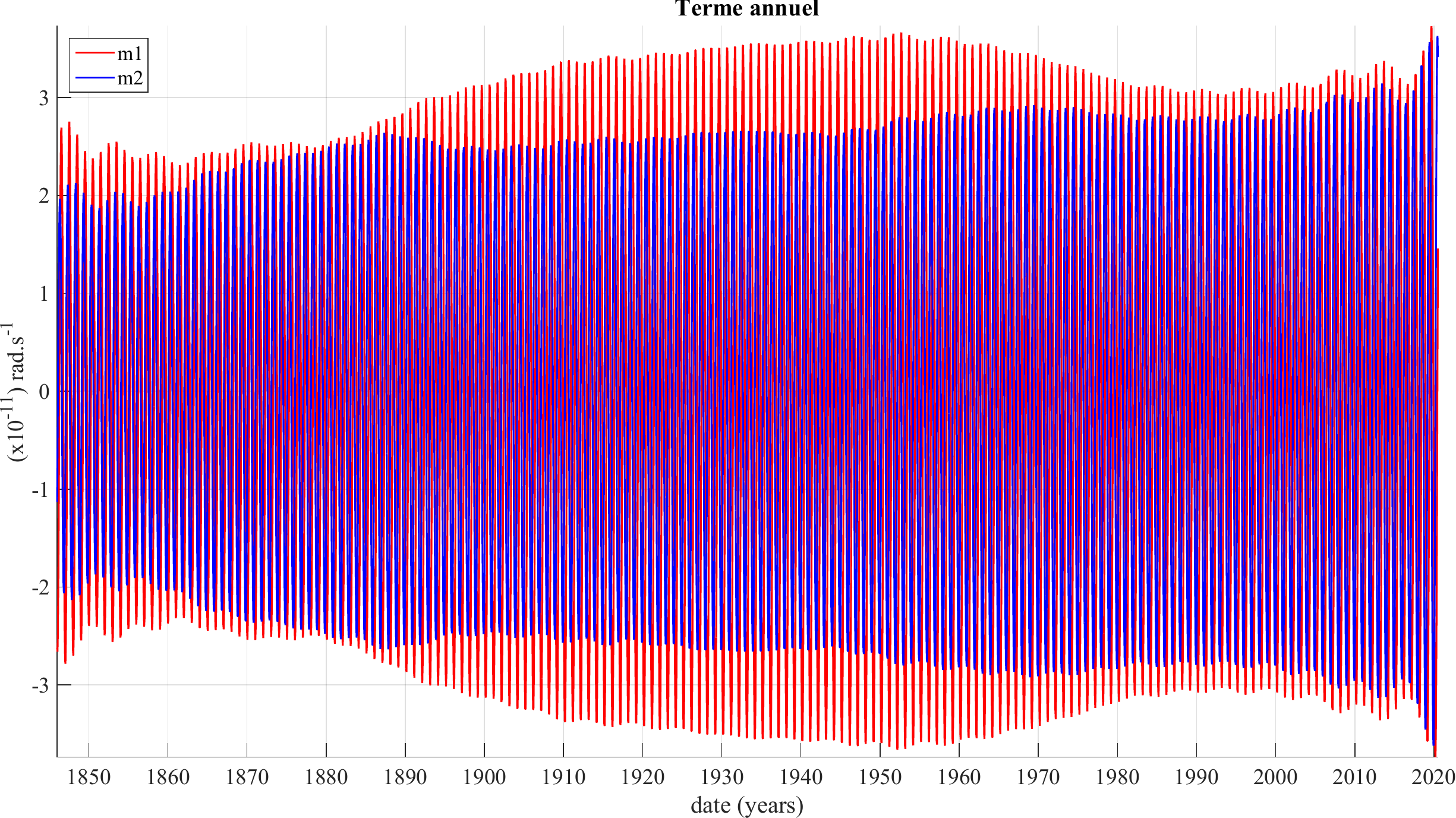}}
	\end{center}
	\caption{Forced pseudo-cycles of the pair (m$_1$, m$_2$) extracted by SSA}
	\label{m1m2_annual}
\end{figure}

\begin{figure}[H]
	\begin{center}
		\tcbox[colback=white]{\includegraphics[width=16cm]{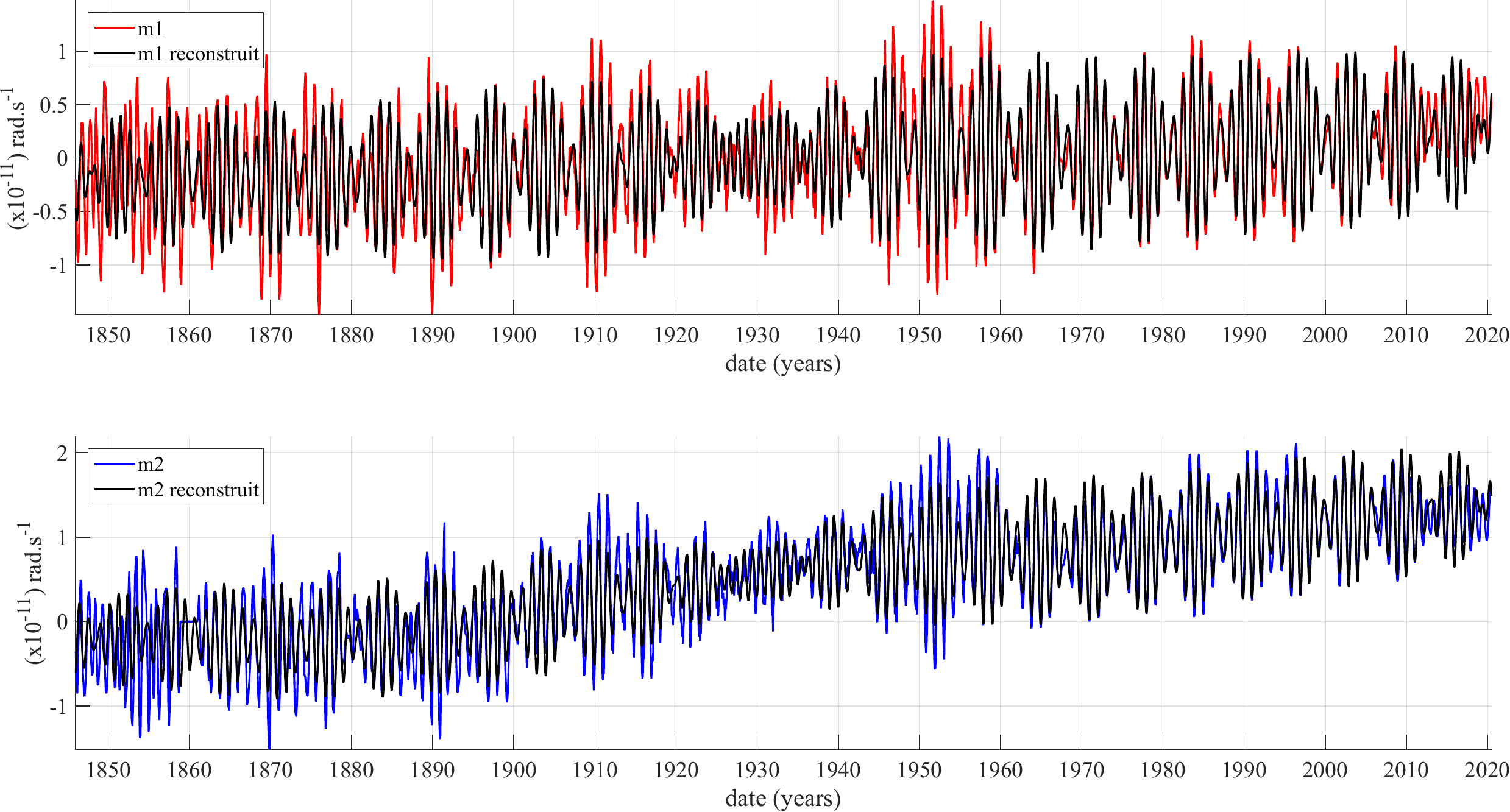}}
	\end{center}
	\caption{Comparison between the raw component m$_1$ (red curve) and the sum of the extracted components (black curve). Below, the same comparison for m$_2$.}
	\label{m1m2_comparaison}
\end{figure}

	Since SSA operates only in the data space and not in the transform space, unlike \couleur{Fourier} or \couleur{Wavelets}, it is possible to reconstruct a signal from the extracted cycles and trends that best fits a theory such as \couleur{Liouville-Euler}. SSA thus allows us to discard any information that is not accounted for by a system of equations and that could complicate its resolution or inversion. The SSA acts as a non-linear physical filter. In the problem of interest here, namely how much the pole drift and the free and forced oscillations contribute to the original signal, we simply need to sum them up and compare them with the originals (\cf. Figure (\ref{m1m2_comparaison})).

	We can see here that considering only the first 5 components obtained from the decomposition of \textbf{X}, grouped in 3 sets, largely explains this polar motion signal.

\subsection{Nonlinear Filtering}
\begin{figure}[H]
	\begin{center}
		\tcbox[colback=white]{\includegraphics[width=16cm]{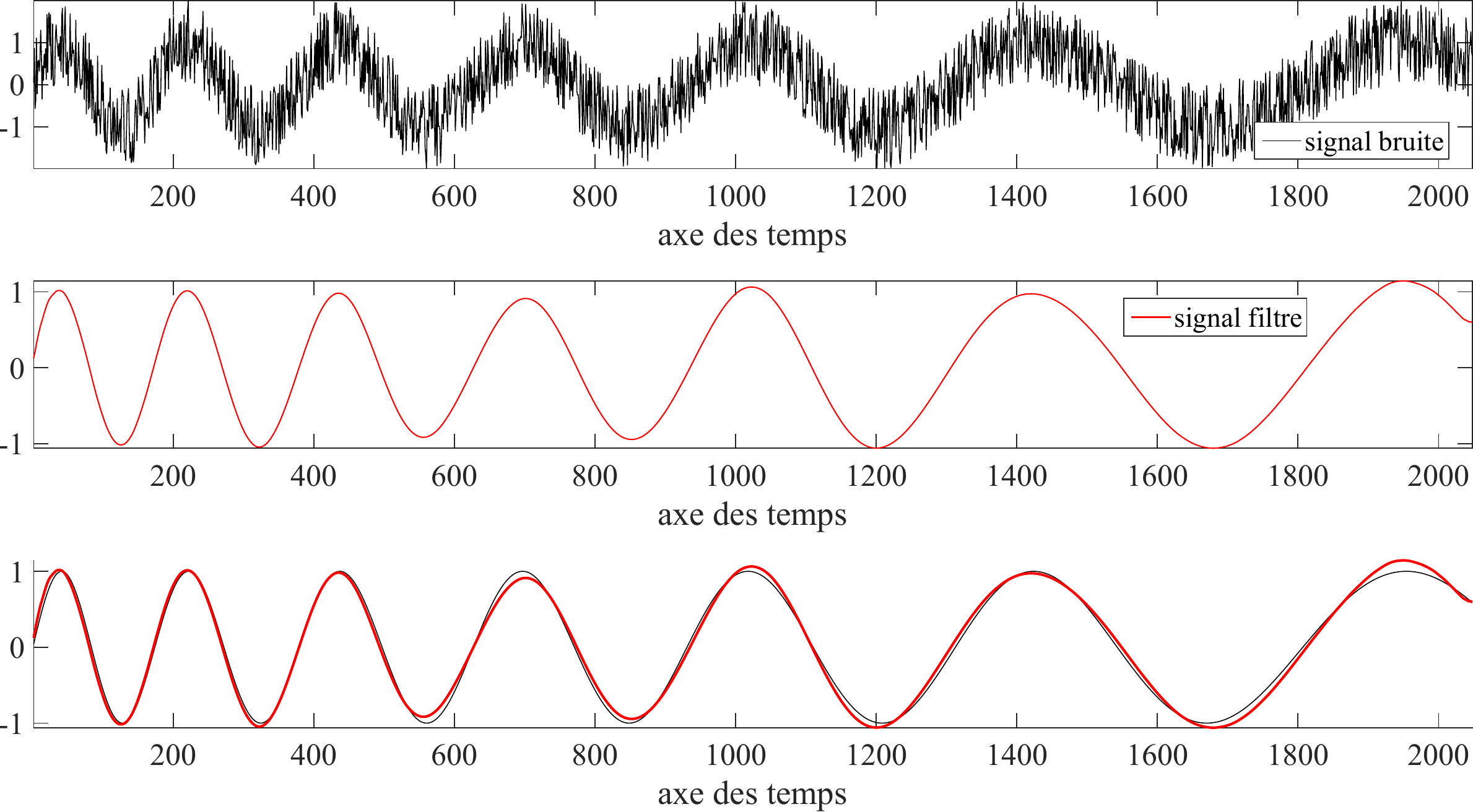}}
	\end{center}
	\caption{At the top, in black, is a sine wave with an increasing phase over time, to which we have added noise. In the middle, in red, is the result of the SSA filtering. Below, superimposed, are the original signal (unnoised, in black) and the filtered signal (in red).}
	\label{ssa_filter1}
\end{figure}

\begin{figure}[H]
	\begin{center}
		\tcbox[colback=white]{\includegraphics[width=16cm]{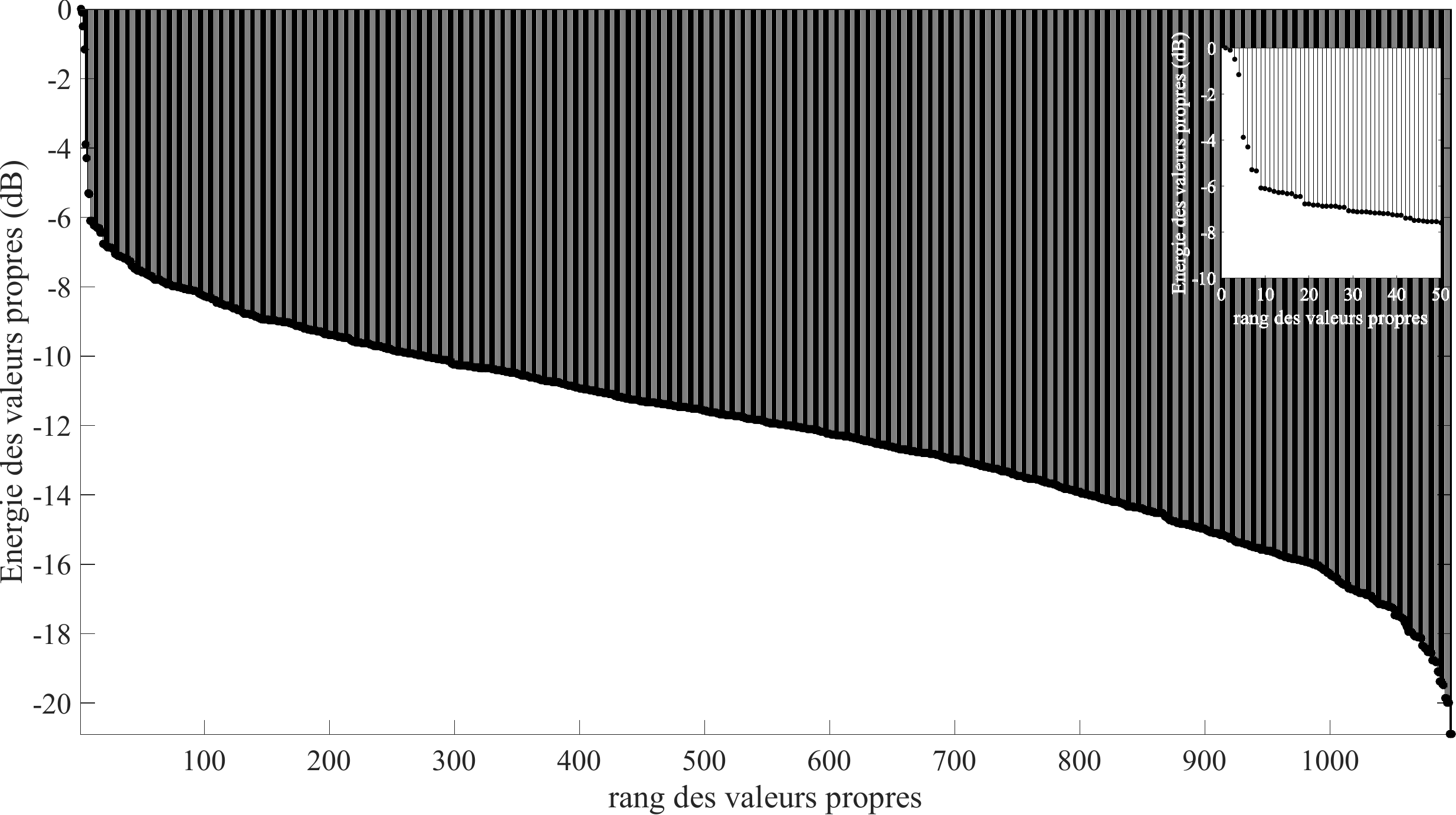}}
	\end{center}
	\caption{Eigenvalues in dB of the noisy signal from figure (\ref{ssa_filter1}). On the top right is a zoom of these eigenvalues between ranks 1 and 50.}
	\label{ssa_filterdb1}
\end{figure}

\begin{figure}[H]
	\begin{center}
		\tcbox[colback=white]{\includegraphics[width=16cm]{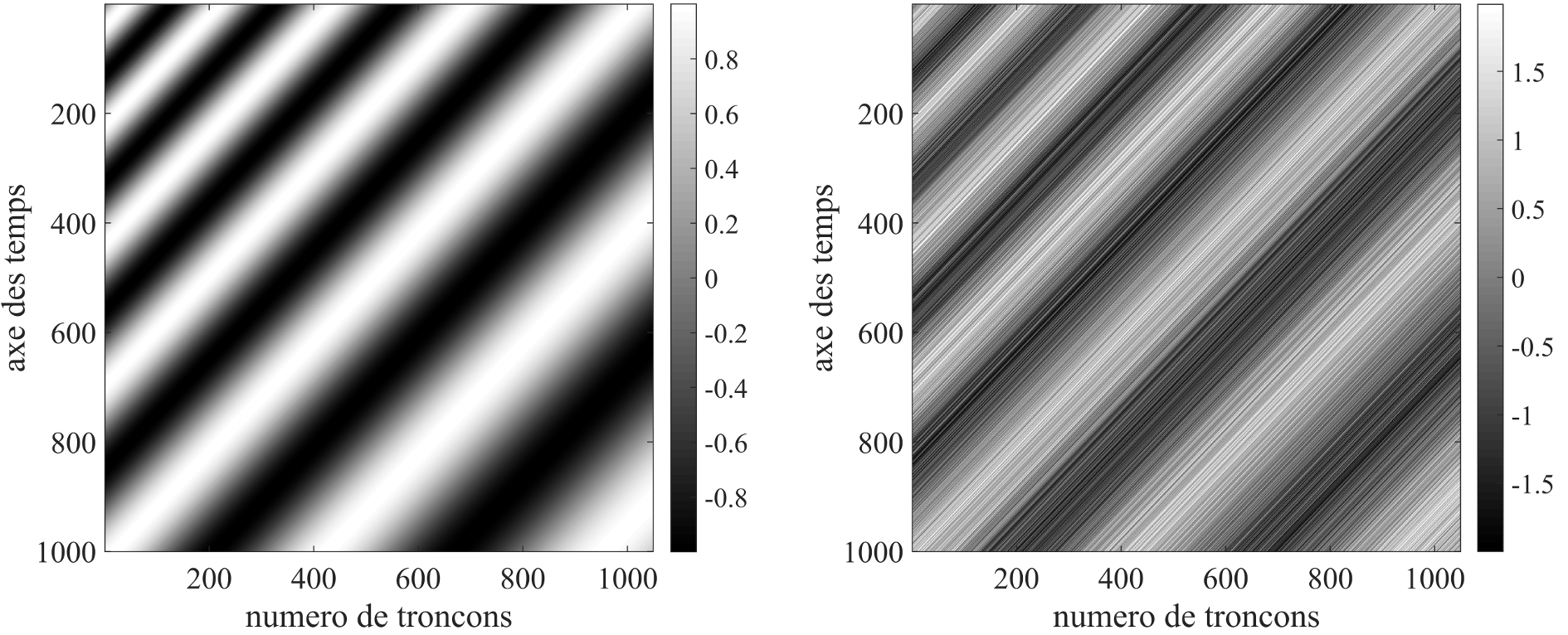}}
	\end{center}
	\caption{Hankel matrices: on the left, the signal before adding noise; on the right, the noisy signal.}
	\label{ssa_filter_hankel}
\end{figure}

	We will revisit the example from Figure (\ref{ondelettes08}), where we filtered a noisy signal with a phase that increased over time using wavelets. The aim here is not to determine whether wavelets or SSA provide better filtering, but rather to compare the implementation of the two approaches on the same signal. This signal has two very interesting advantages for us. Firstly, this sine wave has a phase that varies linearly with time, and secondly, the additive noise is of a magnitude significantly greater than that of the sine wave itself. Using the \couleur{ssa\_04.m} script, we obtain the results shown in the figure (\ref{ssa_filter1}). First, at the top, we see the signal to be analysed, followed in the middle by its filtered version obtained by SSA. This almost perfect filter was expected. The Hankel matrix, unlike the EOF autocorrelation matrix, is composed of segments of the signal to be analysed; therefore, if the signal is only the sum of a first-order predictable signal with white noise, then the pattern of the corresponding Hankel matrix will be that of the first-order signal alone (i.e. without the noise, \cf Figure (\ref{ssa_filter_hankel})). It will then be easy for the SVD to isolate this noise into low energy eigenvalues (see Figure (\ref{ssa_filterdb1})). It is clear that after the eighth eigenvalue, the energy drops below -6 dB, which is less than 50\% of the signal amplitude.
\newpage

\chapter{\titrechap{Inverse Problem}}
\minitoc
\section{Introduction}
	The theory of inverse problems often has a bad reputation. Among other things, it is considered to be too mathematical, detached from reality, and impractical. For these reasons, it is seen by many as the preserve of a community admired for its intellectual achievements but not taken seriously when it comes to practical applications with real data collected from the field. While it is true that some geophysicists working on inverse problems deserve such criticism, it is unfair to generalise this negative impression. In fact, the last decade has seen numerous successes of inverse methods. In geophysics, seismic tomography inversions have provided images of the Earth's mantle. In seismology, the most advanced 3D migration methods are based on nonlinear inversion techniques. In geomagnetism, magnetotelluric inversions and those reconstructing the flow of liquid iron at the surface of the outer core can be cited. Meteorology and oceanography have also seen significant progress in inverse problems, the specificity of which (large volumes of data, spatio-temporal variability of models) requires the development of methods such as data assimilation, which is gaining increasing interest in geophysics. In medicine, inverse problems are increasingly used to analyse electroencephalographic and electrocardiographic data. They are also present in more traditional imaging algorithms such as ultrasound or electrical tomography.

	The applications mentioned above show that the theory of inverse problems now constitutes a \textit{corpus} of considerable volume, resulting from an evolution of research over the last 40 years. In addition to theoretical advances, computational innovations, which have also progressed spectacularly, now allow the implementation of methods that were considered inapplicable only 20 years ago. Before delving into the history, it is useful to define what an inverse problem is by describing a very simple case that will allow us to illustrate the various concepts we will be exploring throughout this course.

\section{An inverse problem example}
	The task is to locate a tunnel by measuring the gravity field along a profile perpendicular to the tunnel axis. We will therefore work in the two-dimensional approximation $\left(x,z\right)$ and assume that the tunnel has a circular cross-section. The vertical component of the gravitational anomaly generated by the tunnel and calculated along the profile $\left(x,z=0\right)$ is given by
\begin{equation} 
	g_{z}\left(x\right)=\dfrac{2\pi .  G . \rho . r_t^2 . z_t}{\left(x-x_{t}\right)^{2}+z_{t}^{2}},				\label{FFF}
\end{equation}

\begin{figure}[H]
	\begin{center}
		\tcbox[colback=white]{\includegraphics[width=16cm]{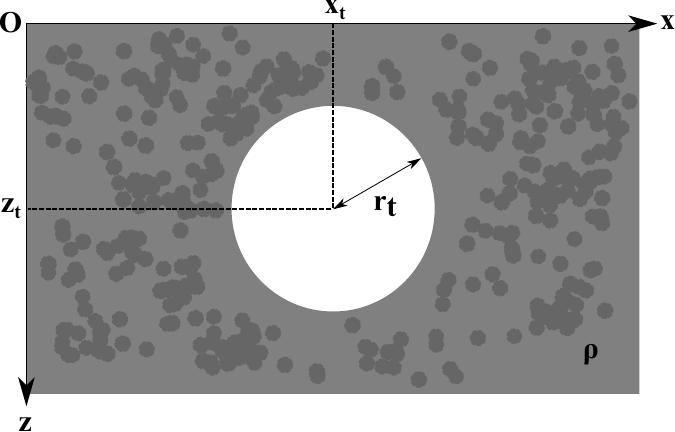}}
	\end{center}
	\caption{Diagram of a classic subsurface geophysical problem: detection of a gravimetric anomaly due to the presence of a tunnel.}
	\label{tunnel01}
\end{figure}

	$\rho$ is the density of the rock in which the tunnel of radius $r_{t}$ is located. The coordinates of the tunnel axis are $x_{t}$ and $z_{t}<0$. The left-hand side of the equation represents the gravimetric anomaly, which is comparable to the data that will allow us to determine the parameters. This data is the primary information for the inverse problem, as it will allow us to improve our understanding of the tunnel model (see figure \ref{tunnel01}). However, this primary information is not the only information available to us, as some symbols on the right hand side of the equation can be considered as more or less known. This is another source of information, known as \apriori information.

	Depending on our level of knowledge, the symbols on the right can be either data or unknowns - parameters of the problem. For example, we might assume that the density, radius and horizontal position are known and that the only parameter of the inverse problem is the depth $z_{t}$. In this case we face a non-linear problem because, for example, if $z_{t}$ is multiplied by $2$, the gravimetric anomaly is certainly not multiplied by $2$. It is also possible that the only unknown parameter is $\rho$, and then the problem is linear, because when $\rho$ is multiplied by $2$, the gravimetric anomaly is doubled. In the most general case, we can assume that we are solving for the four parameters $\left\{ \rho, r_{t}, x_{t}, z_{t} \right\}$.

	Depending on the \apriori information available - or believed to be available \thinspace ! - the inverse problem will take different analytical forms. It is clear that the initial parameterisation of an inverse problem implicitly contains a lot of information. The last symbol we have not yet discussed is the position $x$ at which the gravitational measurements are made. It is usually assumed that this position, known as the independent variable, is perfectly known. We will see that even this variable can be considered as imperfectly known when working within the most general framework of inverse problem theory. An ultimate complication can be added if the accuracy of the equation itself is questioned by considering that the tunnel may not necessarily have a circular cross section, but rather a "potato-like" shape.

	As we can see, the inverse problem of locating a tunnel can be made as complex as desired to illustrate many aspects of inverse problem theory. We will reformulate it throughout the course, gradually increasing its complexity until we reach its most general form.

\section{General structure of inverse problems}
\begin{figure}[H]
	\begin{center}
		\tcbox[colback=white]{\includegraphics[width=16cm]{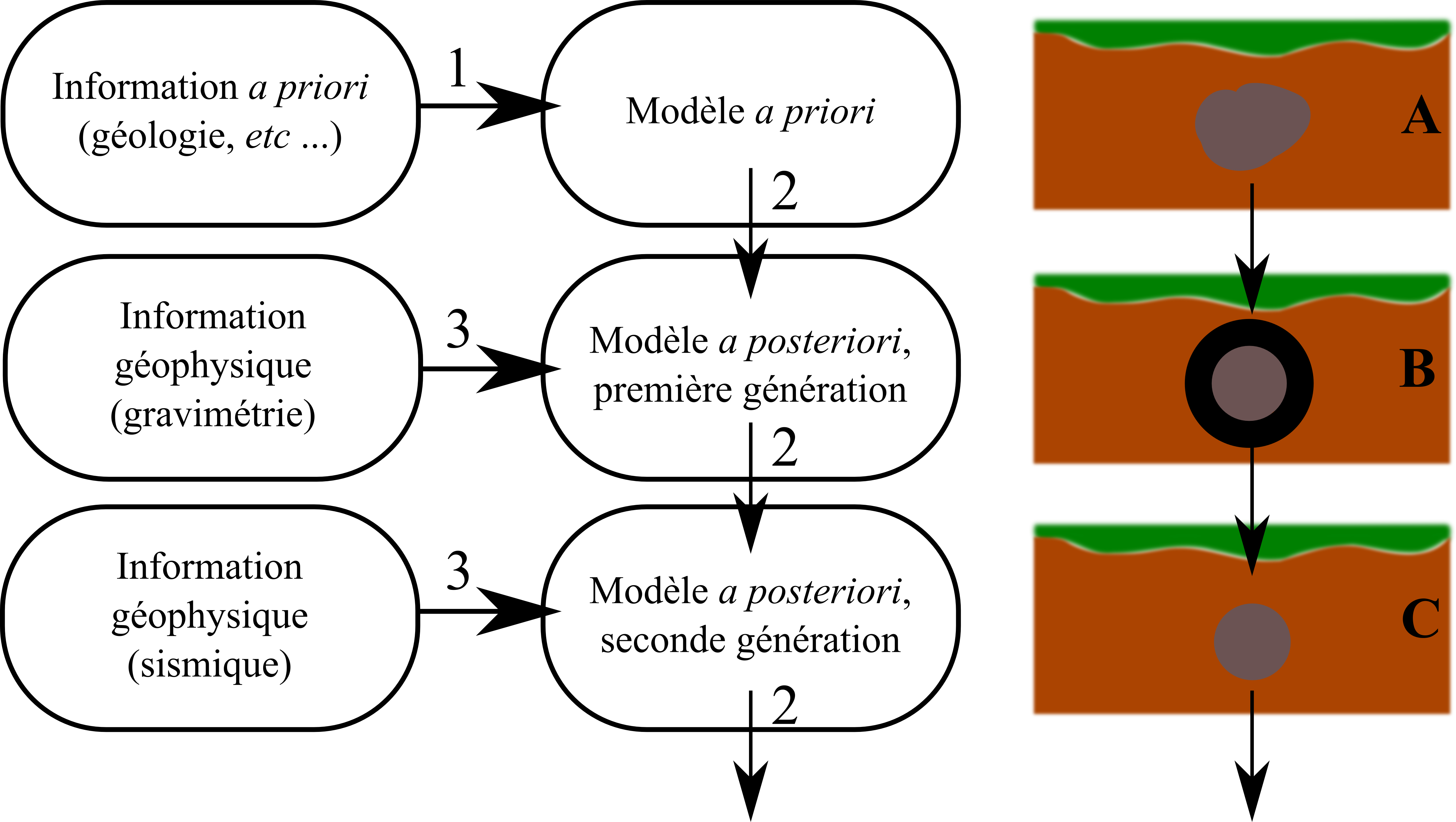}}
	\end{center}
	\caption{General structure of inverse problems}
	\label{principe1}
\end{figure}

	The example presented earlier has characteristics found in most inverse problems, the general structure of which is summarised in Figure (\ref{principe1}). The goal of inverse problems is to improve our understanding of an object or phenomenon by using more and more information about it. For example, general geological knowledge about a region (top left case) might indicate the presence of a fossil magma chamber underground. The same general knowledge indicates that a magma chamber is a more or less spherical structure composed of rocks with densities within a certain range. This leads to a set of \apriori models - top, center - an example of which is shown in box A. These models are infinite in number and are often described in vague and non-numerical terms, which means that they are difficult to manipulate on a computer. However, the geologist's expertise allows the design of a gravimetric experiment based on these models, which will provide information-middle left box-that will refine our knowledge of the magmatic chamber in such a way that the set of acceptable models-centre box-is smaller than the \apriori models. The links 2 and 3 leading to these sets are an inverse problem. The resulting \aposteriori models are more accurate. For example, the models in box B are spherical, with a possible radius within a relatively narrow range and a depth that is fairly well defined. One could then carry out a second geophysical experiment, such as a seismic test, to provide new information - lower left box - that would allow the positioning of the roof of the magmatic chamber - box C - and thus significantly reduce the set of acceptable models - lower middle box. The links 2 and 3 leading to this new set of \aposteriori models form a second inverse problem, where the \apriori models are the \aposteriori models from the first inverse problem. In this way, inverse problems can be linked sequentially to improve our understanding of the magmatic chamber.

\begin{figure}[H]
	\begin{center}
		\tcbox[colback=white]{\includegraphics[width=16cm]{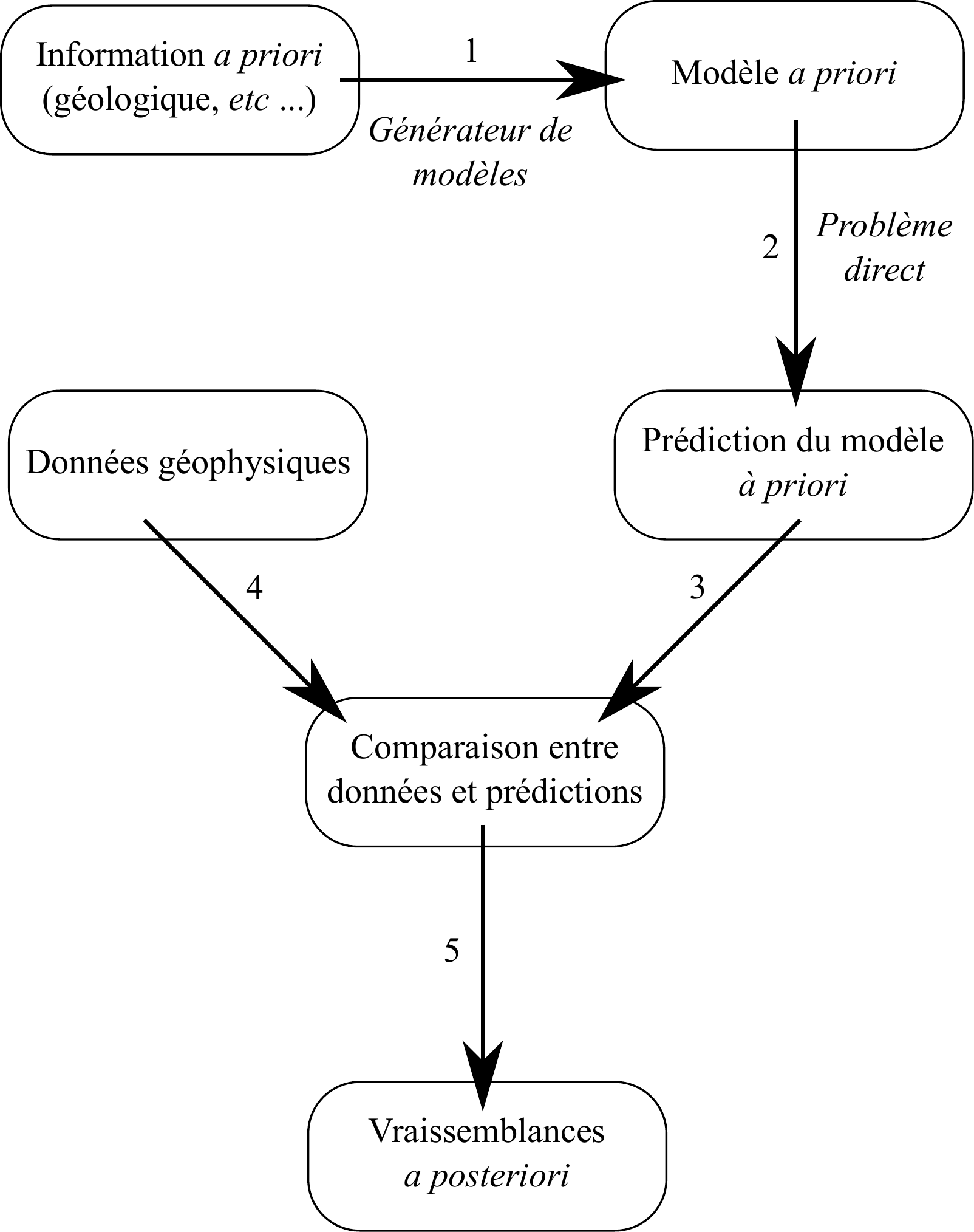}}
	\end{center}
	\caption{General Structure of Inverse Problems}
	\label{principe2}
\end{figure}

	Figure (\ref{principe2}) details the structure of a specific inverse problem, specifically the links 1, 2, and 3 from Figure (\ref{principe1}). Link 1 is a model generator that produces \apriori models compatible with the initial information available before acquiring geophysical data. The model generation step is crucial in this general framework and represents one of the significant challenges in inverse problem theory. This difficulty arises because geological information is often vague and non-numeric, making it challenging to generate \apriori models in a computer that adequately cover the wide range of models envisioned by the expert geologist. In this course, we will explore partial solutions to this problem (geostatistics, projection onto convex sets, \ldots). The next step is arrow 2, which represents the forward problem. This involves selecting \apriori models and calculating their geophysical response to compare with the data collected in the field \via arrows 3 and 4, which lead to the decision box. The forward problem lies at the heart of the inverse problem and often needs to be solved many times. Therefore, it is crucial that the forward problem can be solved as quickly as possible on the computer, which sometimes necessitates the use of approximate solutions. For example, in seismics, asymptotic methods (ray tracing) are often faster than wave equation methods (finite differences or finite elements). In the case of the magma chamber, a simplified forward problem might be to assume that the chamber is spherical. It is up to the expert geologist to decide whether such an approximation is acceptable given his \apriori knowledge; if the magma chamber could be oblong, it is clear that the spherical approximation does not allow proper exploration of the range of models the geologist has in mind. In such cases, the forward problem must be adapted to use, for example, ellipsoidal shapes, which may better suit the geologist's ideas. The decision step allows the selection of \apriori models that are acceptable and will belong to the set of \aposteriori models. This set is the solution to the inverse problem, and it can sometimes be a challenge to present it in a way that is simple and easy for the user to understand. It is sometimes possible to create a visual representation, such as a film showing the \aposteriori models in proportion to their likelihood, but this is not always very meaningful.

\section{A little bit of history}
\subsection{The 1960s}
	Inverse problems were introduced into geophysics towards the end of the 1960s when \shortciteN{backus1967numerical}, \shortciteN{backus1968resolving}, \shortciteN{backus1970uniqueness} published a series of theoretical papers laying the foundations of the theory. Numerous papers followed, either within the same theoretical framework or focusing on specific applications. The 1970s was thus a period of considerable development in the theory of inverse problems. This was particularly true for the theory of linear and linearised problems. Several fundamental principles were established, such as that the statistical uncertainty in the parameters of a model decreases as the resolution of the model - its "fineness" - increases. Although the notion of an ill-posed problem dates back to the early twentieth century (\couleur{Hadamard}), it became commonplace, and it was recognised that geophysical data alone are generally not sufficient to produce an unambiguous model. Basic algorithms were proposed to deal with these challenges as efficiently as possible, such as inversion by singular value decomposition and singular vectors \shortciteN{jackson1972interpretation}, which had been proposed much earlier in applied mathematics \shortciteN{levenberg1944method}, \shortciteN{penrose1955generalized}.

\subsection{The 1970s}
	It was in the early 1970s that a different perspective was proposed by \shortciteN{franklin1970well}, who published a paper explaining that certain ill-posed linear inverse problems become well-posed when formulated in probabilistic terms. The title of this paper, \textit{"Well-posed stochastic extension of ill-posed linear problems"}, implies that if the inverse problem is formulated in terms of finding the probability density of various models within the \apriori\ model space, then the solution sought (i.e. the probability density) is unique. In this sense, the problem is well-posed. At the same time, there has been a growing recognition of the importance of \textit{a priori} information that helps to reduce or even eliminate the ill-posed nature of an inverse problem. This information enhances the data provided by geophysical measurements, as if the data were more abundant, of a different nature and less noisy. \apriori\ information is also used to introduce constraints on the parameters being sought. For example, in gravimetry the constraint that density must be positive can be applied. Unfortunately, \apriori\ information proved difficult to incorporate into the formalisms of the time \shortciteN{jackson1979use}.
  
\subsection{The 1980s}
	The 1980s saw many developments in theory. Applications were also plentiful, but many remained unconvincing, mainly because they produced results that were difficult to integrate into broader frameworks. For example, certain electromagnetic inversion results were difficult to reconcile with geological interpretations, while other seismic inversion results provided little guidance for decisions such as whether to drill an oil well. The main problems in the early 1980s were that inverse methods often paid too little attention to \apriori\ information, the geological nature of which did not fit easily into highly mathematical formalisms. Even the joint inversion of different geophysical data (seismic + gravimetry, \etc.) remained rare \shortciteN{vozoff1975joint}, \cite{lines1988cooperative}. Another major drawback was that inversion methods often failed to account for the multiplicity of possible solutions resulting from insufficient and noisy data. This was a significant handicap when the inversion was intended to inform decision making. The decade of the 1980s is important because inverse problems began to be developed in other scientific fields such as astrophysics, meteorology and medical imaging. Each of these scientific fields contributed to the improvement of the techniques. For example, medicine made significant innovations in imaging dynamic media and developed methods suitable for inverse problems where parameters vary over time. Meteorology invented data assimilation methods, useful when new data are constantly arriving and need to be incorporated into an inversion.

\subsection{The 1990s}
	The article by \textit{Franklin} \shortciteN{franklin1970well} laid the foundations for a stochastic approach to inverse problems, but it took about twenty years for this approach to become commonplace. Among the foundational papers in the probabilistic approach to inverse problems is that of \shortciteN{tarantola1982generalized}, published in 1982, where the authors establish the basis for inversion in terms of probability densities of parameter values. This perspective has its roots in the work of \couleur{Bayes} (1702-1761) \shortcite{barnard1958studies} and has been the subject of numerous publications. In this course we will see that the \textit{Bayesian} approach to inverse problems is very flexible and allows \apriori\ information to be explicitly considered. However, it is only recently that this approach has become popular, largely because we now have sufficiently powerful computers to take full advantage of its benefits. Advances in computing have revived algorithms published in the 1950s that were impractical for intensive use at the time. This is the case of the \couleur{Metropolis} algorithm \shortcite{metropolis1953equation}, proposed in 1953, shortly after the advent of the first computers.

\subsection{The 2000s}
	Paradoxically, after a long period of heavy mathematisation accompanying the development of approximate methods (gradient methods, perturbations, \etc), work on inverse problems has become more refined and is currently focused on the challenging problem of incorporating \apriori\ information and solving highly nonlinear problems. Several global solution search algorithms, such as simulated annealing, which we will explore later, can now be implemented for inverse problems of realistic complexity (i.e. combinatorial complexity). These methods have been applied in geophysics to solve inverse problems in seismology, seismic imaging, electrical tomography, \etc. Considerable effort is devoted to improving these algorithms (simulated annealing, genetic algorithms, neural networks, \etc.), which are based on intensive computations and require immense computing power. Continuing advances in computing now make it possible to access affordable computing power via PC \textit{clusters} or the Internet, allowing inverse problems to be formulated realistically, i.e. using realistic models. Within the next decade, computers will be powerful enough to solve many inverse problems, and the next decade should also see inversion methods becoming more widely used and integrated into the geophysicist's toolbox alongside signal processing techniques.

\section{Our philosophy}
	The series of examples we will see is merely an introduction. We have chosen to focus more specifically on the underlying philosophy of inverse problems, and to describe only a few techniques that are both easy to implement and general enough to be applicable in a wide range of cases. We have chosen to frame the inverse problem in terms of information theory because we believe this is the most general way to approach the subject. Indeed, one could say that solving an inverse problem involves transporting information. The transport of information can be subtle and may not only rely on physics and mathematics, but also require considerable expertise. An example of this is the inverse problems in palaeontology, which involve reconstructing the life history of an animal from an incomplete skeleton. The approach we will take is useful for understanding the difficulty of palaeontologists' tasks, but the mathematics we develop will certainly not be of much help. A major problem with inverse problem theory is that it relies on a mathematical formulation that is difficult to apply to the natural sciences. As a result, many geological inverse problems still defy rigorous theoretical approaches. Some attempts have been made through geostatistics, which has gained prominence for its ability to 'mathematise' geological information. However, a true theory of inverse problems applicable to geology remains to be established.

\chapter{\titrechap{Information \& Inverse Problems}}
\minitoc
\section{The definition of information}
\subsection{Information and Complexity}
	We will concentrate on defining information quantitatively, so that it can be treated as a measurable quantity. The definition of information that we will adopt is the one proposed by \couleur{Léon Brillouin} in 1959 \shortcite{brillouin1959science}, which is based on statistical considerations. Consider a problem with an a priori number of possible answers equal to $N$, for which we have no information. Under these conditions, all possible answers are equally probable, and we will say that the information $I$ needed to uniquely determine the number of a posteriori answers is defined by
\begin{equation}
	I=\ln N
\end{equation}

	The greater the number of \apriori answers, the more information is needed to obtain a unique \aposteriori answer. This is intuitive. The unit of information is the $nep$ when the natural logarithm is used in the definition above; it becomes the $digit$ for the decimal logarithm and the $bit$ for the base-2 logarithm. Consider the example of a problem where the number of a priori answers is limited to $N=2$. The information needed to solve this problem is $I=\ln2\simeq0.693 \ nep$.

	The choice of a logarithmic function is due to the desire for information to have the property of additivity. For example, consider two independent problems with a priori numbers of answers $N_{1}$ and $N_{2}$ respectively. The number of answers to the combined problems is therefore,
\begin{equation}
	N_{1,2}=N_{1}\times N_{2},
\end{equation}

which gives,
\begin{equation}
	\begin{split}	
		I_{1,2} & = \ln\left(N_{1}\times N_{2}\right)\\
				& = \ln\left(N_{1}\right)+\ln\left(N_{2}\right)\\
				& = I_{1}+I_{2}
	\end{split}	
\end{equation}

	The information needed to solve both problems simultaneously is simply the sum of the individual pieces of information. This property also corresponds to our intuition. If the number of \aposteriori answers is no longer 1, but $N'$, then the information gained is given by,
\begin{equation}
	\begin{split}
		I^{^{\prime}} & =  \ln\left(\frac{N}{N'}\right)\\
					  & =  \ln N-\ln N'\\	
					  & <  I.
	\end{split}	
\end{equation}

	We can verify that this expression correctly reduces to the one previously discussed when the  \aposteriori answer is unique. It also shows that the information needed to partially solve a problem is less than the information needed to fully solve it.

\subsection{Information and Probabilities}
	Let us now consider the case where the possible \apriori answers are no longer equally probable. Each answer $R_{i}$ is associated with a probability - a likelihood - $p_{i}$. Of course we do,
\begin{equation}
	\sum_{i}p_{i}=1.
\end{equation}

	Let us return to the simple problem. The set of \apriori answers contains only two elements,
\begin{equation}
	\mathcal{R}=\left\{ R_{1},R_{2}\right\}.
\end{equation}

	We know that the information needed to solve this problem is about 0.693 nep if the two \apriori answers are equally probable. Let us express the probabilities as,
\begin{equation}
	p_{1}=\frac{N_{1}}{N_{1}+N_{2}},\;\; p_{2}=\frac{N_{2}}{N_{1}+N_{2}},
\end{equation}

	where $N_{1}$ and $N_{2}$ are positive integers. The complexity $N$ of the problem, whose \apriori answers have probabilities $p_{i}$, is equal to the number - divided by $N_{1}+N_{2}$ - of ways in which a sequence of $N_{1}+N_{2}$ symbols $R_{i}$ can be formed, knowing that there are $N_{1}$ equal to $R_{1}$ and, obviously, $N_{2}$ equal to $R_{2}$. A simple counting calculation shows that the complexity,
\begin{equation}
	\begin{split}
		N & =  \frac{\left(N_{1}+N_{2}\right)\left(N_{1}+N_{2}-1\right)\left(N_{1}+N_{2}-2\right)\times\cdots\times\left(N_{2}+1\right)}{2\times3\times\cdots\times N_{1}}\\
 		  & =  \frac{\left(N_{1}+N_{2}\right)!}{N_{1}!\times N_{2}!}
	\end{split}
\end{equation}

	where the division by $N_{1}!$ is due to the fact that the $N_{1}$ symbols $R_{1}$ are interchangeable. If we compute the information from $N$, we get,
\begin{equation}
	\begin{split}
	I & =  \ln N\\
	  & =  \left[\ln\left(N_{1}+N_{2}\right)!-\ln N_{1}!-\ln N_{2}!\right].
	\end{split}
\end{equation}

	If $N_{1}$ and $N_{2}$ are chosen large enough - that is, $>100$ - we can use \couleur{Stirling}'s formula,
\begin{equation}
	\ln Q!\simeq Q\left(\ln Q-1\right),
\end{equation}

to find,
\begin{equation}
	\begin{split}
		I & \simeq \left(N_{1}+N_{2}\right)\ln\left(N_{1}+N_{2}\right)-N_{1}\ln N_{1}-N_{2}\ln N_{2}\\
		  & = -\left(N_{1}+N_{2}\right)\left[\frac{N_{1}}{N_{1}+N_{2}}\ln\frac{N_{1}}{N_{1}+N_{2}}+\frac{N_{2}}{N_{1}+N_{2}}\ln\frac{N_{2}}{N_{1}+N_{2}}\right]\\
		  & = -\left(N_{1}+N_{2}\right)\left[p_{1}\ln p_{1}+p_{2}\ln p_{2}\right].
	\end{split}
\end{equation}

	The last expression still depends on $N_{1}$ and $N_{2}$, which is problematic because these numbers are not uniquely determined. For example, the probabilities $1/3$ and $2/3$ can be represented either by $N_{1}=1$ and $N_{2}=2$ or by $N_{1}=2000$ and $N_{2}=4000$. However, the information should not depend on any particular choice. Therefore the complexity $N$ has to be normalised to get an acceptable expression. To do this, it is sufficient to divide the above information by the number of realisations, so that,
 \begin{equation}
	I=\frac{1}{N_{1}+N_{2}}\left[\ln\left(N_{1}+N_{2}\right)!-\ln N_{1}!-\ln N_{2}!\right],
\end{equation}

	which results in a measure of information that depends only on the laws of probability,
\begin{equation}
	I=-\left[p_{1}\ln p_{1}+p_{2}\ln p_{2}\right].
\end{equation}

	At the level of complexity, this renormalisation amounts to a choice,
\begin{equation}
	N=\left[\frac{\left(N_{1}+N_{2}\right)!}{N_{1}!\times N_{2}!}\right]^{1/\left(N_{1}+N_{2}\right)}.
\end{equation}

	Although obtained using \couleur{Stirling}'s approximation, the expression for the information can be considered exact in the sense that $N_{1}$ and $N_{2}$ can always be chosen to be as large as desired. If you consider one of the answers to be certain,
\begin{equation}
p_{1}=1\; et\; p_{2}=0\Rightarrow I=0\; nep\Rightarrow N=1,
\end{equation}

	we find that the information needed to solve the problem is zero, which is obvious since the answer to the problem is known \apriori. The complexity is then $1$. Other examples,
\begin{eqnarray}
	p_{1} & = & \frac{4}{10}\; et\; p_{2}=\frac{6}{10}\Rightarrow I\simeq0.673\; nep\Rightarrow N=1.96,\\
	p_{1} & = & \frac{2}{10}\; et\; p_{2}=\frac{8}{10}\Rightarrow I\simeq0.500\; nep\Rightarrow N=1.65,\\
	p_{1} & = & \frac{1}{10}\; et\; p_{2}=\frac{9}{10}\Rightarrow I\simeq0.325\; nep\Rightarrow N=1.38,
\end{eqnarray}

	show that the information required decreases as the probability of one of the answers decreases. The associated complexity then varies from $2$ to $1$.

	The calculations just performed can be generalised to any number of \apriori\ answers $R_{i}$ associated with probabilities $p_{i}$. We then obtain the definition originally proposed by \couleur{Claude Shannon},
\begin{equation}
	I=-\sum_{i}p_{i}\ln p_{i}
	\label{AAA}
\end{equation}

\subsection{Equally likely answers = maximum information}
	We will now show that the information required to answer a question is maximized when the $N$ \apriori answers $R_{i}$ are equally probable. Referring to equation (\ref{AAA}), we seek the probabilities $p_{i}$ such that,
\begin{equation}
	\dfrac{\partial I}{\partial p_{i}}=0,\;\;\;\left(i=1,\cdots,N\right),
\label{hrg}
\end{equation}

	which returns to the definition of information,
\begin{equation}
	\dfrac{\partial}{\partial p_{i}}\sum_{k}p_{k}\ln p_{k}=0,\;\;\;\left(i=1,\cdots,N\right),
\label{equi2}
\end{equation}

	which, when expanded, yields,
\begin{equation}
	\ln p_{i}+1+\sum_{k\neq i}\dfrac{\partial p_{k}}{\partial p_{i}}\ln p_{k}+\sum_{k\neq i}\dfrac{\partial p_{k}}{\partial p_{i}}=0,\;\;\;\left(i=1,\cdots,N\right).
\label{equi3}
\end{equation}

	Note that the probabilities are normalised, \ie
\begin{equation}
	p_{k}=1-\sum_{i\neq k}p_{i},
\end{equation}

	which implies,
\begin{equation}
	\dfrac{\partial p_{k\neq i}}{\partial p_{i}}=-1
\end{equation}

and consequently the condition (\ref{equi3}) becomes,
\begin{equation}
	\ln p_{i}-\sum_{k\neq i}\ln p_{k}-N+2=0,\;\;\;\left(i=1,\cdots,N\right).
	\label{equi4}
\end{equation}

	By evaluating the equation (\ref{equi4}) for two different indices $i$ and $j$ and performing the subtraction, we obtain \[ \ln p_{i}-\sum_{k\neq i}\ln p_{k}-\ln p_{j}+\sum_{k\neq j}\ln p_{k}=0,\;\;\;\left(i=1,\cdots,N\right).\] Finally, by cancelling the opposing terms, we find \[2\ln p_{i}-2\ln p_{j}=0\] which is satisfied when the two probabilities are equal,
\begin{equation}
	p_{i}=p_{j}
\end{equation}

	Strictly speaking, the proof should be completed with an analysis of the signs of the second partial derivatives to establish that the \textit{extremum} identified is indeed a \textit{maximum}.

\subsection{About the Tunnel}
	Let's illustrate equation (\ref{AAA}) using our tunnel-finding problem. Suppose the only unknown parameter is the depth $z_{t}$ of the centre of the tunnel. One way of framing the problem is to say that we need to find $z_{t}$ within a set of possible values,
\begin{equation}
	\left\{ R_{i}\right\} =\left\{ 5,6,7,8,9,10,11,12,13,14,15,16,17,18,19,20\right\} .
\end{equation}

	If all depths are \apriori equally likely, the information needed to find the correct depth is given by,
\begin{equation}
	I=2.773\; nep.
\end{equation}

	Geological information may lead us to believe that depths below 10 metres or above 18 metres are unlikely. We can express this using the following probability table,
\begin{equation}
	\left\{ \dfrac{1}{34},\dfrac{1}{34},\dfrac{1}{34},\dfrac{1}{34},\dfrac{1}{34},\dfrac{3}{34},\dfrac{3}{34},\dfrac{3}{34},\dfrac{3}{34},\dfrac{3}{34},\dfrac{3}{34},\dfrac{3}{34},\dfrac{3}{34},\dfrac{3}{34},\dfrac{1}{34},\dfrac{1}{34}\right\} .
\end{equation}

	The information required is now given by,
\begin{equation}
	I=2.654\; nep
\end{equation}

	and we can therefore say that the geological information provided is equivalent to
\begin{equation}
	I_{g\acute{e}ol}=2.773-2.654=0.119\; nep.
\end{equation}

\section{Mutual Information}
\subsection{Coupling Information}
	Suppose the problem to be solved involves finding two answers from two \apriori sets of answers, $A_{i}$ and $B_{j}$. Let $p\left(A_{i}, B_{j}\right)$ denote the probabilities of all possible \apriori pairs $\left(A_{i}, B_{j}\right)$. We have,
\begin{equation}
	\sum_{i}\sum_{j}p\left(A_{i},B_{j}\right)=1
\end{equation}

	and we have the marginal probabilities,
\begin{equation}
	\begin{split}
		p\left(A_{i}\right) & =  \sum_{j}p\left(A_{i},B_{j}\right),\\
		p\left(B_{j}\right) & =  \sum_{i}p\left(A_{i},B_{j}\right).
	\end{split}
\end{equation}

	It can be shown in a direct way that,
\begin{equation}
	\sum_{i}p\left(A_{i}\right)=\sum_{j}p\left(B_{j}\right)=\sum_{i}\sum_{j}p\left(A_{i}\right)\times p\left(B_{j}\right)=1.
\end{equation}

The coupling information is given by,
\begin{equation}
	I\left(A,B\right)=-\sum_{i}\sum_{j}p\left(A_{i},B_{j}\right)\ln p\left(A_{i},B_{j}\right),
\end{equation}

	and the marginal information,
\begin{equation}
	\begin{split}
	I\left(A\right) & = -\sum_{i}p\left(A_{i}\right)\ln p\left(A_{i}\right)\\
			   \,\, & = -\sum_{i}\sum_{j}p\left(A_{i},B_{j}\right)\ln p\left(A_{i}\right),
	\end{split}
\end{equation}

\begin{equation}
	\begin{split}
		I\left(B\right) & = -\sum_{j}p\left(A_{j}\right)\ln p\left(B_{j}\right)\\
				   \,\, & = -\sum_{j}\sum_{i}p\left(A_{i},B_{j}\right)\ln p\left(B_{j}\right).
	\end{split}
\end{equation}

	Note that, 
\begin{equation}
	\begin{split}
		I\left(A\right)+I\left(B\right) & = -\sum_{i}\sum_{j}p\left(A_{i},B_{j}\right)\ln\left[p\left(A_{i}\right)\times p\left(B_{j}\right)\right]\\
			   \,\, & = -\sum_{i}\sum_{j}p\left(A_{i},B_{j}\right)\ln\left[p\left(A_{i},B_{j}\right)+q\left(A_{i},B_{j}\right)\right]\\
\,\, & = -\sum_{i}\sum_{j}p\left(A_{i},B_{j}\right)\ln\left[p\left(A_{i},B_{j}\right)\left(1+\frac{q\left(A_{i},B_{j}\right)}{p\left(A_{i},B_{j}\right)}\right)\right]\\
\,\, & = I\left(A,B\right)-\sum_{i}\sum_{j}p\left(A_{i},B_{j}\right)\ln\left[1+\frac{q\left(A_{i},B_{j}\right)}{p\left(A_{i},B_{j}\right)}\right].
	\end{split}
\end{equation}

	To proceed further, it is necessary to prove a useful result. Let us consider the function,
\begin{equation}
	f\left(x\right)=x-\ln\left(1+x\right)
\end{equation}

	defined in the interval $\left]-1,+\infty\right[$. It is easily verified that $f\left(0\right)=0$, $f^{\prime}\left(0\right)=0$ et $f^{\prime}\left(x\right)=x/\left(1+x\right)$, which gives,
\begin{equation}
	\begin{split}
		f^{\prime}\left(x>0\right) & >  0\Longrightarrow f\left(x>0\right)>f\left(0\right)\\
		f^{\prime}\left(-1<x<0\right) & <  0\Longrightarrow f\left(-1<x<0\right)>f\left(0\right).
	\end{split}
\end{equation}

	The function $f$ is therefore minimal at $x=0$ and strictly positive everywhere else. Hence we have,
\begin{equation}
	x\geq\ln\left(1+x\right)
\end{equation}

	in the whole domain of $f$. Now set
\begin{equation}
	x\equiv\frac{q\left(A_{i},B_{j}\right)}{p\left(A_{i},B_{j}\right)},
\end{equation}

	we have,
\begin{equation}
	-q\left(A_{i},B_{j}\right)\leq-p\left(A_{i},B_{j}\right)\ln\left[1+\frac{q\left(A_{i},B_{j}\right)}{p\left(A_{i},B_{j}\right)}\right],\end{equation}
	
	and it follows,
\begin{equation}
	I\left(A,B\right)-\sum_{i}\sum_{j}q\left(A_{i},B_{j}\right)\leq I\left(A,B\right)-\sum_{i}\sum_{j}p\left(A_{i},B_{j}\right)\ln\left[1+\frac{q\left(A_{i},B_{j}\right)}{p\left(A_{i},B_{j}\right)}\right].\end{equation}

	Furthermore, the very definition of $q\left(A_{i},B_{j}\right)$ implies that,
\begin{equation}
	\sum_{i}\sum_{j}q\left(A_{i},B_{j}\right)=0.
\end{equation}

	Combining the various results obtained, we then find that,
\begin{equation}
	I\left(A,B\right)\leq I\left(A\right)+I\left(B\right),
\end{equation}

	\ie the coupling information is less than or equal to the sum of the marginal information. If the two answers to be found are independent, the probability is
\begin{equation}
	p\left(A_{i},B_{j}\right)=p\left(A_{i}\right)\times p\left(B_{j}\right)
\end{equation}

	and equality holds,
\begin{equation}
	I\left(A,B\right)=I\left(A\right)+I\left(B\right).
\end{equation}

\subsection{Conditional information}
	The probability of the pairs can be expressed in the form,
\begin{equation}
	\begin{split}
		p\left(A_{i},B_{j}\right) & =  p\left(A_{i}\right)\times p\left(B_{j}|A_{i}\right)\\ 
								  & =  p\left(B_{j}\right)\times p\left(A_{i}|B_{j}\right)
		\label{condi2}
	\end{split}
\end{equation}

	where $p\left(B_{j}|A_{i}\right)$ is the conditional probability of $B_{j}$ given that $A_{i}$ has occurred. The coupling information then takes the form,
\begin{equation}
	\begin{split}
		I\left(A,B\right) & =  -\sum_{i}\sum_{j}p\left(A_{i},B_{j}\right)\ln p\left(A_{i},B_{j}\right) \\ 
						  & =  -\sum_{i}\sum_{j}p\left(A_{i},B_{j}\right)\ln\left[p\left(A_{i}\right)\times p\left(B_{j}|A_{i}\right)\right]\\ 
						  & =  -\sum_{i}\sum_{j}p\left(A_{i},B_{j}\right)\ln p\left(A_{i}\right)-\sum_{i}\sum_{j}p\left(A_{i},B_{j}\right)\ln p\left(B_{j}|A_{i}\right)\\ 
						  & =  -\sum_{i}p\left(A_{i}\right)\ln p\left(A_{i}\right)-\sum_{i}\sum_{j}p\left(A_{i},B_{j}\right)\ln p\left(B_{j}|A_{i}\right)\\ 
						  & =  I\left(A\right)+I\left(B|A\right),
	\label{condi7}
	\end{split}
\end{equation}

	where the conditional information has been introduced
\begin{equation}
	\begin{split}
		I\left(B|A\right) & =  -\sum_{i}\sum_{j}p\left(A_{i},B_{j}\right)\ln p\left(B_{j}|A_{i}\right)\\ 
						  & =  -\sum_{i}\sum_{j}p\left(A_{i},B_{j}\right)\ln\frac{p\left(A_{i},B_{j}\right)}{p\left(A_{i}\right)}.
	\label{condi9}
 	\end{split}
\end{equation}

	This information is furthermore such that,
\begin{equation}
	I\left(A\right)+I\left(B\right)\geq I\left(A,B\right)=I\left(A\right)+I\left(B|A\right),\label{condi10}
\end{equation}

	which implies that,
\begin{equation}
	I\left(B|A\right)\leq I\left(B\right).\label{condi11}
\end{equation}

	This expression means that the information about the responses $B_{j}$ given the response $A$ is reduced compared to the information about the responses $B_{j}$ alone. If the responses $A_{i}$ and $B_{j}$ are independent --- that is, if knowing $A$ does not provide any additional information about $B$ --- the conditional information is equal to the marginal information.

\subsection{About the Tunnel}
Let us illustrate equation (\ref{condi11}) using our tunnel search problem, assuming that we want to determine the depth $z_{t}$ of the centre of the tunnel at two locations separated by several tens of metres. So the problem is to find the two depths $z_{t}$,
\begin{equation}
	\begin{split}
		\left\{ A_{i}\right\}  & =  \left\{ 5,6,7,8,9,10,11,12,13,14,15,16,17,18,19,20\right\} \\
		\left\{ B_{j}\right\}  & =  \left\{ 5,6,7,8,9,10,11,12,13,14,15,16,17,18,19,20\right\} .
	\end{split}
\end{equation}

	As we have shown, if all depths are equally likely, the information is \[I\left(A\right)=I\left(B\right)=2.773\; nep.\]  If the depth estimates are independent, then \[ I\left(A,B\right)=I\left(A\right)+I\left(B\right)=5.546\; nep.\]  On the other hand, if technical information indicates that the difference between the two depths should not exceed 3 metres, because it is known that the tunnel does not have a slope greater than a certain value, then it becomes clear that knowing the first depth provides information about the second, still unknown, depth. This information limits the number of a priori possible depths, reducing the number of \apriori answers from $N=16^{2}$ to only $N=100$. This reduces the coupling information \[I\left(A,B\right)=\ln100=4.605\; nep,\] and the conditional information in this case is \[I\left(B|A\right)=I\left(A,B\right)-I\left(A\right)\,\,=4.605-2.773\,\,=1.832\; nep.\] The information related to the tunnel slope constraint is equal to the difference between the coupling information calculated with and without the constraint, \[I_{pente}=5.546-4.605=0.941\; nep.\]

\section{Case of continuous distributions}
\subsection{There is a problem !}
	The definition of information given by \couleur{Shannon} and that we have seen so far can be generalised to the case of probability densities $\rho\left(x\right)$ where $x$ can vary continuously. This is, for example, the case of the tunnel depth, which we initially assumed to take discrete values, whereas in reality it can take any value within a given interval \apriori. The \couleur{Shannon} information for a probability density $\rho$ is given by
\begin{equation}
	I_{Shannon}\equiv-\int\rho\left(x\right)\ln\rho\left(x\right)dx.
	\label{conti1}
\end{equation}

	Note by the way that the density $\rho$ is such that,
\begin{equation}
	\int\rho\left(x\right)dx=1,
	\label{condi2}
\end{equation}
	and that we can have $\rho\left(x\right)>1$ for certain values of $x$.

Now consider the case of determining the tunnel depth when the depth is \apriori contained within the interval $S=\left[z_{inf},z_{sup}\right]$. If we assume that the depths are equally likely, then,
\begin{equation}
	\rho\left(z_{t}\in S\right)=\dfrac{1}{z_{sup}-z_{inf}},\;\;\rho\left(z_{t}\notin S\right)=0.
	\label{conti3}
\end{equation}

	The information associated with this probability density is given by,
\begin{equation}
	I_{Shannon}=-\ln\frac{1}{z_{sup}-z_{inf}}.
	\label{conti4}
\end{equation}

	If $z_{sup}-z_{inf}=1$ metres, we find that $I_{Shannon}=0$, which, according to what we have seen so far, implies that we have the answer to the question of determining the depth of the tunnel. However, this is not the case, since the depth is contained within an interval of one metre in width. Worse still, if we now express the distances in centimetres, we find that the associated information is $I_{Shannon}=\ln 100$! This means that the quantification of information depends on the choice of units, which means that information loses the absolute character we had previously ascribed to it.

	In his 1948 paper, \shortciteN{shannon1948mathematical} notes this problem and points out that it is not serious, since what really matters is the variation of information for a fixed choice of units.

\subsection{A new definition of Information}
	In their paper, \couleur{Albert Tarantola} and \couleur{Bernard Valette} \cite{tarantola1982generalized} propose a definition of information that is invariant under changes in coordinate systems or units. They propose,
\begin{equation}
	I_{TarVal}\equiv\int\rho\left(x\right)\ln\frac{\rho\left(x\right)}{\mu\left(x\right)}dx,
	\label{conti5}
\end{equation}

	where the probability density $\mu$ represents the \textit{maximum} state of ignorance about the variable $x$. Note that with this definition, the information obtained no longer represents the information needed to answer the question posed, but rather the information available to answer the question. It is, in a sense, the complementary information to that considered previously.

	How should the \textit{maximum} state of ignorance be chosen? The idea is that this state should be the one that provides the least information about the answer to the question posed. A natural choice is to use a uniform distribution over the \textit{a priori} interval, since we have seen that the case of equally probable outcomes corresponds to the state that requires the most information to answer the question. However, this choice is not always appropriate. Consider, for example, the problem of locating an earthquake on the Earth's surface. If we work in Cartesian coordinates $\left(x,y\right)$, the natural choice is,
\begin{equation}
	\mu\left(x,y\right)=constant.
	\label{conti6}
\end{equation}

	However, if we work in spherical coordinates $\left(\theta,\phi\right)$, where the surface element is $ds=R\sin\theta,d\theta,d\phi$, the probability density corresponding to an equally probable distribution with respect to the surface is given by,
\begin{equation}
	\mu\left(\theta,\phi\right)=constante.R.\sin\theta.
	\label{conti7}
\end{equation}

\subsection{Information conjunction}
	The introduction of the \textit{maximum} state of ignorance requires an adaptation of the formula for the conjunction of information seen in the case of discrete events. Let $\sigma$ be the probability density associated with the information $I_{\sigma}$ corresponding to the conjunction of two pieces of information $I_{1}$ and $I_{2}$, whose respective probability densities are $\rho_{1}$ and $\rho_{2}$. We then have,
\begin{equation}
	\begin{split}
		I_{\sigma} & = \int\sigma\left(x\right)\ln\frac{\sigma\left(x\right)}{\mu\left(x\right)}dx \\ 
		I_{1} & =  \int\rho_{1}\left(x\right)\ln\frac{\rho_{1}\left(x\right)}{\mu\left(x\right)}dx\\ 
		I_{2} & =  \int\rho_{2}\left(x\right)\ln\frac{\rho_{2}\left(x\right)}{\mu\left(x\right)}dx.\label{conti10}
	\end{split}
\end{equation}

	These different pieces of information must be combined according to logical rules that take into account the existence of the \textit{maximum} state of ignorance. These rules are,
\begin{equation}
	I_{\sigma} =\left(I_{1}\& I_{2}\right)=\left(I_{2}\& I_{1}\right)\;\;\; commutation
	\label{regle1}
\end{equation}
\begin{equation}
	\rho_{1}\left(x\right) =0  \Rightarrow  \sigma\left(x\right)=0\;\;\; absorption
	\label{regle2}
\end{equation}
\begin{equation}
	\rho_{1}\left(x\right) =\mu\left(x\right)   \Rightarrow  \sigma\left(x\right)=\rho_{2}\left(x\right)\;\;\; non-information
	\label{regle3}
\end{equation}

	The condition \ref{regle1} simply states that the conjunction of information must be commutative. This requires a symmetric form of $\sigma$ with respect to the densities $\rho_{1}$ and $\rho_{2}$. The condition \ref{regle2} corresponds to the fact that if one of the probability densities is zero for certain values of $x$, then the density $\sigma$ must also be zero for those values. This absorption property is analogous to multiplication, which implies that $\sigma$ must be a function of the product $\rho_{1}\times\rho_{2}$, which automatically satisfies the commutativity imposed by the first condition. The third condition \ref{regle3} takes into account the maximum ignorance $\mu$. Finally, considering the form of $\sigma$ dictated by the first two conditions, we find that,
\begin{equation}
	\sigma\left(x\right)=\dfrac{\rho_{1}\left(x\right)\rho_{2}\left(x\right)}{\mu\left(x\right)}.
	\label{conjonc1}
\end{equation}

\section{Direct problem $=$ information}
\subsection{Still in the tunnel}
	We will start with our favourite example to illustrate and intuitively grasp the developments that will follow. To do this, we will rephrase it slightly to introduce the concept of a direct problem. We have seen that knowing one depth can provide information to determine a second depth. That is, providing a piece of data - the first depth - can improve the information we have about an unknown - the second depth. Building on this observation, it is easy to modify the formulation of the problem slightly and assume that the data is no longer the first depth, but a measurement of the gravitational field. Similarly, the constraint on the slope of the tunnel - which allowed us to 'connect' the two depths - can be replaced by \couleur{Newton}'s law, which relates the tunnel depth to the gravitational anomaly. You might think that \couleur{Newton}'s law is perfectly known and, unlike the slope constraint, leaves no room for tolerance. This is incorrect; there are many reasons why \couleur{Newton}'s law is 'fuzzy' when applied to our tunnel! For example, we do not know the exact density of the surrounding rock, we are not sure if the tunnel is perfectly cylindrical, etc. In short, the direct problem of calculating the gravitational anomaly as a function of depth is an imprecise law that can be described by a distribution of conditional probabilities, which we will denote by,
\begin{equation}
	p\left(g|z_{t}\right).
\end{equation}

\subsection{Direct problem $=$ conditional probability}
	The perspective we have just illustrated with the tunnel example is extremely powerful because it allows not only to relax the rigidity of the mathematical relations describing the direct problem, but also to take measurement uncertainties into account. This is certainly what makes the information-theoretic approach to inverse problems so attractive. From the most general point of view, the direct problem, which relates the data to the parameters that are the unknowns of the inverse problem, is thus expressed in terms of a conditional probability density,
\begin{equation}
	\textrm{DIRECT PROBLEM}=\textrm{PROBABILITY}\left(\textrm{DATA}|\textrm{PARAMETERS}\right).
	\label{direct1}
\end{equation}

\section{Inverse problem = information transfer}
\subsection{\aposteriori conditional information}
	We have seen that the coupling information is given by,
\begin{equation}
	I\left(A,B\right)=I\left(A\right)+I\left(B|A\right)=I\left(B\right)+I\left(A|B\right),
	\label{conjonc2}
\end{equation}

	which implies,
\begin{equation}
	I\left(A|B\right)=I\left(A\right)-I\left(B\right)+I\left(B|A\right).
	\label{conjonc3}
\end{equation}

	This equation provides the solution to an inverse information transfer problem: the \aposteriori conditional information we can obtain about the answer $A$ is equal to the \aposteriori information about $A$ minus the \aposteriori information about $B$ and plus the conditional information about $B$ given $A$. As we have already shown,
\begin{equation}
	I\left(A|B\right)\leq I\left(A\right),
	\label{conjonc4}
\end{equation}

	\ie the \aposteriori information needed to know $A$ is less than the \apriori information we had. In other words, the \aposteriori knowledge we have is greater than the \apriori knowledge. Seen in this way, solving the inverse problem involves increasing our knowledge about the answer $A$.

\subsection{The \couleur{Bayes} formula (discrete events)}
	We will now bridge to the next chapter concerning probabilities. Let's begin with the case of discrete events by explaining solution \ref{conjonc3},
\begin{equation}
	-\sum_{i}\sum_{j}p\left(A_{i},B_{j}\right)\ln p\left(A_{i}|B_{j}\right)=-\sum_{i}\sum_{j}p\left(A_{i},B_{j}\right)\ln\frac{p\left(A_{i}\right)p\left(B_{j}|A_{i}\right)}{p\left(B_{j}\right)},\label{conjonc5}
\end{equation}

	or, in equivalent terms,
\begin{equation}
	\sum_{i}\sum_{j}p\left(A_{i},B_{j}\right)\ln\left[\frac{p\left(A_{i}|B_{j}\right)p\left(B_{j}\right)}{p\left(A_{i}\right)p\left(B_{j}|A_{i}\right)}\right]=0.
\label{conjonc6}
\end{equation}

If we want this relationship to hold in general, the logarithmic term must be identically zero, so we have,
\begin{equation}
	p\left(A_{i}|B_{j}\right)=\frac{p\left(A_{i}\right)p\left(B_{j}|A_{i}\right)}{p\left(B_{j}\right)}.
	\label{bayes1}
\end{equation}

	This relation is known as the \couleur{Bayes} formula. It plays a very important role in probability theory. In the following chapters we will see how this formula can be used to solve inverse problems. 

	By recalling this, a slightly different form of the \couleur{Bayes} formula can be obtained,
\begin{equation}
	p\left(B_{j}\right)=\sum_{k}p\left(A_{k},B_{j}\right)=\sum_{k}p\left(A_{k}\right)p\left(B_{j}|A_{k}\right).
\end{equation}

	We then find that,
\begin{equation}
	p\left(A_{i}|B_{j}\right)=\frac{p\left(A_{i}\right)p\left(B_{j}|A_{i}\right)}{\sum_{k}p\left(A_{k}\right)p\left(B_{j}|A_{k}\right)}.
\label{bayes2}
\end{equation}

We have just established that the manipulation of information can be reduced to the manipulation of probability laws.

\subsection{The generalised \couleur{Bayes} formula (continuous case)}
	Although the \couleur{Bayes} formula (\ref{bayes2}) is indeed used to solve many inverse problems, it is important to remember that its derivation is within the framework of discrete event theory. In the continuous case, one must use the information conjugate seen earlier, for which the resulting probability density is given by the equation \ref{conjonc1}, repeated here,
\begin{equation}
	\sigma\left(x\right)=\frac{\rho_{1}\left(x\right)\rho_{2}\left(x\right)}{\mu\left(x\right)},
	\label{conjonc1b}
\end{equation}

	which is the continuous case equivalent of the \couleur{Bayes} formula.

\chapter{\titrechap{Bayesian inversion}}
\minitoc
\section{Probabilities \& Inverse Problems}
\subsection{Probabilities, Frequencies, and Information}
	It is useful to begin with some thoughts on the concept of probability. In its purest sense, the concept of probability is associated with the idea of repeating an experiment in which the outcome is not identical but, on the contrary, varies from trial to trial. The most common example is throwing a dice. The result of a single throw is an integer between 1 and 6. The number obtained from one roll to the next is not necessarily the same. In signal theory, a stochastic process refers to the system under consideration in the experiments. In our example, the stochastic process is the system consisting of the dice, the receiving surface and the thrower. Each throw is a realisation of the stochastic process. The characterisation of a process is done in terms of statistics and in particular probabilities. In the case of dice, we calculate the frequency of occurrence of each possible number. If this frequency of occurrence is calculated from a very large number of throws, to the point where the number can be considered infinite, the frequency of occurrence is called a probability. In this case, the notion of probability is clearly defined and is based on counting within a set of realisations of a stochastic process with a finite number of possible outcomes.

	Inverse problem theory uses a notion of probability that is sometimes different from what we have just discussed. Here, probabilities are used to quantify the likelihood of an event. For example, a certain possible depth of the tunnel might be considered unlikely if engineers or geologists consider it unlikely. This is rarely a probability calculated in the same way as a die, i.e. by running statistics on a large number of tunnel depths. Probability is a more ambiguous concept that can, of course, include objective statistical data, but also subjective and more difficult to define information. In fact, many of the probabilities dealt with in inverse problem theory are actually likelihoods. This creates a gap in the theory because the foundations on which our rigorous calculations are based can be questioned. For example, one could move away from traditional probabilities in favour of fuzzy logic, which combines information differently.

\subsection{Probability Densities}
	Until now, we have only discussed discrete probabilities calculated for a finite number of possible outcomes. For example, in the case of dice, where the number of outcomes is limited to 6.  In this context, probability is a measure that involves counting the elements of the sets under consideration. A measure must satisfy the following basic properties,
\begin{itemize}
	\item the measure is always positive,, $M\left(\cdot\right)\geq0$;
	\item the measure of the empty set is zero, $M\left(\emptyset\right)=0$;
	\item the measure of the entire space is 1, $M\left(\Omega\right)=1$;
	\item the measure satisfies the additivity property for a collection of disjoint sets, $M\left(\Omega_{1}\cup\Omega_{2}\right)=M\left(\Omega_{1}\right)+M\left(\Omega_{2}\right)$ if $\Omega_{1}\cap\Omega_{2}=\emptyset$. 
\end{itemize}

	When working with continuous random variables, a different measure must be adopted, which we will define over an interval $\mathcal{I}$, so that for any interval $\mathcal{A} \subset \mathcal{I}$ we have,
\begin{equation}
	M\left(\mathcal{A}\right)\equiv\int_{\mathcal{A}}\rho\left(x\right)\, dx
\end{equation}

	We have a valid measure if $\rho\left(x\right) \geq 0$. We will say that $\rho\left(x\right)$ is a probability density function if,
\begin{equation}
	\int_{\mathcal{I}}\rho\left(x\right)\, dx=1
\end{equation}

	The probability density function allows you to calculate the probability that a realisation $x^{\prime}$ of the random variable $x$ lies within a given interval $\mathcal{A}\subset\mathcal{I}$,
\begin{equation}
	P\left(x^{\prime}\in\mathcal{A}\right)=\int_{\mathcal{A}}\rho\left(x\right)\, dx
\end{equation}

	We can calculate the mathematical expectation value -- that is, the mean -- of the random variable $x$,
\begin{equation}
	\overline{x}\equiv E\left[x\right]=\int_{\mathcal{I}}x\rho\left(x\right)\, dx
\end{equation}

	and the variance,
\begin{equation}
	\sigma_{x}^{2}\equiv E\left[\left(x-\overline{x}\right)^{2}\right]=\int_{\mathcal{I}}\left(x-\overline{x}\right)^{2}\rho\left(x\right)\, dx
\end{equation}

	By generalisation we will define the $n$-th central moment as,
\begin{equation}
E\left[\left(x-\overline{x}\right)^{n}\right]=\int_{\mathcal{I}}\left(x-\overline{x}\right)^{n}\rho\left(x\right)\, dx
\end{equation}

	An example of a probability density is given by $\rho\left(x\right)=\pi^{-1/2}\exp\left(-x^{2}\right)$ where the interval $\mathcal{I}=\Bbb{R}$. 
\newline	

	Indeed, one can verify that: \newline

	\dag \ the normalisation condition is satisfied,
\begin{equation}
M\left(\mathcal{I}\right)=\frac{1}{\sqrt{\pi}}\int_{\Bbb{R}}\exp\left(-x^{2}\right)\, dx=1
\end{equation}
\newline

	\dag \ the positivity of the measure,
\begin{equation}
M\left(\mathcal{A}\subset\mathcal{I}\right)=\frac{1}{\sqrt{\pi}}\int_{\mathcal{A}}\exp\left(-x^{2}\right)\, dx\geq0,
\end{equation}
\newline

	\dag \ the additivity of the measure,
\begin{equation}
M\left(\mathcal{A}\cup\mathcal{A}^{\prime}\right)=M\left(\mathcal{A}\right)+M\left(\mathcal{A}^{\prime}\right)
\end{equation}
\newline

	\dag \ when $\mathcal{A}\cap\mathcal{A}^{\prime}=\emptyset$ we finally have,
\begin{equation}
	M\left(\emptyset\right)=0.
\end{equation}

	When probability densities are used, \couleur{Bayes}' formula (\ref{bayes2}) takes the form,
\begin{equation}
	\rho\left(y|x\right)=\frac{\rho\left(y\right)p\left(x|y\right)}{\int_{\mathcal{I}}\rho\left(y\right)p\left(x|y\right)\, dy}
	\label{EEE}
\end{equation}

	The function $\rho\left(y\right)$ is the \apriori probability density, and $\rho\left(y|x\right)$ is the \aposteriori probability density.

\subsection{Mathematical expectation value of a function}
	\couleur{Bayes}' formula for probability densities involves the integral,
\begin{equation}
	\int_{\mathcal{I}}\rho\left(y\right)p\left(x|y\right)\, dy
\end{equation}

	More generally, you will often encounter integrals of the form,
\begin{equation}
	E\left[f\left(x\right)\right]\equiv\int_{\mathcal{I}}f\left(x\right)\rho\left(x\right)\, dx
\end{equation}

	which, as an extension of what we saw in the previous section, we will define as the mathematical expectation value of the function $f\left(x\right)$ with respect to the probability density $\rho\left(x\right)$.

\subsection{Multivariate probabilities}
	The generalisation to the case of multivariate probability densities is immediate by introducing the vector random variable $\mathbf{x}$ and the function $\rho\left(\mathbf{x}\right)$. As before, we define the mean by,
\begin{equation}
	\overline{\mathbf{x}}\equiv\int_{\mathcal{I}}\mathbf{x}\rho\left(\mathbf{x}\right)\, d\mathbf{x}
\end{equation}

	and the covariance matrix by,
\begin{equation}
C_{ij}\left(\overline{\mathbf{x}}\right)\equiv\int_{\mathcal{I}}\left(x_{i}-\overline{x}_{i}\right)\left(x_{j}-\overline{x}_{j}\right)\rho\left(\mathbf{x}\right)\, d\mathbf{x}
\end{equation}

	The marginal probability allows us to determine the probability of finding a realisation of a component $x_{i}$ of the random variable within an interval $\mathcal{A}_{i}$,
\begin{equation}
	P\left(x_{i}\in\mathcal{A}_{i}\right)=\iint_{\mathcal{A}_{i}}\rho\left(\mathbf{x}\right)\, d\mathbf{x}
\end{equation}

	where inner integration is performed over the complete intervals corresponding to the components of $\mathbf{x}$ except $x_{i}$.

	An example of a bivariate probability density defined on $\mathcal{I}=\Bbb{R}\times\Bbb{R}$ is,
\begin{equation}
	\rho\left(x,y\right)=\frac{1}{\pi}\exp\left[-\left(x^{2}+y^{2}\right)\right]
\end{equation}

	One verifies that the marginal probability density for $x$ actually gives a univariate probability density,
\begin{equation}
	\begin{split}
	\rho\left(x\right) & =  \frac{1}{\pi}\int_{\Bbb{R}}\exp\left[-\left(x^{2}+y^{2}\right)\right]\, dy\\
 & =  \frac{1}{\sqrt{\pi}}\exp\left(-x^{2}\right)
	\end{split}
\end{equation}

\section{A few common probability distributions}
\subsection{The normal distribution (\couleur{Gauss})}
	The normal distribution is given by,
\begin{equation}
\rho\left(\mathbf{x}\right)=\frac{1}{\left(2\pi\right)^{N/2}\sqrt{\det\mathbf{C}}}\exp\left[-\frac{1}{2}\left(\mathbf{x}-\overline{\mathbf{x}}\right)^{t}\mathbf{C}^{-1}\left(\mathbf{x}-\overline{\mathbf{x}}\right)\right]
\end{equation}

	where $N$ is the dimension of the vector $\mathbf{x}$ and $\mathbf{C}$ is the covariance matrix, which is symmetric and positive definite. If the components of $\mathbf{x}$ are independent variables, this matrix is diagonal, and its elements are the variances associated with each component of $\mathbf{x}$.

\subsection{Generalised Gaussian distributions}
	Generalised Gaussian distributions are defined by the family,
\begin{equation}
\rho_{p}\left(x\right)\equiv\frac{p^{1-1/p}}{2\sigma_{p}\Gamma\left(1/p\right)}\exp\left(-\frac{\left|x-\overline{x}\right|^{p}}{p\left(\sigma_{p}\right)^{p}}\right)\;\;\; p\geq1,
\end{equation}

where,
\begin{equation}
\sigma_{p}\equiv\left(\int_{\mathcal{I}}\left|x-\overline{x}\right|^{p}\,\rho\left(x\right)dx\right)^{1/p},
\end{equation}

	is a generalised measure of the dispersion of a probability density $\rho\left(x\right)$.

\subsection{The \textit{log-normal} distribution}
	The log-normal distribution is defined for $x \geq 0$,
\begin{equation}
	\frac{1}{\sqrt{2\pi}x \sigma}\exp\left[\frac{-\ln\left(x/m\right)^{2}}{2\sigma^{2}}\right]
\label{def-log-normale}
\end{equation}

	and has a mean of $m\exp\left(\sigma^{2}/2\right)$, a median of $m$, and a variance of $m^{2}\exp\left(\sigma^{2}\right)\exp\left(\sigma^{2}-1\right)$. Each log-normally distributed variable $x$ is associated with a variable $\ln \left(x\right)$ that follows a normal distribution. Similar to how the normal distribution is often obtained by adding random variables, the log-normal distribution is often obtained by multiplying random variables. As a result, the log-normal distribution is often useful for representing fluctuations due to multiplicative effects. More formally, the log-normal distribution is used to represent variables that are subject to proportional changes, where the resulting value is obtained by applying a random factor to the previous value.

\subsection{The \couleur{Poisson} disribution}
	The \couleur{Poisson} distribution is defined for positive integer variables $x$,
\begin{equation}
\frac{\lambda^{x}}{x!}\exp\left(-\lambda\right).
\label{def-poisson}
\end{equation}

	The mean and the variance are both equal to $\lambda$, which must be positive. The \couleur{Poisson} distribution is often used to represent rare random events, such as earthquakes in intra-plate zones. Surprisingly, the \couleur{Poisson} distribution also accurately represents the sequence of fatal accidents caused by horse kicks in the Prussian army in the 19th century!

\subsection{The \textit{gamma} ($\Gamma$-) distribution}
	The $\Gamma$-distribution is defined for $x \geq 0$ and is given by
\begin{equation}
	\frac{\lambda^{k}}{\Gamma\left(k\right)}x^{k-1}\exp\left(-\lambda x\right)
\label{def-gamma}
\end{equation}

	where $\lambda$ and $k$ are two positive parameters representing the scale and shape of the distribution, respectively. The name of the distribution comes from its denominator $\Gamma\left(k\right)$, which ensures the normalization of the distribution. The mean is $k/\lambda$ and the variance is $k/\lambda^{2}$. When $k = 1$, the distribution simplifies to the exponential distribution. The Gamma distribution is similar to the \couleur{Poisson} distribution but has a lighter tail, resulting in lower probabilities for extreme values.

\subsection{The beta ($\beta$-) distribution}
	The $\beta$-distribution is defined for $0 \leq x \leq 1$,
\begin{equation}
	\frac{\Gamma\left(a+b\right)}{\Gamma\left(a\right)\Gamma\left(b\right)}x^{a-1}\left(1-x\right)^{b-1}
\label{def-beta}
\end{equation}

	The mean is $a/(a+b)$, and the variance is $ab/\left[(a+b)^2(a+b+1)\right]$. The two shape parameters, $a$ and $b$, must be positive.

\subsection{The \couleur{Pareto} distribution}
	This distribution, also known as the hyperbolic or power law, is named after the Italian economist \couleur{Vilfredo Pareto}, who used it in the late 19th century to describe personal wealth in certain societies. It is defined for $x \geq a$ with positive shape parameters $a$ and $b$,
\begin{equation}
	\frac{ba^b}{x^{1+b}}.
\label{def-pareto}
\end{equation}

	The mean is given by $ab/(b-1)$ for $b > 1$. For $b > 2$, the variance is,
\begin{equation}
	\frac{a^2b}{\left[(b-1)^2(b-2)\right]}
\end{equation}

	and is infinite for $b \leq 2$. The \couleur{Pareto} distribution is often used to represent scale laws found in nature. In this distribution, the probability that the variable $x > u > a$ is given by $(a/u)^b$. A particular application of the \couleur{Pareto} distribution is in modelling flood peaks.

\subsection{The \textit{binomial} distribution}
	The binomial distribution is defined for positive integer values of $x$,
\begin{equation}
	\binom{n}{x}p^{x}(1-p)^{n-x}
\label{def-binomiale}
\end{equation}

	The mean of this distribution is $np$ and the variance is given by $np(1-p)$. This distribution gives the probability of $x$ events occurring in a series of length $n$, given that the probability of an event occurring is $p$. The binomial distribution can be used to calculate the probabilities of events occurring that do not respond systematically to a given cause. For example, what is the probability of a seismological station being struck by lightning in a year with 45 thunderstorms?

\subsection{The \couleur{Cauchy} distribution}
	The \couleur{Cauchy} distribution is defined by,
\begin{equation}
	\frac{1}{\pi b\left[1+\left(\frac{x-a}{b}\right)^2\right]}
\label{def-cauchy}
\end{equation}

	where the parameter $b > 0$. The \couleur{Cauchy} distribution has a slow-decaying tail, which assigns a relatively high probability to extreme values. As a result, the mean and variance are not defined. However, the median is equal to $a$. The \couleur{Cauchy} distribution is a \couleur{Lévy}-stable distribution, meaning that the sum of variables drawn from a \couleur{Cauchy} distribution will also follow a \couleur{Cauchy} distribution.

\subsection{The \couleur{Weibull} distribution} 
	This distribution is defined for positive integers of $x$,
\begin{equation}
	(\dfrac{a}{b^a})x^{a-1}\exp\left[-\left(\dfrac{x}{b}\right)^a\right],
\label{def-weibull}
\end{equation}

	where $a$ is the shape parameter of the distribution and $b$ is the scale parameter. The mean is given by $b\Gamma(1+1/a)$ and the variance by $b^2\left[\Gamma(1+2/a) - \Gamma^2(1+1/a)\right]$. When $a = 1$, the distribution reduces to the exponential distribution, and to the \couleur{Rayleigh} distribution when $a = 2$.
	
\section{\couleur{Bayes}' formula and inversion"}
\subsection{General solution}
	Let us revisit the information conjunction formula \ref{conjonc1} established for continuous variables, which is equivalent to \couleur{Bayes}' formula,
\begin{equation}
	\sigma\left(z\right)=\frac{\rho\left(z\right)\theta\left(z\right)}{\mu\left(z\right)}
	\label{conjonc1c}
\end{equation}

	and examine the meaning of the different probability densities that make it up. First, it is important to note that, in an inverse problem, we traditionally have data $x$ and parameters $y$ that form the random variable $z$ in the formula above, and so the formula can be rewritten in a more explicit form as follows,
\begin{equation}
	\sigma\left(x,y\right)=\frac{\rho\left(x,y\right)\theta\left(x,y\right)}{\mu\left(x,y\right)}.
	\label{inverse1}
\end{equation}

	The probability density $\rho$ can be considered as the \apriori probability on the parameters and data, while $\theta$ represents the probabilistic version of the forward problem, \ie the probability density relating the data to the parameters within the framework of a physical law or, in the absence of a law, \via statistical relationships. By integrating \ref{inverse1} with respect to $x$, one obtains the marginal probability density for $y$,
\begin{equation}
	\sigma\left(y\right)=\int\frac{\rho\left(x,y\right)\theta\left(x,y\right)}{\mu\left(x,y\right)}dx
	\label{inverse2}
\end{equation}

	which is the most general solution to an inverse problem \cite{TarVal82}. Note that the marginal probability over the data $x$ can also be evaluated to obtain the \aposteriori probability over the measured values of the data,
\begin{equation}
	\sigma\left(x\right)=\int\frac{\rho\left(x,y\right)\theta\left(x,y\right)}{\mu\left(x,y\right)}dy
	\label{inverse2b}
\end{equation}

\subsection{Solution for \apriori independent data and parameters}
	If the data and parameters are \apriori independent, the probability densities take the form,
\begin{eqnarray}
	\rho\left(x,y\right) & = & \rho_{x}\left(x\right).\rho_{y}\left(y\right)\label{inverse3}\\
	\theta\left(x,y\right) & = & \theta\left(x|y\right).\mu_{y}\left(y\right).\label{inverse4}
\end{eqnarray}

The probability $\theta\left(x,y\right)$ given by equation \ref{inverse4} is a conditional probability that contains no information about the parameters since their marginal probability represents the maximum ignorance $\mu_{y}$. Substituting these expressions into the general solution \ref{inverse2}, and assuming that $\mu\left(x,y\right)=\mu_{x}\left(x\right)\cdot\mu_{y}\left(y\right)$, gives the solution to the inverse problem when the data and parameters are \apriori independent,
\begin{equation}
	\sigma\left(y\right)=\rho_{y}\left(y\right)\int\frac{\rho_{x}\left(x\right)\theta\left(x|y\right)}{\mu_{x}\left(x\right)}dx
	\label{inverse5}
\end{equation}

	The very existence of the marginal probability $\sigma\left(y\right)$ depends on whether the various probabilities that make up the equation \ref{inverse5} are consistent with each other. If the \apriori probabilities are inconsistent with the measured data and the forward problem, then it is possible that the marginal probability at $y$ is zero everywhere. The marginal probability over the data $x$ can be obtained in a similar way by substituting \ref{inverse3} and \ref{inverse4} in \ref{inverse2b},
\begin{equation}
	\sigma\left(x\right)=\frac{\rho_{x}\left(x\right)}{\mu_{x}\left(x\right)}\int\rho_{y}\left(y\right)\theta\left(x|y\right)dy
	\label{inverse5b}
\end{equation}

\subsection{Solution for an exact physical law}
	If there is an exact physical law for predicting the data from the parameters, then there exists a function $g$ such that,
\begin{equation}
	x=g\left(y\right),
	\label{direct2}
\end{equation}

and the conditional probability associated with the forward problem can be written as,
\begin{equation}
	\theta\left(x|y\right)=\delta\left(x-g\left(y\right)\right)
	\label{direct3}
\end{equation}

	where $\delta$ is the \couleur{Dirac} distribution. In this case, equation \ref{inverse5} becomes,
\begin{eqnarray}
	\sigma\left(y\right) & = & \rho_{y}\left(y\right)\int\frac{\rho_{x}\left(x\right)\delta\left(x-g\left(y\right)\right)}{\mu_{x}\left(x\right)}dx
	\label{inverse6}\\
 & = & \rho_{y}\left(y\right)\frac{\rho_{x}\left(g\left(y\right)\right)}{\mu_{x}\left(g\left(y\right)\right)}.\label{inverse6a}
\end{eqnarray}

\subsection{Solution using \couleur{Bayes}' formula}
	Let us recall Bayes' formula, from which we will derive the developments that follow,
\begin{equation}
	\rho\left(y|x\right)=\frac{\rho\left(y\right)p\left(x|y\right)}{\int_{\mathcal{I}}\rho\left(y\right)p\left(x|y\right)\, dy}.
	\label{GGG}
\end{equation}

	It is important to clearly define the role of each term in this equation. The probability density $\rho\left(y|x\right)$ is what we are looking for - it is the \aposteriori conditional probability of having $y$ given that $x$ has occurred, and it is the most general answer to an inverse problem. The probability density $\rho\left(y\right)$ is called the \apriori because it is assumed to contain all the information available about $y$ before the experiment was performed, \ie before $x$ was known. Finally, the conditional probability $p\left(x|y\right)$ - also called the likelihood - takes into account the fact that the data $x$ is uncertain and that models $y$ that do not perfectly reproduce the data (i.e. the particular realisation of the random variable $x$) are acceptable within limits defined by $p\left(x|y\right)$. Bayesian inversion depends critically on how the likelihood $p\left(x|y\right)$ is defined, and a significant part of the expertise in physics lies in determining the likelihood accurately. This requires calibration of the method used, as well as the most credible simulations where the true response $y$ is known, and so on. The establishment of a scientific fact can be considered achieved when the relevant community is convinced. In scientific debates, new or surprising results are typically challenged for their reliability, which in the language of \couleur{Bayesian} inversion amounts to debating the choice of probability $p\left(x|y\right)$.

\section{The tunnel again}
\subsection{Example 1: one data and one parameter}
	Let us return to our favourite example and illustrate the use of Bayes' formula (\ref{GGG}) to estimate the depth $z_{t}$ from a measurement $g_{1} = -62;\mu Gal$ of the gravitational field taken at $x_{1}=x_{t}$. In this particular case, we assume that all other parameters in equation (\ref{FFF}) are sufficiently well known and do not need to be determined in the inverse problem. So, let us assume that $x_{t}$ is known, and that $r_{t} = 3; m$ and $\rho = 2700; kg/m^{3}$. The forward problem is then reduced to,
\begin{equation}
	g_{1}^{\prime}\left(z_{t}\right)=\frac{\alpha}{z_{t}}\;\;\left(\mu Gal\right)
	\label{LDH}
\end{equation}

	where $\alpha \simeq -1018.84$. Suppose the measurement is accompanied by a Gaussian uncertainty with a standard deviation of $\sigma_{g} = 5;\mu Gal$. The probability of the measurement with respect to the true value -- which is unknown to us, but in this example where we have taken $z_{t} = 16; m$ is $58.96;\mu Gal$ -- of the gravity $g_{1}^{\prime}$ is given by,
\begin{equation}
	p\left(g_{1}|g_{1}^{\prime}\right)=\frac{1}{\sqrt{2\pi}\sigma_{g}}\exp\left[-\frac{\left(g_{1}-g_{1}^{\prime}\left(z_{t}\right)\right)^{2}}{2\sigma_{g}^{2}}\right]
\end{equation}

La dépendance de $g_{1}^{\prime}$ par rapport à $z_{t}$ permet d'obtenir la vraisemblance nécessaire pour la formule de \couleur{Bayes},
\begin{equation}
	p\left(g_{1}|z_{t}\right)=p\left[g_{1}|g_{1}^{\prime}\left(z_{t}\right)\right]
\end{equation}

	Implementing Bayes' formula requires defining the \apriori probability density on $z_{t}$. If we assume that all depths between $z_{\min} = 5; m$ and $z_{\max} = 25; m$ are equally probable, then we have,
\begin{equation}
	\rho\left(z_{t}\right)=\frac{1}{\Delta z}\Pi\left(\frac{z_{t}-z_{moy}}{\Delta z}\right)
	\label{DKB}
\end{equation}

where $\Delta z = z_{\max} - z_{\min}$ and $z_{moy} = \left(z_{\max} + z_{\min}\right)/2$. The \aposteriori probability density then becomes,
\begin{equation}
	\begin{split}
\rho\left(z_{t}|g_{1}\right) & = \frac{\rho\left(z_{t}\right)p\left(g_{1}|z_{t}\right)}{\int_{z_{\min}}^{z_{\max}}\rho\left(z_{t}\right)p\left(g_{1}|z_{t}\right)\, dz_{t}}\\
\,\, & = \frac{\Pi\left(\frac{z_{t}-z_{moy}}{\Delta z}\right)\exp\left[-\left(g_{1}-\alpha/z_{t}\right)^{2}/\left(2\sigma_{g}^{2}\right)\right]}{\int_{z_{\min}}^{z_{\max}}\exp\left[-\left(g_{1}-\alpha/z_{t}\right)^{2}/\left(2\sigma_{g}^{2}\right)\right]\, dz_{t}}
	\end{split}
\end{equation}

\begin{figure}[!]
	\begin{center}
		\tcbox[colback=white]{\includegraphics[width=16cm]{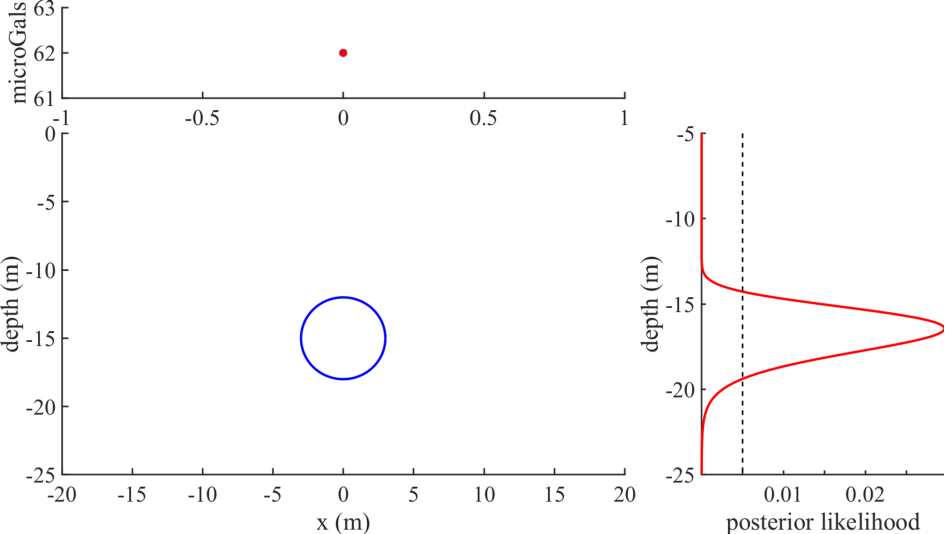}}
	\end{center}
	\caption{\apriori probabilities (dashed line, uniform distribution) and \aposteriori probabilities (solid curve) of the tunnel depth when a single gravimetric measurement is taken directly above the tunnel. \label{tunnel1}}
\end{figure}

	It can be observed that the \aposteriori probability density is no longer a uniform distribution and has a maximum relatively localised within the interval $\left[z_{\min},z_{\max}\right]$ (Figure \ref{tunnel1}, generated using the script \couleur{ex\_tunnel\_01.m}). We say that the parameter $z_{t}$ is resolved, which means that the information provided by the data is useful in determining the unknown parameter. It is possible to compute the \apriori and \aposteriori information on $z_{t}$ to see the effect of the data $g_{1}$. The information needed to determine the depth before using the gravimetric measurement is given by	
	
\begin{equation}
	\begin{split}
	I_{priori}\left(z_{t}\right) & = -\int_{z_{\min}}^{z_{\max}}\rho\left(z_{t}\right)\ln\rho\left(z_{t}\right)\, dz_{t}\\
\,\, & = \frac{1}{\Delta z}\int_{z_{\min}}^{z_{\max}}\ln\Delta z\, dz_{t}\\
\,\, & = \ln\Delta z\\
\,\, & = 3.00\; nep
	\label{LZB}
	\end{split}
\end{equation}

The \aposteriori information is given by,
\begin{equation}
	\begin{split}
	I_{posteriori}\left(z_{t}|g_{1}\right) & = -\int_{z_{\min}}^{z_{\max}}\rho\left(z_{t}|g_{1}\right)\ln\rho\left(z_{t}|g_{1}\right)\, dz_{t}\\
\,\, & \simeq 1.74\; nep
	\end{split}
\end{equation}

	Thus, one can calculate the information provided by the gravimetric measurement,
\begin{equation}
	I_{gravi}\left(g_{1}\right)=I_{priori}-I_{posteriori}\simeq1.26\; nep.
\label{MBR}
\end{equation}

	This information is not zero, which means that our knowledge of the depth $z_{t}$ has increased. We say that the parameter is resolved. Consistent with intuition, the previous expressions show that the \aposteriori information continues to decrease as the standard deviation of the measurement uncertainty decreases. There is a threshold beyond which the \aposteriori information is almost equal to the \apriori information, at which point the gravimetric measurement becomes essentially useless. 
 
\begin{figure}[!]
	\begin{center}
		\tcbox[colback=white]{\includegraphics[width=16cm]{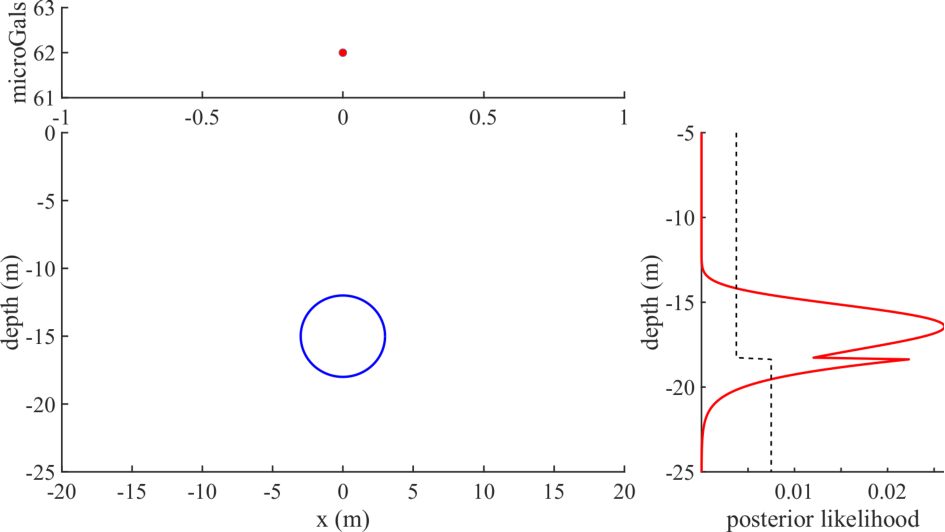}}
	\end{center}
	\caption{\apriori (dashed line) and \aposteriori (solid line) probabilities of the tunnel depth when a single gravimetric measurement is taken directly above the tunnel. In this example, the \apriori probability is not uniform, resulting in a significant change in the \aposteriori probability (see figure \ref{tunnel1}).\label{tunnel4}}
\end{figure}

	It is of course possible to start with a non-uniform prior probability, as in the example shown in figure \ref{tunnel4}. In this case, the \aposteriori probability changes significantly, highlighting the importance of \apriori information in solving inverse problems,
	
\subsection{Example 2: two data and one parameter}
	Let us revisit Example 1 by adding a second gravimetric measurement and see what this means for our knowledge of the tunnel depth. Suppose the data are,
\begin{center}
\begin{tabular}{|c|c|c|}
\hline \hline $x_{i}-x_{t}\ (m)$&  $g_{i}\ (\mu Gal)$&  $\sigma_{g}\ (\mu Gal)$ \\
\hline \hline
\ 	&   &  \    \\
0   &  -62. &  5. \\
\ 	&   &  \    \\
10. & -44  & 5. \\
\ 	&   &  \    \\
\hline \hline
\end{tabular}
\end{center}

	Using the vector notation, $\mathbf{g}$, to represent the data, the probability is then given by
\begin{equation}
	p\left(\mathbf{g}|z_{t}\right)=\frac{1}{\left(\sqrt{2\pi}\sigma_{g}\right)^{2}}\exp\left[-\frac{\| \mathbf{g}-\mathbf{g}^{\prime}\left(z_{t}\right)\|^{2}}{2\sigma_{g}^{2}}\right],
\end{equation}

	where $\mathbf{g}^{\prime}\left(z_{t}\right)$ represents the forward problem, that is, the calculation of the theoretical gravity as a function of the depth $z_{t}$ that we wish to test. We have,
\begin{equation}
	g_{i}^{\prime}\left(z_{t}\right)=\frac{\alpha z_{t}}{\left(x_{i}-x_{t}\right)^{2}+z_{t}^{2}}.
\end{equation}

\begin{figure}[!]
	\begin{center}
		\tcbox[colback=white]{\includegraphics[width=16cm]{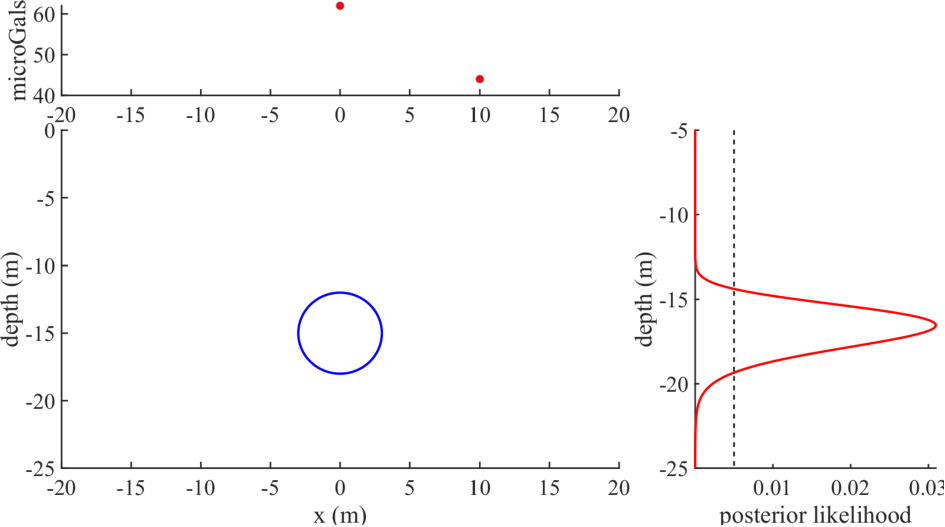}}
	\end{center}
	\caption{\apriori probabilities (dashed line, smooth curve) and \aposteriori probabilities (solid line) from two gravimetric measurements, one directly above the tunnel and the other 10 metres above.\label{tunnel2}}
\end{figure}

	Using the same \apriori probability density (\ref{DKB}) as in Example 1, we find that (Figure \ref{tunnel2})
\begin{equation}
	\begin{split}
\rho\left(z_{t}|\mathbf{g}\right) & =  \frac{\rho\left(z_{t}\right)p\left(\mathbf{g}|z_{t}\right)}{\int_{z_{\min}}^{z_{\max}}\rho\left(z_{t}\right)p\left(\mathbf{g}|z_{t}\right)\, dz_{t}}\\
\,\, & =  \frac{\Pi\left(\frac{z_{t}-z_{moy}}{\Delta z}\right)\exp\left[-\left[\mathbf{g}-\mathbf{g}^{\prime}\left(z_{t}\right)\right]^{2}/\left(2\sigma_{g}^{2}\right)\right]}{\int_{z_{\min}}^{z_{\max}}\exp\left[-\left[\mathbf{g}-\mathbf{g}^{\prime}\left(z_{t}\right)\right]^{2}/\left(2\sigma_{g}^{2}\right)\right]\, 
dz_{t}}
	\end{split}
\end{equation}

\newpage

	Of course, the prior information remains unchanged compared to Example 1 and is given by equation (\ref{LZB}). However, the posterior information is given by,
\begin{equation}
	\begin{split}
	I_{posteriori}\left(z_{t}|\mathbf{g}\right) & = & -\int_{z_{\min}}^{z_{\max}}\rho\left(z_{t}|\mathbf{g}\right)\ln\rho\left(z_{t}|\mathbf{g}\right)\, dz_{t}\\
\,\, & \simeq & 1.69\; nep
	\end{split}
\end{equation}

	This allows us to calculate the information provided by the gravimetric measurements,
\begin{equation}
	I_{gravi}\left(\mathbf{g}\right)=I_{priori}-I_{posteriori}\simeq1.31\; nep
\end{equation}

\begin{figure}[!]
	\begin{center}
		\tcbox[colback=white]{\includegraphics[width=16cm]{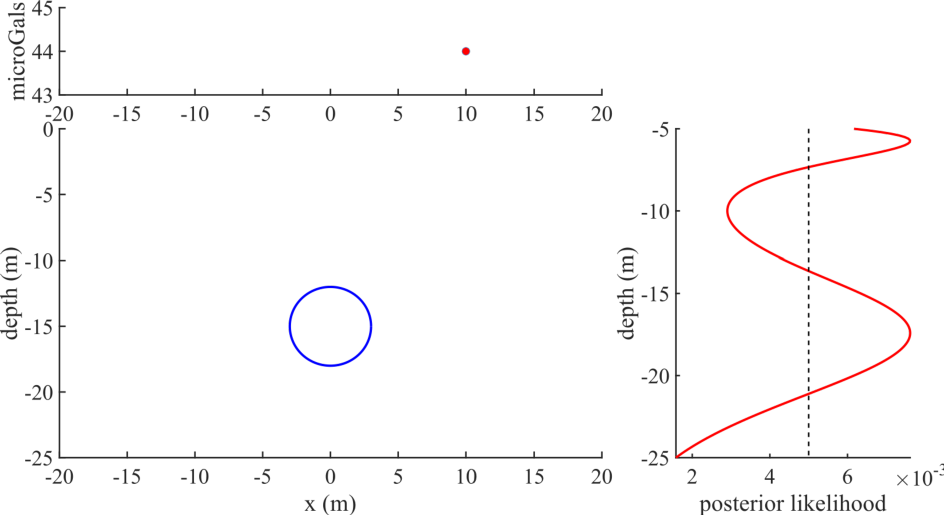}}
	\end{center}
	\caption{\apriori probabilities (uniform distribution, dashed line) and \aposteriori probabilities (solid curve) corresponding to the inverse problem solved with only the gravimetric data located 10 metres from the tunnel axis. It can be seen that the posterior probability is not well localised, indicating that the data do not effectively resolve the tunnel depth (\cf \couleur{ex\_tunnel\_02.m}).\label{tunnel3}}
\end{figure}

This information is only slightly less than that obtained in the previous example, indicating that the parameter $z_{t}$ is not better resolved and that the gravimetric data $g_{2}$ has contributed negligible additional information.Let's examine this situation more closely by calculating the solution to the inverse problem using only the measurement $g_{2}$ taken at $x_{2} - x_{t} = 10; m$. The \aposteriori probability density (Figure \ref{tunnel3}) is, in contrast to the previous case, poorly localised and has two \textit{maxima}. The posterior information associated with this probability density is given by,
\begin{equation}
	I_{posteriori}\left(z_{t}|g_{2}\right)=2.93\; nep
\end{equation}

	and so we have,
\begin{equation}
	I_{gravi}\left(g_{2}\right)=0.07\; nep
\end{equation}

	which confirms that the data provide little additional information. The depth parameter is poorly resolved in this case. This can be understood by noting that the function $g_{2}^{\prime}\left(z_{t}\right)$ is equal to $44;\mu Gal$ at two relatively different depths. This explains the presence of two maxima in the \aposteriori probability density.

\subsection{Example 3: One data and two parameters}
	We can complicate the inverse problem by assuming that the horizontal position $x_{t}$ of the tunnel is poorly determined and is included as one of the parameters. In this example we will only use the gravimetric data $g_{1}$ from example 1. Under these conditions,
\begin{equation}
	p\left(g_{1}|x_{t},z_{t}\right)=\frac{1}{\sqrt{2\pi}\sigma_{g}}\exp\left[-\frac{1}{2\sigma_{g}^{2}}\left(g_{1}-g_{1}^{\prime}\left(x_{t},z_{t}\right)\right)^{2}\right]
\end{equation}

	where the forward problem is given by,
\begin{equation}
	g_{i}^{\prime}\left(x_{t},z_{t}\right)=\frac{\alpha z_{t}}{\left(x_{i}-x_{t}\right)^{2}+z_{t}^{2}}.
\end{equation}

For example, we can set the \apriori probability density as,
\begin{equation}
	\rho\left(x_{t},z_{t}\right)=\frac{1}{\Delta x\Delta z}\Pi\left(\frac{x_{t}}{\Delta x}\right)\Pi\left(\frac{z_{t}-z_{moy}}{\Delta z}\right)
	\label{WHZ}
\end{equation}

	which indicates that the horizontal position is \apriori within an interval of length $\Delta x=50; m$ centered on the measurement location. Thus, we have,
\begin{equation}
	\rho\left(x_{t},z_{t}|g_{1}\right)=\frac{\Pi\left(\frac{x_{t}}{\Delta x}\right)\Pi\left(\frac{z_{t}-z_{moy}}{\Delta z}\right)\exp\left[-\frac{1}{2\sigma_{g}^{2}}\left(g_{1}-g_{1}^{\prime}\left(x_{t},z_{t}\right)\right)^{2}\right]}{\int_{x_{\min}}^{x_{\max}}\int_{z_{\min}}^{z_{\max}}\exp\left[-\frac{1}{2\sigma_{g}^{2}}\left(g_{1}-g_{1}^{\prime}\left(x_{t},z_{t}\right)\right)^{2}\right]\, dx_{t}dz_{t}}
\end{equation}

\begin{figure}[!]
	\begin{center}
		\tcbox[colback=white]{\includegraphics[width=16cm]{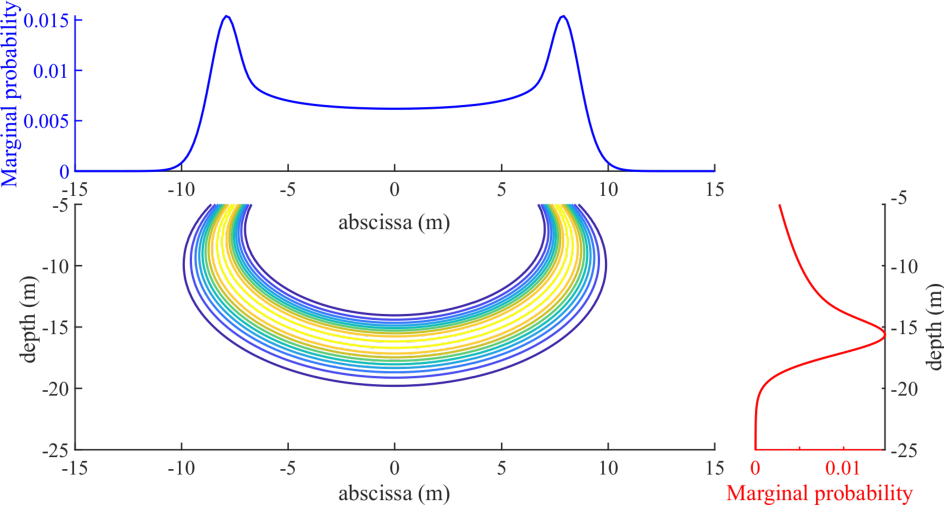}}
	\end{center}
	\caption{\aposteriori probability (contour plots, bottom left) of the horizontal position and depth of the tunnel obtained from a single gravimetric measurement (taken directly above the tunnel, although this is not known). The marginal probability for the depth is significantly less localised than when the data were used to determine the depth alone (see figure \ref{tunnel1}), indicating that the addition of parameters in an inverse problem affects the determination of the \textbf{OTHER} parameters (\cf \couleur{ex\_tunnel\_03.m}).\label{tunnel5}}
\end{figure}

	This \aposteriori probability density is relatively complex (Figure \ref{tunnel5}) and has a horseshoe shape, indicating the correlation between the two parameters $x_{t}$ and $z_{t}$. The \apriori information is given by,
\begin{equation}
	\begin{split}
	I_{priori}\left(x_{t},z_{t}\right) & = -\int_{x_{\min}}^{x_{\max}}\int_{z_{\min}}^{z_{\max}}\rho\left(x_{t},z_{t}\right)\ln\rho\left(x_{t},z_{t}\right)\, dx_{t}dz_{t}\\
\,\, & = \ln\left(\Delta x\Delta z\right)\simeq6.91\; nep
	\label{KCA}
	\end{split}
\end{equation}

	and the \aposteriori information is given by,
\begin{equation}
	\begin{split}
I_{posteriori}\left(x_{t},z_{t}|g_{1}\right) & =  -\int_{x_{\min}}^{x_{\max}}\int_{z_{\min}}^{z_{\max}}\rho\left(x_{t},z_{t}|g_{1}\right)\ln\rho\left(x_{t},z_{t}|g_{1}\right)\, dx_{t}dz_{t}\\
\,\, & \simeq  1.89\; nep	
	\end{split}
\end{equation}

	The information provided by the data $g_{1}$ is therefore $I_{gravi}\left(g_{1}\right)\simeq 5.02\ nep$. The marginal probability densities are respectively,
\begin{equation}
	\rho\left(x_{t}|g_{1}\right)=\frac{\Pi\left(\frac{x_{t}}{\Delta x}\right)\int_{z_{\min}}^{z_{\max}}\exp\left[-\frac{1}{2\sigma_{g}^{2}}\left(g_{1}-g_{1}^{\prime}\left(x_{t},z_{t}\right)\right)^{2}\right]\, dz_{t}}{\int_{x_{\min}}^{x_{\max}}\int_{z_{\min}}^{z_{\max}}\exp\left[-\frac{1}{2\sigma_{g}^{2}}\left(g_{1}-g_{1}^{\prime}\left(x_{t},z_{t}\right)\right)^{2}\right]\, dx_{t}dz_{t}}
\end{equation}

	and,
\begin{equation}
	\rho\left(z_{t}|g_{1}\right)=\frac{\Pi\left(\frac{z_{t}-z_{moy}}{\Delta z}\right)\int_{x_{\min}}^{x_{\max}}\exp\left[-\frac{1}{2\sigma_{g}^{2}}\left(g_{1}-g_{1}^{\prime}\left(x_{t},z_{t}\right)\right)^{2}\right]\, dx_{t}}{\int_{x_{\min}}^{x_{\max}}\int_{z_{\min}}^{z_{\max}}\exp\left[-\frac{1}{2\sigma_{g}^{2}}\left(g_{1}-g_{1}^{\prime}\left(x_{t},z_{t}\right)\right)^{2}\right]\, dx_{t}dz_{t}}
\end{equation}

	The marginal informations are respectively,
\begin{equation}
	\begin{split}
	I_{priori}\left(x_{t}\right) & = \ln\Delta x\simeq3.91\; nep\\
	I_{priori}\left(z_{t}\right) & =  \ln\Delta z\simeq3.00\; nep
	\end{split}
\end{equation}
\begin{equation}	
	\begin{split}
	I_{posteriori}\left(x_{t}|g_{1}\right) & =  -\int_{x_{\min}}^{x_{\max}}\rho\left(x_{t}|g_{1}\right)\ln\rho\left(x_{t}|g_{1}\right)\, dx_{t}\simeq2.93\; nep\\
	I_{posteriori}\left(z_{t}|g_{1}\right) & = -\int_{z_{\min}}^{z_{\max}}\rho\left(z_{t}|g_{1}\right)\ln\rho\left(z_{t}|g_{1}\right)\, dz_{t}\simeq2.60\; nep
	\end{split}
\end{equation}

	It can be seen that the $g_{1}$ data did not provide the same amount of information about the two parameters,
\begin{equation}
	\begin{split}
	I_{gravi}\left(g_{1}\rightsquigarrow x_{t}\right) & \equiv  I_{priori}\left(x_{t}\right)-I_{posteriori}\left(x_{t}|g_{1}\right)\simeq0.98\; nep\\
	I_{gravi}\left(g_{1}\rightsquigarrow z_{t}\right) & \equiv  I_{priori}\left(z_{t}\right)-I_{posteriori}\left(z_{t}|g_{1}\right)\simeq 0.40\; nep
	\end{split}
\end{equation}

An important observation can already be made by comparing these results with those from the first example. It can be seen that in example 1 the data $g_{1}$ contributed an information value of $1.26; nep$ to our knowledge of the depth $z_{t}$. In contrast, in example 3, the same data contributes only $0.40; nep$, which is three times less. This illustrates a universal principle in inverse problem theory, which contrasts with the popular notion that data, often referred to as 'information', provides unchanging knowledge about a parameter.

\subsection{Example 4: Two data and two parameters}
	Let us add the second gravimetric measurement $g_{2}$ to the inverse problem introduced in Example 3. The likelihood is given by the formula,
\begin{equation}
	p\left(\mathbf{g}|x_{t},z_{t}\right)=\frac{1}{\left(\sqrt{2\pi}\sigma_{g}\right)^{2}}\exp\left[-\frac{\| \mathbf{g}-\mathbf{g}^{\prime}\left(x_{t},z_{t}\right)\|^{2}}{2\sigma_{g}^{2}}\right]
\end{equation}

	where $\mathbf{g}^{\prime}\left(x_{t},z_{t}\right)$ represents the direct problem,
\begin{equation}
	g_{i}^{\prime}\left(x_{t},z_{t}\right)=\frac{\alpha z_{t}}{\left(x_{i}-x_{t}\right)^{2}+z_{t}^{2}}.
\end{equation}

\begin{figure}[!]
	\begin{center}
		\tcbox[colback=white]{\includegraphics[width=16cm]{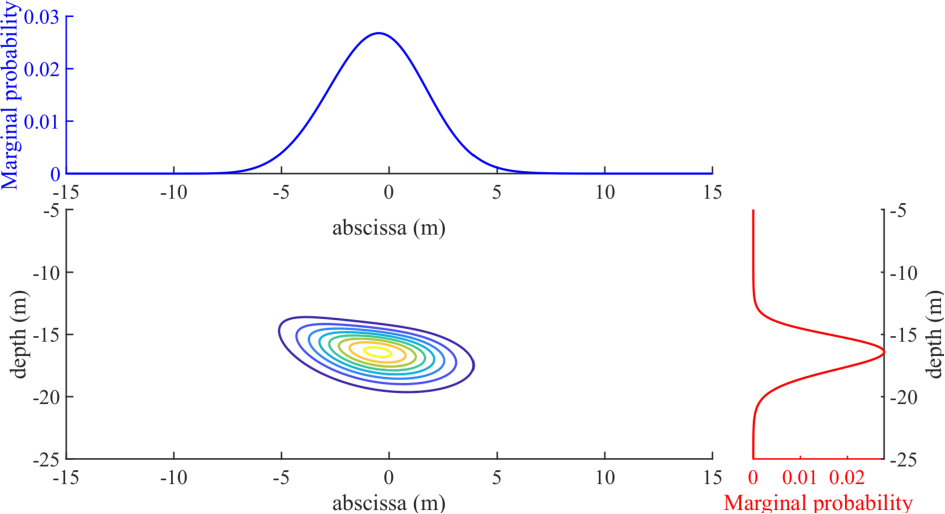}}
	\end{center}
	\caption{\aposteriori probability (contour plots, bottom left) for the horizontal position and depth of the tunnel using two gravimetric measurements. The marginal probabilities indicate that the two parameters are fairly well resolved (\cf \couleur{ex\_tunnel\_04.m}).).\label{tunnel6}}
\end{figure}

	Using the same \apriori probability density (\ref{WHZ}) as in example 3, we find (Figure \ref{tunnel6}),
\begin{equation}
	\rho\left(x_{t},z_{t}|\mathbf{g}\right)=\frac{\rho\left(x_{t},z_{t}\right)p\left(\mathbf{g}|x_{t},z_{t}\right)}{\int_{x_{\min}}^{x_{\max}}\int_{z_{\min}}^{z_{\max}}\rho\left(x_{t},z_{t}\right)p\left(\mathbf{g}|x_{t},z_{t}\right)\, dx_{t}dz_{t}}
\end{equation}

	The prior information remains unchanged from Example 3 and is given by equation (\ref{KCA}), while the \aposteriori information is,
\begin{equation}
	\begin{split}
		I_{posteriori}\left(x_{t},z_{t}|\mathbf{g}\right) & =  	-\int_{x_{\min}}^{x_{\max}}\int_{z_{\min}}^{z_{\max}}\rho\left(x_{t},z_{t}|\mathbf{g}\right)\ln\rho\left(x_{t},z_{t}|\mathbf{g}\right)\, dx_{t}dz_{t}\\
\,\, & \simeq  0.01\; nep
	\end{split}
\end{equation}

	The information provided by the gravimetric measurements,
\begin{equation}
	I_{gravi}\left(\mathbf{g}\right)=I_{priori}-I_{posteriori}\simeq 6.90\; nep
\end{equation}

	The prior marginal information is the same as in example 3, and the \aposteriori marginal information is
\begin{equation}
	\begin{split}
I_{posteriori}\left(x_{t}|\mathbf{g}\right) & \simeq & 2.22\; nep,\\
I_{posteriori}\left(z_{t}|\mathbf{g}\right) & \simeq & 1.82\; nep
	\end{split}
\end{equation}

	The information provided about the two parameters is,
\begin{equation}
	\begin{split}
	I_{gravi}\left(\mathbf{g}\rightsquigarrow x_{t}\right) & \equiv  I_{priori}\left(x_{t}\right)-I_{posteriori}\left(x_{t}|\mathbf{g}\right)\simeq1.69\; nep,\\
	I_{gravi}\left(\mathbf{g}\rightsquigarrow z_{t}\right) & \equiv  I_{priori}\left(z_{t}\right)-I_{posteriori}\left(z_{t}|\mathbf{g}\right)\simeq1.18\; nep
	\end{split}
\end{equation}

\section{Summary of examples 1, 2, 3 and 4}
	It is time to make some summary remarks on the examples we have just discussed and to draw some conclusions that will guide the following sections. The main observations we can make are the following,
\begin{enumerate}
	\item the addition of an extra data point may not improve our knowledge of a parameter (example 2),
	\item the addition of a parameter can significantly reduce the knowledge of another parameter that was previously well resolved (example 3),
	\item a data may improve our knowledge of one parameter, but not another (example 4),
	\item the \aposteriori probability density often contains several relative maxima that do not necessarily correspond to the true solution, which may, in contrast, correspond to a relative minimum (example 3).
\end{enumerate}

	In a more general sense, it was observed that the information provided by the data $g_{2}$ was used in very different ways from one inverse problem to another. In example 2 this information was used very sparingly and it can be said that the data $g_{2}$ was practically useless. In contrast, in example 4 this data proved to be important, where it contributed significantly to the knowledge of the parameter $x_{t}$. This reflects a very classic behaviour of information in human contexts, where, for example, a message may be revealing to one person but meaningless to another. In the realm of inverse problems, different parameters have different sensitivities, or 'resolutions', to different pieces of data. Perhaps even more surprisingly, two gravimetric measurements that might initially be expected to play similar roles can have such different degrees of importance \aposteriori.

\chapter{\titrechap{Monte Carlo Methods}}
\minitoc
\section{Introduction}
	The \textit{Bayesian} solution to an inverse problem is the \aposteriori\ probability density
\begin{equation}
	\rho\left(\mathbf{y}|\mathbf{x}\right)=\dfrac{\rho\left(\mathbf{y}\right)p\left(\mathbf{x}|\mathbf{y}\right)}{\int_{\mathcal{I}}\rho\left(\mathbf{y}\right)p\left(\mathbf{x}|\mathbf{y}\right)\, d\mathbf{y}}
	\label{monte1}
\end{equation}

	where $\mathbf{x}$ and $\mathbf{y}$ represent the data and parameter vectors of the problem, respectively. In general, the dimensions $D_{x}$ and $D_{y}$ of these vectors are very large, and as soon as $D_{y}>3$ one encounters difficulties in visualising the function $\rho\left(\mathbf{y}|\mathbf{x}\right)$. Moreover, the exhaustive and systematic exploration of the \apriori solution space $\mathcal{I}$, which was feasible for the examples concerning the tunnel, is no longer possible because the number of computations required is immense.

	The visualisation problem can be partially solved in several ways. For example, several authors, including \couleur{Albert Tarantola}, advocate the creation of films in which the images consist of several acceptable \aposteriori\ solutions. The more an \aposteriori solution is probable, the more often its image appears in the film. However, this approach is rarely used because the practical realisation of these films must take into account the fact that the frequency of appearance is not necessarily a linear function of the probability, if one wants to take into account physiological factors such as retinal persistence and mental factors such as memory retention. For example, the images may need to be sorted in a certain way to help the viewer better grasp the different classes of solutions. We have already used this technique, which has proved very useful in certain cases, and we have found that a random appearance of the images makes the film very difficult to use. Although the film technique remains experimental and unusual for now, we believe it may become more important in the future as visualisation methods continue to improve.

	Another solution to the visualisation problem is to visualise only the marginal probability densities,
\begin{equation}
	\rho\left(y_{i}|\mathbf{x}\right)=\int\rho\left(\mathbf{y}|\mathbf{x}\right)\, d\mathbf{y}_{\neq i}
	\label{monte2}
\end{equation}

	where $y_{i}$ represents the parameter for which the marginal probability is computed, and $d\mathbf{y}{\neq i}$ is the vector of dimension $D{y}-1$, excluding the dimension corresponding to $y_{i}$. The representation of marginal probabilities is, of course, very simple since we are dealing with functions that depend only on the single variable $y_{i}$

\section{Integration by the \couleur{Monte Carlo} method}
	The calculation of the marginal probabilities \ref{monte2} requires the integration of the \aposteriori probability density $\rho\left(\mathbf{y}|\mathbf{x}\right)$. However, we have found that it is practically impossible to evaluate this probability density systematically and uniformly over the entire \apriori solution space. Therefore, the integral \ref{monte2} cannot be evaluated by numerical methods that require systematic knowledge of $\rho\left(\mathbf{y}|\mathbf{x}\right)$, but it is possible to use integration by the \textit{Monte Carlo} method based on random sampling of the \apriori space. Let $\left\{ \mathbf{y}{1},\mathbf{y}{2},\cdots,\mathbf{y}{n},\cdots,\mathbf{y}{N}\right\} $ be a collection of $N$ models, all with the same component $y_{i}$ and with the other $D_{y}-1$ components randomly drawn from the \apriori space of volume $V$. Then we have,
\begin{equation}
	\rho\left(y_{i}|\mathbf{x}\right)=\int\rho\left(\mathbf{y}|\mathbf{x}\right)\, d\mathbf{y}_{\neq i}\approx V\left\langle \rho\right\rangle \pm V\sqrt{\frac{\left\langle \rho^{2}\right\rangle -\left\langle \rho\right\rangle ^{2}}{N}}
	\label{monte3}
\end{equation}

	where,
\begin{equation}
	\left\langle \rho\right\rangle \equiv\frac{1}{N}\sum_{n=1}^{N}\rho\left(\mathbf{y}_{n}|\mathbf{x}\right)\;\;\;\;\left\langle \rho^{2}\right\rangle \equiv\frac{1}{N}\sum_{n=1}^{N}\rho^{2}\left(\mathbf{y}_{n}|\mathbf{x}\right)
	\label{monte4}
\end{equation}

	The term $\pm$ in the equation \ref{monte3} is an estimate of the uncertainty in the value of the integral.

\begin{figure}[H]
	\begin{center}
		\tcbox[colback=white]{\includegraphics[width=16cm]{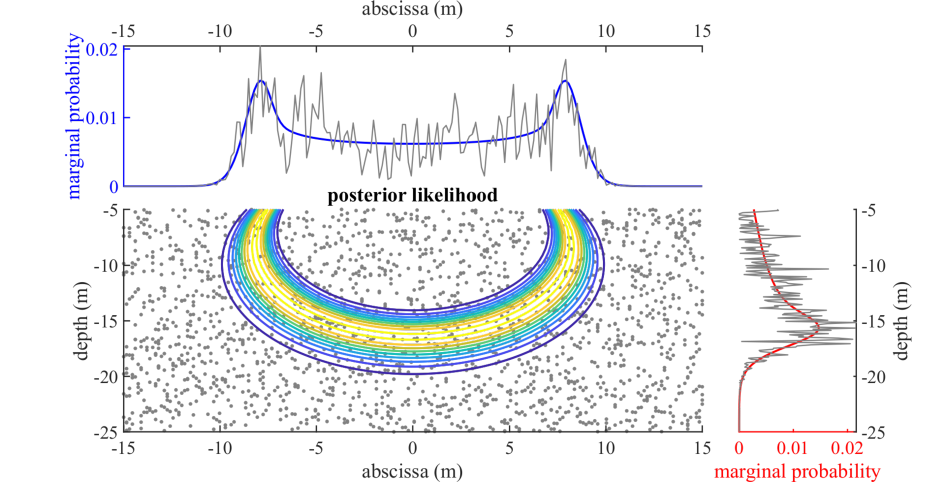}}
	\end{center}
	\caption{\aposteriori probability (contour plots at bottom left) for the horizontal position and depth of the tunnel when only a single gravimetric measurement is used. The marginal probabilities obtained by the \couleur{Monte Carlo} method are shown with dashed lines. Ten samples were taken for each value of $x_{t}$ (top curve) or $z_{t}$ (right curve). \label{carlo1}}
\end{figure}

\begin{figure}[H]
	\begin{center}
		\tcbox[colback=white]{\includegraphics[width=16cm]{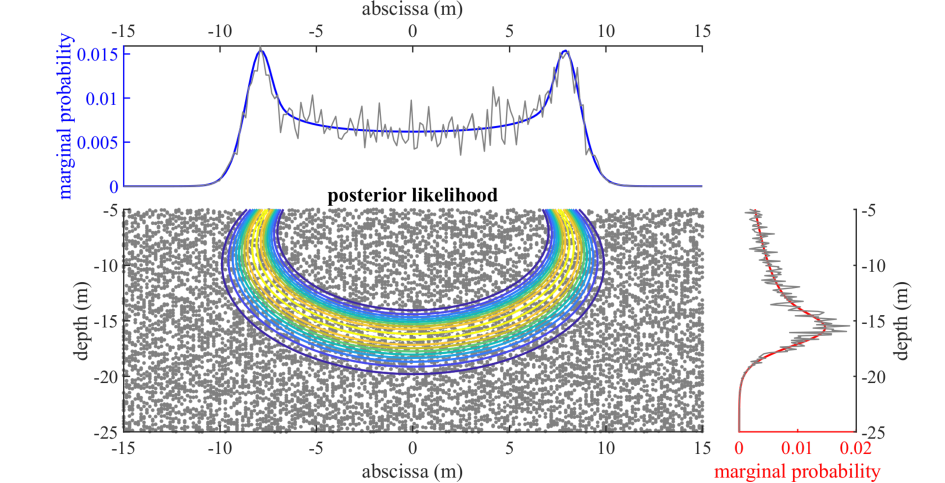}}
	\end{center}
	\caption{Similar to Figure \ref{carlo1}, but for 50 \textit{Monte Carlo} samples.\label{carlo2}}
\end{figure}

\begin{figure}[H]
	\begin{center}
				\tcbox[colback=white]{\includegraphics[width=16cm]{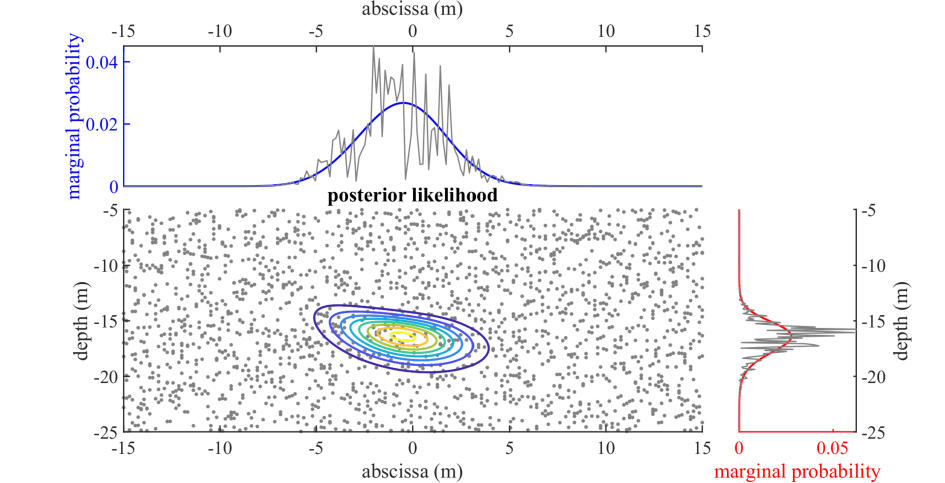}}
	\end{center}
	\caption{Similar to Figure \ref{carlo1}, but for two gravimetric measurements.\label{carlo3}}
\end{figure}

\begin{figure}[H]
	\begin{center}
		\tcbox[colback=white]{\includegraphics[width=16cm]{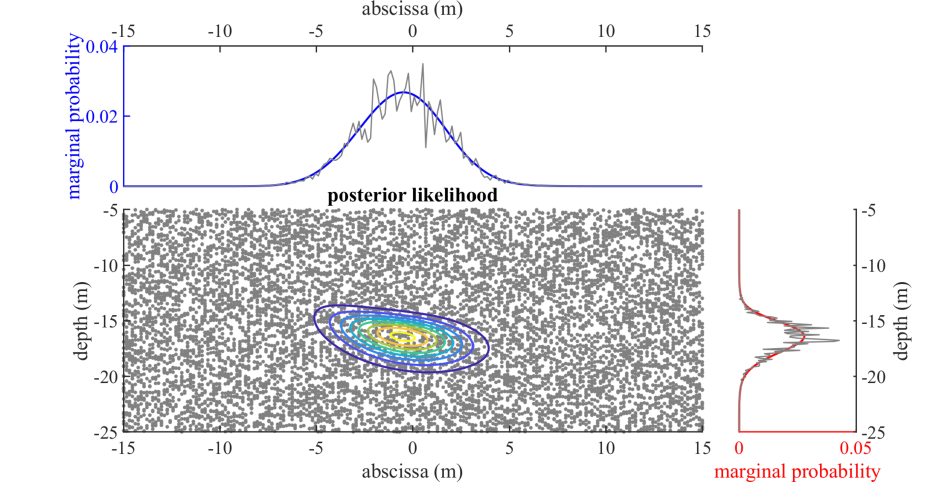}}
	\end{center}
	\caption{Similar to Figure \ref{carlo3}, but for 50 \textit{Monte Carlo} samples..\label{carlo4}}
\end{figure}

	Figures \ref{carlo1}, \ref{carlo2}, \ref{carlo3} and \ref{carlo4} show the marginal probabilities obtained by \couleur{Monte Carlo} integration. These probabilities differ quite significantly from the curves (grey lines) obtained by regular sampling of the \apriori\ model space. Only a large number of samples, greater than that used for regular sampling (50 for figures \ref{carlo2} and \ref{carlo4}), allows the recovery of curves that appear correct. Note that for figures \ref{carlo3} and \ref{carlo4}, the error is significant in the region of maximum probability because most of the integral is contributed by a small region of the integration domain that is not properly sampled by the randomly drawn models.

\begin{figure}[H]
	\begin{center}
		\tcbox[colback=white]{\includegraphics[width=16cm]{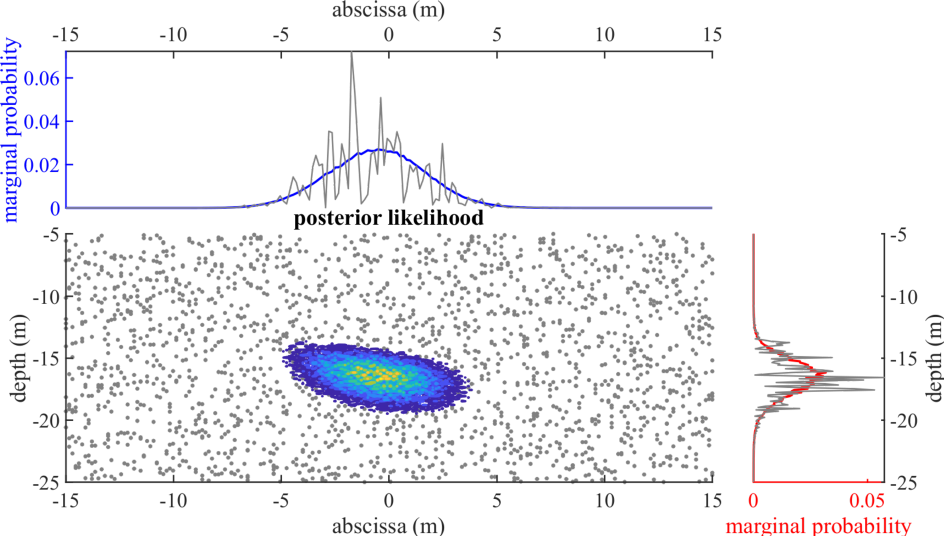}}
	\end{center}
	\caption{Similar to Figure \ref{carlo3}, but for 10 samples from a \couleur{Sobol} sequence. The more regular sampling reduces the integration error.\label{carlo5}}
\end{figure}

	Figures \ref{carlo1} and \ref{carlo3} were created with the script \couleur{ex\_tunnel\_05.m}; images \ref{carlo2}, \ref{carlo4} and \ref{carlo5} were created with the script \couleur{ex\_tunnel\_06.m}.

\newpage

\newpage

	The integration error caused by random sampling decreases as $N^{-1/2}$, whereas the error caused by regular sampling decreases as $N^{-1}$. It is therefore tempting to perform a random sampling that has the advantage of regular sampling, \ie one that is random but distributes the points relatively evenly. This can be achieved using quasi-random sequences, such as those of \couleur{Sobol}, which produce values with a quasi-uniform density that improves as the sequence lengthens (see figures \ref{carlo3} and \ref{carlo5}). This type of sampling gives better numerical integrations, but is limited to a small number of parameters (typically less than 10) and is not significantly more efficient than regular and systematic sampling.

\section{\couleur{Metropolis} algorithm}
\subsection{Importance sampling}
	Integration using the \couleur{Monte Carlo} method does not correctly integrate the marginal probabilities because the random sampling does not give sufficient weight to small regions where the probability density is significant. One way to overcome this is to generate a sequence of random models whose distribution is $\rho\left(\mathbf{y}|\mathbf{x}\right)$ to reduce the error in the mean of equation \ref{monte4}
	
	The \couleur{Metropolis} algorithm, invented in 1953 \shortcite{metropolis1953equation} at the dawn of the computer age, enables this particular type of random sampling, known in the Anglo-Saxon literature as \textit{"importance sampling"}. Basically, the \couleur{Metropolis} algorithm is a \couleur{Markov} chain in which a model is replaced by a successor under the control of a process that is partly random and partly guided. It is this process that constrains the set of generated models to conform to the imposed probability density.

\subsection{\couleur{Markov} chain} 
	A \couleur{Markov} chain is defined by a transition probability law,
\begin{equation}
	P\left(\mathbf{y}_{i}^{0}\rightarrow\mathbf{y}_{j}^{1}\right),
\end{equation}

	which generates a set $\mathcal{M}{1}=\left\{ \mathbf{y}{j}^{1}\right\}$ from a set $\mathcal{M}{0}=\left\{ \mathbf{y}{i}^{0}\right\}$. That is to say, when the law $P$ is applied to each element of $\mathcal{M}{0}$, it results in $\mathcal{M}{1}$. If we group the sets of solutions $\mathbf{y}{i}^{0}$ and $\mathbf{y}{j}^{1}$ into vectors $\mathbf{Y}_{0}$ and matrix $\mathbf{Y}_{1}$,
\begin{equation}
	\mathbf{P}\cdot\mathbf{Y}_{0}=\mathbf{Y}_{1}.
\end{equation}

	We want to repeat the transformation procedure by iteratively applying $\mathbf{P}$ starting from the initial set $\mathbf{Y}_{0}$, so that, after a large number of iterations, the population of the final set satisfies the probability law $\rho\left(\mathbf{y}|\mathbf{x}\right)$. This iterative process has the form
\begin{equation}
	\lim_{n\rightarrow\infty}\mathbf{P}^{n}\cdot\mathbf{Y}_{0}=\mathbf{Y},
\end{equation}

	where the final set $\mathbf{Y}$ consists of models $\mathbf{y}_{i}$ with an appearance frequency of $\rho\left(\mathbf{y}{i}|\mathbf{x}\right)$.

	It is necessary for the algorithm to be stable, \ie the point $\mathbf{Y}$ must be the only fixed point of the flow,
\begin{equation}
	\mathbf{P}\cdot\mathbf{Y}=\mathbf{Y}.
\end{equation}

	Three conditions are necessary to ensure the uniqueness of the fixed point. The first is to state that every initial model $\mathbf{y}_{i}^{0}$ must have an image in the set $\mathcal{M}_{1}$. This amounts to saying that the sum of the transformation probabilities is 1,
\begin{equation}
	\sum_{\mathcal{M}_{1}}P\left(\mathbf{y}_{i}^{0}\rightarrow\mathbf{y}_{j}^{1}\right)=1.
	\label{M1}
\end{equation}

	This equation simply means that it is certain that an element of the target set can be obtained by applying the transformation rule to the initial set. It also requires that any initial model $\mathbf{y}_{i}^{0}$ can be transformed, even with a very small probability, into one of the models in the target set $\mathcal{M}_{1}$,
\begin{equation}
	P\left(\mathbf{y}_{i}^{0}\rightarrow\mathbf{y}_{j}^{1}\right)>0,\;\forall\left(\mathbf{y}_{i}^{0},\mathbf{y}_{j}^{1}\right)\in\mathcal{M}_{0}\times\mathcal{M}_{1}.
	\label{M2}
\end{equation}

	This condition is known as the strong \textit{ergodicity} condition. The third condition is sufficient, but not necessary, to ensure that the transformation $P$ satisfies the desired properties. This is the microscopic equilibrium condition (detailed balance condition),
\begin{equation}
	P\left(\mathbf{y}_{i}^{0}\rightarrow\mathbf{y}_{j}^{1}\right)\rho\left(\mathbf{y}_{i}^{0}|\mathbf{x}\right)=P\left(\mathbf{y}_{j}^{1}\rightarrow\mathbf{y}_{i}^{0}\right)\rho\left(\mathbf{y}_{j}^{1}|\mathbf{x}\right),\;\forall\left(\mathbf{y}_{i}^{0},\mathbf{y}_{j}^{1}\right)\in\mathcal{M}_{0}\times\mathcal{M}_{1},
	\label{M3}
\end{equation}

	which we will see is satisfied by the \couleur{Metropolis} algorithm.

	If conditions (\ref{M1}), (\ref{M2}), and (\ref{M3}) are satisfied, then we have,
\begin{equation}
	\begin{split}
\sum_{\mathcal{M}_{0}}P\left(\mathbf{y}_{i}^{0}\rightarrow\mathbf{y}_{j}^{1}\right)\rho\left(\mathbf{y}_{i}^{0}|\mathbf{x}\right) & =  \sum_{\mathcal{M}_{0}}P\left(\mathbf{y}_{j}^{1}\rightarrow\mathbf{y}_{i}^{0}\right)\rho\left(\mathbf{y}_{j}^{1}|\mathbf{x}\right)\\
 & =  \rho\left(\mathbf{y}_{j}^{1}|\mathbf{x}\right),
 \end{split}
 \label{FIX}
\end{equation}

	where the properties (\ref{M3}) and (\ref{M1}) have been used successively. The relation (\ref{FIX}) shows that the probability density $\rho\left(\mathbf{y}|\mathbf{x}\right)$ is indeed a fixed point of the \textit{Markov} chain.

\subsection{The \couleur{Metropolis} algorithm} 
	Relation (\ref{FIX}) can be satisfied in several ways, among which the \couleur{Metropolis} algorithm uses the transformation law defined by,
\begin{eqnarray}
	P\left(\mathbf{y}_{i}^{0}\rightarrow\mathbf{y}_{j}^{1}\right) & = & 1\;\;\textrm{if}\;\;\rho\left(\mathbf{y}_{i}^{0}|\mathbf{x}\right)<\rho\left(\mathbf{y}_{j}^{1}|\mathbf{x}\right)\label{MT1}\\
	P\left(\mathbf{y}_{i}^{0}\rightarrow\mathbf{y}_{j}^{1}\right) & = & \frac{\rho\left(\mathbf{y}_{j}^{1}|\mathbf{x}\right)}{\rho\left(\mathbf{y}_{i}^{0}|\mathbf{x}\right)}\;\;\textrm{if}\;\;\rho\left(\mathbf{y}_{i}^{0}|\mathbf{x}\right)\geq\rho\left(\mathbf{y}_{j}^{1}|\mathbf{x}\right).\label{MT2}
\end{eqnarray}

	Equation (\ref{MT1}) shows that the transformation $\mathbf{y}_{i}^{0} \rightarrow \mathbf{y}_{j}^{1}$ is accepted whenever the proposed image model has a higher probability than the previous model. In contrast, equation (\ref{MT2}) shows that the transformation is possible, but not certain, if the image model is less probable than the previous model. Clearly then,
\begin{equation}
	P\left[\rho\left(\mathbf{y}_{i}^{0}|\mathbf{x}\right)<\rho\left(\mathbf{y}_{j}^{1}|\mathbf{x}\right)\right]+P\left[\rho\left(\mathbf{y}_{i}^{0}|\mathbf{x}\right)\geq\rho\left(\mathbf{y}_{j}^{1}|\mathbf{x}\right)\right]=1,
\end{equation}

	which is the particular event, the algorithm defined by (\ref{MT1}) and (\ref{MT2}) satisfies condition (\ref{M1}).

	Relation (\ref{MT1}) implies,
\begin{equation}
	P\left(\mathbf{y}_{i}^{0}\rightarrow\mathbf{y}_{j}^{1}\right)\rho\left(\mathbf{y}_{i}^{0}|\mathbf{x}\right)=\rho\left(\mathbf{y}_{i}^{0}|\mathbf{x}\right)\;\;\textrm{si}\;\;\rho\left(\mathbf{y}_{i}^{0}|\mathbf{x}\right)<\rho\left(\mathbf{y}_{j}^{1}|\mathbf{x}\right),
\end{equation}

	and relation (\ref{MT2}) provides,
\begin{equation}
	P\left(\mathbf{y}_{j}^{1}\rightarrow\mathbf{y}_{i}^{0}\right)\rho\left(\mathbf{y}_{j}^{1}|\mathbf{x}\right)=\rho\left(\mathbf{y}_{i}^{0}|\mathbf{x}\right)\;\;\textrm{si}\;\;\rho\left(\mathbf{y}_{i}^{0}|\mathbf{x}\right)<\rho\left(\mathbf{y}_{j}^{1}|\mathbf{x}\right)
\end{equation}

	Combining these two results, we find that
\begin{equation}
	P\left(\mathbf{y}_{i}^{0}\rightarrow\mathbf{y}_{j}^{1}\right)\rho\left(\mathbf{y}_{i}^{0}|\mathbf{x}\right)=P\left(\mathbf{y}_{j}^{1}\rightarrow\mathbf{y}_{i}^{0}\right)\rho\left(\mathbf{y}_{j}^{1}|\mathbf{x}\right)\;\textrm{si}\;\rho\left(\mathbf{y}_{i}^{0}|\mathbf{x}\right)<\rho\left(\mathbf{y}_{j}^{1}|\mathbf{x}\right)
	\label{RS1}
\end{equation}

	Furthermore, the relation (\ref{MT2}) allows us to write that,
\begin{equation}
	P\left(\mathbf{y}_{i}^{0}\rightarrow\mathbf{y}_{j}^{1}\right)\rho\left(\mathbf{y}_{i}^{0}|\mathbf{x}\right)=\rho\left(\mathbf{y}_{j}^{1}|\mathbf{x}\right)\;\textrm{si}\;\rho\left(\mathbf{y}_{i}^{0}|\mathbf{x}\right)\geq\rho\left(\mathbf{y}_{j}^{1}|\mathbf{x}\right)
\end{equation}

	while relation (\ref{MT1}) implies that,
\begin{equation}
	P\left(\mathbf{y}_{j}^{1}\rightarrow\mathbf{y}_{i}^{0}\right)\rho\left(\mathbf{y}_{j}^{1}|\mathbf{x}\right)=\rho\left(\mathbf{y}_{j}^{1}|\mathbf{x}\right)\;\textrm{si}\;\rho\left(\mathbf{y}_{i}^{0}|\mathbf{x}\right)\geq\rho\left(\mathbf{y}_{j}^{1}|\mathbf{x}\right)
\end{equation}

	These two formulas provide the condition,
\begin{equation}
	P\left(\mathbf{y}_{i}^{0}\rightarrow\mathbf{y}_{j}^{1}\right)\rho\left(\mathbf{y}_{i}^{0}|\mathbf{x}\right)=P\left(\mathbf{y}_{j}^{1}\rightarrow\mathbf{y}_{i}^{0}\right)\rho\left(\mathbf{y}_{j}^{1}|\mathbf{x}\right)\;\textrm{si}\;\rho\left(\mathbf{y}_{i}^{0}|\mathbf{x}\right)\geq\rho\left(\mathbf{y}_{j}^{1}|\mathbf{x}\right)
	\label{RS2}
\end{equation}

	Combining the conditional results (\ref{RS1}) and (\ref{RS2}), it follows that,
\begin{equation}
	P\left(\mathbf{y}_{i}^{0}\rightarrow\mathbf{y}_{j}^{1}\right)\rho\left(\mathbf{y}_{i}^{0}|\mathbf{x}\right)=P\left(\mathbf{y}_{j}^{1}\rightarrow\mathbf{y}_{i}^{0}\right)\rho\left(\mathbf{y}_{j}^{1}|\mathbf{x}\right)
\end{equation}

	which is nothing other than the microscopic balance condition (\ref{M3}) seen earlier.

\subsection{Example}
	Let's examine how the (\ref{MT1}) and (\ref{MT2}) algorithms work on a binary example where there are two possible models. In this case, the set of allowed models is,
\begin{equation}
	\mathcal{I}=\left\{ 0,1\right\}.
\end{equation}

	Suppose we want to generate a set of models with the respective probabilities given by,
\begin{equation}
	\rho\left(0\right)=\frac{1}{3}\;\;\textrm{et}\;\;\rho\left(1\right)=\frac{2}{3}.
\end{equation}

	Let the initial set be,
\begin{equation}
	\mathcal{M}_{0}=\left\{ 0,0,0,0,0,0,0,0,0,0,0,0,0,0,0,0,0,0,0,0,0\right\}.
\end{equation}

	The first iteration of the algorithm yields the set,
\begin{equation}
	\mathcal{M}_{1}=\left\{ 1,1,1,1,1,1,1,1,1,1,1,1,1,1,1,1,1,1,1,1,1\right\}
\end{equation}

	since only transformation (\ref{MT1}) was successful, since model $0$ is less likely than model $1$. The second iteration tests the transformation $1 \rightarrow 0$, which will include the equation (\ref{MT2}),
\begin{equation}
	P\left(1\rightarrow0\right)=\frac{\rho\left(0\right)}{\rho\left(1\right)}=\frac{1}{2}
\end{equation}

This transition is therefore random and indicates that the probability of performing the transformation is $0.5$. The practical implementation of this transition is to use a random number generator to produce a number $r \in [0,1]$ drawn from a uniform distribution. If,
\begin{eqnarray}
r & \leq & \frac{\rho\left(0\right)}{\rho\left(1\right)}\Rightarrow P\left(1\rightarrow0\right)\;\textrm{accepts},\\
r & > & \frac{\rho\left(0\right)}{\rho\left(1\right)}\Rightarrow P\left(1\rightarrow0\right)\;\textrm{rejects}.\end{eqnarray}

	Using the random number generator on our calculator, we could find that,
\begin{equation}
	\mathcal{M}_{2}=\left\{ 1,1,0,1,1,0,0,1,1,1,0,0,1,0,0,0,1,0,1,1,1\right\}
\end{equation}

	The next iteration involves testing both transformations $1 \rightarrow 0$ and $0 \rightarrow 1$. The first is random, as we observed in the second iteration, while the second is certain, as in the first iteration. Still using our pocket calculator, we obtained,
\begin{equation}
	\mathcal{M}_{3}=\left\{ 0,1,1,0,1,1,1,1,0,1,1,1,1,1,1,1,1,1,1,0,0\right\}
\end{equation}

	A similar calculation provides,
\begin{eqnarray}
	\mathcal{M}_{4} & = & \left\{ 1,1,1,1,1,0,1,0,1,1,0,0,0,1,0,1,1,1,0,1,1\right\} ,\nonumber \\
	\mathcal{M}_{5} & = & \left\{ 0,0,0,0,0,1,0,1,1,1,1,1,1,1,1,1,0,0,1,1,1\right\} ,\nonumber \\
	\mathcal{M}_{6} & = & \left\{ 1,1,1,1,1,0,1,0,0,1,1,1,1,0,0,1,1,1,0,1,0\right\} ,\nonumber \\
	\mathcal{M}_{7} & = & \left\{ 0,1,0,1,1,1,0,1,1,0,1,1,1,1,1,0,1,1,1,1,1\right\} ,\nonumber \\
	\mathcal{M}_{8} & = & \left\{ 1,1,1,0,1,0,1,1,1,1,1,0,0,0,0,1,0,0,0,0,0\right\} ,\nonumber \\
	\mathcal{M}_{9} & = & \left\{ 1,1,1,1,0,1,0,0,0,1,1,1,1,1,1,1,1,1,1,1,1\right\} ,\nonumber \\
	\mathcal{M}_{10} & = & \left\{ 1,1,1,1,1,0,1,1,1,0,1,0,1,0,0,0,1,0,1,1,1\right\}.
\end{eqnarray}

	Given the fixed probabilities $\rho\left(0\right)$ and $\rho\left(1\right)$, the sets generated should ideally contain seven $0$s and fourteen $1$s. The counts obtained are,
\begin{center}
	\begin{tabular}{|cc|c|c|c|c|c|c|c|c|c|c|c|}
\hline
\hline
 \ & $\mathcal{M}_{0}$&  $\mathcal{M}_{1}$&  $\mathcal{M}_{2}$&  $\mathcal{M}_{3}$&  $\mathcal{M}_{4}$&
 $\mathcal{M}_{5}$&  $\mathcal{M}_{6}$&  $\mathcal{M}_{7}$&  $\mathcal{M}_{8}$&  $\mathcal{M}_{9}$&
 $\mathcal{M}_{10}$ \\
\hline
\hline 
 0 & 21 & 0 & 9 & 5 & 7 & 8 & 7 & 5 & 11 & 4 & 7 \\
 \hline
 1 & 0 & 21 & 12 & 16 & 14 & 13 & 14 & 16 & 10 & 17 & 14 \\
\hline
\hline  
	\end{tabular}
\end{center}

	Aggregating the results for the last 9 sets, the average is as follows,
\begin{equation}
	N\left(0\right)=\frac{63}{9}\Rightarrow\rho\left(0\right)=\frac{1}{3}
\end{equation}

	which is excellent. Of course, in practice, it is necessary to work with a sufficient number of iterations and with sets containing many elements, but it is remarkable to observe that even with small samples like those in our example, the \couleur{Metropolis} algorithm already yields good results.

\subsection{Example of the tunnel}
\begin{figure}[H]
	\begin{center}
		\tcbox[colback=white]{\includegraphics[width=16cm]{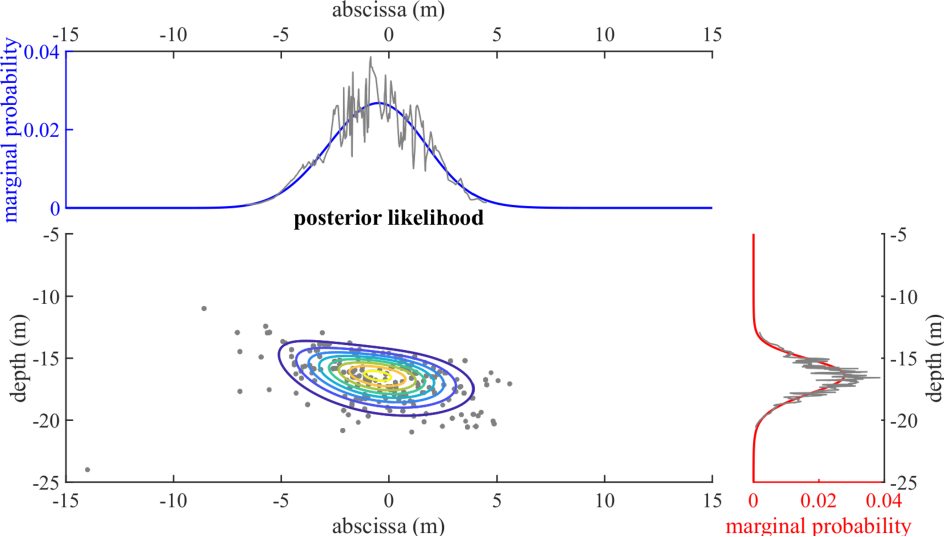}}
	\end{center}
	\caption{Similar to Figure \ref{carlo4}, but for 50 samples from a \couleur{Metropolis} chain. Importance sampling helps to reduce the integration error. The dots represent the 2500 models from a \couleur{Metropolis} chain. Their distribution density follows the probability density. The small number of points compared to the 2500 models indicates that many models are duplicated many times, which means that the acceptance of new models was very rare. This is explained by the fact that each new model was randomly generated in the \apriori space, independently of the previous model.\label{carlo6}}
\end{figure}

	Let us return to the tunnel example used previously for integration with the \couleur{Monte Carlo} method. Figure \ref{carlo6}, generated with the script \couleur{ex\_tunnel\_07.m}, shows the result of the integration obtained with \couleur{Metropolis} sequences of 50 terms. A reduction of the error is observed, especially in the marginal probability concerning $x_{t}$ (see figure \ref{carlo4}).

\chapter{\titrechap{Simulated Annealing}}
\minitoc
\section{Aim of the method}
	We have seen that the \couleur{Metropolis} algorithm can, in principle, generate a sequence of models $\mathbf{y}$ according to a probability density $\rho\left(\mathbf{y}|\mathbf{x}\right)$. However, this algorithm alone does not quickly yield good estimates of marginal probabilities when the probability density consists of local lobes in model space. In such cases, many tested models are rejected if they lie within one of the lobes, because most new models fall into areas of very low probability and therefore have little chance of being accepted. This leads to significant waste in the computation of direct problems, making the method inefficient. If the probability density is multimodal, the chances of exploring all the high-probability lobes are very low, leading to poor assessment of marginal probabilities.

	The idea of simulated annealing \shortcite{kirkpatrick1983} is to guide the models towards the lobes of maximum probability density using the \couleur{Metropolis} algorithm, with two important modifications, which are,
\begin{enumerate}
	\item the  use of a model generation process that has a 'memory' so that the new models are in some sense close to the previous models,
	\item a deformation of the probability density that gradually reveals the lobes during the \couleur{Metropolis} process.
\end{enumerate}

	The conjunction of these two aspects allows the generation of models that are confined to the vicinity of the probability density lobes, resulting in greater efficiency.

\section{Control temperature}
	Fundamentally, Simulated Annealing is the \couleur{Metropolis} algorithm implemented with a probability density whose topology evolves over the course of iterations.  This evolution allows, as mentioned above, a gradual transition from an almost uniform distribution to the \aposteriori density $\rho\left(\mathbf{y}|\mathbf{x}\right)$, which can potentially be multimodal. The evolution law assumes the following obvious relationship,
\begin{equation}
	\rho\left(\mathbf{y}|\mathbf{x}\right)=\exp\left[\ln\left(\rho\left(\mathbf{y}|\mathbf{x}\right)\right)\right].
	\label{recuit1}
\end{equation}

	Let's rewrite this formula by including a parameter $T$, which we will call the temperature
\begin{equation}
	\rho_{T}\left(\mathbf{y}|\mathbf{x}\right)=k_{T}\exp\left[\frac{\ln\left(\rho\left(\mathbf{y}|\mathbf{x}\right)\right)}{T}\right],\;\; T\geq0,
	\label{recuit2}
\end{equation}
 
	where $k_{T}$ is a normalization constant. Evidently,
\begin{equation}
	\rho_{1}\left(\mathbf{y}|\mathbf{x}\right)=\rho\left(\mathbf{y}|\mathbf{x}\right),
	\label{recuit3}
\end{equation}

	and also,
\begin{equation}
	\rho_{\infty}\left(\mathbf{y}|\mathbf{x}\right)=k_{\infty},
	\label{recuit4}
\end{equation}

	that is, at infinite temperature, $\rho_{T}\left(\mathbf{y}|\mathbf{x}\right)$ approaches a uniform probability density. Thus, as $T$ varies from $1$ to infinity, the probability density $\rho_{T}\left(\mathbf{y}|\mathbf{x}\right)$ gradually deforms, providing a means to control the topology of the probability density that guides the \couleur{Metropolis} algorithm.

\section{Perturbing the models}
	The second crucial aspect of Simulated Annealing is the memory of the process, \ie the fact that the models generated retain certain parameters from previous models while modifying others. There is no precise mathematical rule to describe this process, but rather principles that should be followed by defining rules specific to the particular inverse problem at hand.

	The primary principle is that the transition from one model to the next should not disrupt the guidance towards the modes of the probability density provided by the \couleur{Metropolis} process. For this reason, a completely random generation of models is not suitable, as it would result in a path through model space without memory. However, it is also essential that the path allows the exploration of large "territories" within this space in relatively few iterations to avoid algorithmic stagnation and the confinement of the series of models to a very limited volume. It is therefore clear that the model generation process must have somewhat contradictory properties: a substantial degree of movement similar to \couleur{Monte-Carlo} methods and a perturbative memory similar to gradient-based methods.

	Depending on whether the inverse problem involves discrete variables, as in the case of the travelling salesman problem that we will discuss later, or continuous variables, as in the tunnel example, the model generation process may differ significantly. Indeed, even for a discrete problem like the travelling salesman problem, which can have a very large combinatorial space (\eg $2^{32}$), it remains finite and it is possible to design model generation processes where the distance, measured in terms of the number of random draws to move from one model to another, remains small (\eg $100$). In contrast, when dealing with continuous variables, the distance between two models becomes infinite, even if they vary over a finite interval. In such cases, the model generation process may need to adapt as the temperature decreases during the iterations of \couleur{Metropolis}.

\section{The Simulated Annealing algorithm}
	Considering the above, the main steps of Simulated Annealing are as follows 
\begin{enumerate}
	\item loop $j$ over the temperature
	\begin{enumerate}
		\item function defining the temperature $T_{j}$
		\item \couleur{Metropolis}loop $i$  
		\begin{enumerate}
			\item generation of the model to be tested $\mathbf{y}_{i+1}^{T_{j}}$
			\item evaluation of the \aposteriori probability $\rho\left(\mathbf{y}_{i+1}^{T_{j}}|\mathbf{x}\right)$
			\item acceptance or rejection of the transition $\mathbf{y}_{i}^{T_{j}}\rightarrow\mathbf{y}_{i+1}^{T_{j}}$
		\end{enumerate}
		\item end of the $i$-th \couleur{Metropolis} loop
		\item convergence test
	\end{enumerate}
	\item end of the $j$-th loop over the temperature
\end{enumerate}

	As we have already noted, the temperature control and model generation steps are particularly crucial and determine the success or failure of the method. Unfortunately, there are no precise and universal rules for the development of these steps, as their form depends on the specific inverse problem at hand. It is also worth noting that the sequential nature of the above algorithm poses challenges that we will address by proposing a modified algorithm in which several \couleur{Metropolis} loops operate in parallel. To the best of our knowledge, this new algorithm is very similar to genetic algorithms.

\section{Example: the traveling salesman problem}
\subsection{Introduction}
	The travelling salesman problem is famous as a typical case of an optimisation problem with extremely high combinatorial complexity, and its efficient solution has been one of the reasons for the success of simulated annealing. The problem is to determine the order in which a travelling salesman should visit a given number of cities, exactly once, in order to minimise his travel distance. If $\textrm{N}$ is the number of cities to visit, then the number of possible \apriori solutions is $\textrm{N!}$, which quickly leads to an extremely large combinatorial space. For example, if $N=32$, the combinatorial space already exceeds $10^{35}$. Therefore, an exhaustive exploration of the solution space to find optimal solutions is out of the question. 

\subsection{Generation of models}
	For this problem, a model is an ordered list of cities, and the model generation process produces a list from another list. To satisfy both the memory constraints of the algorithm and the ability to explore the \apriori model space quickly, the processes typically used for this problem involve randomly selecting a small number of cities from the list and permuting them, which can be deterministic. Often only 2 cities are selected and swapped, resulting in \eg ,
\begin{equation}
	\left\{ \textrm{Nice},\textrm{Rennes},\textrm{Brest},\textrm{Lille},\textrm{Caen}\right\} \mapsto\left\{ \textrm{Nice},\textrm{Caen},\textrm{Brest},\textrm{Lille},\textrm{Rennes}\right\} .
	\label{recuit5}
\end{equation}

	It is observed that such a process has a clear memory effect, as a new list differs only slightly from the previous one. At the same time, it allows for rapid movement through the model space, since at most $N-1$ permutations are required to move from one list to any other.
	
\subsection{Example of how to operate}
\begin{figure}[H]
	\begin{center}
		\tcbox[colback=white]{\includegraphics[width=16cm]{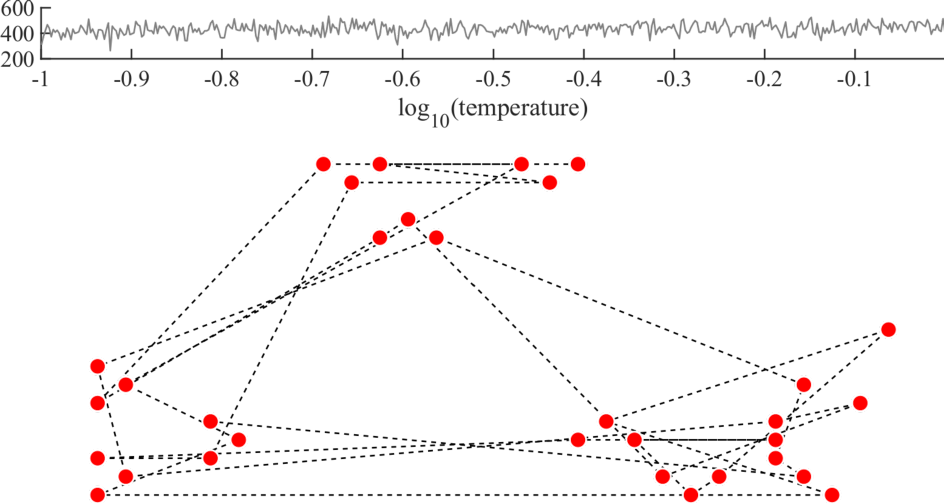}}
	\end{center}
	\caption{Path of the traveling salesman at the end of the \textit{Metropolis} loop for $T=10^{-1}$. The total distance is 309.\label{voyage1}}
\end{figure}

\begin{figure}[H]
	\begin{center}
		\tcbox[colback=white]{\includegraphics[width=16cm]{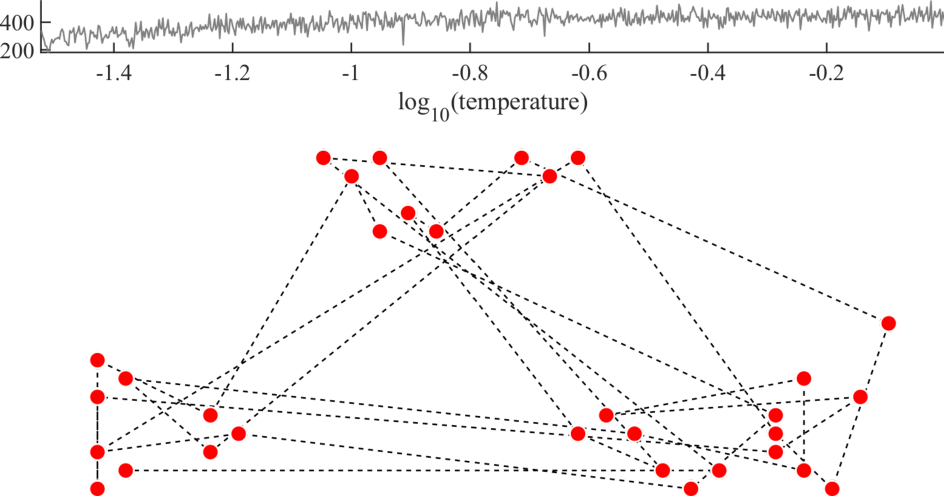}}
	\end{center}
	\caption{Path of the traveling salesman at the end of the \textit{Metropolis} loop for $T=10^{-1.5}$. The beginning of the cost reduction (path length) is noted. The total distance is 203. \label{voyage2}}
\end{figure}

\begin{figure}[H]
	\begin{center}
		\tcbox[colback=white]{\includegraphics[width=16cm]{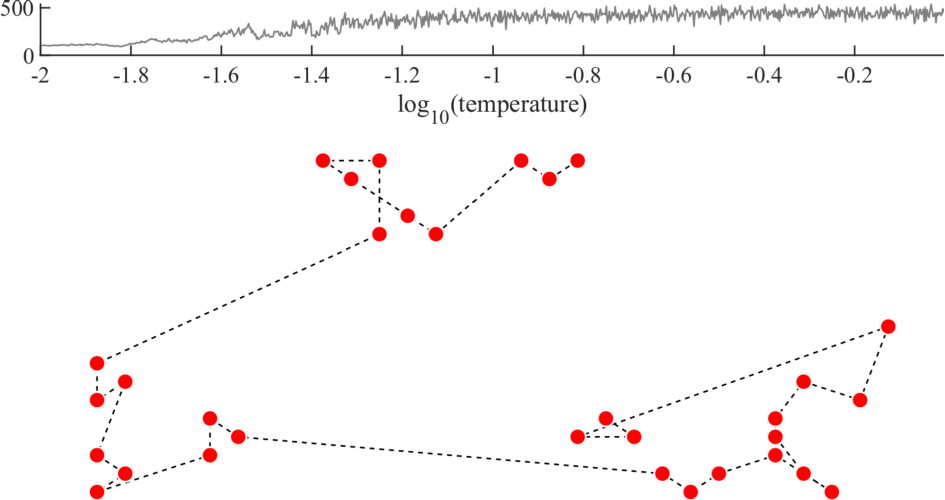}}
	\end{center}
	\caption{Path of the traveling salesman at the end of the \textit{Metropolis} loop for $T=10^{-2}$. The cost decreases in steps due to the hierarchical geography of the cities. The total distance is 104.\label{voyage3}}
\end{figure}

\begin{figure}[H]
	\begin{center}
		\tcbox[colback=white]{\includegraphics[width=16cm]{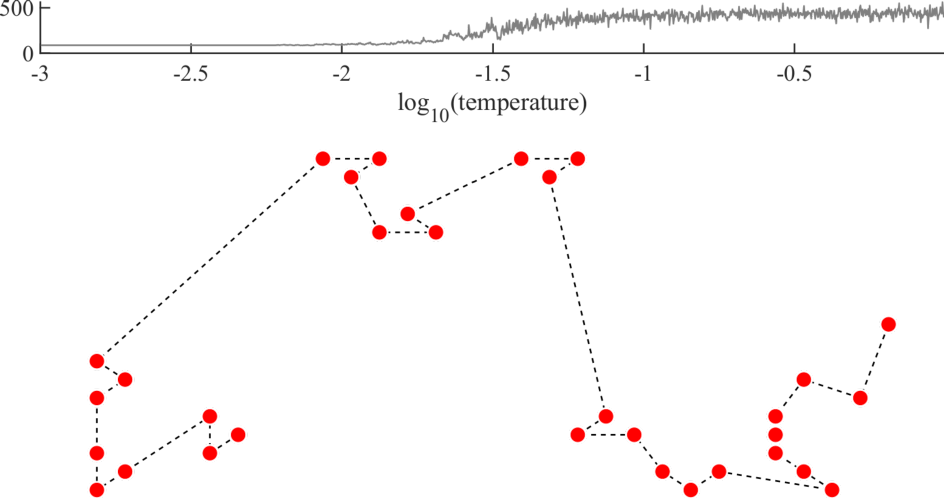}}
	\end{center}
	\caption{Path of the traveling salesman at the end of the \textit{Metropolis} loop for $T=10^{-3}$. This is the end of the cooling phase, and the cost no longer decreases significantly. The total distance is 88.\label{voyage4}}
\end{figure}

	The following example involves 32 cities whose geographical distribution follows a hierarchy of 'countries,' 'regions,' and 'municipalities.' For this example, the temperature was controlled \via a geometric sequence,
\begin{equation}
	T_{j+1}=0.995T_{j},
	\label{recuit6}
\end{equation}

	and 100 iterations were performed for each \couleur{Metropolis} loop (\ie for a given temperature $T_{j}$). Figures \ref{voyage1}, \ref{voyage2}, \ref{voyage3} and \ref{voyage4} illustrate the evolution of the path during the cooling process. It can be observed (curves at the top of each figure) that the cost decreases very rapidly when the temperature is around $10^{-1.5}$ (figure \ref{voyage2}) and stabilises when the temperature drops below $10^{-2}$. At the end of the run (Figure \ref{voyage4}), the total number of models generated is only 137,900, which is very small compared to the combinatorial complexity of the problem, which is $32!$.

\chapter{\titrechap{Methods of Least Squares}}
\minitoc
\section{Introduction}
	Let us revisit the \textit{Bayesian} solution to an inverse problem,
\begin{equation}
	\rho\left(\mathbf{y}|\mathbf{x}\right)=\frac{\rho\left(\mathbf{y}\right)p\left(\mathbf{x}|\mathbf{y}\right)}{\int_{\mathcal{I}}\rho\left(\mathbf{y}\right)p\left(\mathbf{x}|\mathbf{y}\right)\, d\mathbf{y}},
	\label{carre1}
\end{equation}

	and develop it for the particular case where the \textit{a priori} probability on the models is uniform, $\rho\left(\mathbf{y}\right) = \rho_{y}$, and where the errors on the data $\mathbf{x}$ are distributed according to a centred normal (Gaussian) distribution with zero mean and covariance matrix $\mathbf{C}$. We then have,
\begin{equation}
	p\left(\mathbf{x}|\mathbf{y}\right)=\frac{1}{\left(2\pi\right)^{N/2}\sqrt{\det\mathbf{C}}}\exp\left[-\frac{1}{2}\left(\mathbf{x}-\mathbf{x}_{\mathbf{y}}\right)^{t}\mathbf{C}^{-1}\left(\mathbf{x}-\mathbf{x}_{\mathbf{y}}\right)\right],
	\label{carre2}
\end{equation}

	where $\mathbf{x}_{\mathbf{y}}$ represents the predictions (synthetic data) corresponding to the model $\mathbf{y}$, and $N$ is the number of data points considered, which form the components of the vector $\mathbf{x}$. When the errors are uncorrelated, the covariance matrix is a diagonal matrix with elements $\sigma_{i}$, and by expanding equation \ref{carre2} above, we obtain,
\begin{equation}
	p\left(\mathbf{x}|\mathbf{y}\right)=\frac{1}{\left(2\pi\right)^{N/2}\prod_{i=1}^{N}\sigma_{i}}\exp\left[-\sum_{i=1}^{N}\frac{\left(x_{i}-x_{\mathbf{y},i}\right)^{2}}{2\sigma_{i}^{2}}\right].
	\label{carre3}
\end{equation}

	The \textit{a posteriori} probability of the models $\mathbf{y}$ given the data $\mathbf{x}$ is thus such that,
\begin{equation}
	\rho\left(\mathbf{y}|\mathbf{x}\right) = \frac{\rho_{y}}{\left(2\pi\right)^{N/2}\prod_{i=1}^{N}\sigma_{i}}\exp\left[-\sum_{i=1}^{N}\frac{\left(x_{i}-x_{\mathbf{y},i}\right)^{2}}{2\sigma_{i}^{2}}\right],
	\label{carre4}
\end{equation}

	and we see that the model $\mathbf{y}_{mc}$ corresponding to the maximum probability density is such that,
\begin{equation}
	S\left(\mathbf{y}=\mathbf{y}_{mc}\right)\equiv\sum_{i=1}^{N}\frac{\left(x_{i}-x_{\mathbf{y}_{mc},i}\right)^{2}}{\sigma_{i}^{2}}\;\;\;\textrm{MINIMUM}.
	\label{carre5}
\end{equation}

	The model $\mathbf{y}_{mc}$ is therefore the one that minimises the sum of the squared differences between the data and the predictions. For this reason, this model is called the least squares solution to the inverse problem.

\section{Linear problem: the normal equations}
	When the forward problem is linear, the relationship between the predictions $\mathbf{x}_{\mathbf{y}}$ and the model parameters $\mathbf{y}$ is of the form,
\begin{equation}
	\mathbf{x}_{\mathbf{y}}=\mathbf{L}.\mathbf{y}
	\label{carre6}
\end{equation}

	where the matrix $\mathbf{L}$ contains the coefficients $L_{i,j}$ such that the predictions $x_{\mathbf{y},i}$ are obtained as a linear combination of the $M$ parameters $y_{j}$
\begin{equation}
	x_{\mathbf{y},i}=\sum_{j=1}^{M}L_{i,j}y_{j}.
	\label{carre7}
\end{equation}

	Inserting this equation into the equation \ref{carre5}, we obtain,
\begin{equation}
	S\left(\mathbf{y}\right)=\sum_{i=1}^{N}\frac{\left(x_{i}-\sum_{j=1}^{M}L_{i,j}y_{j}\right)^{2}}{\sigma_{i}^{2}},
	\label{carre8}
\end{equation}

	and, for $S$ to be minimised, its partial derivatives with respect to the parameters $y_{k}$ must be set to zero, which means that,
\begin{equation}
	\frac{\partial S}{\partial y_{k}}=-2\sum_{i=1}^{N}\frac{1}{\sigma_{i}^{2}}\left(x_{i}-\sum_{j=1}^{M}L_{i,j}y_{mc,j}\right)L_{i,k}=0\;\; k=1,\cdots,M.
	\label{carre9}
\end{equation}

	By changing the order of summation and eliminating some multiplicative factors, the equation \ref{carre9} takes the following form,
\begin{equation}
	\sum_{j=1}^{M}\left(\sum_{i=1}^{N}\frac{L_{i,k}L_{i,j}}{\sigma_{i}^{2}}\right)y_{mc,j}=\sum_{i=1}^{N}\frac{L_{i,k}x_{i}}{\sigma_{i}^{2}}\;\; k=1,\cdots,M.
	\label{carre10}
\end{equation}

	Notice that the $M$ terms on the right side of the above equation are components of a vector such that,
\begin{equation}
	\left(\begin{array}{c}
\vdots\\
\sum_{i=1}^{N}\frac{L_{i,k}x_{i}}{\sigma_{i}^{2}}\\
\vdots\end{array}\right)=\left[\begin{array}{ccccc}
 &  & \vdots\\
\frac{L_{1,k}}{\sigma_{1}} & \cdots & \frac{L_{i,k}}{\sigma_{i}} & \cdots & \frac{L_{N,k}}{\sigma_{N}}\\
 &  & \vdots\end{array}\right]\left(\begin{array}{c}
x_{1}/\sigma_{1}\\
\vdots\\
x_{i}/\sigma_{i}\\
\vdots\\
x_{N}/\sigma_{N}
	\end{array}\right).
	\label{carre11}
\end{equation}

	By introducing the matrix,
\begin{equation}
	\mathbf{L}_{\sigma}\equiv\left[
		\begin{array}{ccccc}
\frac{L_{1,1}}{\sigma_{1}} & \cdots & \frac{L_{1,k}}{\sigma_{1}} & \cdots & \frac{L_{1,M}}{\sigma_{1}}\\
\vdots &  & \vdots &  & \vdots\\
\frac{L_{i,1}}{\sigma_{i}} & \cdots & \frac{L_{i,k}}{\sigma_{i}} & \cdots & \frac{L_{i,M}}{\sigma_{i}}\\
\vdots &  & \vdots &  & \vdots\\
\frac{L_{N,1}}{\sigma_{N}} & \cdots & \frac{L_{N,k}}{\sigma_{N}} & \cdots & \frac{L_{N,M}}{\sigma_{N}}
		\end{array}\right],
	\label{carre12}
\end{equation}

	and the vector,
\begin{equation}
	\mathbf{x}_{\sigma}\equiv\left(
		\begin{array}{c}
x_{1}/\sigma_{1}\\
\vdots\\
x_{i}/\sigma_{i}\\
\vdots\\
x_{N}/\sigma_{N}
		\end{array}\right),
	\label{carre13}
\end{equation}

	equation \ref{carre11} can be written in compact form as
\begin{equation}
	\left(
	\begin{array}{c}
\vdots\\
\sum_{i=1}^{N}\frac{L_{i,k}x_{i}}{\sigma_{i}^{2}}\\
\vdots
	\end{array}
	\right)=\mathbf{L}_{\sigma}^{T}.\mathbf{x}_{\sigma}.
	\label{carre14}
\end{equation}

	Applying the same procedure to the left-hand side of the equation \ref{carre10}, this equation becomes,
\begin{equation}
	\left(\mathbf{L}_{\sigma}^{T}.\mathbf{L}_{\sigma}\right).\mathbf{y}_{mc}=\mathbf{L}_{\sigma}^{T}.\mathbf{x}_{\sigma}
	\label{carre15}
\end{equation}

	The least squares solution $\mathbf{y}_{mc}$ is formally obtained by solving the equation \ref{carre15},
\begin{equation}
	\mathbf{y}_{mc}=\left(\mathbf{L}_{\sigma}^{T}.\mathbf{L}_{\sigma}\right)^{-1}.\mathbf{L}_{\sigma}^{T}.\mathbf{x}_{\sigma}.
	\label{carre16}
\end{equation}

	Unfortunately, the direct solution of this equation does not generally give an acceptable solution for various reasons that we will examine later, and it is preferable to work directly with the system \ref{carre6}.
\begin{equation}
	\mathbf{x}_{\mathbf{y}}=\mathbf{L}.\mathbf{y}.
	\label{carre17}
\end{equation}

	This system is generally rectangular, since the matrix $\mathbf{L}$ has $N$ rows and $M$ columns, and its solution must be obtained formally by,
\begin{equation}
	\mathbf{y}_{mc}=\mathbf{L}_{\sigma}^{\dag}.\mathbf{x}_{\sigma},
	\label{carre18}
\end{equation}

	where the matrix $\mathbf{L}{\sigma}^{\dag}$ is an operator known as the generalised inverse of $\mathbf{L}{\sigma}$.

\section{Singular Value Decomposition \& Singular Vectors}
	The decomposition of matrices into singular values and vectors (\textbf{SVD}) has its roots in the work of \couleur{Eugenio Beltrami} (1835-1899) published in 1873, which considered the decomposition of real square matrices. A year later, the mathematician \couleur{Camille Jordan} independently made the same discoveries. It was not until 1936 that \couleur{Eckart and Young} established the decomposition of complex rectangular matrices. The most widely used algorithm for performing the \textbf{SVD} decomposition of matrices is due to \couleur{Gene Golub} and \couleur{Christian Reinsch} (\shortciteNP{golub1971singular}).

	The \textbf{SVD} decomposition theorem states that any matrix $\mathbf{L}\in\mathbf{R}^{N\times M}$ can be factored in the form,
\begin{equation}
	\mathbf{L}=\mathbf{U}.\boldsymbol{\Lambda}.\mathbf{V}^{T},
	\label{IG1}
\end{equation}

	where $\mathbf{U}\in\mathbf{R}^{N\times K}$, $\mathbf{V}\in\mathbf{R}^{M\times K}$, and $\boldsymbol{\Lambda}\in\mathbf{R}^{K\times K}$. The $K$ columns $\mathbf{u}_{i}$ of the matrix $\mathbf{U}$ are the singular vectors of the matrix $\mathbf{L}.\mathbf{L}^{T}$, and those $\mathbf{v}_{i}$ of the matrix $\mathbf{V}$ are the singular vectors of the matrix $\mathbf{L}^{T}.\mathbf{L}$. So, we have,
\begin{equation}
	\mathbf{U}.\mathbf{U}^{T}=\mathbf{U}^{T}.\mathbf{U}=\mathbf{I}_{N},
	\label{IG2}
\end{equation}

and,
\begin{equation}
	\mathbf{V}.\mathbf{V}^{T}=\mathbf{V}^{T}.\mathbf{V}=\mathbf{I}_{M}.
	\label{IG3}
\end{equation}

	The matrix $\boldsymbol{\Lambda}$ is diagonal and its $K$ elements are the square roots of the singular values of the matrices $\mathbf{L}.\mathbf{L}^{T}$ and $\mathbf{L}^{T}.\mathbf{L}$. In general, there are $K\leq\min\left(M,N\right)$ non-zero singular values, where $K$ is the rank of the matrix $\mathbf{L}$. So, we have,
\begin{equation}
	\mathbf{L}^{T}.\mathbf{u}_{i}=\lambda_{i}\mathbf{v}_{i}\;\;\; i=1,\cdots,K,
	\label{IG4}
\end{equation}

and,
\begin{equation}
	\mathbf{L}.\mathbf{v}_{i}=\lambda_{i}\mathbf{u}_{i}\;\;\; i=1,\cdots,K.
	\label{IG5}
\end{equation}

	The decomposition given by the equation \ref{IG1} yields two sets of orthogonal and normalised vectors $\mathbf{u}_{i}$ and $\mathbf{v}_{i}$. The vectors $\mathbf{u}{i}$, of which there are $N$ and of dimension $N$, form a basis for a subspace of the vector space containing the data vector $\mathbf{x}$. The vectors $\mathbf{v}_{i}$, of which there are $M$ and of dimension $M$, form a basis for a subspace of the vector space of the parameters $\mathbf{y}$. It is important to note that the matrix $\mathbf{L}$ is reconstructed using the $K$ vectors $\mathbf{u}_{i}$ and $\mathbf{v}{i}$ associated with the $K$ non-zero singular values $\lambda_{i}$. To construct bases for the vector spaces of the data $\mathbf{x}$ and the parameters $\mathbf{y}$, it is necessary to complete the bases formed by the vectors $\mathbf{u}_{i}$ and $\mathbf{v}_{i}$ by adding other orthonormal vectors $\mathbf{u}_{0,i}$ and $\mathbf{v}_{0,i}$. These vectors, which can be considered as singular vectors corresponding to a zero eigenvalue of the matrices $\mathbf{L}$ and $\mathbf{L}^{T}$, form bases for the zero subspaces of the vector spaces of dimensions $N$ and $M$ containing the data and the parameters, respectively. For the basis vectors of the null subspaces, the equations \ref{IG4} and \ref{IG5} are simplified to,
\begin{equation}
	\mathbf{L}^{T}.\mathbf{u}_{0,i}=\mathbf{O}\;\;\; i=K+1,\cdots,N,
	\label{IG6}
\end{equation}

	and,
\begin{equation}
	\mathbf{L}.\mathbf{v}_{0,i}=\mathbf{O}\;\;\; i=K+1,\cdots,M.
	\label{IG7}
\end{equation}

	Similar to the matrices $\mathbf{U}$ and $\mathbf{V}$, whose columns are the vectors $\mathbf{u}_{i}$ and $\mathbf{v}_{i}$, the vectors $\mathbf{u}_{0, i}$ and $\mathbf{v}_{0,i}$ can be grouped to form matrices denoted $\mathbf{U}_{0}$ and $\mathbf{V}_{0}$ of dimensions $N\times\left(N-K\right)$ and $M\times\left(M-K\right)$ respectively.

\section{Solution provided by the spectral decomposition}
	The spectral decomposition of the matrix $\mathbf{L}$, as discussed in the previous section, allows to represent the vectors $\mathbf{x}$ and $\mathbf{y}$ in the bases of the singular vectors,
\begin{eqnarray}
\mathbf{x} & = & \sum_{i=1}^{K}b_{i}\mathbf{u}_{i}+\sum_{i=K+1}^{N}b_{0,i}\mathbf{u}_{0,i}\label{IG8}\\
 & = & \mathbf{U}.\mathbf{b}+\mathbf{U}_{0}.\mathbf{b}_{0},\label{IG8b}\\
\mathbf{y} & = & \sum_{i=1}^{K}a_{i}\mathbf{v}_{i}+\sum_{i=K+1}^{M}a_{0,i}\mathbf{v}_{0,i}\label{IG9}\\
 & = & \mathbf{V}.\mathbf{a}+\mathbf{V}_{0}.\mathbf{a}_{0},\label{IG9b}
\end{eqnarray}

	where the vectors $\mathbf{b}$, $\mathbf{b}_{0}$, $\mathbf{a}$ and $\mathbf{a}_{0}$ are the components of the data $\mathbf{x}$ and the parameters $\mathbf{y}$. Using these notations and introducing the decomposition \ref{IG1}, the system becomes \ref{carre17},
\begin{equation}
	\mathbf{U}.\boldsymbol{\Lambda}.\mathbf{V}^{T}\left[\mathbf{V}.\mathbf{a}+\mathbf{V}_{0}.\mathbf{a}_{0}\right]=\mathbf{U}.\mathbf{b}+\mathbf{U}_{0}.\mathbf{b}_{0},
	\label{IG10}
\end{equation}

	and finding $\mathbf{y}$ amounts to finding the components $\mathbf{a}$ and $\mathbf{a}_{0}$.\ Rewriting the equation \ref{IG10} and premultiplying each term by $\boldsymbol{\Lambda^{-1}}.\mathbf{U}^{T}$, we get,
\begin{eqnarray}
	\mathbf{a} & = & \boldsymbol{\Lambda^{-1}}.\mathbf{b}\label{IG11}\\
 & = & \boldsymbol{\Lambda^{-1}}.\mathbf{U}^{T}.\mathbf{x}\label{IG11b}
\end{eqnarray}

	The vector $\mathbf{a}_{0}$ cannot be determined in the same way from the equation \ref{IG10} and must be set arbitrarily or determined using additional information to that contained in $\mathbf{x}$. Therefore the solution $\mathbf{y}$ is given by
\begin{equation}
	\mathbf{y}=\mathbf{V}.\boldsymbol{\Lambda^{-1}}.\mathbf{U}^{T}.\mathbf{x}+\mathbf{V}_{0}.\mathbf{a}_{0}
	\label{IG12}
\end{equation}

\section{Obtained solution properties}
	The solution given by the equation \ref{IG12} has certain properties which we will now examine. The first of these is that the components $\mathbf{a}_{0}$ are arbitrary, which means that the solution is not unique. Uniqueness is only achieved when the number of singular values is equal to the dimension $M$, \ie the number of parameters, because in this case the base $\mathbf{V}_{0}$ is empty and $\mathbf{y}$ is uniquely defined by the components $\mathbf{a}$, which are themselves determined by the data $\mathbf{x}$ in equation \ref{IG11b}.

	The second important property is that the solution obtained via equation \ref{IG12} is a least squares solution in the sense that the residual vector, which contains the discrepancies between the data $\mathbf{x}$ and the model predictions $\mathbf{x}_{\mathbf{y}}$, is such that,

\begin{eqnarray}
	\mathbf{e} & \equiv & \mathbf{L}.\mathbf{y}-\mathbf{x}\label{IG13}\\
 & = & \mathbf{U}.\boldsymbol{\Lambda}.\mathbf{V}^{T}.\left[\mathbf{V}.\mathbf{a}+\mathbf{V}_{0}.		\mathbf{a}_{0}\right]-\mathbf{U}.\mathbf{b}-\mathbf{U}_{0}.\mathbf{b}_{0}\label{IG13b}\\
 & = & \mathbf{U}.\left[\boldsymbol{\Lambda}.\mathbf{a}-\mathbf{b}\right]-\mathbf{U}_{0}.\mathbf{b}_{0}.		\label{IG13c}
\end{eqnarray}

	The norm of this vector is,
\begin{equation}
	\mathbf{e}^{T}.\mathbf{e}=\left\Vert \boldsymbol{\Lambda}.\mathbf{a}-\mathbf{b}\right\Vert ^{2}+\left\Vert \mathbf{b}_{0}\right\Vert ^{2},
	\label{IG14}
\end{equation}

	and is minimal if $\mathbf{a}=\boldsymbol{\Lambda^{-1}}.\mathbf{b}$ (equation \ref{IG11}), i.e. if the solution $\mathbf{y}$ is the one given by equation \ref{IG12}. In this case the quadratic error is,
\begin{equation}
	\mathbf{e}^{T}.\mathbf{e}=\left\Vert \mathbf{b}_{0}\right\Vert ^{2},
	\label{IG15}
\end{equation}

	and is solely controlled by the projection of the data $\mathbf{x}$ onto the vectors $\mathbf{u}_{0,i}$ of the null subspace. Based on this result, the equation \ref{IG12} can be rewritten using the notation denoting the least squares solution,
\begin{equation}
\mathbf{y}_{mc}=\mathbf{V}.\boldsymbol{\Lambda^{-1}}.\mathbf{U}^{T}.\mathbf{x}+\mathbf{V}_{0}.\mathbf{a}_{0}.\label{IG12b}
\end{equation}

\section{Example: Signal deconvolution}
	We will now illustrate the previous sections with an example commonly encountered in signal processing: deconvolution. It is indeed common, as in seismology, to try to recover the input signal $y(t)$ of a system (assumed to be linear and stationary) from the output signal $x(t)$ and the impulse response of the system $l(t)$. The relationship is given by,
\begin{equation}
	x\left(t\right)=l\left(t\right)*y\left(t\right).
	\label{DS1}
\end{equation}

	In practice, the convolution $*$ is applied to discrete and truncated signals ($\mathbf{x}$, $\mathbf{y}$, and $\mathbf{l}$) \via the Z-transform,
\begin{equation}
	\sum_{n=1}^{N}x_{n}Z^{n}=\left(\sum_{j=1}^{J}l_{j}Z^{j}\right)\left(\sum_{m=1}^{M}y_{m}Z^{m}\right)
	\label{DS2}
\end{equation}

In this example, we will consider a system where the output $\mathbf{x}$ is the second derivative of the input signal $\mathbf{y}$, taking,
\begin{equation}
	\mathbf{l}=\left(
		\begin{array}{c}
-1\\
+2\\
-1
		\end{array}\right)
	\label{DS3}
\end{equation}

	The application of this filter to an input signal $\mathbf{y}$ can be written in matrix form, revealing the matrix $\mathbf{L}$ as discussed in the previous sections.
\begin{equation}
	\left(
		\begin{array}{c}
x_{1}\\
x_{2}\\
x_{3}\\
x_{4}\\
\vdots
		\end{array}\right)=\left(
		\begin{array}{ccccc}
-1 & 2 & -1 & 0 & \cdots\\
0 & -1 & 2 & -1 & \cdots\\
0 & 0 & -1 & 2 & \cdots\\
0 & 0 & 0 & -1 & \cdots\\
\vdots & \vdots & \vdots & \vdots & \ddots
		\end{array}\right)\left(
		\begin{array}{c}
y_{1}\\
y_{2}\\
y_{3}\\
y_{4}\\
\vdots
		\end{array}\right)
	\label{DS4}
\end{equation}

	where the matrix, which has a very specific structure, is called a \couleur{Toeplitz} matrix and in this case has been written by removing the edge effects of the convolution given by the equation \ref{DS2}.

\begin{figure}[H]
	\begin{center}
		\tcbox[colback=white]{\includegraphics[width=16cm]{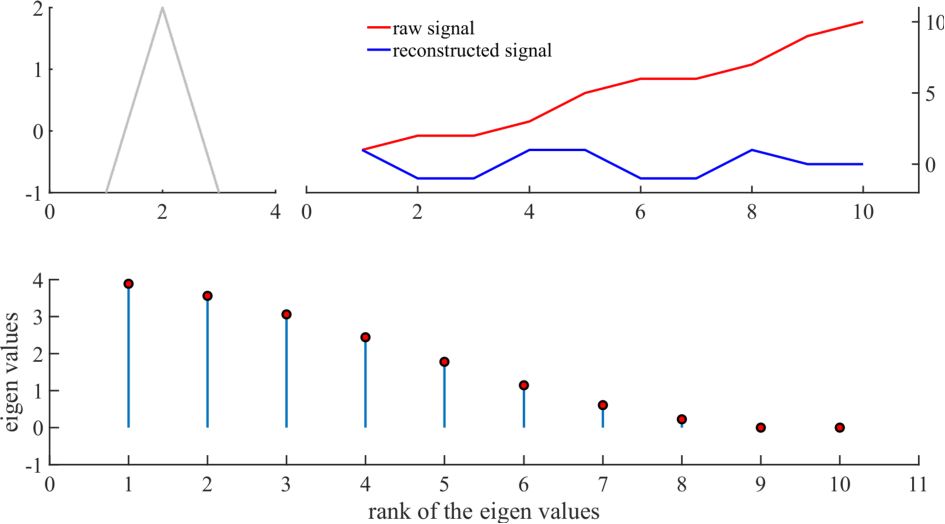}}
	\end{center}
	\caption{Results of signal deconvolution using a second-order finite difference filter.\label{deconv1}}
\end{figure}

\begin{figure}[H]
	\begin{center}
		\tcbox[colback=white]{\includegraphics[width=16cm]{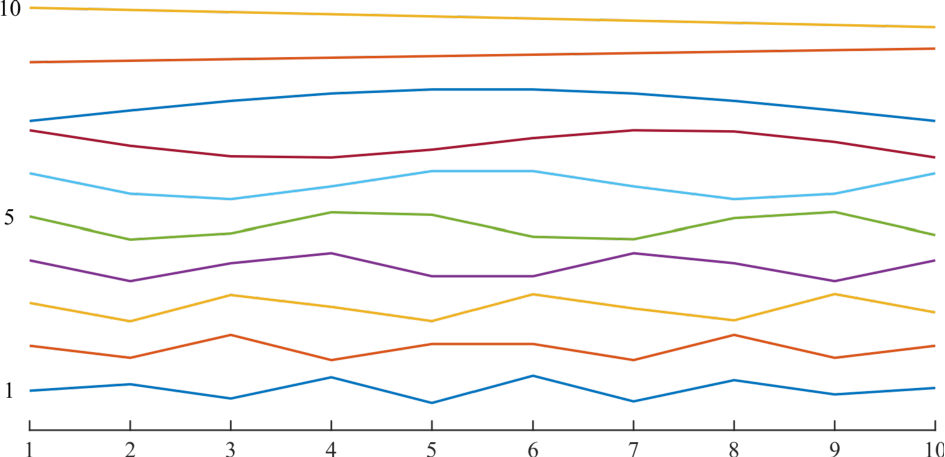}}
	\end{center}
	\caption{Singular vectors corresponding to the singular values shown at the bottom of Figure \ref{deconv1}.\label{deconv2}}
\end{figure}

	Figure \ref{deconv1} shows the results of the inversion by singular value decomposition (SVD) and singular vectors of the \couleur{Toeplitz} matrix for an input signal $\mathbf{y}$ with $M=10$ values and an output signal $\mathbf{x}$ with $N=8$ values. This is an example of an underdetermined problem, since the number of unknowns exceeds the number of data points. The \couleur{Toeplitz} matrix therefore has at most $K=8$ non-zero singular values. This is illustrated in the lower part of figure \ref{deconv1}, which shows the spectrum of the singular values of the \couleur{Toeplitz} matrix. Since the signal $\mathbf{y}$ belongs to a vector space of dimension $M=10$, there are two basis vectors corresponding to zero singular values, which form a basis for the zero subspace. The solution $\mathbf{y}_{mc}$ obtained from equation \ref{IG12b} by setting $\mathbf{a}_{0}=\boldsymbol{0}$ is shown in the top right of figure \ref{deconv1} (solid line). It can be seen that this solution differs significantly from the theoretical solution, shown as a dashed line, and contains a significant trend that is not captured in the $\mathbf{y}_{mc}$ solution. Figure \ref{deconv2} shows the singular vectors that form the basis of the $\mathbf{y}$ solution space. The vectors corresponding to zero singular values are numbers 8 and 10 from the bottom, and it is evident that these vectors model a linear trend and a constant value. This explains why the trend is not found in the solution; it belongs to the zero subspace. This is logical since the filter is a second derivative operator that cancels constant or linear functions. It is of course possible to obtain a solution identical to the theoretical one, but this requires the choice of the correct vector $\mathbf{a}{0}$, which can only be done using \apriori information provided in addition to the data $\mathbf{x}$.

\chapter{\titrechap{Generation of \apriori models}}
\minitoc
\section{Introduction}
We have seen the benefit of being able to generate models according to the \apriori probability density $\rho\left(\mathbf{y}\right)$. This capability significantly improves  the efficiency of simulated annealing in the sense that more models constructed in this way are retained. Furthermore, generating models according to the \apriori probability allows for more easily escaping from local minima when dealing with a multimodal \aposteriori density.

	It is therefore interesting to have methods capable of producing models that immediately satisfy a set of constraints, which may be of a highly variable nature. Such techniques exist, and one of the most popular in the geosciences is undoubtedly geostatistics, whose success is based on its ability to incorporate qualitative and disparate geological information quantitatively. Another way of incorporating a priori constraints relatively easily is to use the method of projection onto convex subspaces. This technique allows an arbitrary model to be modified into one that comes close to satisfying the required constraints.

\section{Convex sets: Definitions}
	Let us first establish the basic mathematical concepts. We say that a set of models $\mathcal{E}=\left\{ y_{i}\right\} $ is convex if,
\begin{equation}
	\forall\left(\mathbf{y}_{i},\mathbf{y}_{j}\right)\in\mathcal{E}\times\mathcal{E}\;\mathrm{et}\;\lambda\in\left[0,1\right],\;\mathrm{alors}\;\lambda\mathbf{y}_{i}+\left(1-\lambda\right)\mathbf{y}_{j}\in\mathcal{E}.
\end{equation}

	It is easy to show that,
\begin{equation}
	\mathcal{E}\;\mathrm{et}\;\mathcal{E}^{\prime}\;\mathrm{convexes}\Rightarrow\mathcal{E}\cap\mathcal{E}^{\prime}\;\mathrm{convexe.}
\end{equation}

	We will say that a convex set is a cone if,
\begin{equation}
	\forall\mathbf{y}\in\mathcal{E}\;\mathrm{et}\;\mu>0,\;\mathrm{alors}\;\mu\mathbf{y}\in\mathcal{E}.
\end{equation}

	The sets $\Bbb{R}^{n}$ and $\Bbb{R}_{+}^{n}$ are convex. A closed interval $\left[a,b\right]$ in $\Bbb{R}$ is convex, as well as the set of positive continuous functions.

\section{Projections onto convex sets}
	We will now consider some convex sets that are particularly interesting for generating models in the simulated annealing algorithm. For each of these sets, we will also show how to project an arbitrary model onto these convex sets.

\subsection{Imposed Values}
	We will denote by $\mathcal{C}$ the set of models $\mathbf{y}{i}$ for which certain components have known and fixed values $c{k}$, that is to say,
\begin{equation}
	\mathcal{C}\equiv\left\{ \mathbf{y}_{i}\,;\; y_{i,k}=c_{k}\right\} .
\end{equation}

	It is easy to show that this set is convex. The projection of any model $\mathbf{y}$ onto $\mathcal{C}$ is obtained by assigning the fixed values $c_{k}$ to the corresponding components,
\begin{equation}
	\mathbf{y}\rightarrow\mathbf{y}_{\mathcal{C}}\;;\; y_{\mathcal{C},k}=c_{k}.
\end{equation}

\subsection{Valeurs bornées}
	The set $\mathcal{B}$ denotes the class of models whose components are bounded,
\begin{equation}
	\mathcal{B}\equiv\left\{ \mathbf{y}_{i}\,;\; a_{k}\leq y_{i,k}\leq b_{k}\right\} .
\end{equation}

Projection onto this convex set involves adjusting the components whose values are outside the allowed interval,
\begin{equation}
	\mathbf{y}\rightarrow\mathbf{y}_{\mathcal{B}}\;;\; y_{\mathcal{B},k}=\max\left[\min\left(y_{i,k},b_{k}\right),a_{k}\right].
\end{equation}

\subsection{Discontinuity}
	The set $\mathcal{D}$ denotes the class of models with a discontinuity of amplitude $\mathbf{d}\left(x\right)$ along a boundary defined by $f\left(x\right)=0$. The set is thus defined by,

\begin{equation}
\mathcal{D}\equiv\left\{ \mathbf{y}_{i}\,;\;\mathbf{y}_{i}\left(x^{+}\right)-\mathbf{y}_{i}\left(x^{-}\right)=\mathbf{d}\left(x\right)\right\} ,.\end{equation}

	and the projection onto this convex set is such that,
\begin{equation}
	\begin{split}
\mathbf{y} & \rightarrow  \mathbf{y}_{\mathcal{D}}=\mathbf{y}+\frac{1}{2}\mathbf{d}\left(x\right)\;;\; f\left(x\right)>0\\
\mathbf{y} & \rightarrow  \mathbf{y}_{\mathcal{D}}=\mathbf{y}-\frac{1}{2}\mathbf{d}\left(x\right)\;;\; f\left(x\right)<0.
	\end{split}
\end{equation}

\subsection{Sequencing}
	The set $\mathcal{S}$ denotes the class of models with a specified ordering along. We have,
 \begin{equation}
\mathcal{S}\equiv\left\{ \mathbf{y}_{i}\,;\;\mathbf{y}_{i}\left(x_{1}\right)\geq\mathbf{y}_{i}\left(x_{2}\right)\right\}
\end{equation}

	and the projection onto this convex set is achieved by,
\begin{equation}
\mathbf{y}\left(x_{1}\right)=\mathbf{y}\left(x_{2}\right)=\frac{1}{2}\left(\mathbf{y}\left(x_{1}\right)+\mathbf{y}\left(x_{2}\right)\right)
\end{equation}

	This constraint allows, for example, for the imposition of rivers when generating fractal terrains.

\subsection{Imposed Mean}
	The set $\mathcal{M}$ denotes the class of models with an imposed mean $y_{moy}$. We have,
\begin{equation}
\mathcal{M}\equiv\left\{ \mathbf{y}_{i}\,;\;\left\langle \mathbf{y}_{i}\right\rangle =y_{moy}\right\}
\end{equation}

	and the projection is performed by,
\begin{equation}
\mathbf{y}\rightarrow\mathbf{y}_{\mathcal{M}}=\mathbf{y}-\left\langle \mathbf{y}\right\rangle +y_{moy}.\end{equation}

\subsection{Maximum Energy}
	The set $\mathcal{E}$ denotes the class of models with energy $e$ less than or equal to a certain value $e_{0}$. We have,
\begin{equation}
E\equiv\left\{ \mathbf{y}_{i}\,;\; e=\int y^{2}\left(x\right)dx\leq e_{0}\right\}
\end{equation}

	and the projection is performed by,
\begin{equation}
\mathbf{y}\rightarrow\mathbf{y}_{E}=\mathbf{y}*\sqrt{e_{0}/e}
\end{equation}

\bibliographystyle{fchicago}
\bibliography{fourier_biblio,pb_inverse}

\listoffigures

\printindex

\newpage
\thispagestyle{empty} 
\begin{tikzpicture}[remember picture,overlay]
      
 \fill[top color=blue!70!black!20, bottom color=cyan!20!blue!60] 
        (current page.south west) rectangle (current page.north east);

    \draw (current page.center) node [inner sep=1cm]{
    
    }; 
\end{tikzpicture}

\end{document}